\documentclass[rmp,aps,twocolumn,showpacs,amsmath,floatfix]{revtex4-1}
\pdfoutput=1 
\usepackage{graphics}
\usepackage{graphicx,color}% Include figure files
\usepackage{bm}% bold math
\usepackage{dcolumn}% Align table columns on decimal point
\usepackage[T1]{fontenc}

\def\lamb#1#2{$^{#1}_{\Lambda}${#2}}
\def\lam#1#2{$^{#1}_{~\Lambda}${#2}}
\def\lamc#1#2{$^{#1}_{~~\Lambda}${#2}}
\def\Kpi{($K^-,\pi^-$)}
\def\Kpig{($K^-,\pi^- \gamma$)}
\def\piK{($\pi^+,K^+$)}
\def\eeK{($e,e'K^+$)}
\def\piKg{($\pi^+,K^+\gamma$)}

\begin{document}
\title{Strangeness in nuclear physics}

\author{A. Gal}
\email{avragal@vms.huji.ac.il}
\affiliation{Racah Institute of Physics, The Hebrew University,
Jerusalem 91904, Israel}
\author{E.~V. Hungerford}
\email{hunger@uh.edu}
\affiliation{University of Houston, Houston, TX 77204, USA}
\author{D.~J. Millener}
\email{millener@bnl.gov}
\affiliation{Brookhaven National Laboratory, Upton, NY 11973, USA}

\begin{abstract} 
Extensions of nuclear physics to the strange sector are reviewed, 
covering data and models of $\Lambda$ and other hypernuclei, 
multistrange matter, and antikaon bound states and condensation. 
Past achievements are highlighted, present unresolved problems discussed, 
and future directions outlined.
\end{abstract}                                                                 


%13.75.Ev: hyperon-nucleon interactions
%13.75.Jz: kaon-baryon interactions
%21.80.+a: hypernuclei
%25.80.-e: meson- and hyperon-induced reactions
%26.60.+c: nuclear matter aspects of neutron stars
%97.60.Jd: neutron stars

%\date{May 2001}
\maketitle
\newpage 
\tableofcontents

\section{Introduction}
\label{sec:intro}

\subsection{Brief historical overview}
\label{subsec:history}

In the early 1950s a quantum number, conserved under the strong
interaction, was introduced ~\cite{gellmann53,nakano53} in order
to explain the behavior of the ``strange'' particles which had been
observed in emulsions exposed to cosmic rays. Almost
simultaneously, the first hypernucleus, formed by a $\Lambda$ hyperon
bound to a nuclear fragment, was observed in an emulsion
exposed to cosmic rays ~\cite{danysz53}. For the next 20 years or so,
hypernuclei were explored using emulsion detectors, first with cosmic
rays, and then with beams from existing accelerators.  
Within the last 40 years, modern particle accelerators and electronic
instrumentation has increased the rate and breadth of the experimental 
investigation of strangeness in nuclei. As always, theoretical interest 
has closely followed the experimental development. 

The behavior of a $\Lambda$ in a nuclear system is a nuclear many-body 
problem, since the forces between the baryons are predominantly hadronic 
and the time scale of the strong interaction is about $10^{-23}$ s 
compared to the weak-interaction lifetime of a $\Lambda$ in the nuclear 
medium~\cite{bhang98,park00} of approximately $10^{-10}$ s. Therefore, 
the combined hypernuclear system can be treated using well developed 
nuclear-theory models such as the shell or mean-field models with an 
effective $\Lambda$-nucleus interaction. New dynamical symmetries may also 
arise in hypernuclei, {\it e.g.} by treating the $\Lambda$ hyperon shell-model 
orbitals on par with those of nucleons within the Sakata version of SU(3) 
symmetry \cite{sakata56}. This approach was found useful in hypernuclear 
spectroscopic studies \cite{auerbach81,auerbach83}. Furthermore, by coupling 
SU(3)-Sakata with SU(2)-spin, the resulting SU(6) symmetry group presents 
a natural extension of Wigner's SU(4) spin-isospin symmetry group in light 
nuclei \cite{dg81}. 

$\Lambda$ hypernuclei also offer a test-ground for microscopic approaches to 
the baryon-baryon interaction. Thus, since one-pion exchange (OPE) between 
a $\Lambda$ hyperon and a nucleon is forbidden by isospin conservation, the 
$\Lambda N$ interaction has shorter range, and is dominated by higher mass 
(and multiple) meson exchanges when compared to the $NN$ interaction. For 
example, two-pion exchange between a $\Lambda$ hyperon and a nucleon proceeds 
through intermediate $\Sigma N$ states ($\Lambda N\rightarrow \Sigma N
\rightarrow \Lambda N$), potentially leading to non-negligible three-body 
$\Lambda NN$ forces \cite{gibson88}. The analogous mechanism of intermediate 
$\Delta N$ states ($NN\rightarrow\Delta N \rightarrow NN$) in generating 
three-body $NNN$ forces in two-pion exchange~\cite{epelbaum09} seems to be 
less important in nuclear physics, not only because the $NN$ interaction is 
dominated by OPE, but also because of the considerably higher excitation
mass of the $\Delta$ resonance with respect to that of the $\Sigma$ hyperon. 
Such theoretical expectations may be explored in hypernuclear few-body and 
spectroscopic calculations. 

Finally, the $\Lambda$ can be used as a selective probe of the nuclear medium, 
providing insight into nuclear properties that cannot be easily addressed 
by other techniques. Thus, from a hadronic as opposed to a quark 
perspective, the $\Lambda$ remains a distinguishable baryon within the 
nucleus, and samples the nuclear interior where there is little direct 
information on the single-particle structure of nuclei. Because of this, 
various aspects of hypernuclear studies such as $\Lambda$ decay, or the 
spectra of heavy hypernuclear systems, can illuminate nuclear features 
which would be more obscured in conventional nuclei. 

 Useful material on the subject of this review can be found in the
proceedings of the recent triennial conferences on Hypernuclear and
Strange Particle Physics (Gal and Hungerford, 2005; Pochodzalla and
Walcher, 2007; Gibson \textit{et al.}, 2010; Juli{\'{a}}-Diaz \textit{et al.},
2013), special volumes (Motoba, Akaishi, and Ikeda, 1994; Gal and Hayano,
2008; Hiyama, Motoba, and Yamamoto, 2010b; Gal, Hashimoto, and
Pochodzalla, 2012), schools~\cite{REZ2006}, and several review 
articles (Hashimoto and Tamura, 2006; Botta, Bressani, and Garbarino, 2012;
Feliciello and Nagae, 2015).

\subsection{General features of hypernuclear structure}
\label{subsec:general} 

To review nomenclature, a hypernucleus is constructed from a normal nucleus, 
with atomic weight A and atomic number Z, by adding one or more bound 
hyperons ($\Lambda$, $\Sigma$, $\Xi$, and perhaps $\Omega$). For example, 
the hypernucleus \lam{12}{C} consists of 12 baryons, one of these being 
a $\Lambda$ hyperon. It has atomic number 6, as noted by the label C. However 
for a general hypernucleus, the atomic number identifier is a measure of the 
system charge, and not necessarily the number of protons, since hyperons can 
carry charge. 

\begin{table}[thb]
\caption{Experimental $\Lambda$ separation energies, $B_\Lambda$, of light 
hypernuclei from emulsion studies. These are taken from a compilation 
\cite{davis86} of results from \cite{juric73,cantwell74}, omitting 
\lam{15}{N}~\cite{davis90}. A reanalysis for \lam{12}{C}~\cite{dluzewski88}
gives 10.80(18) MeV. \label{tab:blambda}} 
\begin{ruledtabular} 
\begin{tabular}{lcc} 
Hypernucleus & Number of events & $B_{\Lambda}\pm\Delta B_{\Lambda}$ (MeV) \\
\hline 
  \lamb{3}{H}  & \phantom{1}204 & $0.13\pm 0.05$ \\
  \lamb{4}{H}  & \phantom{1}155 & $2.04\pm 0.04$ \\
  \lamb{4}{He} & \phantom{1}279 & $2.39\pm 0.03$ \\
  \lamb{5}{He} & 1784           & $3.12\pm 0.02$ \\
  \lamb{6}{He} & \phantom{12}31 & $4.18\pm 0.10$ \\
  \lamb{7}{He} & \phantom{12}16 & not averaged \\
  \lamb{7}{Li} & \phantom{1}226 & $5.58\pm 0.03$ \\
  \lamb{7}{Be} & \phantom{12}35 & $5.16\pm 0.08$ \\
  \lamb{8}{He} & \phantom{123}6 & $7.16\pm 0.70$ \\
  \lamb{8}{Li} & \phantom{1}787 & $6.80\pm 0.03$ \\
  \lamb{8}{Be} & \phantom{12}68 & $6.84\pm 0.05$ \\
  \lamb{9}{Li} & \phantom{123}8 & $8.50\pm 0.12$ \\
  \lamb{9}{Be} & \phantom{1}222 & $6.71\pm 0.04$ \\
  \lamb{9}{B}  & \phantom{123}4 & $8.29\pm 0.18$ \\
  \lam{10}{Be} & \phantom{123}3 & $9.11\pm 0.22$ \\
  \lam{10}{B}  & \phantom{12}10 & $8.89\pm 0.12$ \\
  \lam{11}{B}  & \phantom{12}73 & $10.24\pm 0.05$ \\
  \lam{12}{B}  & \phantom{12}87 & $11.37\pm 0.06$ \\
  \lam{12}{C}  & \phantom{123}6 & $10.76\pm 0.19$ \\
  \lam{13}{C}  & \phantom{123}6 & $11.69\pm 0.12$ \\
  \lam{14}{C}  & \phantom{123}3 & $12.17\pm 0.33$ \\
 %\lam{15}{N}  & \phantom{12}14 & $13.59\pm 0.15$ \\
\end{tabular}
\end{ruledtabular}
\end{table}

A hypernucleus is characterized by its spin, isospin, and in the case of 
$\Lambda$ hypernuclei, a strangeness of $-$1. If the $\Lambda$ is injected 
into the nuclear system, the resulting hypernucleus will normally de-excite 
by a nuclear Auger process, or by $\gamma$ emission. The resulting ground 
state then decays by the weak interaction, emitting $\pi$ mesons as in 
the free $\Lambda$ decay, and also nucleons in a four-fermion in-medium 
interaction $\Lambda N\to NN$. Therefore, observation of the energetics of 
hypernuclear formation and decay can provide information on binding energies 
and spins of hypernuclear ground states. To conserve baryon number, a reaction 
producing a hypernucleus commonly replaces a nucleon with a $\Lambda$. In 
terms of the shell model, a hypernucleus is then described by a set of 
$\Lambda N^{-1}$ particle-hole excitations of the target nucleus which are 
coupled to specific values of spin and isospin. 

The acquisition of hypernuclear binding energies, well-depths, and positions 
of the hypernuclear levels began in the 1960s. Early work included $K^-$ 
absorption in emulsions and bubble chambers, where hyperfragments were 
identified by their mesonic decays. These efforts successfully established 
the binding energies of a number of light hypernuclei in their ground 
states (g.s.) where the $\Lambda$ is in the lowest $s_{1/2}$ orbit, 
as summarized in Table~\ref{tab:blambda}. In 1972, the existence of 
a \lam{12}{C} particle-unstable state with a $\Lambda$ in the $p$ orbit 
was confirmed \cite{juric72}, and the reaction 
$K^- +{^{12}\text{C}}\rightarrow\pi^- +p$~+~\lam{11}{B} in emulsion was used 
to study excited states of \lam{12}{C}. Beginning in the mid 1970's, the 
structure of $p$-shell hypernuclei was further explored via \Kpi\ reactions
using accelerated beams of kaons and magnetic spectrometers. Binding 
energies of heavier hypernuclear systems were extracted from spectra 
obtained using the \piK\ reaction. This reaction has greater probability 
to populate interior states. 
\begin{figure}[thb]
\center
\includegraphics[height=6cm]{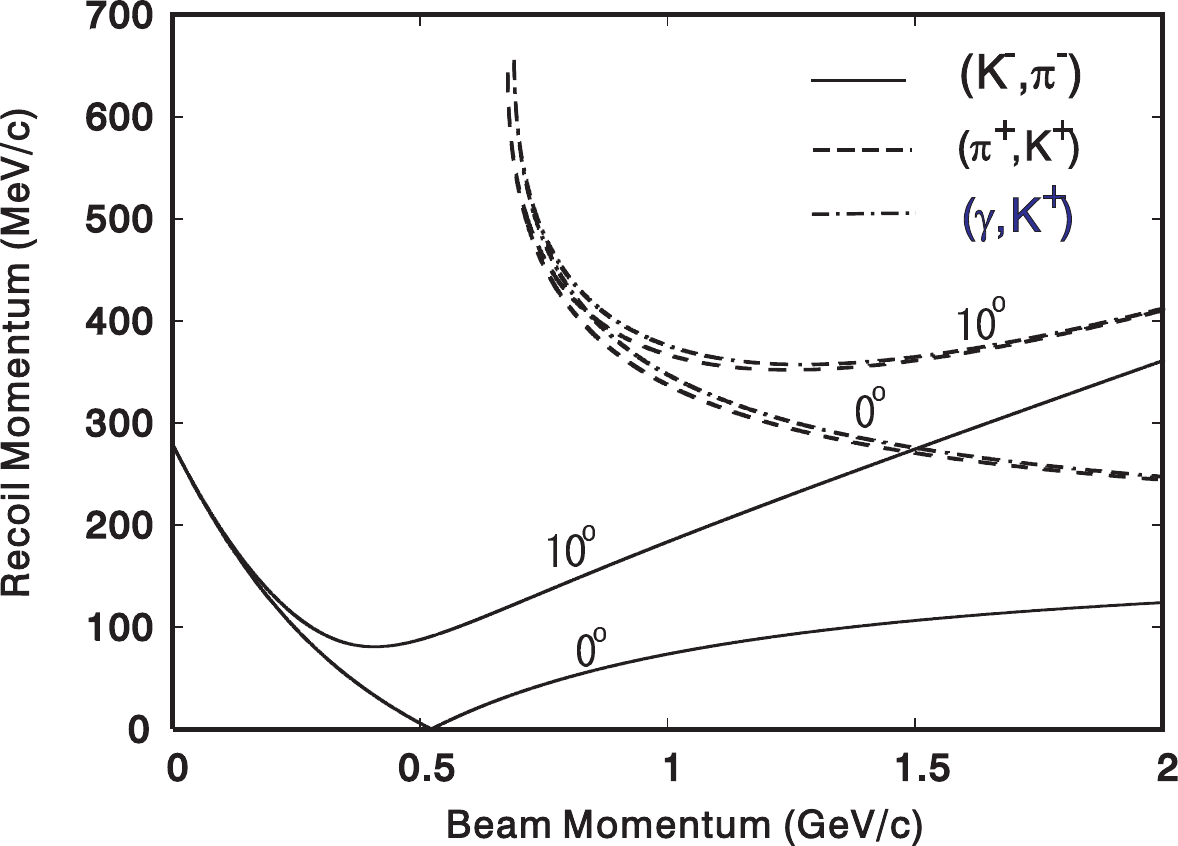}
\caption{The recoil momentum of the $\Lambda$ hypernucleus produced from 
a $^{12}$C target, as a function of the incident particle momentum and angle
for several production reactions. \label{fig:kinematics}}
\end{figure}
Unfortunately, the mass (or binding energy) scale for most of the data 
was normalized to the emulsion $B_\Lambda$ value (Table~\ref{tab:blambda}) 
for \lam{12}{C} that is determined by only a few events. This, coupled 
with resolution issues in the reaction spectra, lead to some uncertainties 
in binding energies. Some of the binding-energy uncertainties have been sorted 
out in recent years by comparing with ($e,e'K^+$) electroproduction 
measurements (see Sec.~\ref{subsubsec:spe}). 

\subsubsection{Kinematics}
\label{subsubsec:kinem} 

The kinematics for several elementary reaction processes are shown in 
Fig.~\ref{fig:kinematics}. As indicated in the figure, the $(K^{-},\pi^{-})$ 
reaction can have low, essentially zero momentum transfer to the produced 
$\Lambda$ hypernuclei. This also holds for $\Sigma$ hypernuclei. Thus the 
probability is large that $\Lambda$ and $\Sigma$ hyperons, when produced at 
low momentum transfer, will interact with, and bind to, the residual spectator 
nucleus. On the other hand reactions such as ($\pi^+,K^+$) or ($\gamma,K^+$) 
have high momentum transfer with respect to the nuclear Fermi momentum, 
producing recoil hyperons that have a high probability of escaping the 
nucleus. Such reactions are loosely termed ``quasi-free" (QF) processes, 
although the hyperon actually experiences continuum, final-state interactions
(FSI). Obviously, in the case of higher momentum transfer, cross sections 
to bound states are significantly reduced. 

\begin{figure} 
\center 
\includegraphics[width=6.5cm]{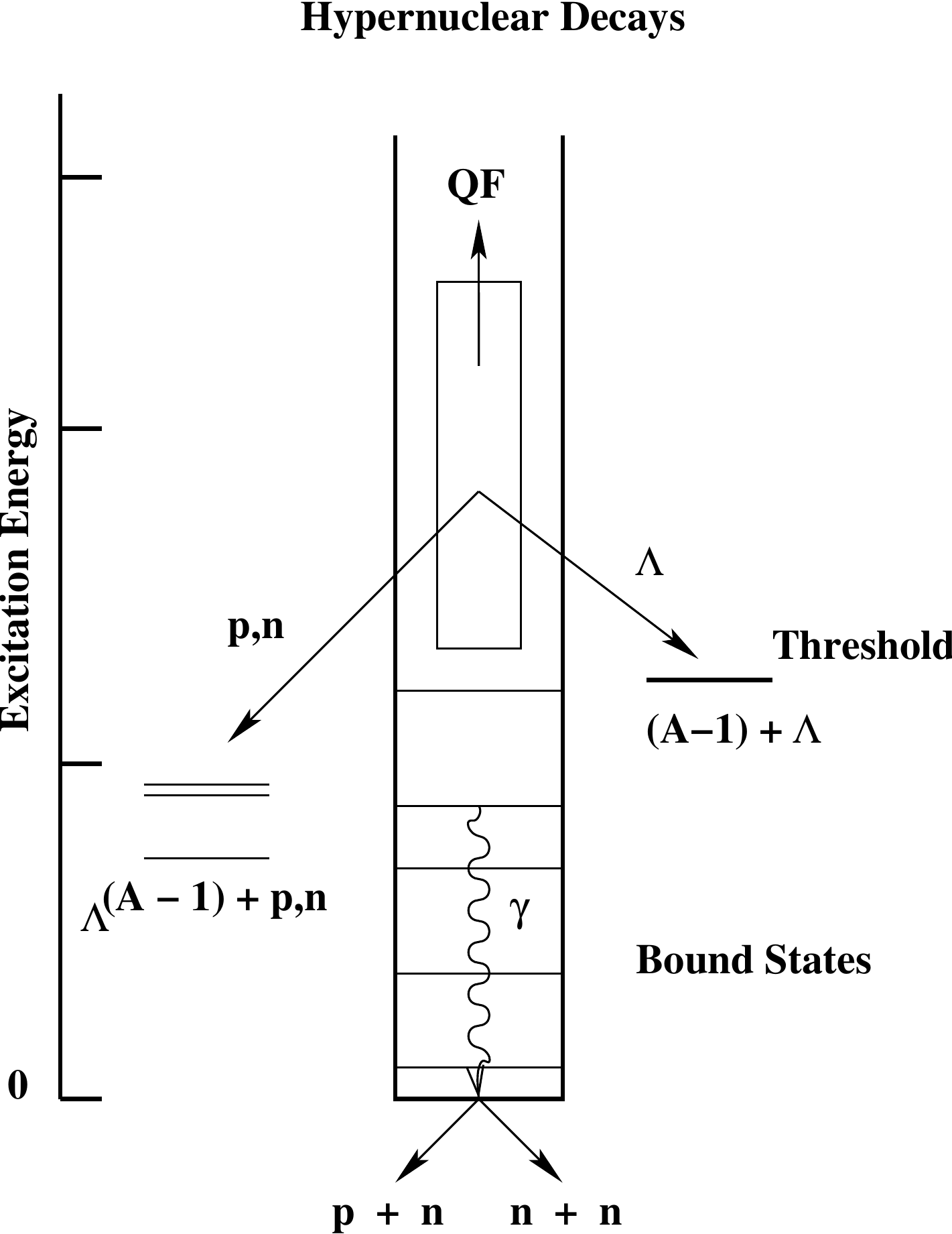} 
\caption{A schematic representation of the decays of an excited 
hypernucleus, showing in particular the decay of highly excited 
states by Auger and $\gamma$-ray  transitions.} 
\label{fig:auger} 
\end{figure} 

Furthermore, a $K^{-}$ strongly interacts with nucleons through various 
resonant states. Thus incident kaons in a $(K^{-},\pi^{-})$ reaction 
attenuate rapidly in nuclear matter, and the transition density should peak 
at the nuclear surface to maximise the cross section. Combining this 
with low momentum transfer as discussed above, the \Kpi\ reaction 
most likely involves an outer shell neutron, simply replacing this neutron 
with a $\Lambda$ having the same single-particle quantum numbers. 
On the other hand, energetic $\pi^{+}$ and $K^{+}$ particles have longer 
mean-free paths in nuclear matter, and give larger momentum transfer to 
the hyperon. Thus they can interact with interior nucleons, and can impart 
significant angular momentum transfer. However, such reactions have greater 
quasi-free strength.

After production, a bound hypernucleus generally deexcites to the state 
in which all the baryons reside in their lowest single-particle levels, 
from which the hypernuclear ground state then decays via the weak interaction. 
The energy released in the nuclear transitions is removed by 
gamma rays, or Auger neutron emission (see Fig.~\ref{fig:auger}) because the 
neutron (or proton) emission threshold can be lower than the $\Lambda$ 
emission threshold. Above the $\Lambda$ threshold, $\Lambda$ as well as 
nucleon emission can occur. It is interesting to note that particle-unstable 
hypernuclear levels near $B_{\Lambda}=0$ are experimentally observed to have 
narrow widths. Nuclear states at comparable excitation energies
would be broad. However, the narrow width of $\Lambda$-nuclear states is 
due to the weakness of the $\Lambda N$ interaction relative to the $NN$ 
interaction \cite{likar86}.

\subsubsection{Examples - kinematic considerations of hypernuclear production}  
\label{subsubsec:examples} 

The $(K_\text{stop}^{-},\pi^{-})$ reaction was the first reaction used for 
hypernuclear production, as kaon beams, particularly those produced in 
early accelerator experiments, were weak and the intensity of pions in the 
beams obscured the production reaction pions. Thus, it was easier to identify 
a stopped $K^{-}$, and stopping the $K^{-}$ assured that essentially all the 
kaons interacted with the target. As discussed earlier, this 
reaction was used with an emulsion detector, to produce an excited complex 
of states in \lam{12}{C} which decayed by proton emission to \lam{11}{B}. 
In this case, the emitted proton energy was measured in the emulsion, 
and the level structure interpreted in terms of three $p$-shell $\Lambda$ 
states located at about 11 MeV excitation energy \cite{dalitz86}. These 
included a narrow state with width equal to the experimental resolution 
($\approx$100 keV) just 140 keV below the $^{11}$C+$\Lambda$ threshold. This 
state was assigned as the expected $0^{+}$ state \cite{dalitz86,davis08}. 
Beneath this state was a broader level with a width of $\approx$600 keV 
which was interpreted as one of the expected $2^{+}$ states. The third state, 
750 keV below the second state, had a width of $\approx$150 keV and was also 
interpreted as another $2^{+}$ state. 

Later, it was recognized that the incident momentum of the in-flight 
$^{A}Z(K^{-},\pi^{-})$\lamb{A}{Z} reaction could be chosen so that the 
momentum transferred to the hypernucleus is close to zero, 
Fig.~\ref{fig:kinematics}, and that kaon beams near 750 MeV/c provide 
a maximum in the elementary cross section. Thus, using this reaction, 
a series of experiments were initiated at CERN \cite{povh80} and then 
at BNL \cite{,chrien79,may81}. The spectra produced by the $(K^{-}, \pi^{-})$ 
experiments show peaks for substitutional states near the nuclear surface 
(i.e., a neutron replaced by a $\Lambda$ with the same quantum numbers). 

In the case of $\Sigma$ production \cite{dover84a}, the $N(K^{-},\pi)\Sigma$ 
differential cross section in the forward direction shows two enhancements, 
one at about 400 MeV/c and a smaller one of different isospin at about 750 
MeV/c. A 400 MeV/c momentum is generally too low to be useful, since the 
intensity of secondary kaon beams drops rapidly below 600 MeV/c. On the 
other hand, zero momentum transfer occurs at an incident kaon momentum 
of about 300 MeV/c, and QF production is significantly enhanced if the 
incident momentum is greater than 600 MeV/c. However, there have been 
several searches for $\Sigma$ hypernuclei using very low momentum kaon 
beams \cite{bertini80,bertini84,bertini85}. Finally, there is another 
enhancement in the elementary $N(K^{-},\pi)\Lambda$ cross section at 
about 1.7 GeV/c. This momentum range, bearing some promise of appreciable 
polarization, has been used recently in \Kpig\ experiments (J-PARC E13)
using a 1.5 GeV/c beam from the J-PARC K1.8 beam line.

\subsection{The distorted wave impulse approximation}
\label{subsec:DWIA} 
Obviously to produce a hypernucleus, one needs to bind the hyperon in 
a nuclear potential well. This potential is usually generated by fitting 
its depth to some known $\Lambda$ single-particle binding energy in 
a Woods-Saxon shaped well with geometry derived from nuclear 
phenomenology. Potential wells for nucleons are often obtained from 
density-dependent mean-field calculations. 

 In a simple single-particle model, a production reaction removes a 
nucleon from a nuclear level (nuclear shell) replacing it with a 
$\Lambda$ in a $\Lambda$ level (shell). Thus, for a closed-shell target 
nucleus, the structure developed in a $\Lambda N^{-1}$ particle-hole model 
provides an obvious basis for a theoretical description of the production 
process. In this model, the production reaction can be described by the 
distorted wave impulse approximation (DWIA) (H\"{u}fner, Lee, and Weidenmuller,
1974; Bouyssy, 1977).
This formulation views the target as a collection of nucleons in 
single-particle levels, with the amplitude for production occurring 
between the incident projectile and a nucleon in the target. In this
most straightforward treatment of the hypernuclear production cross section,
the laboratory cross section for \Kpi\ (or similarly \piK) reactions 
can be written in terms of the two-body cross section on a nucleon 
in a factorized form as~\cite{itonaga94,motoba10}
\begin{equation}
 {d\sigma(\theta)\over d\Omega_L} = \alpha_\text{kin} 
{d\sigma_{el}(\theta)\over d\Omega_L} N_\text{eff}(if;\theta)
\label{eq:xsn}
\end{equation}
where $\alpha_\text{kin}$ is a kinematic factor involving the energies and 
momenta of the participants and $N_\text{eff}(if;\theta)$ is the distorted-wave 
integral, known as the effective neutron number, defined by
\begin{eqnarray}
&& {1\over 2J_i+1}\sum_{M_iM_f}|\langle J_fM_fT_f\tau_f|\int d\bm{r}
\chi^{(-)*}_{\pi}\left(\bm{k}_{\pi},\frac{M_A}{M_H}\bm{r}\right)
 \nonumber \\
&&\times \sum_{j=1}^A U_{-}(j)\delta \left(\bm{r}-\frac{M_C}{M_A}\bm{r}_j\right)|
J_iM_iT_i\tau_i\rangle|^2\chi^{+}_{K}(\bm{k}_K,\bm{r})\, .
\label{eq:neff}
\end{eqnarray}
$M_H$ ($M_A$) is the hypernuclear (target) mass and $M_C$ refers to the 
nuclear core of the hypernucleus. A zero-range interaction is assumed 
and the operator $U_{-}$ converts a neutron into a $\Lambda$ hyperon.
In a more sophisticated treatment that also enables the calculation of
hypernuclear polarizations~\cite{itonaga94}, a term 
$f+g(\bm{\sigma}_j\cdot\widehat{\bm{n}})$ is included under the summation 
over $j$ rather than using the two-body cross section in the factorized 
form of Eq.(\ref{eq:xsn}). Here, $f$ and $g$ denote the two-body 
spin-nonflip and spin-flip amplitudes and $\widehat{\bm{n}}$ is a unit 
vector perpendicular to the reaction plane. The BNL group~\cite{auerbach83} 
factors out the square of the Fermi-averaged amplitude $f$.

In Eq.(\ref{eq:neff}), the $\chi$'s are the distorted incident and final wave 
functions for the kaon and pion obtained from the nuclear optical potentials. 
Motoba and collaborators use eikonal distorted waves based on the
elementary $KN$ and $\pi N$ cross sections while the BNL group fits the
elastic scattering of 800 MeV/c $\pi$'s and $K$'s on $^{12}$C. For cross 
sections, the results from the two groups are in quite good agreement.

As for any inelastic scattering involving a one-body transition, 
the nuclear structure information is encoded in one-body density-matrix 
elements (OBDME), namely the matrix elements between the initial nuclear 
and final hypernuclear states of a coupled product of an annihilation 
operator for the nucleon and a creation operator for the 
$\Lambda$~\cite{auerbach83}.  An instructive example occurs when the 
hypernuclear wave function represents a simple weak-coupling state 
(this is a reasonable approximation because the $\Lambda N$ interaction 
is quite weak). Then, the OBDME that governs the cross section is
\[ \langle \alpha_c J_c T_f, j_\Lambda 0;J_fT_f||
\left(a^+_{j_\Lambda}\widetilde{a}^{ }_{j_N}\right)^{\Delta J 1/2}||
\alpha_i J_i T_i \rangle \vspace{-15pt}\]
\begin{eqnarray}
%\langle p^{n-1}\alpha_c J_c T_f, j_\Lambda 0;J_fT_f||
%\left(a^+_{j_\Lambda}\widetilde{a}^{ }_{j_N}\right)^{\Delta J 1/2}||
%p^n\alpha_i J_i T_i \rangle = & \nonumber \\
& = & (-)^{j_N +j_\Lambda -\Delta J}\, U(J_ij_NJ_fj_\Lambda,J_c\Delta J)
\nonumber \\
& & \times \mbox{} \langle \alpha_c J_c T_f||\widetilde{a}_{j_N}
||\alpha_i J_i T_i \rangle \; . 
\label{eq:formation}
\end{eqnarray} 
Here, $\alpha_c$ denotes a specific core state, $U$ is a unitary Racah 
coefficient for the recoupling of three angular momenta, $\Delta J$ is the 
angular momentum transfer, and the isospin transfer is $1/2$. 

The radial part of the transition density is given by products of the 
$\Lambda$ and nucleon radial wave functions. Also, an overall isospin 
Clebsch-Gordan coefficient factors out of Eq.(\ref{eq:neff}) leaving a 
reduced matrix element in isospace. Finally, one is left with a sum over 
products of radial integrals and OBDME for each permitted angular 
momentum transfer $\Delta J$. To see the consequences of the spin-flip 
characteristics of the reaction used to produce the hypernuclear states, 
it is useful to change the coupling from $(j_Nj_\Lambda)\Delta J$ to 
$(l_Nl_\Lambda)\Delta L\Delta S\Delta J$. For \Kpi\ reactions near 800 MeV/c 
and \piK\ reaction at 1.04 GeV/c, $\Delta S\!=\!0$ dominates. On the other 
hand, for \eeK\ reactions $\Delta S\!=\!1$ dominates (especially for the 
favored high $\Delta J$); see the appendix of Millener (2012) for a 
discussion of the combinations of OBDME that govern the various production 
reactions.

As Eq.~(\ref{eq:formation}) shows, the OBDME is proportional to the 
spectroscopic amplitude for the removal of the struck nucleon from the 
target. This leads to the intuitive, and important result in the 
weak-coupling limit, that the total strength for forming the states in a 
weak-coupling multiplet (summing over $J_fj_\Lambda$, with $j_\Lambda$ denoting
the members of a $\Lambda$ spin-orbit doublet) is proportional to 
the pickup spectroscopic factor, $\sum_{j_N}C^2S_{j_N}(c)$, from the 
target. Here, the isospin Clebsch-Gordan coefficient $C$ 
is obtained by changing the order of 
coupling in the overall isospin Clebsch-Gordan coefficient and 
$S_{j_N}(c)\!=\!<i||a^+_{j_N}||c>^2$. Failure to resolve
the states in a multiplet strongly limits the information that can be
obtained on the spin-dependence of the $\Lambda N$ interaction and 
underlines why high-resolution $\gamma$-ray detection is so important.

Distortions of the incident and exit waves generally do not change the shape 
of the angular distributions, but can reduce the reaction amplitudes by 
up to an order of magnitude [see Table 2 of Millener (1990)]. The 
factorized two-body amplitude must be averaged over the Fermi momentum of 
the participating nucleons in the medium. This can reduce the cross section 
typically by $10-20$\%. Finally, the DWIA approximation assumes that the 
reaction amplitude can be expressed by a two-body on-shell $t$-matrix. 
Corrections to this approximation, and reaction processes that include 
instantaneous interactions with more than one nucleon, are expected to be 
small. 

\subsection{Continuum excitations}
\label{subsec:continuum} 

In many situations the $\Lambda$ is produced in unbound, continuum states.  
This especially occurs in high momentum transfer reactions, but even in the 
$(K^{-},\pi^{-})$ substitutional reaction the $\Lambda$ can be unbound, as 
the $\Lambda$-nucleus well depth is approximately half of that of the 
nucleon-nucleus well depth. In hypernuclear production, this leads to the 
creation of a continuum background of excitations above the $\Lambda$-nucleus 
threshold. The continuum is sometimes discussed in terms of a QF 
reaction. In this model, the QF continuum spectrum is obtained by calculating 
the statistical density of states for the reaction on a single-particle 
nuclear state which produces an unbound $\Lambda$ recoiling under the 
influence of a $\Lambda$-nucleus potential. Calculations of the spectrum can 
be undertaken in a Fermi-gas model, so that the shape of the spectrum is 
determined by kinematics and the $\Lambda$-nucleus well depth \cite{dalitz76}. 
Applying this analysis to the continuum data of several medium-mass 
hypernuclei, a $\Lambda$-nucleus well depth of $\approx$30 MeV is extracted.

On the other hand, contributions to the continuum spectrum should also 
include nuclear structure information. Inclusion of nuclear structure 
can be treated by several methods \cite{kishimoto86,motoba88}, 
the most common is the continuum shell model \cite{halderson88}, 
where the QF and resonant behavior are simultaneously calculated. 
The general features of continuum production are best observed by 
comparing the spectra from various reactions \cite{itonaga90}. 
Above the continuum threshold, decay widths and the density of states 
increase rapidly. These appear as a rising, rather featureless background, 
with perhaps a few broad structures lying near threshold.
When modified by final state interactions \cite{watson52}, the QF process 
can be applied to the extraction of the hyperon-nucleus interaction from 
the shape of the continuum spectrum near threshold.

\subsection{The nuclear Auger effect}
\label{subsec:auger} 

 From previous arguments, a hypernucleus can be modeled as a set of 
single-particle nucleon holes and $\Lambda$ states. A reaction can 
place a $\Lambda$ particle in any of the bound or unbound levels 
of the nucleus, from which it may escape the nuclear potential well, 
cascade downward in energy, or become trapped in an isomeric level 
\cite{likar86}. A bound $\Lambda$ eventually reaches the ground state 
from which it weakly decays, see Fig.~\ref{fig:auger}. The energy released 
in these transitions is removed either by $\gamma$ rays, or by Auger 
neutron (or perhaps proton) emission since nucleon emission thresholds 
can be lower than the $\Lambda$ emission threshold. However, nucleon 
emission can also occur from unbound $\Lambda$ states. Thus the final 
hypernuclear species may differ from the one initially produced. 
Indeed, the hypernuclear system may fragment, producing a residual 
hypernucleus much lower in mass. Consequently, hypernuclei can be studied 
not only in production, where the reaction is constrained by a few measured 
particles which completely determine the residual system, but also in decay, 
when the production process may be ill-determined but measurement of the 
decay products is sufficient to determine a specific hypernucleus. Therefore, 
unless some additional information is available, just measuring energies of 
$\gamma$-ray transitions is generally not sufficient to identify a hypernucleus 
or, moreover, the levels involved in the hypernuclear transition.

\subsection{Strangeness production: reactions and experimental
techniques}
\label{subsec:production}

\subsubsection{The $(K^{-}_\text{stop},\pi^{-})$ reaction}
\label{subsubsec:stopminus} 

\begin{table}[b]
\caption{Branching ratios (in \%) for hyperon production using stopped 
$K^{-}$ \cite{vandervelde77}. \label{tab:kpi_branch}}
\begin{ruledtabular}
\begin{tabular}{lccccc} 
Ratio & H & D & He & C & Ne \\
\hline 
$R(\Lambda \pi^{0})$ & 4.9 & 5. & 6.2 & 4.4 & 3.4   \\
$R(\Sigma^{+} \pi^{-})$ & 14.9 & 30. & 37.3 & 37.7 & 37.7   \\
$R(\Sigma^{-} \pi^{+})$ & 34.9 & 22. & 10.9 & 16.8 & 20.4   \\
$R(\Sigma^{0} \pi^{0})$ & 21.4 & 23. & 21.2 & 25.7 & 27.6   \\
$R(\Lambda \pi^{-})$ & 9.7 & 10. & 12.6 & 8.7 & 6.7   \\
$R(\Sigma^{0} \pi^{-})$ & 7.1 & 5. & 5.9 & 3.3 & 2.1   \\
$R(\Sigma^{-} \pi^{0})$ & 7.1 & 5. & 5.9 & 3.3 & 2.1   \\
$R_{n}/R_{p}$ & 0.31 & 0.25 & 0.32 & 0.18 & 0.12   \\
$R_{m}$ &  & 0.01 & 0.16 & 0.19 & 0.23   \\
\end{tabular}
\end{ruledtabular}
\end{table}

Kaon capture at rest generally leads to $\Sigma$ rather than $\Lambda$ 
production. Approximately 5 times as many $\Sigma$'s as $\Lambda$'s are 
produced in $K^{-}$ capture on carbon \cite{tamura94}. A strong $\Sigma$ QF 
production background is present in recent $\Lambda$ hypernuclear production 
experiments by the FINUDA Collaboration \cite{agnello11a}. The prominence of 
$\Sigma\pi$ relative to $\Lambda\pi$ final states in $K^-$ capture at rest is 
demonstrated in Table~\ref{tab:kpi_branch}, taken from $K^{-}_{\rm stop}$ 
reactions in bubble chambers \cite{vandervelde77}. In this table, the $R$ 
factors are the branching fractions to a particular channel upon $K^{-}$ 
capture, and the ratio $R_{n}/R_{p}$ is the ratio of captures on neutrons to 
captures on protons. The ratio $R_{m}$ is the branching ratio for capture on 
multinucleon clusters in the nucleus with no emitted pions, reaching values 
about 20\% in nuclei beyond carbon. Of the several possible two-nucleon 
absorption channels, $\Sigma^-p$ pairs emitted in $K^-$ capture at rest on 
$p$-shell nuclear targets have been observed at rates $\sim$4\% 
\cite{agnello15}. The table shows clearly a reversal of the ratio 
$R(\Sigma^-\pi^+)/R(\Sigma^+\pi^-)$ when going from capture on hydrogen to 
capture on nuclear targets. This reflects the proximity of the $\Lambda(1405)$ 
subthreshold resonance which is more readily accessed kinematically in $K^-$ 
reactions on nuclei, as studied recently in capture at rest experiments on 
$p$-shell nuclear targets \cite{agnello11c}.

The $(K^{-}_\text{stop},\pi^{-})$ reaction proceeds when a kaon is absorbed 
from an atomic orbit into the nucleus \cite{hufner74}. X-ray measurements of 
kaon absorption on $^{12}$C \cite{wiegand67} indicate that 20\% of all the 
kaons are captured from $d$ orbits, while the remaining 80\% are believed to be
captured from low angular momentum, $l_K\!=\!0$ or 1, and large $n_K$ states. 
Kaon absorption at rest provides momentum transfer approximately equal 
to the Fermi momentum of a bound $\Lambda$, and for a carbon target 
angular-momentum transfers $J\leq 4$ are possible. Since the stopped 
reaction has higher momentum transfer than the in-flight reaction, 
it is much less selective. In comparison, the QF process is stronger for 
stopped kaons than in flight, so that it becomes difficult to resolve 
states near $B_{\Lambda}=0$ due to QF background. Therefore the 
effectiveness of the stopped kaon reaction, particularly for the higher 
energy levels, is limited, even with improved energy resolution.

The $(K^{-}_\text{stop},\pi^{-})$ reaction was extensively used to produce 
hypernuclei before separated kaon beam lines were available. During capture, 
a $\Lambda$ hyperon is produced by the reaction, 
$K^-+n\,\rightarrow\,\Lambda +\pi^-$. 
In the first counter experiment of this type at the CERN PS \cite{faessler73}, 
a kaon beam was brought to rest in a carbon target, and following the 
absorption of the kaon, a \lam{12}{C} hypernucleus was formed and identified 
by the emission of the $\pi^-$. Two broad peaks were observed in the pion 
spectrum, one with $B_{\Lambda} = 11 \pm 1$ MeV and the other with 
$B_{\Lambda} = 0 \pm 1$ MeV. The widths were dominated by the experimental 
resolution, $6\pm 1$ MeV, and the two peaks were subsequently identified 
as excitations with the $\Lambda$ residing in the $s$ and $p$ shells. 
Formation rates for these states were estimated as $(2\pm 1)\times 10^{-4}$, 
and $(3\pm 1)\times 10^{-4}$ per stopped kaon, respectively. In another 
$K_\text{stop}^-$ experiment, the $^{12}$C($K^-_\text{stop},\pi^-)$\lam{12}{C} 
reaction was observed~\cite{tamura94}, with rates per stopped kaon for the 
formation of these $s$-shell and $p$-shell states given by $(0.98\pm 0.12) 
\times 10^{-3}$ and $(2.3\pm 0.3)\times 10^{-3}$, respectively. These 
formation probabilities were a factor of 3 larger than those calculated by 
Gal and Klieb (1986) and a factor of 8 larger than the Matsuyama-Yazaki 
values \cite{matsuyama88}. However, the relative strength of the two 
peaks was found to be in better agreement with theory. 

\begin{figure} 
\includegraphics[width=8.0cm]{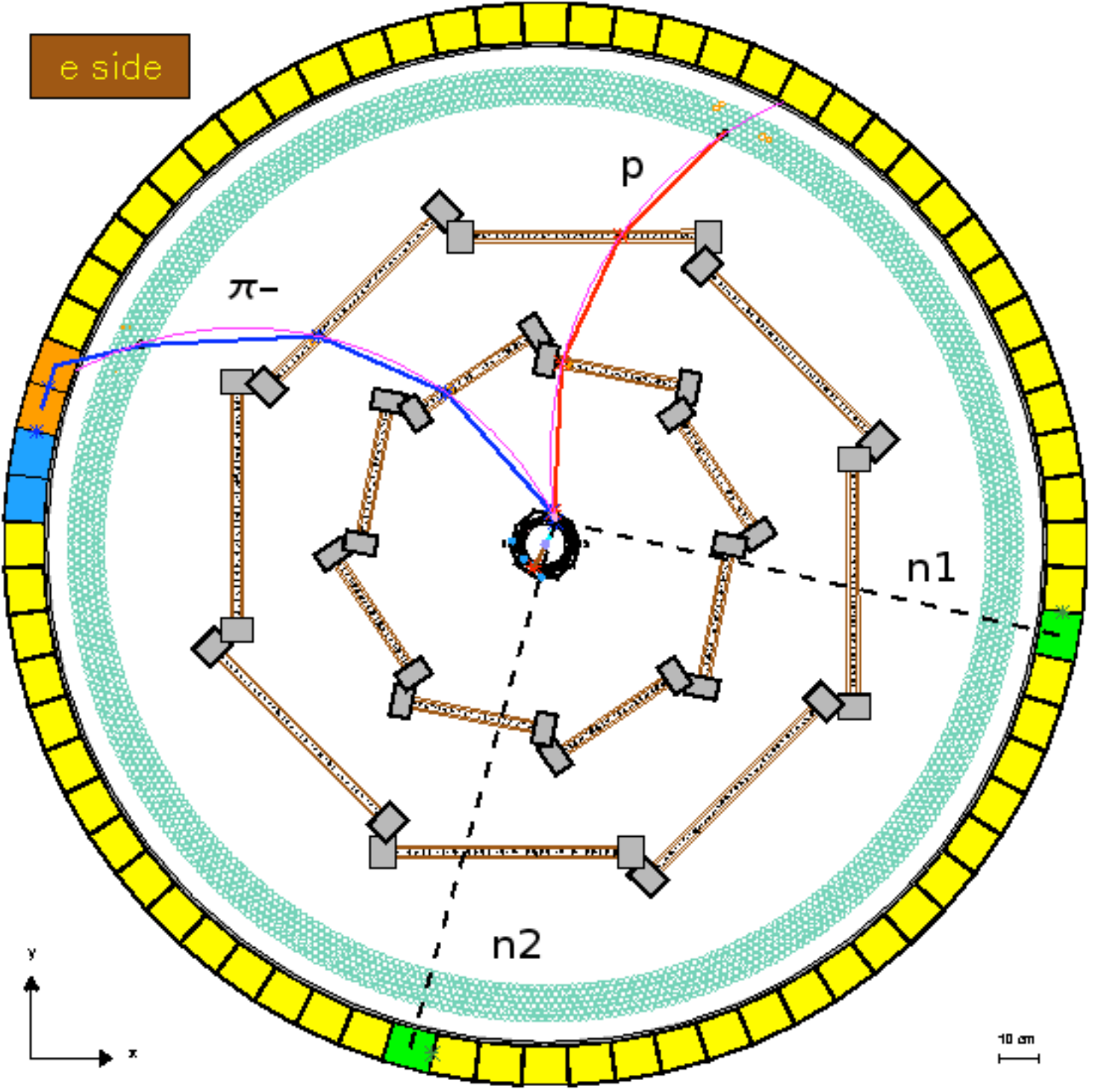} 
\caption{Illustration of a $\Lambda np\to nnp$ event 
observed by FINUDA on a $^7$Li target. The $\pi^-$ track arises from the 
formation of \lamb{7}{Li}. Adapted from Agnello \textit{et al.}, 2012b.} 
\label{fig:2Nfinuda} 
\end{figure} 

More recently, the FINUDA Collaboration at the DA$\Phi$NE $e^+e^-$ colliding 
beam machine in Frascati reported stopped $K^-$ formation rates on several 
$p$-shell targets from $^7$Li to $^{16}$O \cite{agnello05a,agnello11a}. More 
hypernuclear levels in \lam{12}{C} than the two main peaks seen in the earlier 
experiments were observed, with rates consistent for these two peaks with 
the earlier reports. These $p$-shell hypernuclear formation rates were then 
used in a theoretical study of the in-medium modification of the $\bar{K}N$ 
interaction, as derived within a coupled-channel chiral model, concluding 
that the ($K^{-}_\text{stop},\pi^-$) reaction can be used to better 
determine the $K^{-}$-nuclear optical potential depth~\cite{cieply11}. 
FINUDA's special niche in hypernuclear physics was its remarkable performance 
connecting together production and decay of light $\Lambda$ hypernuclei. This 
will become clear in Sec.~\ref{subsec:weakdecay}. 

FINUDA's capabilities are demonstrated in Fig.~\ref{fig:2Nfinuda} by showing 
a complete kinematical reconstruction of a three-nucleon final state in one 
of two \lamb{7}{Li}~$\to {^4{\rm He}}+n+n+p$ decay events observed at 
DA$\Phi$NE~\cite{agnello12b} following stopped-$K^-$ formation of \lamb{7}{Li} 
on a $^7$Li target, with a production $\pi^-$ track clearly visible. 
In another recent example, by correlating $\pi^{+}$ mesons from the 
$(K^{-}_\text{stop},\pi^{+})$ production reaction on a $^6$Li target with 
weak-decay $\pi^-$ mesons, three events were observed that were 
interpreted as production of the neutron-rich exotic hypernucleus 
\lamb{6}{H} that subsequently undergoes a two-body \lamb{6}{H}$\to \pi^- 
+ {^6}$He weak decay (Agnello \textit{et al.}, 2012a, 2012c). 

\begin{table*}[thb]
\caption{$\Lambda$ hypernuclear formation rates in $K^{-}$ capture at rest 
on $^{12}$C, in units of $10^{-3}$ per stopped $K^{-}$. \label{tab:prob}} 
\begin{ruledtabular} 
\begin{tabular}{llcc}  
& \multicolumn{1}{c}{Reference} & $R(s_{1/2})\times 10^{3}$ & 
$R(p_{3/2}+p_{1/2})\times 10^{3}$ \\ 
\hline
\lam{12}{C} theory & Gal and Klieb (1986) & 0.33 & 0.96 \\
\lam{12}{C} theory & Matsuyama and Yazaki (1988) & 0.12 & 0.59 \\
\lam{12}{C} theory & Krej\v{c}i\v{r}\'{i}k, Ciepl{\'{y}}, and Gal 
(2010)\footnotemark [1]  & 0.13--0.43 & 0.43--1.27 \\
\lam{12}{C} experiment & Tamura \textit{et al.} (1994) & $0.98 \pm 0.12$ 
  & $2.3 \pm 0.3$ \\
\lam{12}{B} theory & Krej\v{c}i\v{r}\'{i}k, Ciepl{\'{y}}, and Gal 
(2010)\footnotemark [1]  & 0.06--0.20 & 0.20--0.64 \\
\lam{12}{B} experiment & Ahmed\textit{et al.} (2003)\footnotemark [2] & 
 $0.28 \pm 0.08$  & $0.35 \pm 0.09$ \\
\end{tabular}
\end{ruledtabular} 
\footnotetext [1] {Depending on the $K^-$ nuclear potential, 
from deep to shallow.}
\footnotetext [2] {Multiply by 2 to compare to \lam{12}{C} 
production.} 
\end{table*}

\subsubsection{The $(K^{-}_\text{stop},\pi^{0})$ reaction}
\label{subsubsec:stopzero} 

The $(K^{-}_\text{stop},\pi^{0})$ reaction is an example in which both 
strangeness and charge are exchanged. However, it is expected to have 
the same features as the $(K^{-}_\text{stop},\pi^-)$ reaction, although 
its cross section is reduced by the isospin ratio of 1/2. This reaction 
produces hypernuclear species charge symmetric to those studied by the 
\Kpi\ and \piK\ reactions. In this reaction, the 
two photons from the $\pi^{0}$ decay can be used to identify and measure 
the energy of the outgoing $\pi^{0}$. Thus, not only do all the stopped 
$K^{-}$'s interact in the nuclear target but a thick target can be used 
without degrading the energy resolution because captured $K^{-}$'s have 
essentially zero momentum and the decay photons easily penetrate the 
target without significant energy degradation. 

Comparison of the spectra of charge-symmetric hypernuclei provides information 
that could be helpful to extract the isospin asymmetry of the fundamental 
$\Lambda N$ interaction. This has been studied to some extent in the ground 
states of $s$- and $p$-shell mirror hypernuclear pairs, but aside from binding 
energies, few comparative data are available \cite{gibson95}. However, in 
addition to charge asymmetry in the fundamental $\Lambda N$ interaction, 
Coulomb effects can lead to energy differences between charge-symmetric 
hypernuclei, in part because the added $\Lambda$ compresses its nuclear core, 
thereby increasing its Coulomb energy \cite{hiyama99}. Therefore, a careful 
study of the spectra of several charge symmetric pairs is needed to extract 
both the Coulomb and charge asymmetry effects for the excited, as well as the 
hypernuclear ground states \cite{gal15}. 

In an experiment at BNL \cite{ahmed03}, $\pi^{0}$'s were detected by observing 
the opening angle of the decay photons from the $\pi^{0}$ using a neutral meson
spectrometer (NMS). The NMS \cite{nms89} was a large acceptance photon detector
which measured the total energy of a $\pi^{0}$. It consisted of two arrays of 
60 CsI crystals each fronted by a set of bismuth germanate (BGO) converter and 
wire chamber tracking planes. The CsI crystals provide the photon calorimetry 
to determine the relative energy difference between the decay photons, while 
the BGO and wire chambers determine the location of the photon conversion. A 
dispersed $K^{-}$ beam with a nominal momentum of 690 MeV/c was brought to 
rest in a set of 4 natural graphite targets after it traversed a wedge-shaped, 
brass degrader of central thickness $\approx$141~mm. The degrader compensated 
for the beam dispersion ($\approx$1.2 MeV/c per cm). The energy resolution was 
2.2 MeV (FWHM) which was primarily attributed to problems associated with 
maintaining energy calibrations over the long period of data acquisition. 

In this experiment, the hypernuclear ground-state formation probability 
was found to be $(0.28\pm 0.08)\times 10^{-3}$ and that for the $p$-shell 
states near the $\Lambda$ emisson threshold was $(0.35\pm 0.09)\times 
10^{-3}$. This is compared in Table~\ref{tab:prob} to theoretical and 
experimental values for the $(K^-_\text{stop},\pi^-)$ reaction, that should 
occur twice as often assuming good isospin symmetry. The quoted errors are 
statistical, but because of the difficulty in extracting the yield from 
background the systematic error is somewhat larger for the $p$ shell 
(about 15\%). Following kinematical corrections to isospin conservation, 
the formation probability to the ground state is lower than the previous 
experimental value for \lam{12}{C} formation. However, this result 
still remains higher than the theoretical calculations for the ground 
state (Gal and Klieb, 1986; Matsuyama and Yazaki, 1988; Ciepl{\'{y}} 
\textit{et al.}, 2001, 2003; Krej\v{c}i\v{r}\'{i}k, Ciepl{\'{y}} and Gal,
2010).

\subsubsection{The in-flight {\Kpi}~reaction} 
\label{subsubsec:kpi} 

\begin{figure}[t] 
\includegraphics[width=8cm]{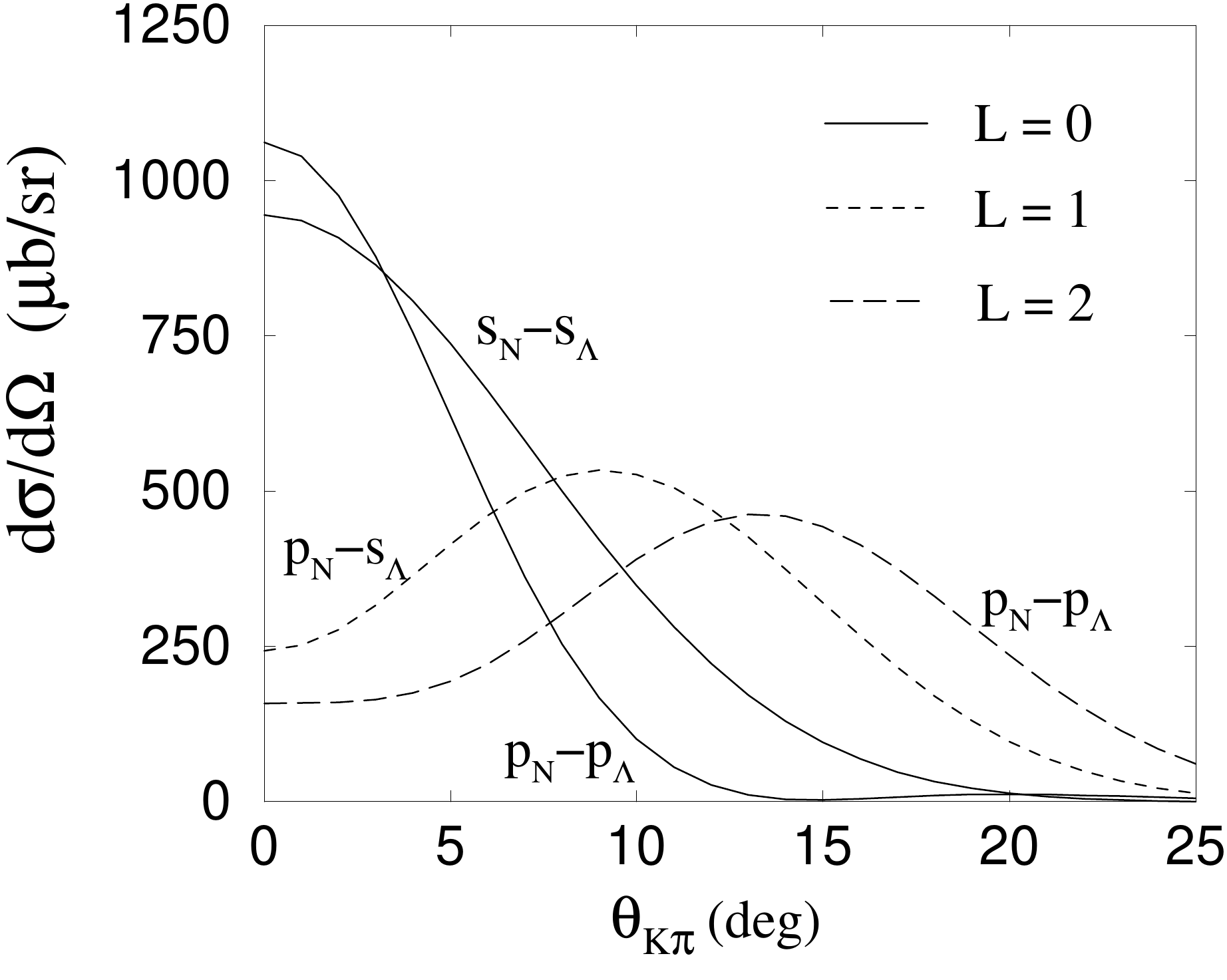} 
\caption{Angular distributions for the \Kpi\ reaction for pure
single-particle transitions on $^{16}$O at $p_K\!=\!900$ MeV/c.
$\Delta L$ is the orbital angular momentum transfer for the
indicated transitions. From Ukai \textit{et al.}, 2008.} 
\label{fig:kpiangdist} 
\end{figure} 

Although hypernuclear spectroscopy was initially studied with stopped 
kaon beams, the in-flight \Kpi\ reaction was introduced to take advantage 
of intense sources of secondary beams and the adoption of modern electronic 
counting to the readout of magnetic spectrometers. The in-flight reaction 
has several advantages as described below.

 The in-flight \Kpi\ reaction was first used at CERN~\cite{bonazzola74,
brueckner75} and then at BNL~\cite{chrien79} for incident kaon momenta in the 
range of 700 to 900 MeV/c where the elementary cross section has a maximum 
[see Fig.~6 of Hashimoto and Tamura (2006)]. Another important feature of the 
elementary reaction at these momenta is that the spin-flip amplitudes are 
small. As Fig.~\ref{fig:kinematics} shows, the momentum transfer to the 
hypernucleus is still small in the forward direction, favoring no transfer of 
orbital (or spin) angular momentum. In this case, the spectra of light 
hypernuclei exhibit peaks when a $\Lambda$ replaces a neutron without changing 
the quantum numbers of the single-particle orbit. This is illustrated in 
Fig.~\ref{fig:kpiangdist} for pure single-particle transitions on $^{16}$O at 
$p_K\!=\!800$ MeV/c. The resulting $\Lambda$ hypernuclear states are called 
``substitutional states'' (populated via ``recoilless'' transitions). The 
strong nuclear absorption of the incident $K^{-}$ and exiting $\pi^-$ limits 
penetration into the nucleus and favors transitions with surface-peaked 
transition densities (generally, between nodeless orbits). 

\begin{figure}[t]
\includegraphics[width=8cm,clip]{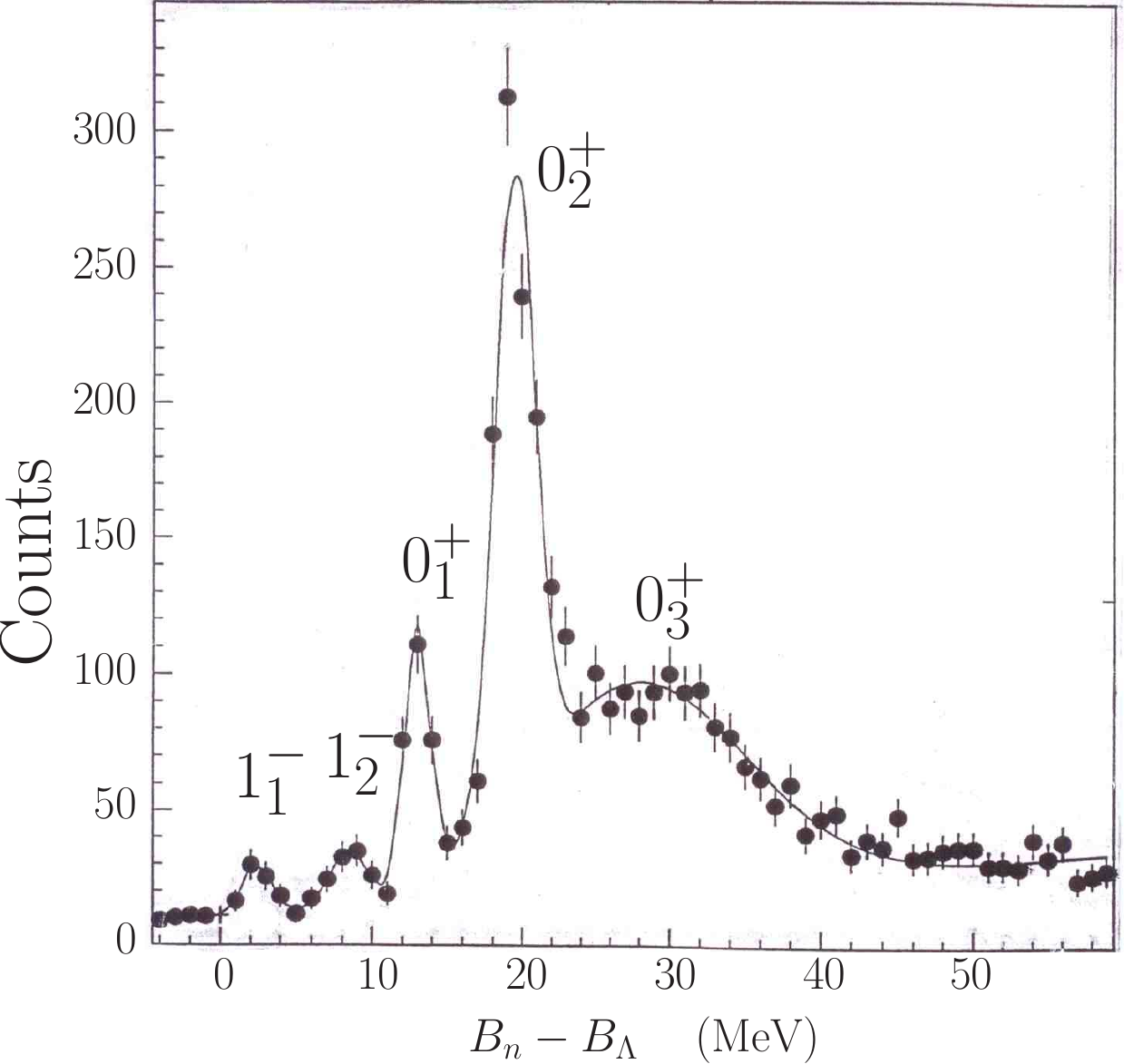} 
\caption{Spectrum for the \Kpi\ reaction on $^{16}$O at incident momentum
$p_K\!=\!715$ MeV/c near $0^\circ$~\cite{brueckner78}. The $1^-$ states are 
$s_\Lambda$ states based on the $p_{1/2}^{-1}$ and  $p_{3/2}^{-1}$ hole states of 
$^{15}$O. The $0^+_{1,2}$ states are $p_{\Lambda}$ substitutional states based on 
the same core states, while the $0^+_3$ state is based on the broad 
$0s$-hole strength in $^{15}$O. For $^{16}$O, $B_n\!=\!15.66$ MeV, so
that $B_\Lambda\!\sim\!13$ MeV for the $1^-_1$ state. The refit of the data
is due to D. H. Davis and D. N. Tovee. Courtesy of D. H. Davis.}
\label{fig:cern16o} 
\end{figure} 

One of the early investigations used the spin splitting of states in 
\lam{16}{O} to obtain a value for the $\Lambda$-nucleus spin orbit 
interaction. In Figure~\ref{fig:cern16o}, the splitting of the two 
$p_\Lambda$ states ($0^+_1$ and $0^+_2$), observed in the 
$^{16}$O\Kpi\lam{16}{O} reaction spectrum, shows that the energy difference 
between the states obtained when replacing a $p_{1/2}$ or $p_{3/2}$ neutron 
by a $\Lambda$ is essentially the same as the energy splitting of the hole 
states in $^{15}$O (6.18 MeV). This indicates that the effective $\Lambda N$ 
spin-orbit splitting is small~\cite{povh80}, a conclusion that remains valid
when the residual $\Lambda N$ interaction is taken into 
account~\cite{bouyssy80}.
A small effective $\Lambda N$ spin-orbit potential was also confirmed 
in the analysis of the angular distribution of the $p_\Lambda$ substitutional 
peak based on the $^{12}$C ground state observed in the $^{13}$C\Kpi\lam{13}{C}
reaction spectrum. In this experiment~\cite{may81}, the $p_{1/2\Lambda}$ state
is formed via a $\Delta L\!=\!0$ transition near $0^\circ$ while the
$p_{3/2\Lambda}$ state is formed via a $\Delta L\!=\!2$ transition near
$15^\circ$ (see Fig.~\ref{fig:kpiangdist}). Therefore, by measuring a shift 
of $0.36\pm 0.3$~MeV in the excitation of the substitutional peak between 
$0^{\circ}$ and $15^{\circ}$, the $\Lambda$ spin-orbit coupling was shown 
to be small~\cite{auerbach81,auerbach83}. Finally, the $\Lambda$ spin-orbit 
splitting in \lam{13}{C} was found to be very small by observing two 
$\gamma$ rays of energy $\approx$11~MeV, taken to be  
$p_{j\Lambda}\to s_{1/2\Lambda}$ $E1$ transitions correlated with two constituent 
states in this substitutional peak, and found to be split by 
152$\pm$54(stat)$\pm$36(syst)~keV \cite{ajimura01,kohri02}. 

After the initial success in applying the $(K^{-},\pi^{-})$ reaction to 
$\Lambda$ hypernuclei, an attempt was made to look for bound $\Sigma$ states 
using the same reaction \cite{bertini80,bertini84,bertini85}. Although it 
was expected that such structures would have a large width due to the strong 
conversion $\Sigma N \rightarrow \Lambda N$, this research remained in 
a confused state for a number of years, limited by the low statistics of 
the experiments which perhaps also encouraged theoretical speculations. 
Experimentally, a number 
of light $\Sigma$-nuclear systems were investigated, particularly for $s$- 
and $p$-shell $\Sigma$ nuclear systems. Attempts were made to use lower 
incident kaon momentum to reduce the QF component in the reaction, and to 
enhance substitutional-state production. All these investigations indicated 
some reaction strength below the $\Sigma$ emission threshold, but the 
interpretation of the observed structure was limited by statistical 
fluctuations (Dover, Millener, and Gal, 1989).    

\begin{figure}[th] 
\includegraphics[width=8cm]{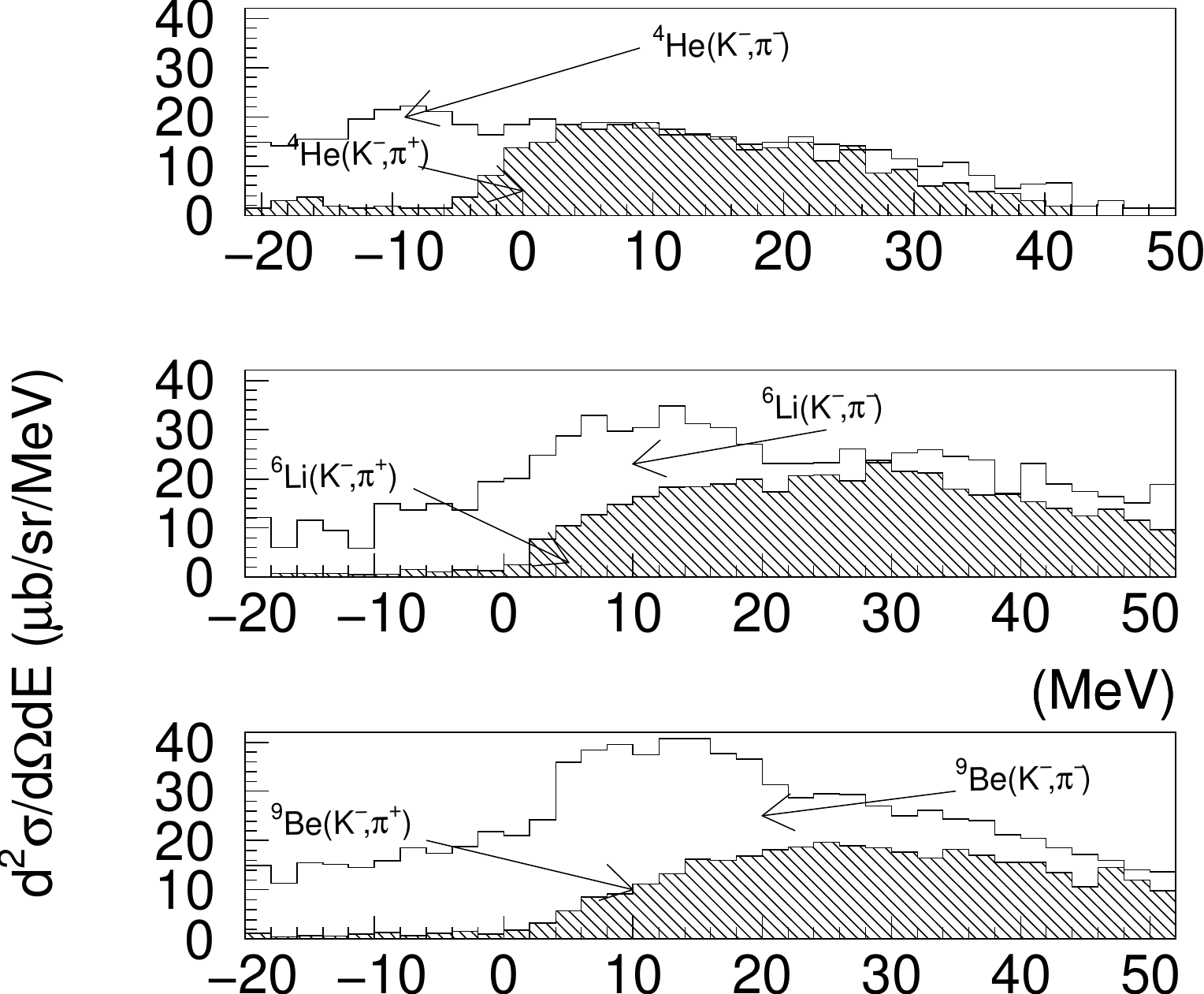} 
\caption{Experimental data from various targets showing the progressive 
changes in the $\Sigma$-nucleus interaction for both $(K^{-},\pi^{-})$ 
and $(K^{-},\pi^{+})$ reactions. From Bart \textit{et al.}, 1999.} 
\label{sigma_states} 
\end{figure} 

Two high-statistics experiments were eventually completed, one \cite{nagae98} 
on a $^{4}$He target and one \cite{bart99} on a series of $p$-shell nuclei. 
The result provided a consistent picture for $\Sigma$ nuclear interactions 
in light nuclear systems. A significant dependence on isospin was found by 
observing production differences in the spectra from \Kpi\ and
$(K^-,\pi^+)$ reactions. This is shown in Fig.~\ref{sigma_states} where 
one sees a progressive shift of the enhancement below threshold to higher 
energies and a broadening of its width. In the specific case of $^{4}$He, 
a broad bound state having a binding energy of $\approx$4.4~MeV with a width 
of $\approx$7.0~MeV was observed. Note that this state must have isospin 1/2, 
as it is seen only in the \Kpi\ reaction.  

The presence of isospin dependence suggests a strong ``Lane'' term in the 
potential which would have a $1/A$ dependence, reducing the possibility 
of $\Sigma$ hypernuclear states of any width for $A>4$. In particular for 
the $A=4$ system, theoretical analysis has shown that the effective $\Sigma$ 
nucleus potential has a small attractive pocket near the nuclear surface, 
and a strong repulsive core which decreases exponentially as the nuclear 
radius increases. A bound $\Sigma$ could reside in this well and, as the 
nuclear surface has lower density, the conversion width of the $\Sigma$ 
is smaller allowing a broad state to form~\cite{harada90}. 

More recently a study of $\Sigma$-nuclear systems was completed using the 
$(\pi^{-},K^{+})$ reaction in flight on several targets (e.g., C, Si). 
This reaction converts in one step a target proton to a $\Sigma^-$ hyperon. 
While the resulting spectra show a non-vanishing residual strength below the 
$\Sigma$ hypernuclear threshold, no evidence for bound states was found. 
Indeed, when analyzed in DWIA, the spectra are reproduced only by using 
a strongly repulsive $\Sigma$-nucleus potential~\cite{saha04}. The 
$(\pi^{-},K^{+})$ reaction was also studied near the $\Lambda$ hypernuclear 
threshold on a $^{10}$B target, searching for bound states in the neutron-rich 
\lam{10}{Li} hypernucleus~\cite{saha05}. Although no clear peaks could be 
resolved in the $\Lambda$ bound region, the size of the deduced cross section 
is consistent with formation of \lam{10}{Li} through a $\Sigma^-$ admixture of 
probability $\approx$0.1\% induced by $\Sigma^- p \leftrightarrow \Lambda n$ 
coupling (Harada, Umeya, and Hirabayashi, 2009). Very recently J-PARC experiment
E10, using the $(\pi^{-},K^{+})$ reaction on a $^6$Li target, did not observe 
any significant strength in the \lamb{6}{H} bound region~\cite{sugimura14}, 
indicating perhaps a weaker appropriate $\Sigma^-$ admixture than in 
\lam{10}{Li}. This leaves open the question of whether or not the exotic 
neutron-rich hypernucleus \lamb{6}{H} is particle stable as indicated by the 
FINUDA experiment using a $(K^{-}_\text{stop},\pi^{+})$ production reaction 
\cite{agnello12c,agnello12a}; see the discussion at the end of 
Sec.~\ref{subsubsec:stopminus} and the recent calculations by
Gal and Millener (2013) and Hiyama \textit{et al.} (2013).

\subsubsection{The {\piK}~reaction}
\label{subsubsec:pik}

The study of hypernuclear spectra using the \piK\ reaction (Dover, Ludeking, 
and Walker, 1980; Thiessen \textit{et al.}, 1980) was first explored at the 
BNL-AGS in a series
of investigations providing spectra across a wide range of hypernuclei. 
Typical energy resolution of  3--4 MeV was obtained \cite{milner85,pile91}. 
The reaction was then explored in detail at KEK with a dedicated beamline and 
a high resolution spectrometer, SKS \cite{fukuda95}, specifically built to 
detect the reaction kaons. Using this system, the resolution improved to 
about 2 MeV \cite{nagae01}.

\begin{figure}[t]
\includegraphics[width=8cm,clip]{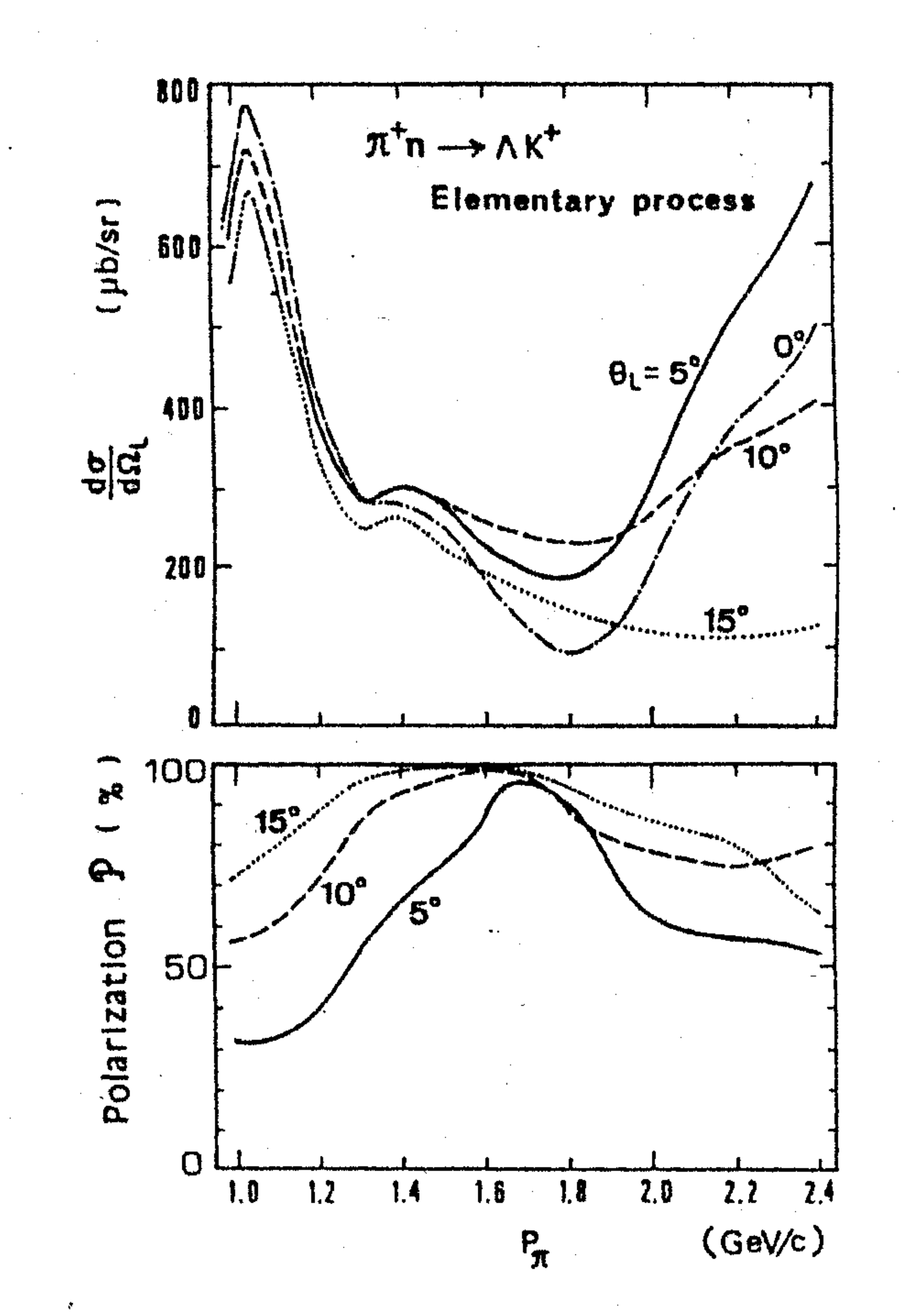}
\caption{The elementary $n(\pi^{+},K^{+})\Lambda$ reaction and the
polarization of the $\Lambda$ as a function of the $\pi$ incident
momentum. From  Band\={o} \textit{et al.}, 1989.}
\label{fig:pk_prod_xtion}
\end{figure}

\begin{figure}[t]
\includegraphics[width=8.3cm,clip]{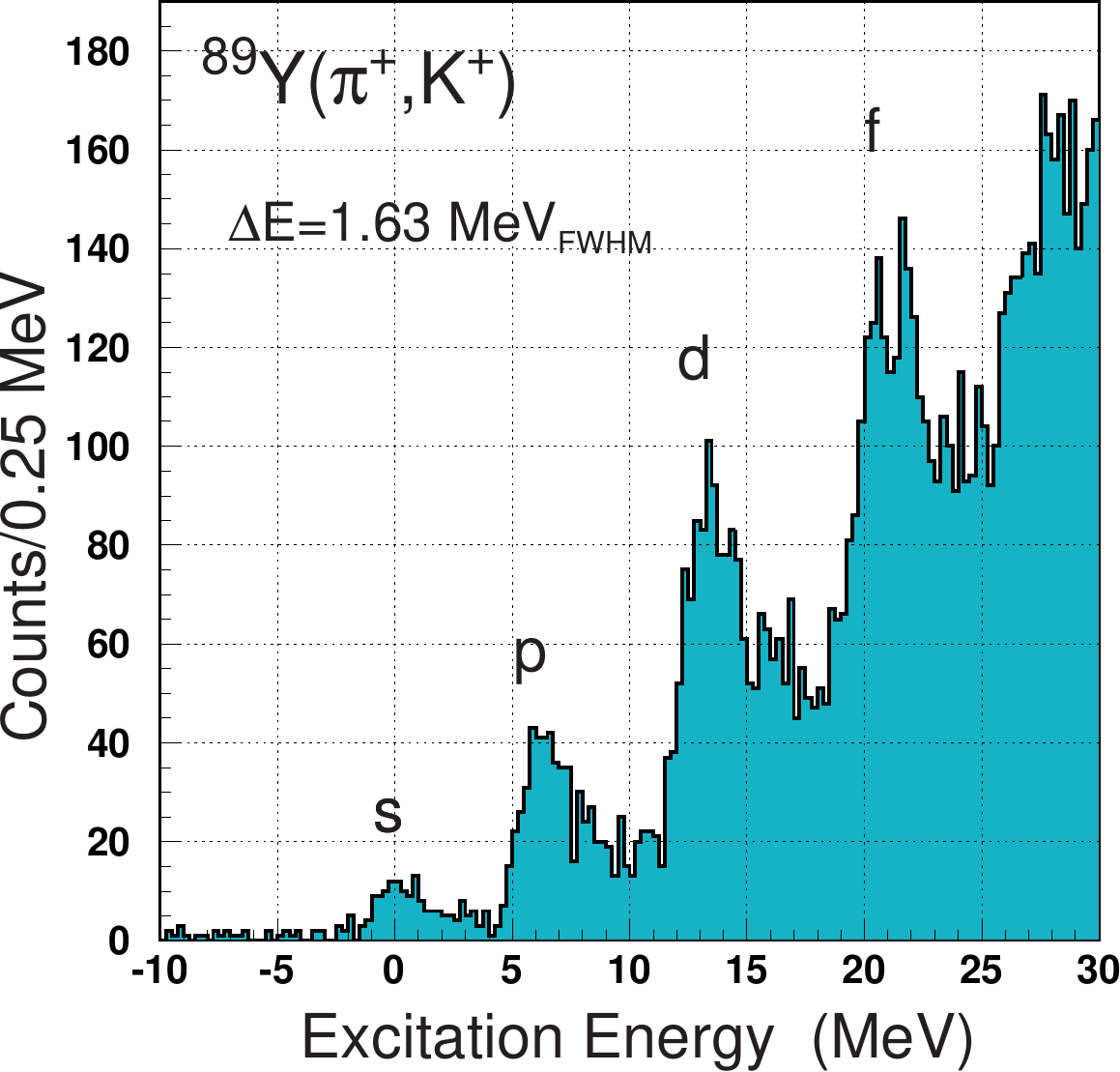}
\caption{The hypernuclear spectrum of \lam{89}{Y} from KEK E369 showing the 
major $\Lambda$ shell structure. From Hotchi \textit{et al.}, 2001.}
\label{fig:L89Y}
\end{figure}

 The elementary reaction $n(\pi^{+},K^{+})\Lambda$ 
peaks at an incident pion momentum near 1.05 GeV/c, as shown in 
Fig.~\ref{fig:pk_prod_xtion}, and all \piK\ experiments have been 
performed at this incident momentum. The outgoing $K^{+}$ 
has a momentum of $\approx$0.7 GeV/c and the momentum and angular-momentum 
transfer to the $\Lambda$ is substantial. The \piK\ reaction then 
preferentially populates spin-stretched states with an 
angular-momentum transfer $\Delta L\!=\!l_n + l_\Lambda$. For nodeless
orbitals, the momentum dependence (form factor) of the transition density 
(product of radial wave functions) is given by $y^{\Delta L/2}e^{-y}$ with
$y\!=\!(bq/2)^2$, where $q$ is the 3-momentum transfer and $b$ is the 
harmonic oscillator parameter ($b^2\!=\!41.5/\hbar\omega$, 
$\hbar\omega\!=\!45A^{-1/3}-25A^{-2/3}$). The maximum of the form factor
occurs for $y\!=\!\Delta L/2$. For light hypernuclei and transitions
to inner $\Lambda$ orbitals in heavier nuclei, the momentum transfer
$q$ is generally over 300 MeV/c which is well past the peak in the form
factor and cross sections are small. However, the \piK\ reaction 
becomes more effective in producing states with large $l_\Lambda$ in heavier 
hypernuclei due to the increasing spin of the valence neutron orbital 
involved in the reaction. Indeed, in Fig.~\ref{fig:L89Y}, the full spectrum of 
nodeless, bound $\Lambda$ orbitals is clearly evident for the \lam{89}{Y} 
hypernucleus \cite{hotchi01}. The main part of the cross section arises 
from associated production on a $g_{9/2}$ neutron, while the origin of 
possible fine structure in the peaks is open to interpretation~\cite{motoba08}.
The $\Delta L\!=\!7$ transition dominating the $f_\Lambda$ peak is well
matched in the sense that the peak of the form factor occurs for 
$q\sim 345$ MeV/c and closely matches the momentum transfer to the
hypernucleus. In general, \piK\ cross sections are found to be roughly a 
factor of 100 below those in the \Kpi\ reactions (different final states are
populated) but, in terms of running time, the decrease in cross section can 
be more than compensated by the increased intensity of pion beams. Because 
the momentum transfer is high, the cross section falls rapidly with angle 
and the angular distribution is not a good indicator of the angular momentum 
transfer.

In contrast to low momentum kaon induced reactions, the $\Lambda$ recoil 
in the \piK\ reaction has substantial polarization at finite forward angles. 
This polarization is due to a combination of the difference the near- and 
far side absorption of the incident pion, and the spin dependence of the 
elementary interaction. With the exception that polarization creates 
specific spin states in the hypernucleus, polarization in the \piK\ reaction 
has not been experimentally used in spectroscopic studies, (e.g., 
angular correlations), as these experiments require coincidence measurements 
at angles where the production rate is low. 

 The \piK\ reaction has so far been the most productive spectroscopic 
reaction across a wide range of nuclei. However, targets are 
large (e.g., several cm$^2$ in area and $\approx$\,gm/cm$^{2}$ thick) 
which is a factor in limiting the energy resolution. The choice of target
is a factor in the selectivity of the reaction. As noted earlier, cross 
sections are proportional to the neutron pickup spectroscopic factor in
the weak-coupling limit. This means that ideally one should choose a target 
with a full shell of a high-$j$ neutron orbit close to the Fermi surface.
At A$\sim$90, this would mean a $^{90}$Zr target but $^{89}$Y has the
advantage that it is a monotope; the $^{88}$Y core nucleus has a $4^-$ ground 
state and a low-lying $5^-$ state (at 232 keV) that are both fed by $g_{9/2}$
neutron removal and a small correction must be made to the extracted
$B_\Lambda$ values~\cite{hasegawa96}. 

The \piK\ reaction provides a textbook example of the single-particle shell 
structure of hypernuclei, with Fig.~\ref{fig:L89Y} showing the prime example.
In Sec.~\ref{subsubsec:spe}, we collect together the $\Lambda$ 
single-particle energies in terms of $B_\Lambda$ values extracted from \piK ,
\eeK , \Kpi , and emulsion studies. Most of the values come from three 
\piK\ experiments at KEK, namely E140a~\cite{hasegawa96} (targets $^{10}$B,
$^{12}$C, $^{28}$Si, $^{89}$Y, $^{139}$La, $^{208}$Pb), E336 (Hashimoto 
\textit{et al.}, 1998; Hashimoto and Tamura, 2006) (targets $^7$Li, $^9$Be, 
$^{12}$C, $^{13}$C, $^{16}$O), and E369~\cite{hotchi01} (targets $^{12}$C, $^{51}$V,
$^{89}$Y). All the targets are largely a single isotope, either because
the natural target is a monotope, or nearly so, or because an enriched
target was used ($^7$Li, $^{10}$B, $^{13}$C, $^{208}$Pb). For the heavier
targets ($^{51}$V, $^{89}$Y, $^{139}$La, $^{208}$Pb), the aim is to identify
peaks due to a series of $\Lambda$ orbitals based on holes in the nodeless
$f_{7/2}$, $g_{9/2}$, $h_{11/2}$, and  $i_{13/2}$ neutron shells. For the
odd-mass targets there is fragmentation of the neutron pickup strength
due to the presence of an odd proton, and this must be accounted for
in the analysis. In addition, other filled neutron orbits can
make substantial contributions to the cross sections, as can be seen from
attempts to analyze the data for \lamc{139}{La} and \lamc{208}{Pb} in
Fig.~27 of Hashimoto and Tamura (2006) . We note that although plane-wave 
impulse approximation calculations seem to capture the essential physics
(Bender, Shyam, and Lenske, 2010), DWIA calculations 
generally give in addition reliable estimates for the cross sections of 
states populated in the \piK\ reaction (Motoba \textit{et al.}, 1988;
Millener, 1990; Motoba, Itonaga, and Yamamoto, 2010). 

\subsubsection{The \eeK\ reaction}
\label{subsubsec:eek}

Traditionally, hypernuclei were produced with secondary beams of kaons or 
pions. Because the \Kpi\ reaction is exothermic, the 3-momentum transfer 
to the $\Lambda$ hypernucleus can be chosen to be small. In the \Kpi\ 
reaction, the cross section to substitutional states (i.e., states 
where the $\Lambda$ acquires the same shell quantum numbers as those of 
the neutron which it replaces) is relatively large. On the other hand, 
the \piK\ reaction has a 3-momentum transfer comparable to the nuclear 
Fermi momentum, and the reaction preferentially populates states with 
high angular momentum transfers (Milner \textit{et al.}, 1985; Band\={o}
and Motoba, 1986). Neither of these two reactions has significant spin-flip 
amplitude at forward angles, and consequently all spectra are dominated by 
transitions to non spin-flip states. Also, aside from early emulsion 
experiments, mesonic-reaction spectroscopy has generally provided hypernuclear 
spectra with energy resolutions $\approx$2~MeV. This is due to the intrinsic 
resolutions of secondary mesonic beamlines, and the target thicknesses required 
to obtain sufficient counting rates. However, one study did achieve 
a spectrum resolution of approximately 1.5 MeV for the \lam{12}{C} 
hypernucleus, using a thin target and devoting substantial time to 
data collection~\cite{hotchi01}. Another significant problem with the 
\Kpi\ and \piK\ reactions is how to fix the absolute scale of hypernuclear
binding energies (no free-neutron target) and this requires normalization
to a known ground-state binding energy, e.g. from emulsion data; 
see the discussion in Sec.~\ref{subsubsec:spe}. 

Electron beams, in comparison, have excellent spatial and energy resolutions, 
and the exchange of a photon can be accurately described by a first order 
perturbation calculation. In addition, electroproduction has been used for 
precision studies of nuclear structure so many experimental techniques are 
well established. Although previous electron accelerators had poor duty 
factors significantly impairing high singles-rate coincidence experiments, 
continuous-beam accelerators have now overcome this limitation. The cross 
section for nuclear kaon electroproduction is smaller than that for 
hypernuclear production by the $(\pi,K)$ reaction for example, but reaction 
rates can be compensated by increased beam intensity. Targets can be 
physically small and thin (10-100 mg cm$^{-2}$), allowing studies of almost 
any isotope. However, a great advantage of the \eeK\  
reaction is the potential to reach energy resolutions of a few 
hundred keV with reasonable counting rates at least up to medium-weight 
hypernuclei~\cite{hungerford94}. Another great advantage
is that the $\Lambda$ and $\Sigma^0$ peaks from the \eeK\ reaction on
hydrogen can be used to calibrate the hypernuclear binding-energy scale.

Furthermore, the \eeK\ reaction proceeds by the absorption 
of a spin--1 virtual photon which carries high spin-flip probability even at 
forward angles. The 3-momentum transfer to a quasi-free $\Lambda$ is high 
(approximately 300 MeV/c at zero degrees for 1.5 GeV incident photons), 
so the resulting reaction is expected to predominantly excite spin-flip 
transitions to spin-stretched states (Motoba, Sotona, and Itonaga, 1994). 
Recall that spin-flip states are not strongly excited in hadronic production, 
and the \eeK\ reaction acts on a proton rather than a neutron, creating 
proton-hole $\Lambda$-particle states which are charge symmetric to those 
studied with meson beams. 

In electroproduction, the $\Lambda$ and $K^{+}$ particles are created 
associatively via an interaction between a virtual photon and a bound 
proton, $p(\gamma,K^{+})\Lambda$. The hypernucleus, \lam{A}{Z}, 
is formed by coupling the $\Lambda$ to the residual nuclear core 
$^{(A-1)}(Z-1)$. In electroproduction, the energy and 3-momentum of 
the virtual photon are defined by $\omega=E_{e}-E^{\prime}_{e}$ 
and $\bm{q}=\bm{p}_{e}-\bm{p}_{e}^{\,\prime}$, respectively. 
The square of the four-momentum transfer of the electron is then 
given by $-Q^{2}=t=\omega^{2}-q^{2}$. As will be shown below, the number 
of (virtual) photons falls rapidly as the scattered electron angle increases 
(increasing $t$), and thus the distribution of (virtual) photons also peaks 
in the forward direction. In addition, the nuclear transition matrix element 
causes the cross section for hypernuclear production to fall rapidly as the
angle between the reaction kaon and the (virtual) photon increases. Thus, 
experiments must be done within a small angular range around the direction 
of the incident electron. To accomplish this, the experimental geometry 
requires two spectrometer arms, one to detect the scattered electron and 
one to detect the kaon, both placed at extremely forward angles.

The electroproduction cross section can be expressed \cite{sotona94} by  
\begin{eqnarray} 
\frac{d^{3}\sigma}{d E^{\prime}_{e}d\Omega^{\prime}_{e} 
d\Omega_{K}} &= & \Gamma\left[\frac{d\sigma_{T}}
{d\Omega_{K}} + \epsilon\frac{d\sigma_{L}}{d\Omega_{K}} 
+ \epsilon \mbox{cos}(2\phi) \frac{d\sigma_{TT}}{d \Omega_{K}} 
\right. \nonumber \\
&& \mbox{} + \left.\mbox{cos}(\phi)\sqrt{2\epsilon(1+\epsilon)}
\frac{d\sigma_{LT}}{d\Omega_{K}}
\right].
\end{eqnarray} 
The factor $\Gamma$ is the virtual flux factor evaluated with electron 
kinematics in the lab frame, and $\phi$ is the angle measuring the 
out-of-plane production of the kaon with respect to the plane containing 
the beam and scattered electron.  The factor $\Gamma$ has the form 
\begin{equation}
\Gamma = \frac{\alpha}{2\pi^{2}Q^2} \,
\left[ \frac{E_\gamma}{1-\epsilon} \right] \,
\frac{E^\prime_e}{E_e}\, .
\end{equation}
In the above equations, $E_{\gamma} = \omega$, $\alpha$ is the fine-structure 
constant and $\epsilon$ is the polarization factor
\begin{equation}
\epsilon = \left[ 1 + \frac{2|\bm{q}|^2}{Q^2}{\mbox{tan}}^2(\theta_e/2)
\right]^{-1}\, .
\end{equation}
The label on each of the cross-section expressions ($T$, $L$, $TT$ and $LT$) 
represent transverse, longitudinal, polarization, and interference terms. 
For real photons of course, $Q^2\!=\!0$, so only the transverse 
cross section is non-vanishing, and for a very forward experimental 
geometry, the virtual photons are almost on the mass shell where 
$Q^2 =\bm{p}^2_{\gamma} - E^2_\gamma \, = \, 0$ so the cross section is 
completely dominated by the transverse component. Thus a good approximation 
replaces electroproduction cross section by the photoproduction cross section 
multiplied by a flux factor. 

\begin{figure}[t]
\includegraphics[width=8.3cm]{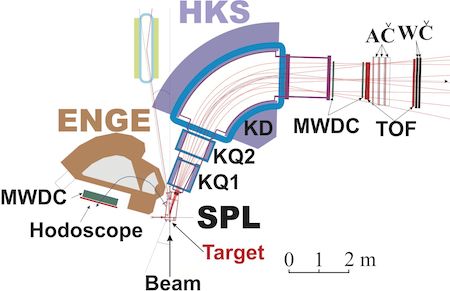}
\includegraphics[width=8.3cm]{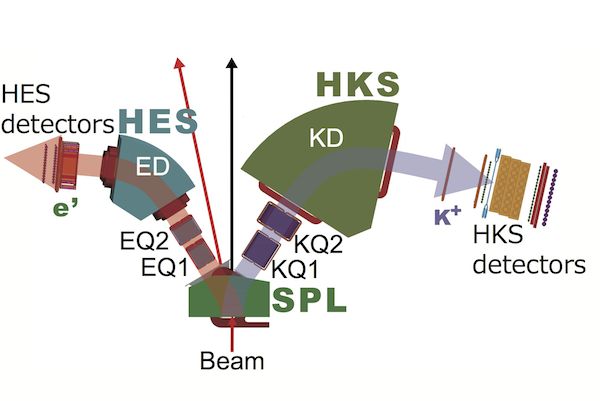}
\caption{Schematic illustration of the experimental setup, 
technique, and upgrades for the Hall C HKS hypernuclear spectroscopy 
experiments E01-011 (upper panel) and E05-115 (lower panel). From 
Tang \textit{et al.}, 2014. \label{fig:layout}} 
\end{figure}

Experimentally, $\Gamma$ is integrated over the angular and momentum 
acceptances of the electron spectrometer. In order to maximize the 
cross section of the elementary $p(\gamma,K^+)\Lambda$ reaction, the 
photon energy is chosen to be $\approx$(1.5-2.5) GeV. To maximize 
the elementary cross section, the virtual photon energy should be near 
1.5 GeV, which determines the scattered electron energy, $E_{e^\prime} = 
E_e - \omega$. Finally, to limit the production of a background of unwanted 
hyperons, the maximum choice for the beam energy should be as close to 1.8 
GeV as possible. The virtual flux factor peaks at zero degrees and falls 
rapidly as the scattering angle increases (Xu and Hungerford, 2003). 
A large percentage of the scattered electrons can be captured in even a 
small solid angle for scattering angles near zero degrees. 

Compared to secondary beam experiments, the magnetic optics of the 
spectrometer systems in electroproduction experiments are less 
complicated because of the small beam spot ($\approx$100$\,\mu$m), 
the $\le 4^{\circ}$ electron scattering angle, and the small momentum 
value of the scattered electron. However, the disadvantage of this 
geometry is a high electron background rate from target bremsstrahlung, 
which ultimately limits the usable beam luminosity and drives the geometry 
away from in-plane scattering. 
  
 Once the choice of the incident and scattered electron momenta is fixed, 
the production kaon momenta are determined by the kaon production angle. 
The kinematics are illustrated in Fig.~\ref{fig:kinematics}. The recoil 
momentum of the $\Lambda$ is comparable to the Fermi momentum and the 
kaon momentum is sufficient to allow a reasonable kaon survival fraction. 
The detector package requires at least a 1000-to-1 kaon-to-pion particle 
identification. Figure~\ref{fig:layout}  shows 
a schematic view of the experimental layout for the JLab Hall C HKS 
hypernuclear spectroscopy experiments E01-011 and E05-115. The splitter SPL 
bends electrons into the high-resolution electron spectrometer (HES) and 
kaons into the high-resolution kaon spectrometer (HKS), so that the 
reaction angles of both the electron and kaon can be observed at very forward 
angles. However, the SPL also bends the incident beam so that the beam must 
be bent back into the beam dump. This is accomplished by bending the incident 
beam before it enters the SPL canceling the bend angle in the SPL. In this 
way, the beam is bent before it is dispersed by the target, producing less 
scattering in magnets and apertures and thus less background. Further, to 
decrease the extremely high electron singles rate, the HES is rotated out 
of the HKS-beam dispersion plane by 7.5$^{\circ}$. This tilt is equivalent 
to a rotation plus a shift of the spectrometer so that scattered electrons 
$\le 4.5^{\circ}$ hit the HES yokes and thus do not enter the spectrometer 
acceptance. This angle was chosen based on a figure of merit optimization 
between hypernuclear yield and accidental background rate. The tilt improved 
the true data rate by an order of magnitude while reducing accidental 
background. The beam and spectrometer parameters are tabulated by Tang
\textit{et al.} (2014). The experimental energy resolution to specific states 
was approximately 600 keV FWHM.

\begin{figure}[t]
\includegraphics[width=8.3cm]{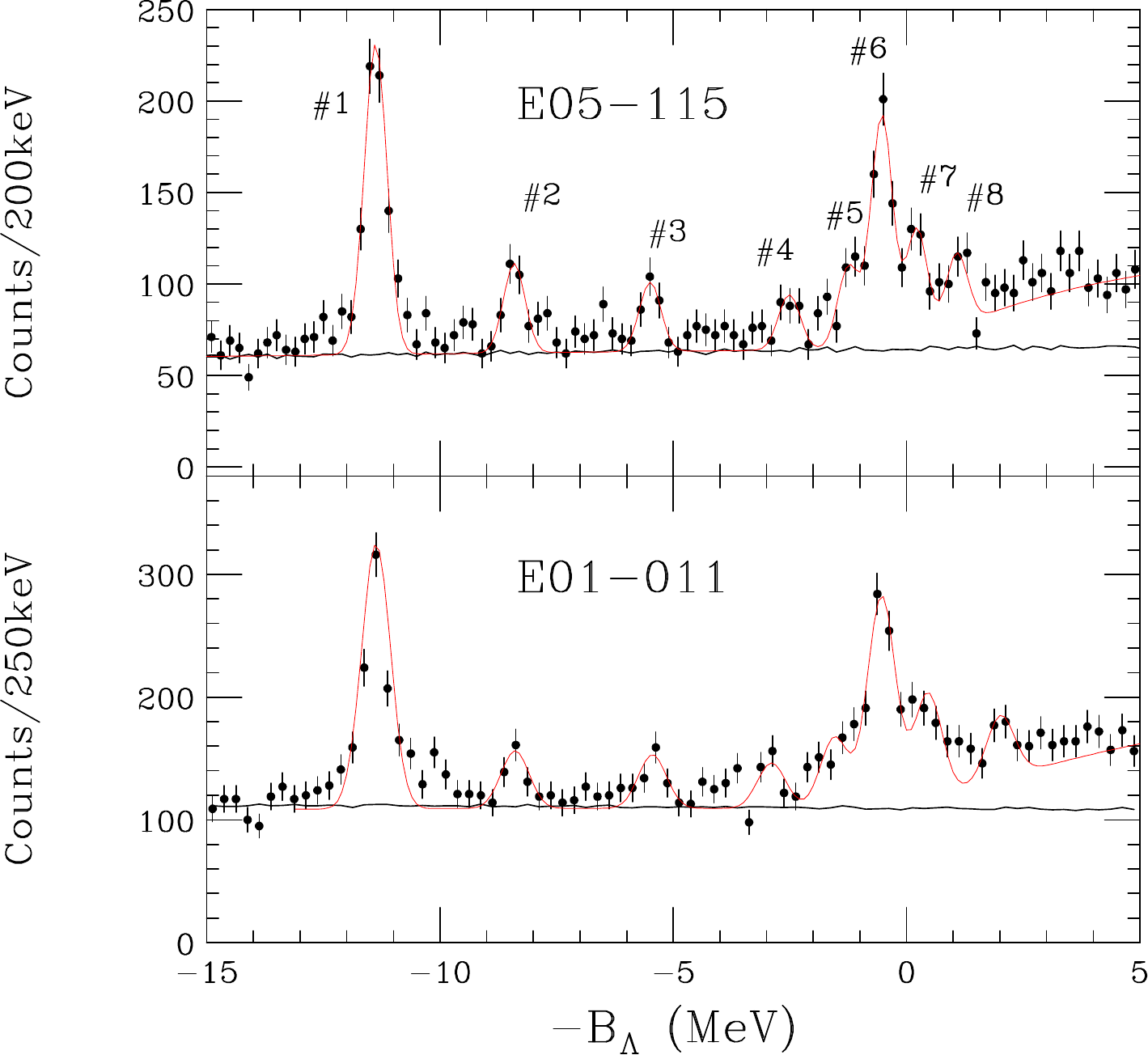}
\caption{Spectroscopy of \lam{12}{B} from the E05-115 and 
E01-011 experiments. The area below the black line is the accidental 
background. From Tang \textit{et al.}, 2014. \label{fig:specb12l}}
\end{figure}

The \lam{12}{B} spectrum obtained in these experiments on a $^{12}$C target is 
shown in Fig.~\ref{fig:specb12l}, demonstrating the improved resolution in the 
more recent E05-115 experiment with respect to that in the older one E01-011 
and also with respect to the Hall A experiment E94-107 \cite{iodice07}. In the 
upper panel of the figure, peaks 1, 2, 3, and 4 result from the 
$p_N\to s_{\Lambda}$ transition strength, with peak 1 standing for the 
\lam{12}{B} g.s. doublet which to a very good approximation is based on the 
$^{11}$B g.s. core state. The other three peaks correspond to coupling the 
$s_{\Lambda}$ hyperon to known excited levels in $^{11}$B. Peaks 5, 6, 7, and 8 
result from the $p_N\to p_{\Lambda}$ transition strength which extends further 
up into the continuum. Similar spectra were reported for the charge-symmetric
hypernucleus \lam{12}{C} in \piK\ and $(K^{-}_\text{stop},\pi^{-})$ experiments 
at KEK \cite{hotchi01} and at DA$\Phi$NE \cite{agnello05a}, respectively. Yet, 
the JLab \eeK\ experiment provides by far the most refined $A\!=\!12$ 
$\Lambda$ hypernuclear excitation spectrum. 

\begin{figure}[thb]
\includegraphics[width=8cm,clip]{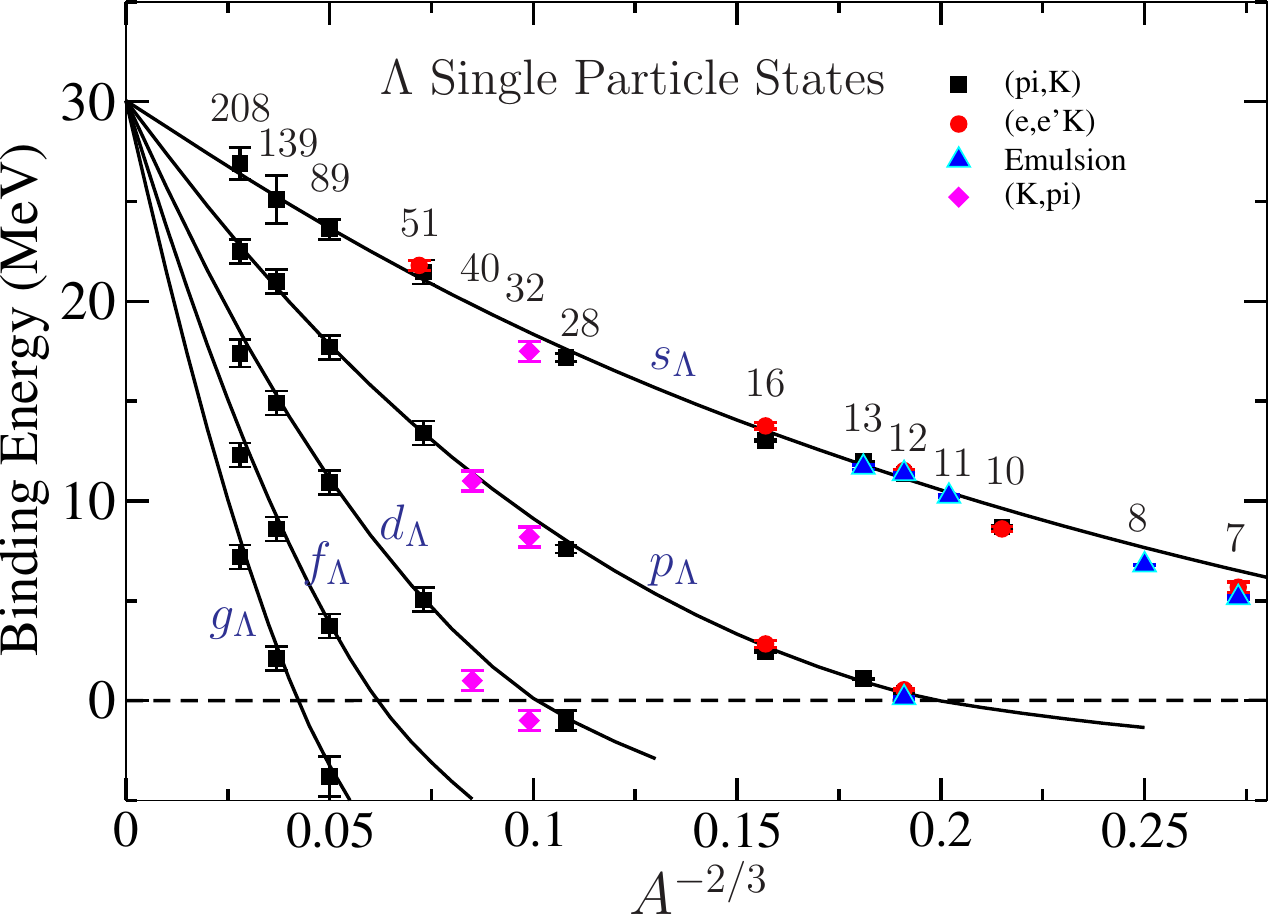} 
\caption{Energy levels of the $\Lambda$ single-particle 
major shells in \lamb{A}{Z} hypernuclei as a function of $A^{-2/3}$. 
The curves are obtained from a standard Woods-Saxon potential $V_{\rm WS}$ 
representing the $\Lambda$-nucleus interaction with depth 
$V_0$=$-$30.05~MeV, radius $R$=$r_0A^{1/3}$, where $r_0$=$1.165$~fm, and 
diffusivity $a=0.6$~fm. Updated from Millener, Dover, and Gal, 1988.} 
\label{fig:spe} 
\end{figure} 

 Very recently, the spectrum of another $p$-shell hypernucleus,
\lam{10}{Be}, was obtained in a JLab Hall C \eeK\ experiment~\cite{gogami16a}.
This experiment gives a $B_\Lambda$ value for a hypernucleus for which
there are only a few emulsion events (see Table~\ref{tab:blambda}). It
shows four clear $s_\Lambda$ peaks as expected from the proton removal
strength from $^{10}$B [see Sec.~\ref{subsec:DWIA} and Fig.~3 of
Millener (2012)].

 The \eeK\ experiments in Hall A were performed using two existing
high-relsolution (long flight path) spectrometers and used a much
higher electron-beam energy of $\sim 3.7$ GeV to increase the $K^+$
survival time. The two essential features of the setup were the
placement of superconducting septum magnets before each spectrometer
to be able to take data at $6^\circ$ and a ring-imaging Cherenkov
detector to provide unambiguous $K^+$ identification. Data were taken
using targets of $^{12}$C~\cite{iodice07}, $^{16}$O~\cite{cusanno09},
and $^9$Be~\cite{urciuoli15}. In particular, $B_\Lambda\!=\!13.76\pm 0.16$
MeV was determined for \lam{16}{N} by using the $\Lambda$ and $\Sigma^0$
peaks from the elementary \eeK\ reaction on the hydrogen in a
waterfall target for calibration.

\subsubsection{Single-particle structure}
\label{subsubsec:spe}

\begin{table}[t]
\caption{$B_\Lambda$ values from a variety of sources for $\Lambda$
single-particle states. \label{tab:spe}}
\begin{ruledtabular} 
\begin{tabular*}{\textwidth}{@{}l@{\extracolsep{\fill}}ccccc}
 Hypernucleus & $s_\Lambda$ & $p_\Lambda$ & $d_\Lambda$ & $f_\Lambda$
& $g_\Lambda$ \\
\hline
  & \multicolumn{5}{c}{\piK} \\
 \lamc{208}{Pb} & 26.9(8) & 22.5(6) & 17.4(7) & 12.3(6) & 7.2(6) \\
 \lamc{139}{La} & 25.1(12) & 21.0(6) & 14.9(6) & 8.6(6) & 2.1(6) \\
 \lam{89}{Y} & 23.6(5) & 17.7(6) & 10.9(6) & 3.7(6) & $-3.8(10)$ \\
 \lam{51}{V} & 21.5(6) & 13.4(6) & 5.1(6) &  &  \\
 \lam{28}{Si} & 17.2(2) & 7.6(2) & $-1.0(5)$ &  &  \\
 \lam{16}{O} & 13.0(2) & 2.5(2) &  &  &  \\
 \lam{13}{C} & 12.0(2) & 1.1(2) &  &  &  \\
 \lam{12}{C} & 11.36(20) & 0.36(20) &  &  &  \\
 \lam{10}{B} & 8.7(3) &  &  &  &  \\
  & \multicolumn{5}{c}{\eeK} \\
 \lam{52}{V} & 21.8(3) &  &  &  &  \\
 \lam{16}{N} & 13.76(16) & 2.84(18) &  &  &  \\
 \lam{12}{B} & 11.52(2) & 0.54(4) &  &  &  \\
 \lam{10}{Be} & 8.55(13) &  &  &  &  \\
 \lamb{7}{He} & 5.55(15) &  &  &  &  \\
  & \multicolumn{5}{c}{Emulsion} \\
 \lam{13}{C} & 11.69(12) & 0.8(3) &  &  &  \\
 \lam{12}{B} & 11.37(6) &  &  &  &  \\
 \lam{12}{C} &          & 0.14(5) &  &  &  \\
 \lamb{8}{Li} & 6.80(3) &  &  &  &  \\
 \lamb{7}{Be} & 5.16(8) &  &  &  &  \\
  & \multicolumn{5}{c}{\Kpi} \\
 \lam{40}{Ca} &  & 11.0(5) & 1.0(5) &  &  \\
 \lam{32}{S} & 17.5(5) & 8.2(5) & $-1.0(5)$ &  &  \\
\end{tabular*}
\end{ruledtabular}
\end{table}

 Taking the positions of the $\Lambda$ major shells as observed in the
\piK\ and other reactions, the $\Lambda$ single-particle energies show a 
very smooth $A$-dependence, which can be reproduced by a simple 
Woods-Saxon potential $V_{\rm WS}$, as shown in Fig.~\ref{fig:spe} for a 
data set that includes information up to \lamc{208}{Pb}~\cite{hasegawa96}. 
The data used in the construction of Fig.~\ref{fig:spe} is given in
Table~\ref{tab:spe}. Because the $B_\Lambda$ values in Table~\ref{tab:spe}
differ in several respects from the values given in the original
papers and reviews [see, e.g., Hashimoto and Tamura (2006)], some 
explanation is needed.

The most important overall change in the tabulated $B_\Lambda$ values arises 
from the fact that the KEK \piK\ data were all normalized to the emulsion 
value of 10.76 MeV for \lam{12}{C}~\cite{hasegawa96}. This differs 
considerably from the emulsion value of 11.37 MeV for \lam{12}{B} that 
is based on a substantial number of events for the characteristic 
$\pi^- + 3\alpha$ decay mode.  It is generally accepted that one should 
not trust the emulsion $B_\Lambda$ values for \lam{12}{C} and beyond because 
of the difficulty of identifying uniquely the decaying hypernucleus and the 
fact that there are very few events in each case~\cite{davis90}. In fact, 
the best determined $B_\Lambda$ value for \lam{12}{C} is 0.14(5) MeV 
based on proton emission from what is interpreted as a $0^+$ with a 
dominant $^{11}$C(g.s.)$\times p_{3/2\Lambda}$ configuration~\cite{davis08}. 
The same analysis gives two $2^+$ states 0.06 and 0.80 MeV below the 
$0^+$ state. These $2^+$ states should be populated in the \piK\ spectrum with 
the upper one dominant. The unresolved $p_\Lambda$ peak from KEK E336 is 
11.00(3) MeV above the ground-state peak~\cite{hashtam06}. Adding 0.14 MeV
and 0.23 MeV for the difference between the $0^+$ state and the $2^+$ centroid
gives 11.37 MeV, the same as the $B_\Lambda$ value for \lam{12}{B}.
Taking into account the fact that different $p_\Lambda$ states are populated
in different reactions, one gets similar values from the \eeK\
\cite{iodice07,tang14} and  $(K_\text{stop}^{-},\pi^{-})$~\cite{agnello05a} 
reactions. Table~\ref{tab:pshell} shows that adding 0.6 MeV to \piK\ 
$B_\Lambda$ values from KEK E336~\cite{hashtam06} gives better agreement 
with the emulsion values. However, for \lam{16}{O} there is still a 
discrepancy with $B_\Lambda\!=\!13.76\pm 0.16$ MeV for 
\lam{16}{N}~\cite{cusanno09}.

 Hasegawa \textit{et al.} (1996) state in their Section II.F
that they apply a shift to the $K^+$ momentum to get the \lam{12}{C} 
ground-state peak at $B_\Lambda\!=\!10.76$ MeV. The relationship between 
$p_K$ and $B_\Lambda$ is linear and nearly independent of the target mass. 
Therefore, the energy shift applied to \lam{12}{C} applies elsewhere.
The numbers for \lam{28}{Si}, \lamc{139}{La}, and \lamc{208}{Pb} in
Table~\ref{tab:spe} are from Table~13 of Hashimoto and Tamura (2006); a 
reanalysis of the KEK E140a data has been made and the errors include an 
estimate for the systematic error associated with the KEK \piK\ experiments.

 Hasegawa \textit{et al.} (1996) made corrections of 0.15, 
0.99, and 1.63 MeV to the extracted $B_\Lambda$ values for 
\lam{89}{Y}, \lamc{139}{La}, and \lamc{208}{Pb}; 0.15 MeV is the centroid 
of the $4^-/5^-$ $\pi p_{1/2}^{-1} \nu g_{9/2}^{-1}$
ground-state doublet of $^{88}$Y, 0.99 MeV is the excitation energy
of the centroid of the $\nu 0h_{11/2}$ pickup strength from $^{139}$La,
and 1.63 MeV is the excitation energy of the  $\nu 0i_{13/2}$ hole state in 
$^{207}$Pb. 

For \lam{89}{Y}, the left-hand peaks in Table VIII of Hotchi \textit{et al.}
(2001) are taken [Motoba \textit{et al.} (2008) argued that the right-hand 
peaks are associated with the $\nu f_{5/2}$ hole state] whereas 
Hashimoto and Tamura (2006) take the centroid of the left-hand 
and right-hand peaks.

\begin{table}[t]
\caption{\piK\ versus emulsion $B_\Lambda$ values for $p$-shell
hypernuclei. The first line contains values from KEK E336~\cite{hashtam06},
the second gives emulsion values from Table~\ref{tab:blambda}, and the last 
is \piK\ plus 0.6 MeV. For comparison, Gogami \textit{et al.} (2016a) have 
averaged the differences for \lamb{7}{Li}, \lamb{9}{Be},  \lam{10}{B}, and 
\lam{13}{C} to obtain a shift of $0.54(5)$ MeV. \label{tab:pshell}}
\begin{ruledtabular} 
\begin{tabular*}{\textwidth}{@{}c@{\extracolsep{\fill}}ccccc}
 \lamb{7}{Li} & \lamb{9}{Be} & \lam{10}{B} & \lam{12}{C} & \lam{13}{C}
 & \lam{16}{O} \\
\hline
 5.22(8) & 5.99(7) & 8.10(10) & 10.76 & 11.38(5) & 12.42(5) \\
 5.58(3) & 6.71(4) & 8.89(12) & 10.76(19) & 11.69(12) & \\
 5.82 & 6.59 & 8.70 & 11.36 & 11.98 & 13.02 \\
\end{tabular*}
\end{ruledtabular}
\end{table}

 For \lam{51}{V}, the left-hand peaks from Table IX of Hotchi \textit{et al.}
(2001) are taken and 0.9 MeV is added because the strongest $\nu f_{7/2}$
pick-up strength from $^{51}$V goes to a closely spaced $7^+$/$5^+$
doublet at 0.9 MeV excitation energy in $^{50}$V (levels up to about 1.3 MeV
are excited by $\nu f_{7/2}$ removal and the $6^+$ ground state is also 
quite strong). This then gives $B_\Lambda\!=\!21.47$ MeV for the $s_\Lambda$ 
state, to which one should add a small amount for the increase in mass by 
one unit to compare with the value of $21.80$ MeV from 
$^{52}$Cr\eeK\lam{52}{V} reaction~\cite{gogami14}.  

For the $p_\Lambda$ energy in \lam{13}{C}, the centroid of
the excitation energies of the $p_{3/2\Lambda}$ (10.83 MeV) and 
$p_{1/2\Lambda}$ (10.98 MeV) states from a $\gamma$-ray 
experiment~\cite{kohri02} is used. 

 Also included in Table~\ref{tab:spe} are recent $B_\Lambda$ values
from JLab; \lam{52}{V}~\cite{gogami14}, \lam{16}{N}~\cite{cusanno09}, 
\lam{12}{B}~\cite{tang14}, \lam{10}{Be}~\cite{gogami16a}, and 
\lamb{7}{He}~\cite{gogami16b}. 
The \Kpi\ values for \lam{32}{S} and \lam{40}{Ca} are CERN 
data~\cite{bertini79}. For \lam{12}{C} and \lam{16}{O}, see 
Br\"{u}ckner \textit{et al.} (1978) and for the summary paper 
(up to a $^{209}$Bi target), see Bertini \textit{et al.} (1981).

\begin{figure*}[th]  
\includegraphics[width=14cm]{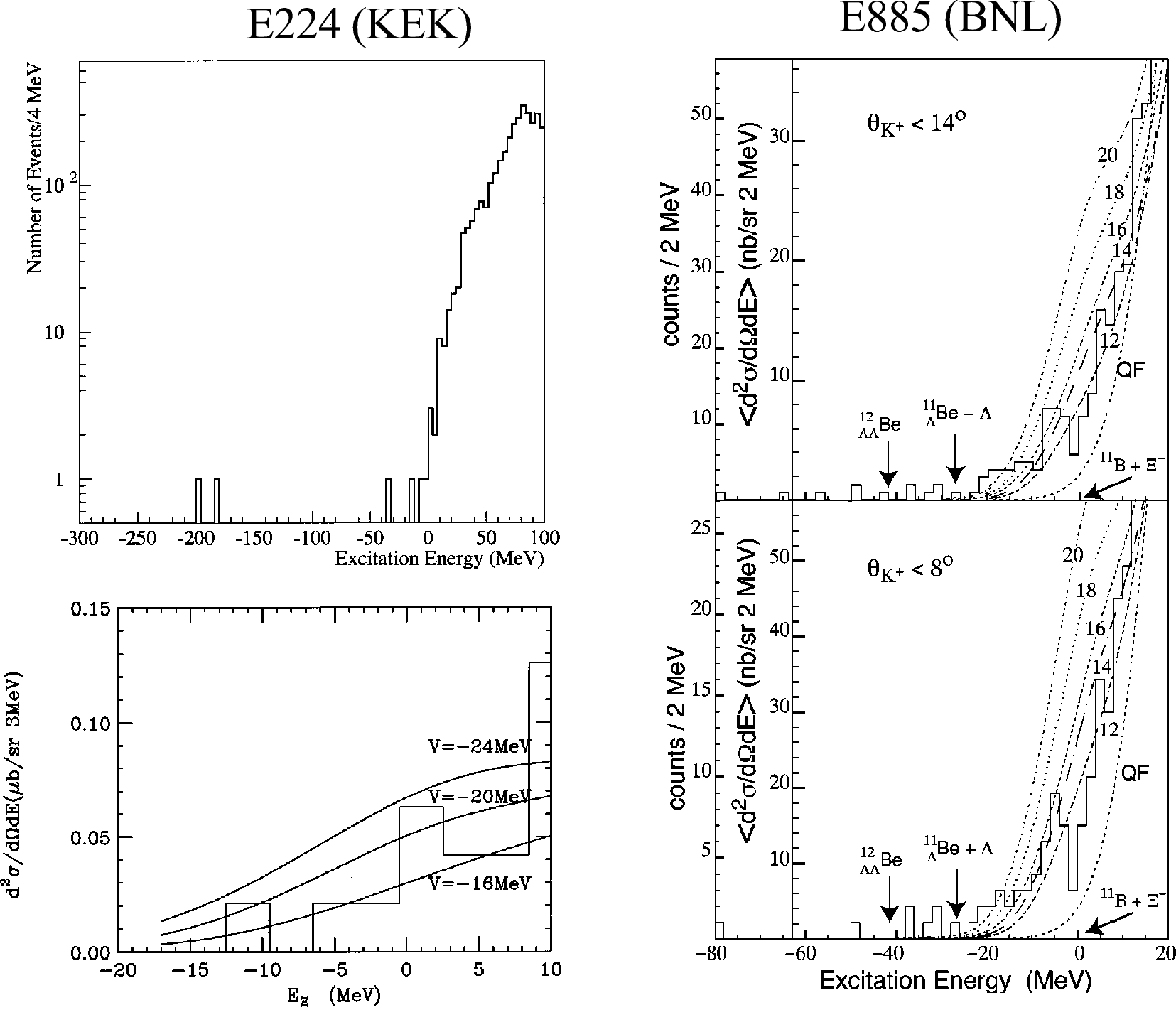} 
\caption{$^{12}{\rm C}(K^-,K^+)$ missing-mass spectra measured in KEK E224 
\cite{fukuda98} (left) and BNL E885~\cite{khaustov00} (right).
The curves correspond to assumptions made on the strength of an attractive 
$\Xi^-$-nucleus potential, folded with the experimental resolution. 
From Nagae, 2007.} 
\label{fig:Xi} 
\end{figure*} 

 The data in Fig.~\ref{fig:spe} are quite well fit by a simple Woods-Saxon
potential. However, when replacing $V_{\rm WS}$ by the 
low-density limit form ${\tilde V}_0\rho_{N}(r)$, with $\rho_{N}$ the nuclear 
density, the fit to the data requires adding a repulsive potential with 
a higher power of $\rho_{N}$ and, obviously, a depth ${\tilde V}_0$ of the 
attractive potential much larger than $V_{\rm WS}$ (Millener, Dover, and Gal,
(1988). The resulting density-dependent $\Lambda$-nucleus potential can be 
traced back within a Skyrme-Hartree-Fock approach to a combination of two-body 
attractive $\Lambda N$ and a three-body repulsive $\Lambda NN$ interaction 
terms. Similar conclusions were also reached by Yamamoto, Band\={o}, and 
\v{Z}ofka (1988). These early papers were based on a \piK\ experiment performed 
at BNL in 1987~\cite{pile91}. Since that time, there have been a large number 
of both non-relativistic and relativistic mean-field calculations that 
reproduce the $\Lambda$ single-particle energies (Mare\v{s} and Jennings,
1994; Cugnon, Lejeune, and Schulze, 2000; Keil, Hoffmann, and Lenske, 2000; 
Vida\~{n}a \textit{et al.}, 2001; Finelli \textit{et al.}, 2009). 
The smooth behavior of the $B_\Lambda$ values is such that it should be 
possible to fit the updated data set in Table~\ref{tab:spe} very well in almost
any model with small adjustments in the parameters. In addition, the
single-particle energies have been fitted using a strongly cancelling
combination of attractive $\Lambda N$ and repulsive $\Lambda NN$ interactions 
(Usmani and Bodmer, 1999; Lonardoni, Pederiva, and Gandolfi, 2014). These 
results are in some tension with the results of recent Nijmegen $YN$ models
(Yamamoto and Rijken, 2013; Nagels, Rijken, and Yamamoto, 2015b)
where G-matrix folding models based just on the $YN$ interaction fit
the $\Lambda$ single-particle energies quite well.

\subsubsection{($K^-,K^+$) and stopped $\Xi^-$ reactions} 
\label{subsubsec:xistop} 

The two-body reaction $K^-p \to K^+\Xi ^-$ is the primary method 
used to produce double strangeness in nuclei. The forward-angle 
cross section of this reaction peaks for incident $K^-$ momentum around 
$p_{\rm lab} = 1.8$~GeV/c, with a value close to $50~\mu{\rm{b/sr}}$. 
The usefulness of the \textit{nuclear} ($K^-,K^+$) reaction in producing $\Xi$ 
hypernuclei was discussed by Dover and Gal (1983). Missing-mass 
spectra on $^{12}$C from experiments done at KEK \cite{fukuda98} and at BNL 
\cite{khaustov00} are shown in Fig.~\ref{fig:Xi}. A full spectrum over a wide 
$\Xi^-$ excitation range is shown in the upper-left diagram, and insets 
centered around the $\Xi^-$ threshold are shown in the rest of the diagrams. 
No conclusive experimental evidence for well defined $\Xi$ hypernuclear 
levels could be determined because of the limited statistics and detector 
resolution of $\approx$10 MeV. However, by fitting to the shape and 
cross-section yield of the spectra in the $\Xi$-hypernuclear region, 
an upper bound of approximately 15 MeV attraction was placed on the $\Xi$ 
hypernuclear potential strength, as shown in the figure by various calculated 
curves. The formation of $\Lambda\Lambda$ hypernuclei via a direct ($K^-,K^+$) 
reaction without intermediate $\Xi$ production is less favorable, requiring 
two steps, each on a different proton, e.g., $K^-p\to\pi^0\Lambda$ followed by 
$\pi^0p \to K^+\Lambda$ (Baltz, Dover, and Millener, 1983). The expected 
position of the $_{\Lambda \Lambda}^{~12}$Be ground state is marked by arrows for 
the BNL E885 experiment. Given the limited statistics, no firm evidence for the 
production of $_{\Lambda \Lambda}^{~12}$Be states was claimed. 

A different class of experiments is provided by stopping $\Xi ^-$ hyperons 
in matter, giving rise to two $\Lambda$'s via the two-body reaction 
$\Xi^-p \to \Lambda\Lambda$ which releases only 23 MeV. Double-$\Lambda$ 
hypernuclei may then be formed in stopped $\Xi ^-$ reactions in a nuclear 
target, after the $\Xi ^-$ hyperons are brought to rest from a ($K^-,K^+$) 
reaction \cite{zhu91}. Calculations by Yamamoto \textit{et al.}, mostly using 
double-$\Lambda$ compound nucleus methodology, provide relative formation 
rates for $\Lambda\Lambda$ hypernuclei (Sano, Wakai, and Yamamoto, 1992;
Yamamoto\textit{et al.}, 1992, 1997; Yamamoto, Sano, and Wakai, 1994).

Dedicated experiments with stopped $\Xi ^-$ hyperons were proposed in 
order to produce some of the lightest $\Lambda \Lambda$ hypernuclei, 
$_{\Lambda \Lambda}^{~~6}$He \cite{zhu91}, $_{\Lambda \Lambda}^{~~4}$H (Kumagai-fuse, 
Koike, and Akaishi, 1995), and $_{\Lambda \Lambda}^{~12}$B \cite{yamada97}, by 
searching for a peak in the outgoing neutron spectrum in the two-body reaction 
\begin{equation} 
\label{eq:lamlam} 
\Xi ^- ~+~ ^AZ ~ \longrightarrow ~ _{\Lambda \Lambda}^{~A}(Z-1) ~+~ n \,. 
\end{equation} 
These proposals motivated the AGS experiment E885~\cite{khaustov00a} which 
used a diamond target ($^{\rm nat}$C) to stop the relatively fast $\Xi ^-$ 
hyperons recoiling from the quasi-free peak of the $p(K^-, K^+)\Xi ^-$ 
reaction in the diamond target. Non-negligible decay losses occur during the 
stopping time of the $\Xi^-$ hyperon, so that a dense target was used to 
produce, stop, and capture the $\Xi^-$ hyperons. An upper bound of a few 
percent was established for the production of the $_{\Lambda\Lambda}^{~12}$Be 
hypernucleus. Experimental evidence for $_{\Lambda \Lambda}^{~~6}$He 
\cite{takahashi01} and $_{\Lambda \Lambda}^{~~4}$H \cite{ahn01a} had to await 
different techniques, although the evidence for the latter species remains 
controversial \cite{randeniya07}. 

The stopped $\Xi ^-$ reaction in deuterium $(\Xi ^- d)_{\rm atom}\rightarrow Hn$ 
was used in AGS experiment E813 to search for the doubly strange $H$ 
dibaryon, yielding a negative result \cite{merrill01}. An earlier search by 
the KEK E224 collaboration, stopping $\Xi ^-$ on a scintillating-fiber active 
carbon target, also yielded a negative result \cite{ahn96}. The ($K^-,K^+$) 
reaction was also used, on a $^3$He target, to establish a stringent upper 
limit on $H$-dibaryon production \cite{stotzer97}. Theoretically, based
on recent lattice QCD calculations by two different groups,
NPLQCD~\cite{beane11} and HALQCD~\cite{inoue11}, and on extrapolation
made to the SU(3)-broken hadronic world (Haidenbauer and Mei{\ss}ner, 2012;
Inoue \textit{et al.}, 2012). the H dibaryon is unbound with respect to the 
$\Lambda\Lambda$ threshold, perhaps surviving in some form near the $\Xi N$ 
threshold.

On the positive side, a double-$\Lambda$ hypernucleus was discovered in light 
emulsion nuclei by the KEK stopped $\Xi^-$ experiment E176 \cite{aoki91} 
and was subsequently interpreted as a $_{\Lambda\Lambda}^{~13}$B hypernucleus 
(Dover \textit{et al.}, 1991; Yamamoto, Takaki, and Ikeda, 1991). This 
experiment produced several events, each showing a decay into a pair of known 
single-$\Lambda$ hypernuclei~\cite{aoki93,aoki95}. Two more events were 
reported by the KEK E373 collaboration \cite{ichikawa01,nakazawa15}, with the 
latter event claimed to imply a lightly bound ${\Xi}^{-}$-${^{14}{\rm N}}$ 
nuclear state. Using these events, one should be able to deduce the properties 
of the initial $\Xi^-$ \textit{atomic} states. However, the 100 keV resolution 
common in emulsion work is 3 orders of magnitude larger than typical 
values anticipated for the strong-interaction shifts and widths of $\Xi^-$ 
atomic levels. This provides a major justification for pursuing a program 
for the measurement of $\Xi^-$ X rays (Batty, Friedman, and Gal, 1999), in 
parallel with strong-interaction reactions involving $\Xi$ hyperons. 

\subsubsection{Hypernuclear lifetime measurements} 
\label{subsubsec:lifetime} 

If the velocity of a hypernucleus recoiling from a production reaction 
is known, its lifetime can be measured by the distance it travels before 
decaying. This recoil-distance technique was used to observe and measure 
the lifetime of many short lived particles. In particular the lifetime of 
a free, unbound $\Lambda$, ($263 \pm 2$)~ps \cite{pdg14}, was determined 
by observing its mesonic decay in a beam of neutrally charged 
hyperons (Poulard, Givernaud, and Borg, 1973; Clayton \textit{et al.}, 1975;
Zech \textit{et al.}, 1977).

Lifetimes of \lamb{3}{H}, \lamb{4}{H} and \lamb{5}{He} measured in emulsion 
were published as early as in 1964 \cite{prem64}, but since hypernuclei are 
generally produced in emulsion with low kinetic energies, only very few 
decayed in flight, incurring relatively large experimental uncertainties 
on the deduced lifetimes. The more precise \lamb{3}{H} lifetime deduced in 
a subsequent emulsion measurement, $\tau$(\lamb{3}{H})=128$^{+35}_{-26}$~ps 
\cite{bohm70a}, is considerably shorter than the one deduced from a helium 
bubble-chamber measurement, $\tau$(\lamb{3}{H})=246$^{+62}_{-41}$~ps 
\cite{keyes73}. The latter is equal to the free $\Lambda$ lifetime within the 
experimental uncertainties. This was explained by Bohm and Wysotski (1970) as 
the possible Coulomb dissociation of the very weakly bound \lamb{3}{H} when 
traversing the high-$Z$ emulsion. Finally, the \lamb{5}{He} lifetime deduced in
that emulsion study \cite{bohm70b} agrees perfectly within its larger 
uncertainties with the lifetime deduced 35 years later in a KEK experiment in 
which \lamb{5}{He} was produced in a \piK\ reaction \cite{kameoka05}. This 
and other lifetimes measured similarly at KEK are listed in 
Table~\ref{tab:lifetimes}, with \lamb{}{Fe} the heaviest $\Lambda$ hypernucleus
for which this information is available. It is clear from the table that 
beginning with \lamb{12}{C} the $\Lambda$ hypernuclear lifetimes saturate at a 
value about 80\% of the free $\Lambda$ lifetime. 

\begin{table} 
\caption{$\Lambda$ hypernuclear lifetimes (in ps) measured at KEK, using 
\piK\ production reactions.} 
\label{tab:lifetimes} 
\begin{ruledtabular}
\begin{tabular}{lcccc} 
$\Lambda$ & $^5_\Lambda$He & $^{12}_{~\Lambda}$C & $^{28}_{~\Lambda}$Si & 
$_\Lambda$Fe \\ \hline 
$263 \pm 2$\footnotemark [1] & $278 \pm 11$\footnotemark [2] & 
$212 \pm 7$\footnotemark [2] & $206 \pm 11$\footnotemark [3] & 
$215 \pm 14$\footnotemark [3] \\ 
\end{tabular}
\end{ruledtabular}
\footnotetext [1] {Olive \textit{et al.} (2014).}
\footnotetext [2] {Kameoka \textit{et al.} (2005).}
\footnotetext [3] {Bhang \textit{et al.} (1998) and Park \textit{et al.} 
(2000).}
\end{table}

The first accelerator experiment to apply the recoil-distance method in 
a hypernuclear experiment used the LBL Bevatron to produce a hypernuclear 
beam by bombarding a polyethylene target with a 2.1 GeV/nucleon $^{16}$O 
beam \cite{nield76}. Spark chamber detectors with photographic readout were 
positioned behind the target and scanned for tracks with a decay vertex. 
The readout trigger required that an interaction occurred in the target and 
a potential decay was observed within a given time delay. These events were 
analyzed by a fit to the form $N(x)=A\exp{(-x/\lambda)}+B$ by varying $A$, 
$B$ and $\lambda$, where $B$ is a constant background, $\lambda$ the mean 
lifetime of the hypernucleus, and $x$ the mesured distance between the vertex 
and the target. Although the actual system which decayed was not directly 
identified, the most likely hypernuclear production reactions were assumed 
to be 
\begin{equation} 
^{16}{\rm O} + p \, \to \, {^{16}_{~\Lambda}{\rm O}} + n + K^{+} \,, 
\label{eq:LBL1} 
\end{equation} 
\begin{equation} 
^{16}{\rm O} + n \, \to \, {^{16}_{~\Lambda}{\rm N}} + n + K^{+} \,.  
\label{eq:LBL2} 
\end{equation} 
The measured mean life was found to be $86^{+33}_{-26}$~ps, which is 
two to three times shorter than lifetimes measured in this hypernuclear 
mass range in more recent, better controlled \piK\ experiments at 
KEK \cite{bhang98,park00}, as demonstrated in Table~\ref{tab:lifetimes}. 

\begin{figure}[t] 
\includegraphics[width=8.0cm]{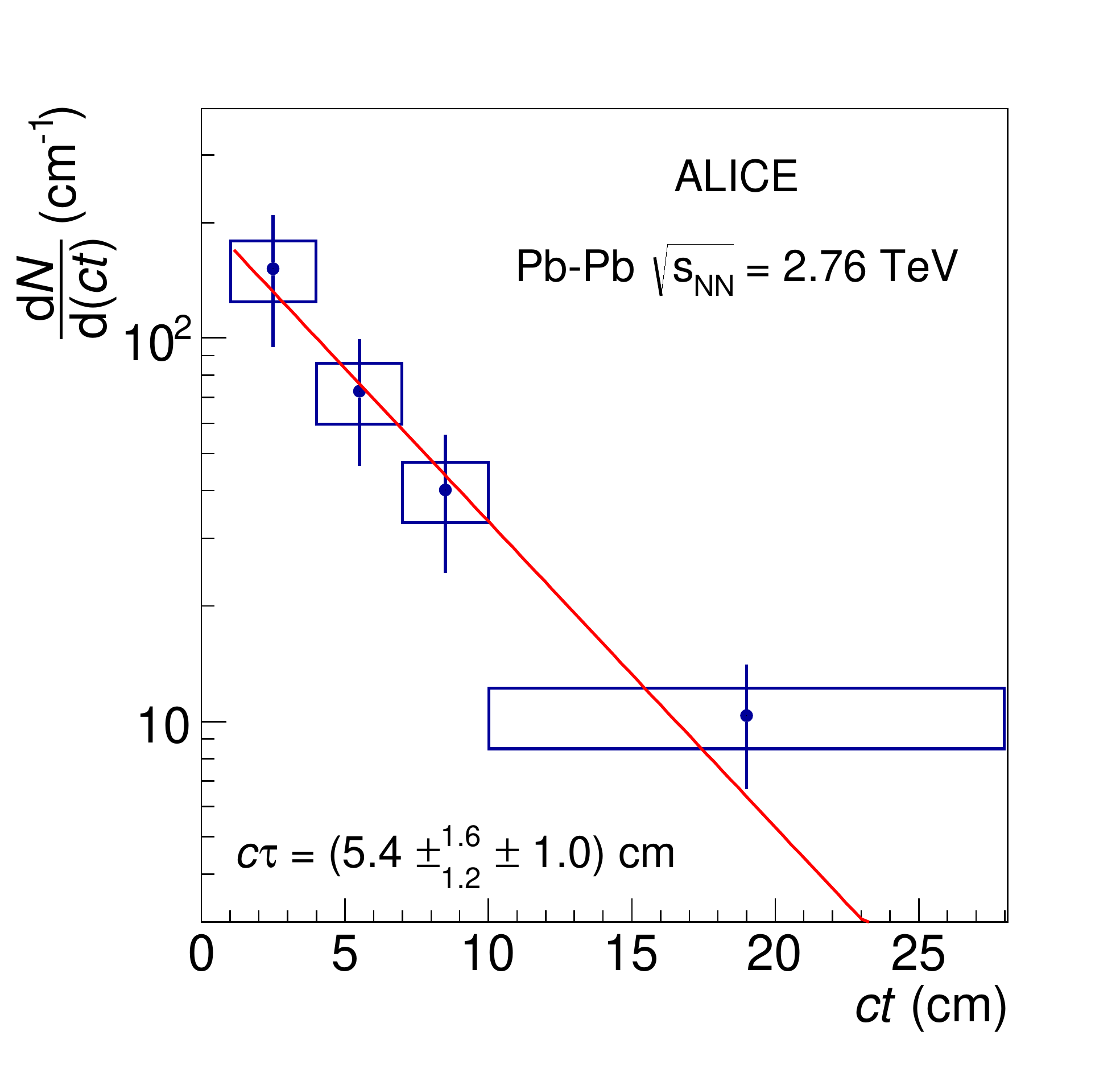} 
\caption{Measured d$N$/d($ct$) distribution and exponential fit used by 
the ALICE Collaboration to determine the lifetime of \lamb{3}{H} produced 
in Pb--Pb central collisions at $\sqrt{s_{NN}}=2.76$~TeV at the CERN-LHC. 
The bars and boxes are statistical and systematic uncertainties, respectively. 
Adapted from Adam \textit{et al.}, 2016a.}
\label{fig:alice} 
\end{figure}

More recently, the HypHI Collaboration at GSI reported lifetimes of 
\lamb{3}{H} and \lamb{4}{H} produced by bombarding a carbon target with a 
2~GeV/nucleon $^6$Li beam \cite{rappold13a}. The lifetime of \lamb{3}{H} has 
also been measured in heavy-ion central collisions, by the STAR Collaboration 
at the BNL-RHIC collider \cite{abelev10} and by the ALICE Collaboration at 
CERN-LHC \cite{adam16a}. These measurements use the time dilation of a Lorentz 
boost to the recoiling hypernucleus produced in the collision, as shown in 
Fig.~\ref{fig:alice} from the ALICE determination of $\tau$(\lamb{3}{H}).  
The values deduced from these measurements for the \lamb{3}{H} lifetime are 
about 25\% shorter than the free $\Lambda$ lifetime, see the latest 
compilation by Rappold \textit{et al.} (2014). This poses a serious 
theoretical challenge as discussed later in Sec.~\ref{subsubsec:lifetimes2}. 

Several programs have attempted to obtain the lifetime of heavy hypernuclei 
using the recoil-distance method for delayed fission after stopping antiprotons 
on Bi and U targets (Bocquet \textit{et al.}, 1987; Armstrong \textit{et al.}, 
1993) or by electroproduction on a Bi target \cite{noga86}. These use 
back-to-back fission fragments from the presumed decay of a recoiling 
hypernucleus to obtain the position of the decay relative to the target. 
As previously, the recoil velocity and decay position provide the 
hypernuclear lifetime. 

As an example, this technique was used by the COSY-13 Collaboration to 
obtain the lifetime of hypernuclei averaged over hypernuclear masses from 
$A$=160--190, 170--200, and 200--230. The data were obtained from the fission 
of nuclear systems recoiling from an approximately 1.9~GeV proton beam 
incident on Au, Bi, and U targets, respectively (Pysz \textit{et al.}, 1999;
Cassing \textit{et al.}, 2003).
Obviously the specific recoiling system was unknown, so the masses and momenta 
of the recoils were obtained from coupled-channel transport and statistical 
evaporation models. In both the COSY-13 and $\bar{p}$ experiments, fragments 
and particles emitted directly from the target were blocked from entering 
the amplitude-sensitive fission detectors -- the recoil shadow method. The 
result of the COSY-13 experiment was a lifetime of ($145\pm 11$)~ps. This is 
significantly shorter than the lifetime expected by extrapolating the measured 
lifetimes listed in Table~\ref{tab:lifetimes} which indicate that saturation 
of hypernuclear lifetimes is achieved already for $A\geq 12$. Cassing
\textit{et al.} (2003) argued that the result shows significant violation of 
the $\Delta I = 1/2$ rule. However, Bauer and Garabarino (2010) pointed out 
that no known mechanism could account for this significant decrease in 
the lifetime compared to ($215\pm 14$)~ps measured for $_{\Lambda}$Fe 
\cite{bhang98,park00,sato05}. Therefore, additional, more constrained 
measurements are needed to resolve this controversy. 

\subsection{Free-space and in-medium $YN$ interactions}
\label{subsec:intmodels}

\subsubsection{The free-space $YN$ interaction}
\label{subsubsec:freespace} 

One of the motivations for the study of hypernuclei was the expectation that 
information on the low-energy $\Lambda N$ interaction could be extracted from 
the spins and binding energies of the $s$-shell hypernuclear systems. Direct 
scattering and reaction measurements involving $\Lambda$'s are extremely 
difficult, since the $\Lambda$ is electrically neutral and its lifetime is 
short, $\approx$263~ps. Thus, production and scattering must be done in the 
same target, and the detector must have sufficient granularity and particle 
identification to analyze scattering events in the presence of a number of 
possible backgrounds. The data that do exist comes mostly from hydrogen 
bubble chambers, and was acquired with a stopping $K^-$ beam. Hence, the 
data analysis must extract the kinematics and rates from tracks in the 
bubble-chamber target as the $\Lambda$ recoils from the $p(K^-_{\rm stop},\pi^0)
\Lambda$ reaction and then scatters from another hydrogen nucleus 
\cite{alexander68,sechizorn68}.    

There are also a few data points for $\Sigma p$ scattering and reactions 
\cite{eisele71} taken using hydrogen bubble chambers. However, a more recent 
technique used a scintillating-fiber target \cite{ahn99}, applying the 
\piK\ reaction to produce and scatter $\Sigma^+$'s in the 
scintillating fiber. This technique tracks the charged $\Sigma^+$'s to, 
and after, their interactions with protons in the fibers by observing 
electronically stored, stereo images of reaction events. The readout 
is triggered by a $(\pi,K)$ spectrometer system that identifies the possible 
production of $\Sigma^+$ recoils that could have re-scattered \cite{ahn99}. 
One might envision using a similar apparatus to obtain $\Lambda N$ scattering 
data, but inferring the energy and scattering angle of a neutral $\Lambda$ 
is not feasible.

Excluding the latest $\Sigma p$ data, there are some 37 $YN$ (hyperon-nucleon) 
data points. Obviously this is insufficient to extract even the scattering 
lengths, so these data are analyzed using models of SU(3)$_{\rm f}$ 
symmetry of the baryon-baryon interaction that make connections with the 
richer $NN$ data. However, SU(3)$_{\rm f}$ is badly broken due to the 
difference in mass between the \textit{s} and (\textit{u},\textit{d}) quarks, 
so that realistic models must include SU(3)$_{\rm f}$ breaking terms. Several 
$YN$ potential models have been developed along these lines for use in 
hypernuclear physics. The most used ones are as follows: 
\begin{itemize} 

\item The Nijmegen models, including the hard-core models D (Nagels, Rijken,
and de Swart, 1977) and F (Nagels, Rijken, and de Swart, 1979), the soft-core 
models NSC89 (Maessen, Rijken, and de Swart, 1989) and NSC97 
(Rijken, Stoks, and Yamamoto, 1999), and the extended soft core models ESC04 
\cite{rijken06a} and ESC08 (Nagels, Rijken, and Yamamoto, 2015b) that, in 
addition to one boson exchange (OBE), also consider pseudoscalar (PS) 
two-meson exchanges and other short-range contributions. These models in 
particular allow extension to hyperon-hyperon ($YY$) potentials where there 
is almost no scattering data (Nagels, Rijken, and Yamamoto, 2015a)
implying unfortunately an increased model dependence. For applications to 
$S=-2$ hypernuclei, see Yamamoto and Rijken (2008). 

\item The Bonn-J{\"{u}}lich multi-meson-exchange models 
(Holzenkamp, Holinde, and Speth, 1989; Reuber, Holinde, and Speth, 1994;
Haidenbauer and Mei{\ss}ner, 2005) that are based on the SU(6) 
symmetry of the quark model. The short-range behavior of the $YN$ interaction 
in these and in the Nijmegen models follows largely from the way scalar-meson 
interactions are introduced, and is therefore necessarily model dependent. 

\item Effective Field Theory (EFT) chiral models, of leading order (LO) 
(Polinder, Haidenbauer and Mei{\ss}ner, 2006) and next to leading order 
(NLO) (Haidenbauer \textit{et al.}, 2013), 
that use regularized PS Goldstone-boson exchange $YN$ potentials, adding
zero-range contact terms to parametrize the short-range behavior of the 
$YN$ coupled-channel interactions. For a recent review see
Haidenbauer (2013).

\end{itemize} 
In addition, a quark-model baryon-baryon potential obeying SU(6) symmetry 
was developed by Fujiwara, Suzuki, and Nakamoto (2007) and used for 
constructing hyperon-nucleus potentials \cite{kohno09}. 

\begin{table*}[t] 
\caption{$\Lambda N$ scattering lengths and effective ranges (in fm) for 
several $YN$ interaction models. For the EFT models, these refer to 
$\Lambda p$ and to cutoff parameter of 600 MeV. \label{tab:LN}}
\begin{ruledtabular}
\begin{tabular}{llcccc}
Model & \multicolumn{1}{c}{Reference} & $a^{s}$ & $r_{0}^{s}$ & 
$a^{t}$ & $r_{0}^{t}$ \\
\hline     
NSC89  & Maessen, Rijken, and de Swart (1989) & -2.79 & 2.89 & -1.36 & 3.18 \\
NSC97e & Rijken, Stoks, and Yamamoto (1999) & -2.17 & 3.22 & -1.84 & 3.17  \\
NSC97f & Rijken, Stoks, and Yamamoto (1999) & -2.60 & 3.05 & -1.71 & 3.33 \\
ESC08c & Nagels, Rijken, and Yamamoto (2015b) & -2.54 & 3.15 & -1.72 & 3.52 \\
J\"{u}lich '04 & Haidenbauer and Mei{\ss}ner (2005) & -2.56 & 2.75 & -1.66 
 & 2.93 \\
EFT (LO) &  Polinder, Haidenbauer, and Mei{\ss}ner (2006) & -1.91 & 1.40 
 & -1.23 & 2.20 \\ 
EFT (NLO) & Haidenbauer \textit{et al.} (2013)  & -2.91 & 2.78 & -1.54 
& 2.72 \\ 
\end{tabular}
\end{ruledtabular}
\end{table*} 

Table~\ref{tab:LN} compares the $\Lambda N$ singlet and triplet scattering
lengths and effective ranges for several models, showing that the $YN$ 
low-energy data cannot determine precisely these low-energy parameters. 
Judging by the $\Lambda N$ scattering lengths, the $\Lambda N$ interaction is 
attractive but is weaker roughly by a factor of 2 than the $NN$ interaction. 
This is consistent with the absence of $\Lambda N$ bound states and with 
the onset of $\Lambda$-hypernuclear binding realized by the 
weakly-bound hypertriton \lamb{3}{H} ($B_{\Lambda}=0.13\pm 0.05$~MeV, see 
Table~\ref{tab:blambda}). The spin dependence of the $\Lambda N$ interaction 
is opposite to that of the $NN$ interaction, with the spin-singlet s-wave 
$\Lambda N$ interaction being stronger than the spin-triplet interaction, 
consistent with the known spin-parity, $J^{\pi}={\frac{1}{2}}^+$, of 
\lamb{3}{H}.

\subsubsection{Extraction of $\Lambda N$ interaction in final-state 
interactions} 
\label{subsubsec:finalstate} 

Extraction of the $NN$ scattering lengths and effective ranges from scattering 
of nucleons in a continuum final state has been throughly explored. The 
technique has been used to compare neutron-neutron to proton-neutron and 
proton-proton scattering in order to obtain charge symmetry breaking 
information (Gross, Hungerford, and Malanify, 1971). The experiments analyze 
the spectrum of a three-body breakup reaction in the region of phase space 
where two final-state nucleons have low relative energy. They require 
excellent energy resolution, but only relative cross sections.

Extension of this technique to obtain the $YN$ scattering lengths 
and effective ranges has also been proposed (Karplus and Rodberg, 1959;
Gibbs \textit{et al.}, 2000). 
Experimentally, one must have an energy resolution $\ll$1~MeV near 
the turning point in phase space where the reduced energy of the 
hyperon and nucleon vanishes. This is not presently possible in mesonic 
production reactions, and while sub-MeV resolution of hypernuclear 
spectra may be obtained in electromagnetic production, quasi-free $\Sigma$ 
production is high, and unfavorable kinematic conditions due to the light 
mass of the recoiling $\Lambda N$ system significantly reduce the resolution. 
The sensitivity of the spectrum shape to the effective range \cite{dohrmann07} 
is thus degraded.

\subsubsection{Comparison of the $\Lambda N$ and $\Sigma N$ interactions}
\label{subsubsec:LNvsSN} 

\begin{figure*}[t]
\centerline{\includegraphics[width=17.0cm]{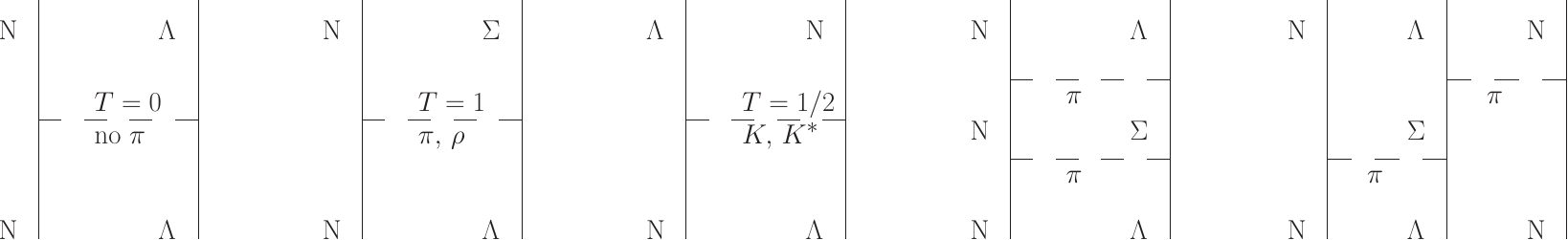}}
\caption{Diagrams showing schematically the important features of the 
coupled $\Lambda N\!-\!\Sigma N$ strangeness $-1$ interaction for 
isospin 1/2. All diagrams are $YN$ Born diagrams. The first diagram represents
generically meson exchanges such as $\eta$ and $\omega$, and the fourth diagram
iterates the one-pion exchange of the second diagram and is included in
the standard coupled-channels approach to the $YN$ interaction. The last 
diagram shows a two-pion-exchange three-body interaction. \label{fig:yn}}
\end{figure*}

The $\Lambda$ has isospin 0, so the $\Lambda N$ interaction occurs only in 
the isospin state $I_{YN}=1/2$ only. On the other hand, the $\Sigma$ hyperon 
has isospin 1, allowing $\Sigma N$ interaction in both isospin states $I_{YN}
=1/2$ and $3/2$. Although $YN$ input data are limited, the Nijmegen, and the
 EFT 
potentials in particular, favor significant $\Sigma N$ spin and isospin 
dependence, yielding strong attraction in the $^{1}S_{0}$, $T\!=\!3/2$ 
and $^{3}S_{1}$, $T\!=\!1/2$ channels and repulsion in the $^{3}S_{1}$, 
$T\!=\!3/2$ and $^{1}S_{0}$, $T\!=\!1/2$ channels. This is rather different 
from the relatively weak spin dependence of the attractive $\Lambda N$ 
interaction in these models. Perhaps the most significant difference between 
the $\Lambda N$ and $\Sigma N$ interactions is the strong conversion $\Sigma 
N\,\to\,\Lambda N$ with energy release of some 80 MeV. This dominates the 
behavior of a $\Sigma$ in the nuclear medium (Dover, Millener, and Gal, 1989), 
implying also appreciable $\Lambda N\,\leftrightarrow\,\Sigma N$ mixing, 
particularly in the $s$-shell hypernuclei, as discussed in 
Sec.~\ref{subsubsec:sshell}. 

\subsubsection{The effective $YN$ interaction}
\label{subsubsec:yneff}

The hyperon-nucleon interaction involves the coupled $\Lambda$N and $\Sigma$N 
channels, as illustrated in Fig.~\ref{fig:yn}. The diagrams in the figure 
make the point that the direct $\Lambda N\!-\!\Lambda N$ interaction does 
not contain a one-pion-exchange contribution because of isospin conservation 
(except for electromagnetic violations via $\Lambda\!-\!\Sigma^0$ mixing) 
while the coupling between the $\Lambda N$ and $\Sigma N$ channels does. 
For this reason alone, the $\Lambda N$ interaction is considerably weaker 
than the $NN$ interaction, and there is reason to believe that the three-body 
$\Lambda NN$ interaction in a hypernucleus could be relatively important.

The free-space interactions are obtained as extensions of meson-exchange 
models for the $NN$ interaction by invoking, e.g., a broken flavor, 
SU(3)$_{\rm f}$, symmetry. The most widely used model is the Nijmegen 
soft-core, one-boson-exchange potential model known as NSC97
(Rijken, Stoks, and Yamamoto, 1999). 
The six versions of this model, labeled NSC97a--f, cover a wide range 
of possibilities for the strength of the central spin-spin interaction 
ranging from a triplet interaction that is stronger than the singlet 
interaction to the opposite situation. More recently, extended soft-core 
versions, ESC04 \cite{rijken06a} and ESC08 (Nagels, Rijken, and
Yamamoto, 2015b), have become available. Effective interactions for use 
in a nuclear medium are then derived through a G-matrix procedure
(Rijken, Stoks, and Yamamoto, 1999; Rijken and Yamamoto, 2006a;
Yamamoto, Motoba, and Rijken, 2010).

The $\Lambda N$ effective interaction can be written (neglecting a quadratic 
spin-orbit component) in the form 
\begin{eqnarray} 
V_{\Lambda N}(r) & =& V_0(r) +V_{\sigma}(r)~ \bm{s}_N\cdot
 \bm{s}_\Lambda +  V_\Lambda (r)~\bm{l}_{N\Lambda}\cdot
\bm{s}_\Lambda  \nonumber\\ 
& & \mbox{} + V_N (r)~\bm{l}_{N \Lambda}\cdot\bm{s}_N +  V_T (r)~S_{12}\; , 
\label{eq:vlam} 
\end{eqnarray} 
where $V_0$ is the spin-averaged central interaction, $V_\sigma$ is the 
difference between the triplet and singlet central interactions, $V_\Lambda$ 
and $V_N$ are the sum and difference of the strengths of the symmetric 
spin-orbit (SLS) interaction 
$\bm{l}_{N\Lambda}\cdot(\bm{s}_{\Lambda} +\bm{s}_N$) 
and antisymmetric spin-orbit (ALS) interaction 
$\bm{l}_{N\Lambda}\cdot(\bm{s}_{\Lambda} -\bm{s}_{N}$), 
and $V_T$ is the tensor interaction with 
\begin{equation} 
S_{12}= 3(\bm{\sigma}_{N}\cdot\hat{\bm{r}})(\bm{\sigma}_{\Lambda}
\cdot\hat{\bm{r}})-\bm{\sigma}_{N}\cdot\bm{\sigma}_{\Lambda}\; . 
\label{eq:tensor} 
\end{equation} 
For the $\Lambda$ in an \textit{s} orbit, $\bm{l}_{N\Lambda}$ is proportional 
to $\bm{l}_{N}$ (Gal, Soper, and Dalitz, 1971). The effective 
$\Lambda N\!-\!\Sigma N$ and $\Sigma N\!-\!\Sigma N$ interactions can be 
written in the same way.

 Effective interactions in common use are the hyperon-nucleon Gaussian (YNG) 
interactions (Yamamoto \textit{et al.}, 1994; Yamamoto, Motoba, and Rijken,
2010) in which each term is represented by an expansion in terms of a 
limited number of Gaussians with different ranges,
\begin{equation}
 V(r) = \sum_i v_i\, e^{-r^2/\beta_i^2} 
\label{eq:gauss} 
\end{equation} 
for the central and spin-orbit components, and 
\begin{equation}
V_T(r) = \sum_i v_i\, r^2\, e^{-r^2/\beta_i^2} 
\label{eq:r2gauss} 
\end{equation} 
for the tensor component. When based on nuclear-matter calculations, the YNG 
matrix elements are made density dependent by parametrizing the coefficients 
$v_i$ through the Fermi momentum $k_F$. 

 Effective interactions for finite nuclei, specifically for \textit{p}-shell 
hypernuclei, have been generated using a Brueckner-Hartree procedure
\cite{halderson08}. These use Yukawa forms in place of the Gaussians above, 
are density-independent, and are available for most of the Nijmegen 
interactions (D. Halderson, private communication). The above Gaussian or 
Yukawa interactions provide a starting point for the interactions that 
give rise to the parameter sets in Eqs.~(\ref{eq:param7})-(\ref{eq:paramls})
describing the energy spectra of $p$-shell hypernuclei. This process is
illustrated in Millener (2010), which also contains some
remarks about the possible role of the double one-pion exchange $\Lambda NN$
interaction (see Fig.~\ref{fig:yn}) introduced long ago for $p$-shell
hypernuclei by Gal, Soper, and Dalitz (1971). Phenomenological, but physically 
motivated, $\Lambda NN$ interactions have been used for the $s$-shell 
hypernuclei and the $\Lambda$ well depth (Bodmer, Usmani, and Carlson, 1984a;
Bodmer and Usmani, 1988). These studies were later extended to a full study of 
$\Lambda$ single-particle energies~\cite{usmani99}, most recently using 
auxilliary-field diffusion Monte Carlo techniques (Lonardoni, Pederiva, and
Gandolfi, 2014). While microscopically
derived $YNN$ interactions have not been available for use in few-body 
calculations~\cite{nogga13}, such interactions have been recently
derived from SU(3) chiral effective field theory~\cite{petschauer16a}
but not yet applied.  

\section{$\Lambda$ Hypernuclei}
\label{sec:lamhyp}

\subsection{Structure calculations}

\subsubsection{$s$-shell hypernuclei}
\label{subsubsec:sshell}

The $s$-shell hypernuclei illustrate many of the features of the $\Lambda N$ 
interaction (Dalitz, Herndon, and Tang, 1972; Nemura, Akaishi, and Suzuki, 
2002; Nogga, Kamada, and Gl\"{o}ckle, 2002). The binding energy 
of the lightest hypernucleus, the hypertriton \lamb{3}{H}, was obtained 
from emulsion (Bohm \textit{et al.}, 1968; Juri\v{c} \textit{et al.}, 1973; 
Davis and Pniewski, 1986). Its spin and parity, 
$J^\pi\!=\!1/2^+$ \cite{dalitz69,keyes70}, was found by analysis of 
its $\pi^-$ weak-decay width~\cite{dalitz58,dalitz59a}. As a consequence, 
one can deduce that the spin-singlet, as opposed to the spin-triplet, 
$\Lambda N$ interaction must be stronger. In addition, as the binding 
energy~\cite{davis86} is only $0.13\pm 0.05$ MeV, there is no bound, excited 
$T=0$ hypertriton state. A bound $T=1$ $\Lambda nn$ was speculated recently 
by the HypHI Collaboration at GSI \cite{rappold13}. However,  $A=3$ 
few-body calculations constrained by the $T=0$ hypertriton (Miyagawa 
\textit{et al.}, 1995; Belyaev, Rakityansky, and Sandhas, 2008; Gal and 
Garcilazo, 2014; Garcilazo and Valcarce, 2014; Hiyama \textit{et al.}, 2014), 
and in some also by the $A=4$ hypernuclei~\cite{gal14a,hiyama14}, rule out a 
bound $\Lambda nn$. The spin-parity and binding energy of the hypertriton 
provide important constraints on the spin components of the central 
$\Lambda N$ potential because of the lack of direct data from low-energy 
$\Lambda N$ scattering  (Downs and Dalitz, 1959; Dalitz, 1969).

\begin{figure}[t] 
\includegraphics[width=8cm]{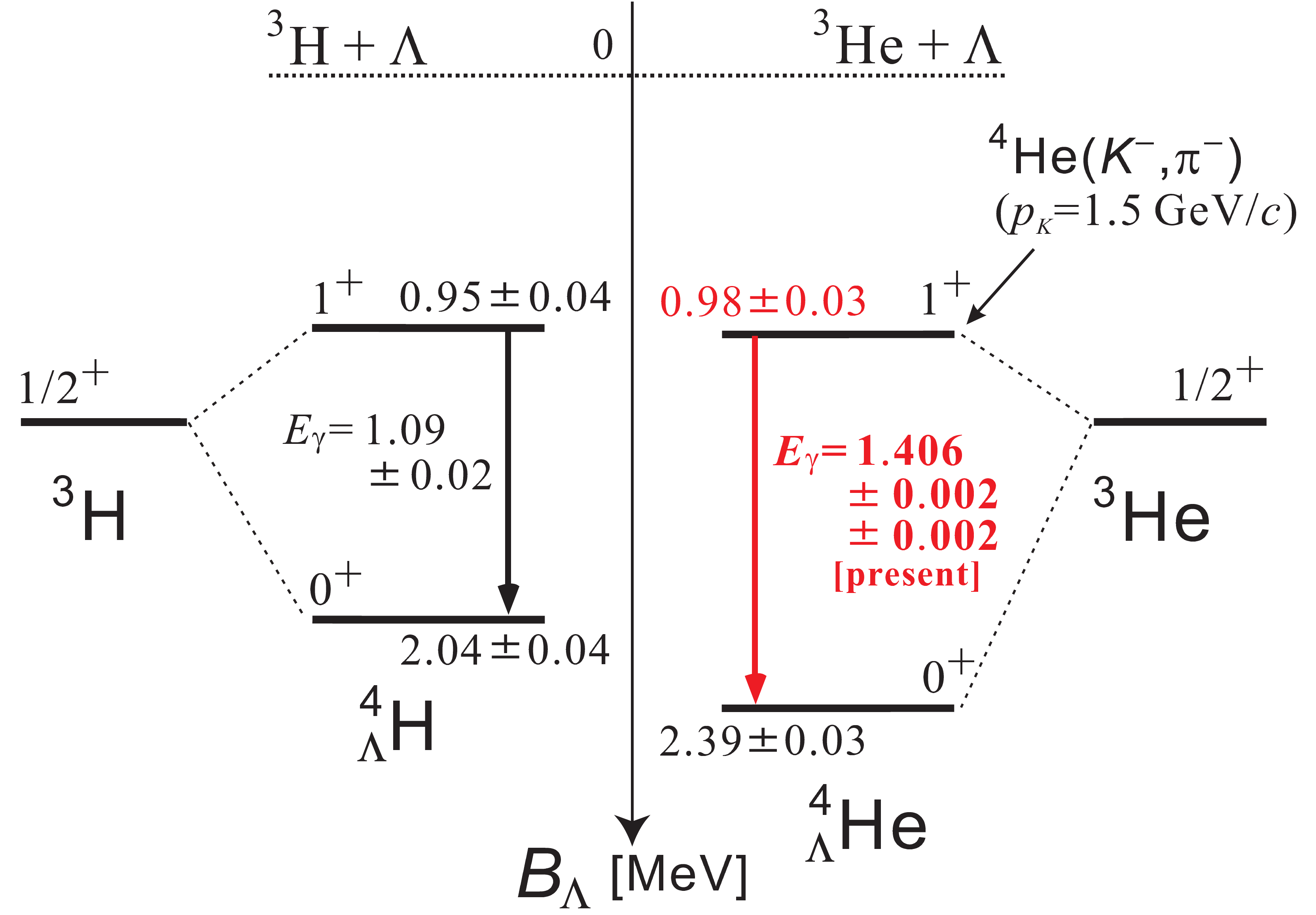} 
\caption{Level diagrams for the $A\!=\!4$ $s$-shell hypernuclei showing the 
ground-state binding energies from emulsion data and incorporating 
information on the \lamb{4}{He} excited state from a $\gamma$-ray experiment 
with the Hyperball-J at J-PARC (energies in MeV). From
Yamamoto {\it et al.}, 2015. \label{fig:fig5}} 
\end{figure}

The binding energies of the $A\!=\!4$ hypernuclei, \lamb{4}{H} and 
\lamb{4}{He}, have been extracted from emulsion data \cite{bohm68,juric73}. 
Spin assignments of these hypernuclei were obtained from analysis of their 
pionic weak decays \cite{dalitz58,downs59}. This isodoublet of hypernuclei 
forms the lightest system of isobaric mirror hypernuclei and provides 
information on charge-symmetry breaking (CSB) in the $\Lambda N$ interaction. 
The excited states of these hypernuclei were observed at $\sim 1.1$ MeV by 
stopping $K^-$ mesons 
in Li isotopes and looking for hypernuclear $\gamma$ transitions to the $A=4$ 
ground states in coincidence with either their $\pi^{-}$ or $\pi^{0}$ weak 
decays \cite{bamberger73,bedjidian79}. A very recent J-PARC experiment, 
E13 \cite{tamura13}, using the in-flight \Kpi\ reaction directly on 
$^4$He, identifies the $M1$ $\gamma$-ray transition in \lamb{4}{He} at 
$E_{\gamma}=1.41$~MeV \cite{yamamoto15}, thereby implying that the $1^+$ 
excitation energy in \lamb{4}{He} is 1.41~MeV, which differs substantially 
from the 1.15~MeV~\cite{bedjidian79} traditionally accepted, 
see Fig.~\ref{fig:fig5}. The resulting CSB in the $1^+$ excited states 
in the $A=4$ hypernuclei is then considerably smaller than in the $0^+$ 
ground states. Also recently, the binding energy of \lamb{4}{H} was
determined to be $B_\Lambda\!=\!2.12\pm 0.01 ({\rm stat})\pm 0.09({\rm syst})$
MeV at Mainz by measuring the momentum of the monochromatic pion from
the two-body decay \lamb{4}{H}$\to {^4}{\rm He}+\pi^-$~\cite{esser15}. This is
consistent, within the systematic error, with the emulsion value shown in 
Fig.~\ref{fig:fig5}. 

The heaviest of the $s$-shell hypernuclei is \lamb{5}{He}. It has 
a spin-parity of $J^\pi\!=\!1/2^+$~\cite{dalitz58} and has no bound excited 
state. Its binding energy \cite{bohm68} of $3.12\pm 0.02$ MeV is anomalously 
low as has been noted for many years (Dalitz, Herndon, and Tang, 1972;
Dalitz, 1973; Hungerford and Biedenharn, 1984). 
A straightforward calculation with just a $\Lambda N$ interaction using the 
measured binding energy values of the $A\!=\!3$, and 4 systems (including 
the excited states) as calibrations, over-binds \lamb{5}{He} by $1-2$~MeV. 
Conversely, fitting to \lamb{3}{H} and \lamb{5}{He} leaves the $A\!=\!4$
hypernuclei underbound. This problem has been attributed to a $\Lambda N$ 
tensor force, a three-body force, $\Lambda-\Sigma$ coupling, and partial 
quark deconfinement. 

 The importance of $\Lambda$-$\Sigma$ coupling in this regard has been simply 
demonstrated by writing two-component wave functions for either the $0^+$ 
or the $1^+$ states of \lamb{4}{He} (or \lamb{4}{H}) with isospin 
$T\!=\!1/2$~\cite{akaishi00}
\begin{equation}
\label{eq:4he}
|{^4_\Lambda\text{He}}\rangle = \alpha s^3s_\Lambda + \beta s^3s_\Sigma \; .
\end{equation}
The $\Sigma$ component is 2/3 $\Sigma^+$ and 1/3 $\Sigma^0$ for \lamb{4}{He}
(2/3 $\Sigma^-$ and 1/3 $\Sigma^0$ for \lamb{4}{H}). The off-diagonal
matrix elements $v(J)$ ($J\!=\!0,1$) between the basis states
can be derived from the $\Lambda$N$-\Sigma$N G matrix for $0s$ 
orbits, giving~\cite{akaishi00,millener07}
\begin{eqnarray}
\label{eq:4he-od}
  v(0) & = & \frac{3}{2}\,{^3g}-\frac{1}{2}\,{^1g}  =  
\overline{\text{V}}'+\frac{3}{4}\Delta' \; , \\
  v(1) & = & \frac{1}{2}\,{^3g}+\frac{1}{2}\,{^1g} 
  =  \overline{\text{V}}'-\frac{1}{4}\Delta' \; , \\
\text{with}& &\overline{\text{V}}'=\frac{1}{4}{^1g} +\frac{3}{4}{^3g}
\quad \text{and}\quad \Delta'={^3g}-{^1g}\, ,
\end{eqnarray}
where the prime on $\overline{\text{V}}'$ and $\Delta'$ is used to
denote the central average and spin-spin matrix elements of the
$\Lambda$-$\Sigma$ coupling interaction. Taking round numbers derived using 
the 10-range Gaussian interaction of Akaishi \textit{et al.} (2000) that 
represents NSC97f yields $^3g\! = \!4.8$\,MeV 
and $^1g\!=\! -1.0$\,MeV, which give $\overline{\text{V}}'\!=\! 3.35$\,MeV 
and $\Delta'\!=\! 5.8$\,MeV. Then, $v(0)\!=\! 7.7$\,MeV and 
$v(1)\!=\! 1.9$\,MeV. In a simple $2\times 2$ problem, the energy shifts 
of the $\Lambda$-hypernuclear states are given by $\sim v(J)^2/\Delta E$ 
with $\Delta E\sim 80$\,MeV [and the admixture $\beta\sim -v(J)/\Delta E$]. 
Thus, the downward energy shift for the $0^+$ state is $\sim 0.74$\,MeV while 
the shift for the $1^+$ state is small. The result is close to that for the 
NSC97f interaction in Fig.~1 of Akaishi \textit{et al.} (2000). 

The observed CSB in the $A\!=\!4$ system is partially due to differences 
in Coulomb energies of the core nuclei, and to the mass difference between 
$\Sigma^{\pm}$ which is $\approx$10\% of the $\Lambda-\Sigma$ mass difference, 
but the fundamental CSB in the $\Lambda N$ interaction is significant, and 
associated primarily with electromagnetic $\Lambda-\Sigma^0$ mixing that 
breaks isospin \cite{dalitz64a}. The CSB of the excited states differs from 
that of the ground states, and obtaining the correct level splittings is not 
trivial. As with the case of the hypertriton discussed above, CSB constrains 
the in-medium $\Lambda N$ interaction, in particular the strong-interaction 
coupling of $\Lambda$'s and $\Sigma$'s in the hypernuclear wavefunction 
(Gibson and Lehman, 1979; Akaishi \textit{et al.}, 2000; Hiyama 
\textit{et al.}, 2001; Nemura, Akaishi, and Suzuki 2002; Nogga, Kamada, 
and Gl\"{o}ckle, 2002; Nogga, 2013; Gal, 2015).

\begin{figure*}[t]
\centerline{\includegraphics[width=17.0cm]{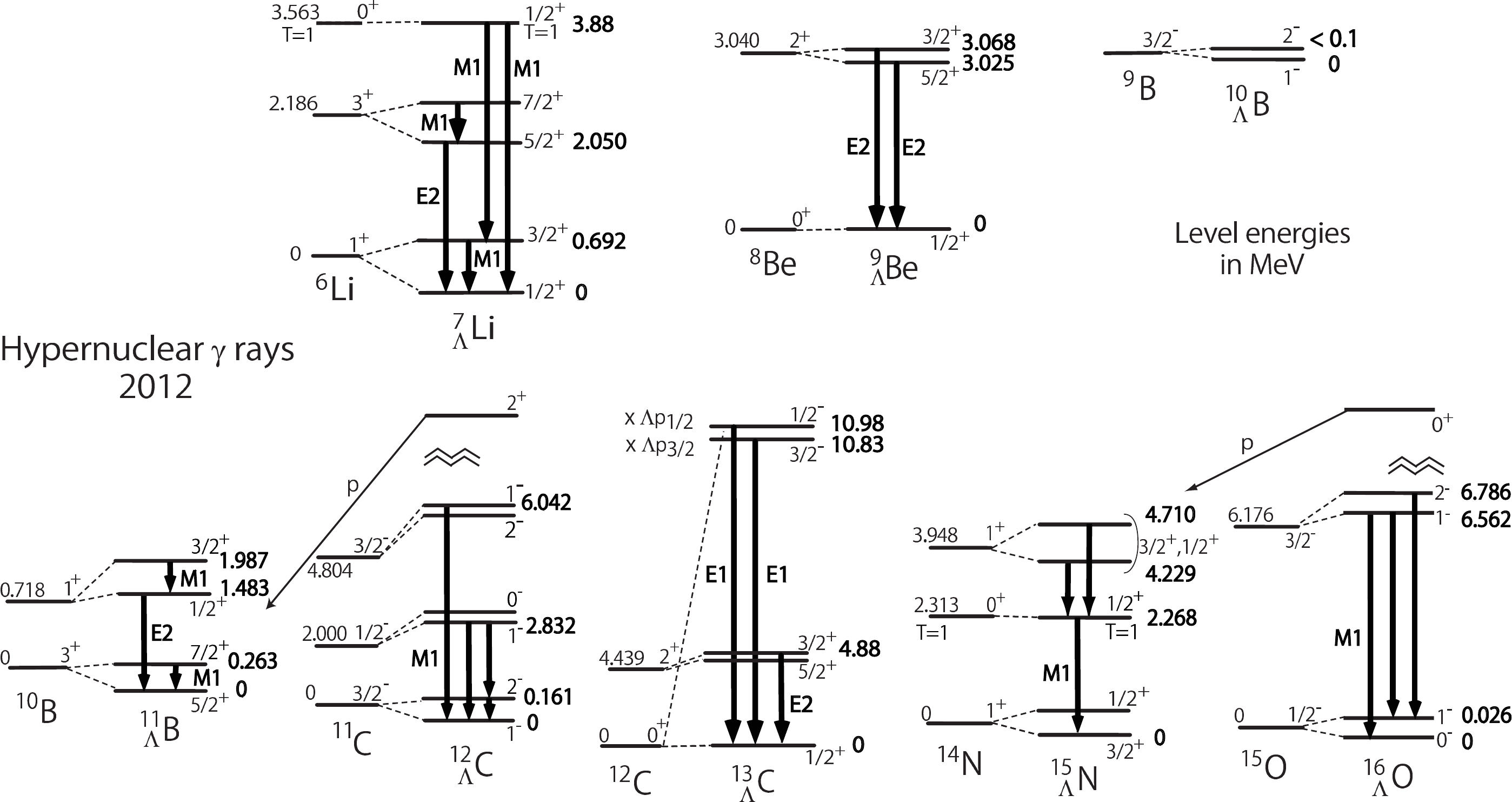}}
\caption{Spectra of $p$-shell hypernuclei showing observed $\gamma$-ray
transitions, all with the Hyperball detector except for the transitions in 
\lam{13}{C}~\cite{ajimura01,kohri02} and \lam{12}{C}, for which the
Hyperball2 detector was used~\cite{hosomi15}. All energies are in MeV.
Adapted from Tamura \textit{et al.}, 2013.}
\label{fig:gamma-spectra}
\end{figure*}

Faddeev-Yakubovsky studies (Nogga, Kamada, and Gl\"{o}ckle, 2002;
Nogga, 2013) of the $A\!=\!4$ hypernuclei using the Nijmegen NSC97 
(Rijken, Stoks, and Yamamoto, 1999) soft-core $YN$ potentials fail to 
reproduce the isodoublet CSB splittings, although NSC97f, in particular, does 
a good job of reproducing the binding energy of \lamb{3}{H} and the 
$0^+ - 1^+$ excitation energy in the $A=4$ hypernuclei. The same study 
also concludes that the probability of finding a $\Sigma$ in the $A\!=\!4$ 
hypernuclei is about 1.5\%, depending on the potential. Higher probability, 
of order 4\% results by using NSC89 (Maessen, Rijken, and de Swart, 1989), and 
the CSB it yields is much larger than for NSC97, but NSC89 is definitely not 
a realistic YN potential for use in hypernuclei. Likewise, the J\"{u}lich'04
interaction~\cite{haidenbauer05} is unsuitable (Nogga, 2013; Gazda 
\textit{et al.}, 2014; Wirth \textit{et al.}, 2014), especially in its 
$\Lambda$-$\Sigma$ coupling characteristics. However, the J\"{u}lich chiral 
$YN$ model at LO (Polinder, Hedenbauer, and Mei{\ss}ner, 2006) shows 
promise~\cite{nogga13,wirth14}, as does~\cite{nogga13} the NLO 
model~\cite{haidenbauer13a}.

The observation of $\pi^{+}$ decay of \lamb{4}{He} 
(see Sec.~\ref{subsubsec:mesonicdecay}) supports the supposition that the 
wave function of this hypernucleus contains a non-negligible $\Sigma$ 
component, although the $\Sigma$ admixture required is considerably beyond 
those provided by the $A\!=\!4$ hypernuclear few-body calculations
(Nemura, Akaishi, and Suzuki, 2002; Nogga, Kamada, and Gl\"{o}ckle, 2002;
Nogga, 2013).

Another few-body, variational calculation (Nemura, Akaishi, and Suzuki, 2002)
attempted to explicitly include three-body forces within a coupled-channel 
approach. This study claims to have obtained reasonable agreement with the 
separation energies for all the $s$-shell hypernuclei, including the excited 
states, by using a NSC97e-simulated potential. However, the genuine NSC97e 
potential in Nogga's calculation (Nogga, Kamada, and Gl\"{o}ckle, 2002;
Nogga, 2013) significantly underbinds \lamb{3}{H}. Therefore, there appears 
sufficient discrepancy between the results of theoretical calculations, and 
also when compared to the data, to warrant a more conservative view that all 
calculations are still missing something.

To summarize the status of \textit{ab initio} calculations, the $\Lambda N$ 
interaction is weaker than the $NN$ interaction, in part because one-pion 
exchange between a $\Lambda$ and a nucleon is forbidden by isospin. 
The inclusion of two-pion exchange introduces coupling of $\Lambda$'s and 
$\Sigma$'s in hypernuclei, in analogy to the coupling of $\Delta$'s with 
nucleons in nuclei. However, $\Lambda-\Sigma$ coupling is much more important 
because of the suppression of the long-range OPE and the smaller mass 
difference between the $\Lambda$ and $\Sigma$. $\Lambda-\Sigma$ coupling 
naturally induces three-body forces as generated by the last diagram in 
Fig.~\ref{fig:yn} (Nemura, Akaishi, and Suzuki, 2002), and 
electromagnetic $\Lambda-\Sigma^0$ mixing generates charge-symmetry breaking 
\cite{gal15,gazda16}. Thus the use of a $\Lambda N$ potential in a many-body 
calculation must include in-medium effects, as these are not included 
in any two-body ``elementary'' potential (Nemura, Akaishi, and Suzuki, 2002; 
Nogga, Kamada, and Gl\"{o}ckle, 2002).

\subsubsection{$p$-shell hypernuclei, $\gamma$-ray measurements, 
and spin dependence of the $\Lambda N$ interaction}
\label{subsubsec:pshell}

\begin{table*}[t]
\caption{Doublet spacings in $p$-shell hypernuclei. $E_c$ identifies 
the core state upon which the doublet is built. Energies are given in keV.  
The entries in the top (bottom) half of the table are calculated using the 
parameters in Eq.~(\ref{eq:param7}) [Eq.~(\ref{eq:param11})]. The individual 
contributions do not sum to exactly $\Delta E^{th}$, which comes
from the diagonalization, because small contributions from the
energies of admixed core states are not included.}
\label{tab:spacings}
\begin{ruledtabular} 
\begin{tabular*}{\textwidth}{@{}l@{\extracolsep{\fill}}ccrrrrrrrr}
\noalign{\smallskip}
 & $J^\pi_u$ & $J^\pi_l$ & $E_c$ &$\Lambda\Sigma$ & $\Delta$ & 
$S_\Lambda$ & $S_N$  & $T$ & $\Delta E^{th}$ & $\Delta E^{exp}$  \\
\noalign{\smallskip}\hline\noalign{\smallskip}
\lamb{7}{Li} & $3/2^+$ & $1/2^+$ & 0 & 72 & 628 & $-1$ & 
$-4$ & $-9$ &  693 & 692 \vspace{1pt}\\
\lamb{7}{Li} & $7/2^+$ & $5/2^+$ & 2186 & 74 & 557 & $-32$ & 
$-8$ & $-71$ &  494 & 471 \vspace{1pt}\\
\lamb{8}{Li} & $2^-$ & $1^-$ & 0 & 149 & 393 & $-14$ & 
$-15$ & $-23$ &  445 & (442) \vspace{1pt}\\
\lamb{9}{Li} & $5/2^+$ & $3/2^+$ & 0 & $116$ & $531$ & 
$-18$ & $-18$ & $-10$ &  $590$ & \vspace{1pt}\\
\lamb{9}{Li} & $3/2^+_2$ & $1/2^+$ & 981 & $-79$ & $229$ & 
$-13$ & $-11$ & $-91$ &  $-13$ & \vspace{1pt}\\
\lamb{9}{Be} & $3/2^+$ & $5/2^+$ & 3030 & $-8$ & $-14$ & 
$37$ & $0$ & $28$ &  $44$ & 43 \vspace{5pt}\\
\lam{10}{Be} & $2^-$ & $1^-$ & 0 & $-10$ & 180 & $-22$ & 
$-4$ & $-33$ &  110 & $<100$ \vspace{1pt}\\
\lam{10}{Be} & $3^-$ & $2^-$ & 2429 & $-19$ & 172 & $-37$ & 
$-5$ & $-10$ &  103 &  \vspace{1pt}\\
\lam{11}{B} & $7/2^+$ & $5/2^+$ & 0 & 56 & 339 & $-37$ & 
$-10$ & $-80$ &  267 & 264 \vspace{1pt}\\
\lam{11}{B} & $3/2^+$ & $1/2^+$ & 718 & 61 & 424 & $-3$ & 
$-44$ & $-10$ &  475 & 505 \vspace{1pt}\\
\lam{12}{C} & $2^-$ & $1^-$ & 0 & 65 & 167 & $-22$ & 
$-12$ & $-42$ &  158 & 161 \vspace{1pt}\\
\lam{15}{N} & $3/2^+_2$ & $1/2^+_2$ & 3948 & 65 & 451 & $-2$ & 
$-16$ & $-10$ &  507 & 481 \vspace{1pt}\\
\lam{15}{N} & $1/2^+_1$ & $3/2^+_1$ & 0 & 45 & 244 & 34 & 
$-8$ & $-214$ &  99 & \vspace{1pt}\\
\lam{16}{O} & $1^-$ & $0^-$ & 0 &$-33$ & $-123$ & $-20$ & 
1 & 188 &  23 & 26 \vspace{1pt}\\
\lam{16}{O} & $2^-$ & $1^-_2$ & 6176 & 92 & 207 & $-21$ & 
1 & $-41$ &  248 & 224 \vspace{1pt}\\
\noalign{\smallskip}
\end{tabular*}
\end{ruledtabular}
\end{table*}

  The results from various production reactions for hypernuclei have
establishd that the $\Lambda$ moves in a potential well about 30 MeV deep 
and that the $\bm{l}_{N\Lambda}\cdot\bm{s}_\Lambda$ spin-orbit term is quite small.
However, multiplets based on particular core levels cannot be resolved.
The splitting of a multiplet is governed by terms in Eq.~(\ref{eq:vlam})
that depend on the spin of the $\Lambda$. In the $p$ shell, the five
$p_N s_\Lambda$ two-body matrix elements depend on the radial integrals 
associated with each component in Eq.~(\ref{eq:vlam}), are conventionally 
denoted by the parameters $\overline{V}$, $\Delta$, $S_\Lambda$, $S_N$ and 
$T$ (Gal, Soper, and Dalitz, 1971)
\begin{equation} 
V_{\Lambda N}  = \overline{V} + \Delta \bm{s}_N\cdot \bm{s}_\Lambda +  
S_\Lambda\bm{l}_{N}\cdot \bm{s}_\Lambda  +  S_N\bm{l}_{N}\cdot\bm{s}_N +  TS_{12} . 
\label{eq:vparam} 
\end{equation} 
Note that the operators associated with $\Delta$ and $S_\Lambda$ are 
$\bm{S}_N\cdot \bm{s}_{\Lambda}$ and $\bm{L}_{N}\cdot \bm{s}_{\Lambda}$ with
$\bm{S}_N$ and $\bm{L}_{N}$ the total nuclear Pauli spin and the total
orbital angular momentum, respectively.
This enables simple estimates for the contributions of $\Delta$ and 
$S_\Lambda$ to be made from the known $LS$ structure of the nuclear-core state.

 The only way to measure the doublet spacings, and hence determine
$\Delta$, $S_\Lambda$, and $T$, is to perform $\gamma$-ray
spectroscopy with high-resolution $\gamma$-ray detectors.
Figure~\ref{fig:gamma-spectra}
shows 20 $\gamma$-ray transitions observed in $p$-shell hypernuclei via
\piKg\ experiments at KEK and \Kpig\ experiments at BNL between 1998
and 2005 using the Hyperball array of 14 large-volume Ge 
detectors~\cite{hashtam06}. It can be seen that the data set includes
the measurement of nine doublet spacings. As will be discussed, the
data for \lamb{7}{Li}, \lamb{9}{Be}, and \lam{16}{O} play an
important role in determining $\Delta$, $S_\Lambda$, and $T$, respectively.
Also, looking ahead, Table~\ref{tab:spacings} shows that all nine doublet 
spacings can be well described in terms of the contributions of these 
three parameters and contributions arising from $\Lambda$-$\Sigma$ mixing. 

 The motivation for including both $\Lambda$ and $\Sigma$ hypernuclear
states in the shell-model basis is provided in the previous subsection
where it is noted that the coupling between these configurations is
necessary to solve the ``overbinding'' problem in the $s$-shell hypernuclei
by providing considerable extra binding energy for the \lamb{4}{H}
and \lamb{4}{He} $0^+$ ground states. This means that the $\Lambda N$
spin-spin interaction and $\Lambda$-$\Sigma$ coupling both contribute 
strongly to the spacing of the $0^+$ and $1^+$ states.

  The $s_Ns_Y$ matrix elements depend entirely on relative $s$ states 
while the central $p_Ns_Y$ matrix elements come from roughly half relative 
$s$ state and half relative $p$ state.
Because the $p$-state matrix elements are much smaller than $s$-state 
matrix elements, the scale for energy shifts from $\Lambda$-$\Sigma$
coupling desreases by factor of four in $p$-shell hypernuclei. This can 
be seen from Fig.~\ref{fig:gamma-spectra} and Table~\ref{tab:spacings} 
but the effects are still significant. 

 The parametrization of Eq.~(\ref{eq:vparam}) applies to the direct 
$\Lambda N$ interaction, the $\Lambda N$-$\Sigma N$ coupling interaction, 
and the direct $\Sigma N$ interaction for both isospin 1/2 and 3/2.
Thus, the input to shell-model calculations is four sets of the
five parameters defined by Eq.~(\ref{eq:vparam}). The parameter values
of most interest are those for the $\Lambda N$ and $\Lambda N$-$\Sigma N$ 
interactions and a prime indicates the $\Lambda N$-$\Sigma N$
parameters. Values for these two parameter sets are based on various Nijmegen 
models  of the $YN$ interactions are given in Section~III of Millener (2010); 
see also Yamamoto, Motoba, and Rijken (2010). The central interactions 
given by $\overline{V}'$ and $\Delta'$ are dominant for the 
$\Lambda N$-$\Sigma N$ interaction. To see which nuclear core states 
contribute to the $\Lambda -\Sigma$ coupling, and make contact with the 
approach of Umeya and Harada (2009, 2011), one can include 
an overall factor $\sqrt{4/3}\,\bm{t}_N\cdot \bm{t}_{\Lambda\Sigma}$ that has a 
value of unity for the two-body matrix elements in Eq.~(\ref{eq:vparam}), 
where $\bm{t}_{\Lambda\Sigma}$ is the operator that converts a $\Lambda$ into a 
$\Sigma$. Then, the core operator associated with  $\overline{V}'$ is 
$\bm{T}_N = \sum_i \bm{t}_{Ni}$. This leads to a non-zero matrix element only 
between $\Lambda$ and $\Sigma$ states that have the same core, with the value
\begin{equation}
 \langle (J_cT,s_\Sigma)JT |V_{\Lambda\Sigma}|(J_cT,s_\Lambda)JT\rangle
 = \scriptstyle{\sqrt{\frac{4}{3}}}\sqrt{T(T+1)}\ \overline{V}' ,
\label{eq:fermi}
\end{equation}
in analogy to Fermi $\beta$ decay of the core nucleus. Similarly,
the spin-spin term involves $\sum_i \bm{s}_{Ni}\bm{t}_{Ni}$ for the core and 
connects core states that have large Gamow-Teller (GT) matrix elements 
between them. This can be the case when the core states are the same
[this has been called coherent $\Lambda$-$\Sigma$ coupling~\cite{akaishi00}]
but, because $\Delta'$ is large, there can be large coupling matrix
elements for other states, often with different isospin but with
the same spatial symmetry. Not surprisingly,
energy shifts due to $\Lambda$-$\Sigma$ coupling grow with the
isospin of the core nucleus and are predicted to be more than 250 keV for 
the ground states of \lamb{9}{He} and \lam{10}{Li} that could be reached
by double-charge-exchange reactions from stable targets~\cite{gal13c}.
 
 Shell-model calculations for $p$-shell hypernuclei start with the
Hamiltonian 
\begin{equation}
 H = H_N + H_Y + V_{NY} ,
\label{eq:hamyn}
\end{equation}
where $H_N$ is an empirical Hamiltonian for the $p$-shell core,
the single-particle $H_Y$ supplies the $\sim 80$\,MeV mass difference
between $\Lambda$ and $\Sigma$, and $V_{NY}$ is the $YN$ interaction.
The shell-model basis states are chosen to be of the form
$|(p^n\alpha_{c}J_{c}T_{c},j_Yt_Y)JT\rangle$,
where the hyperon is coupled in angular momentum and isospin
to eigenstates of the $p$-shell Hamiltonian for the core, with up to three
values of $T_c$ contributing for $\Sigma$-hypernuclear states. This
is known as a weak-coupling basis and, indeed, the mixing of
basis states in the hypernuclear eigenstates is generally
very small. In this basis, the core energies can be taken from
experiment where possible and from the $p$-shell calculation otherwise.

 The technical details of such calculations are quite 
simple~\cite{auerbach83,millener07}. Because the product of creation and 
annilation operators for a two-body $YN$ interaction can written
in terms $a^\dagger a$ pairs for the nucleons and hyperons, we simply need
a complete set of OBDME between $p$-shell
eigenstates (the maximum dimension for a given $JT$ in the $p$ shell is
only 14) to compute matrix elements of the hypernuclear Hamiltonian.
Only isoscalar OBDME are needed in the $\Lambda$ space and isovector
OBDME are needed for the $\Lambda$-$\Sigma$ coupling matrix elements.

Many hypernuclear calculations have used the venerable
Cohen and Kurath (1965) interactions. Here, the $p$-shell
interaction has been refined using the following strategy.
The one-body spin-orbit splitting between the $p_{3/2}$ and
$p_{1/2}$ orbits is fixed to give a good description
of the light $p$-shell nuclei (say for $A\leq 9$). The
overall strength of the tensor interaction is also fixed,
ultimately to produce the cancellation in $^{14}$C $\beta$
decay. The well-determined linear combinations of the central 
and vector $p$-shell interactions are then chosen by fitting
the energies of a large number of states that are known
to be dominantly $p$ shell in character, including the
large spin-orbit splitting at $A\!=\!15$. Some properties
of stable $p$-shell ground states are shown in Table~\ref{tab:par4}
for this interaction in the supermultiplet basis where $[f]K_L L$ label 
representations of SU(3)$\supset$R3 in the orbital space (three single-particle
$p$ states) and $[\widetilde{f}]\beta TS$ label representations of 
SU(4)$\supset$SU(2)$\times$SU(2) in the spin-isospin space (four states); 
$[f] =  [f_1 f_2 f_3]$, with $f_1\ge f_2\ge f_3$ and $f_1+ f_2+ f_3 = n$, 
also labels the spatial symmetry. $K_L$ labels multiple occurrences
of $L$ for a given representation of SU(3) and is obtained by angular-momentum
projection from a specific intrinsic SU(3)$\supset$SU(2) basis state; when
$S\neq 0$, $J$ can be projected from a product of the SU(3)$\supset$SU(2)
intrinsic state and an intrinsic spin state with magnetic quantum number $K_S$
to give a state with $K\!=\!K_L\!+\!K_S$ and a mixture of $L$ values
[see Eqs.(\ref{eq:k3h}) and (\ref{eq:k3})].  The central interaction is 
essentially SU(4) 
conserving and the mixing of different $[f_c]L_cS_c$ is primarily due to
the one-body spin-orbit and two-body SLS and ALS terms in the effective
$p$-shell Hamiltonian. A detailed discussion of $p$-shell nuclei,
including the allowed quantum numbers and spectra, is given in Section 5 of 
Millener (2007). In Table~\ref{tab:par4}
\begin{equation}
|K\!=J\!=\!3/2\rangle =\scriptstyle{\sqrt{\frac{21}{26}}}|L\!=\!1\rangle - 
\scriptstyle{\sqrt{\frac{5}{26}}}|L\!=\!2\rangle,
\label{eq:k3h}
\end{equation}
with $S\!=\!1/2$, while 
\begin{equation}
|K\!=\!J\!=\!3\rangle = \scriptstyle{\sqrt{\frac{6}{7}}}|L\!=\!2\rangle 
-\scriptstyle{\sqrt{\frac{3}{22}}}|L\!=\!3\rangle + \scriptstyle{
\sqrt{\frac{1}{154}}}|L\!=\!4\rangle ,
\label{eq:k3}
\end{equation}
with $S\!=\!1$.

\begin{table}[t]
\caption{Root-mean-square charge radii and dominant wave function components
for the ground states of stable $p$-shell nuclei (par4 interaction). $[f]$
labels the spatial symmetry of the $p$-shell nucleons (see text). The $L$
decomposition of states with good $K$ are given in Eqs.~(\ref{eq:k3h})
and (\ref{eq:k3}). \label{tab:par4}} 
\begin{ruledtabular} 
\begin{tabular}{ccrccl} 
Nucleus & $\langle r^2\rangle^{1/2}_{ch}$ fm & $[f]$ & \% $[f]$ & $J^\pi$ & 
Dominant component\\
\hline 
 $^6$Li & 2.57 & [2] & 98.2 & $1^+$ & $L\!=\!0,S\!=\!1$ \\
 $^7$Li & 2.41 & [3] & 96.6 &  $\frac{3}{2}^-$ & $L\!=\!1,S\!=\!\frac{1}{2}$ \\
 $^9$Be & 2.52 & [41] & 94.7 &  $\frac{3}{2}^-$ & $K\!=\!\frac{3}{2},
         S\!=\!\frac{1}{2}$ \\
 $^{10}$B & 2.45 & [42] & 94.0 &  $3^+$ & $K\!=\!3,S\!=\!1$ \\
 $^{11}$B & 2.42 & [43] & 81.0 &  $\frac{3}{2}^-$ & $K\!=\!\frac{3}{2},
         S\!=\!\frac{1}{2}$ \\
 $^{12}$C & 2.47 & [44] & 79.3 &  $0^+$ & $L\!=\!0,S\!=\!0$ \\
 $^{13}$C & 2.44 & [441] & 66.5 &  $\frac{1}{2}^-$ & $L\!=\!1,
        S\!=\!\frac{1}{2}$ \\
 $^{14}$C & 2.56 & [442] & 59.7 &  $0^+$ & $L\!=\!0,S\!=\!0$ \\
 $^{14}$N & 2.52 & [442] & 94.2 &  $1^+$ & $L\!=\!2,S\!=\!1$ \\
 $^{15}$N & 2.59 & [443] & 100.0 & $\frac{1}{2}^-$ & $L\!=\!1,
        S\!=\!\frac{1}{2}$ \\
\end{tabular}
\end{ruledtabular}
\end{table}

 In the LS basis for the core, the matrix elements of 
$\bm{S}_N\cdot \bm{s}_{\Lambda}$ are diagonal
[similarly for $\bm{L}_{N}\cdot \bm{s}_{\Lambda} = 
(\bm{J}_N -\bm{S}_N)\cdot \bm{s}_{\Lambda}$] and
depend just on the intensities of the total $L$ and $S$ 
for the hypernucleus. Because supermultiplet symmetry 
$[f_c]K_cL_cS_cJ_cT_c$ is generally a good symmetry for $p$-shell core 
states [Table~\ref{tab:par4} and Eqs.~(\ref{eq:k3h}) and (\ref{eq:k3})],
only one or two values of $L$ and $S$ are important. 
Of the remaining $\Lambda N$ parameters, $\overline{V}$ contributes 
only to the overall binding energy; $S_N$ does not contribute to 
doublet splittings in the weak-coupling limit but a negative $S_N$ 
augments the nuclear spin-orbit interaction and contributes to the 
spacings between states based on different core states; in general, 
there are no simple expressions for the coefficients of $T$.

 With reference to Table~\ref{tab:spacings}, the set of $\Lambda N$
parameters used up to \lamb{9}{Be} (chosen to fit the energy spacings 
in \lamb{7}{Li} perfectly) is (parameters in MeV)
\begin{equation}
\Delta= 0.430\ \ S_\Lambda =-0.015\ \ {S}_{N} = -0.390\ \ {T}=0.030 .
\label{eq:param7}
\end{equation}
 The doublet spacings for the heavier $p$-shell hypernuclei 
consistently require a smaller value for $\Delta$
\begin{equation}
\Delta= 0.330\ \ S_\Lambda =-0.015\ \ {S}_{N} = -0.350 \ \ {T}=0.0239 .
\label{eq:param11}
\end{equation}
The  matrix elements for the $\Lambda$-$\Sigma$ coupling interaction, 
based on the G-matrix calculations of Akaishi \textit{et al.} (2000) for the
NSC97$e$,$f$ interactions (Rijken, Stoks, and Yamamoto, 1999), are
\begin{equation}
\overline{V}' = 1.45\ \ \Delta'= 3.04\ \ S_\Lambda' = S_N' = -0.09
\ \ T' = 0.16 .
\label{eq:paramls}
\end{equation}
These parameters are kept fixed throughout the $p$ shell.

 We are now in a position to consider the $\gamma$-ray data
in Fig.~\ref{fig:gamma-spectra} in relation to the breakdown of
doublet spacings in Table~\ref{tab:spacings}. First, on a historical
note, shell-model analyses of $\Lambda$ binding energies for
$p$-shell hypernuclei were attempted long ago, and introduced the
notation still in use for the $\Lambda N$ interaction (Gal, Soper, and
Dalitz, 1971). They also considered a double one-pion-exchange
$\Lambda NN$ interaction. However, progress on characterizing the
$\Lambda N$ interaction was hampered by a lack of data (Gal, Soper, and
Dalitz, 1972, 1978). Nevertheless, the stage was set for studies of 
hypernuclear $\gamma$-rays~\cite{dg78}. The observation of $\gamma$-rays in 
\lamb{7}{Li} and \lamb{9}{Be} at BNL using the \Kpig\ reaction and NaI 
detectors~\cite{may83} finally permitted a convincing shell-model 
analysis~\cite{millener85} with parameters close to those in
Eq.~(\ref{eq:param7}), but without the inclusion of $\Lambda$-$\Sigma$ 
coupling, and inspired other analyses~\cite{fetisov91}. Many of the 
$p$-shell hypernuclei up to \lam{13}{C} have also been studied in
cluster models~\cite{hiyama09}.

 In the first \piKg\ experiment with the Hyperball at KEK in 
1998~\cite{tamura00}, four $\gamma$-rays in \lamb{7}{Li} were seen, 
namely all except the $7/2^+\to 5/2^+$ transiton in 
Fig.~\ref{fig:gamma-spectra}.  Note that the $3/2^+$ ($L\!=\!0,S\!=\!3/2$)
and $7/2^+$ ($L\!=\!2,S\!=\!3/2$) require spin-flip and are not strongly 
populated in the \piK\ reaction~\cite{hiyama99}. The high-energy M1 
transitions from the $1/2^+;T\!=\!1$ level can be seen when the 
Doppler-shift correction is made and their energy difference matches
the 691.7 keV of the transition (peak sharpened by the Doppler correction) 
between the ground-state doublet members. The lineshape for the 
2050-keV $5/2^+\to 1/2^+$ transition gives a lifetime for the $5/2^+$ level
via the Doppler-shift attenuation method~\cite{tanida01}. The derived
reduced electric-quadrupole transition probability B(E2) is considerably 
smaller than expected from the known B(E2) for the $3^+\to 1^+$ transition 
in $^6$Li. The lowest threshold is for \lamb{5}{He}$+d$ at 3.94(4) MeV so that 
the $5/2^+$ state and the $1/2^+$ ground state in \lamb{7}{Li} are considerably
 more bound than the core states in $^6$Li. This entails a shrinkage in the 
size of the radial wave functions, and a
reduction of the B(E2), that is best treated in cluster-model calculations 
for \lamb{7}{Li}~\cite{hiyama99}. The 471-keV M1 $\gamma$-ray in the
upper doublet was seen via $\gamma$-$\gamma$ coincidence with the
$5/2^+\to 1/2^+$ transition in a \Kpig\ experiment on a $^{10}$B target 
at BNL~\cite{ukai06} (following $l\!=\!0$ $^3$He emission from
the $s_N^{-1}s_\Lambda$ substitutional state in \lam{10}{B}).

 From Table~\ref{tab:spacings}, it can be seen that the \lamb{7}{Li}
ground-state doublet spacing comes mostly from the spin-spin interaction 
($\frac{3}{2}\Delta$ in the pure LS limit) with a 10\% assistance from 
$\Lambda$-$\Sigma$ coupling. The situation is similar for the second doublet 
except that contributions from $S_\Lambda$ and $T$ reduce the spacing by 
$\sim 100$ keV. $S_N$ reduces the excitation energies of the $5/2^+;0$ and 
$1/2^+;1$ states by 288 keV and 82 keV, respectively~\cite{millener07}, 
making the $1/2^+$ state just bound.

  In \lamb{9}{Be}, the $^8$Be core states are unbound (by 92 keV for
the ground state) but the presence of the $\Lambda$ raises the $\alpha$
threshold to 3.50 MeV, viz.
\begin{equation}
B_\alpha(^9_\Lambda{\rm Be}) = B_\alpha(^8{\rm Be}) +
B_\Lambda(^9_\Lambda{\rm Be}) -B_\Lambda(^5_\Lambda{\rm He}) ,
\label{eq:balpha}
\end{equation}
meaning that the $\gamma$-rays from the $3/2^+$ and $5/2^+$ states can be
observed. This was achieved using the Hyperball in a \Kpig\ experiment
at BNL~\cite{akikawa02}. With the Doppler correction, peaks were seen at 
3024 and 3067 keV [these are updated energies~\cite{tamura10}].
Only the upper peak is seen following proton emission from \lam{10}{B}
and strong theoretical arguments~\cite{millener05,millener07} indicate 
that this $\gamma$-ray comes from the $3/2^+$ member of the doublet.
Table~\ref{tab:spacings} shows that the small splitting of the doublet
means that $S_\Lambda$ is small (contributions from $\Delta$, $T$, and
$\Lambda$-$\Sigma$ coupling more or less cancel); the splitting is 
$-\frac{5}{2}S_\Lambda$ if the $^8$Be $2^+$ state is pure $L\!=\!2,S\!=\!0$, 
as it is in the $2\alpha + \Lambda$ cluster model~\cite{hiyama00}.

 An earlier experiment with NaI detectors at BNL~\cite{may83} observed
a $\gamma$-ray at 3079(40) keV and put an upper limit of 100 keV
on the doublet splitting. This, and the observation of a 2034(23) keV
$\gamma$-ray in \lamb{7}{Li}~\cite{may83}, revived shell-model studies of
$p$-shell hypernuclei~\cite{millener85}.

 The main objective of a 2001 \Kpig\ experiment at BNL~\cite{ukai04,ukai08}
was to measure the ground-state doublet spacing of \lam{16}{O} that
depends strongly on the matrix element of the $\Lambda N$ tensor interaction
$T$. For a pure $p_{1/2}^{-1}s_\Lambda$ configuration, the spacing 
is~\cite{dg78}
\begin{equation}
 E(1^-_1)-E(0^-) =  -\frac{1}{3}\Delta +\frac{4}{3}S_\Lambda +8\,T .
\label{eq:o16gs}
\end{equation}
Figure~\ref{fig:gamma-spectra} shows that the measured spacing is only
26 keV, derived from the difference in energies of the $\gamma$-rays
from the 6562-keV $1^-$ excited state to the members of the ground-state
doublet. Table~\ref{tab:spacings} shows that the small separation is
the result of a large cancellation between the contributions of $T$ and
the other contributions (mainly $\Delta$). If $\Delta$ is known, this
doublet spacing fixes $T$. The major contributor to the increase in
the spacing between the two doublets relative to the core spacing of 
6.176 MeV is $S_N$ which gives over 500 keV ($\sim -\frac{3}{2}S_N$).

\begin{figure}[t]
\centerline{\includegraphics[width=8.5cm]{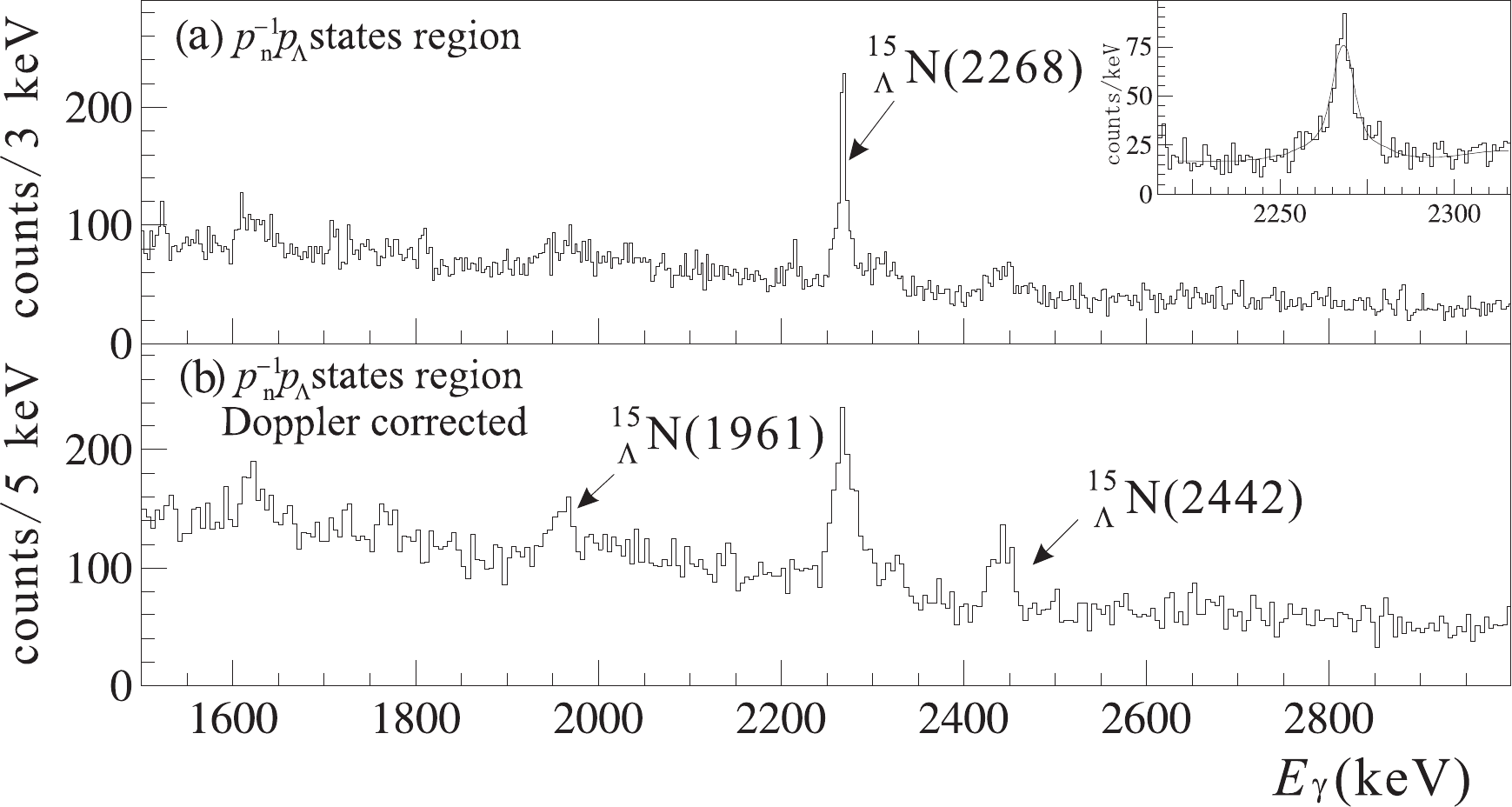}}
\caption{$\gamma$-ray spectra from \lam{15}{N} 
(see Fig.~\ref{fig:gamma-spectra}) following proton emission from 
the $p^{-1}_np_\Lambda$ mass region of \lam{16}{O} populated in the \Kpi\ 
reaction on $^{16}$O (see Fig.~\ref{fig:cern16o}). 
The upper figure shows that the
2268-keV line is sharp without Doppler correction implying a long lifetime
that is obtained from the lineshape analysis shown in the inset. The lower
figure shows the transitions from the upper doublet that appear when the
Doppler correction is made. Adapted from Ukai \textit{et al.}, 2008.}
\label{fig:l15nexp}
\end{figure}

 A weak $\gamma$-ray is also seen in the above experiment~\cite{ukai08}
and is interpreted as a transition from the $2^-$ member of the upper
doublet [the $2^-$ state requires spin-flip to be populated via the
\Kpi\ reaction]. The \lam{15}{N} $\gamma$-rays are seen following proton
emission from the $p_\Lambda$ states of \lam{16}{O} (see 
Fig.~\ref{fig:l15nexp}). The 2268-keV $\gamma$-ray is sharp without
Doppler correction implying a long lifetime [measured at 1.5(4) ps] while
the transitions from the upper doublet are fast and are seen when the
Doppler correction is made. It is interesting that the transition from
the $1/2^+;1$ level to the $1/2^+$ member of the ground-state doublet is 
not seen; in the weak-coupling limit, it should be approximately half
the strength of the 2268-keV transition. We first note that in $^{14}$N
the M1 transition from the 3.498-MeV $1^+$ level (mainly $L\!=\!0,S\!=\!1$)
to the $0^+;1$ level is strong while the M1 transition from the $0^+;1$ level 
to the ground-state is weak because this transtion is the analog of 
$^{14}$C $\beta$ decay and the $\langle\sigma\tau\rangle$ matrix element
essentially vanishes (making the M1 transition mainly orbital). It
turns out~\cite{millener07,ukai08} that small admixtures of the 
$1^+_2;0\times s_\Lambda$ configuration into the wave functions of
the ground-state doublet members produce strong cancellations in the 
hypernuclear M1 matrix elements giving a predicted lifetime of 0.5 ps
for the $0^+;1$ level compared with 0.1 ps for the core transition.
The cancellation is more severe for the $1/2^+;1\to 1/2^+$ transition
but still not quite strong enough because the calculated  
$\gamma$-ray branch to the $1/2^+$ state is 18\% while the experiment
puts an upper limit of $\sim 5$\% at the predicted energy~\cite{ukai08}. 
The upper doublet (the lower member is surely $1/2^+$) is based on an 
$L\!=\!0,S\!=\!1$ core and the splitting is mainly due to the spin-spin 
interaction ($\Delta$) in analogy to the \lamb{7}{Li} ground-state doublet 
and, in fact, the first-excited-state doublet in \lam{11}{B}.

 In \lam{12}{C}, the excitation energies of the excited $1^-$ states provide 
a useful check on the energies of the unresolved peaks in the 
$^{12}$C\eeK\lam{12}{B} reaction~\cite{iodice07,tang14}. The difference in 
the energies of the transitions from the $1^-_2$ level agrees with the
161.5 keV energy measured for the ground-state doublet 
transition~\cite{hosomi15}. This doublet spacing is important because of the
failure to observe the corresponding doublet spacing in \lam{10}{B} in
two \Kpig\ experiments at BNL~\cite{chrien90,ukai04th} that both set an
upper limit of about 100 keV on the doublet spacing. The core nuclei have
similar structures (see Table~\ref{tab:par4}), being essentially particle-hole
conjugates in the $p$ shell (a particle or hole in the Nilsson $K\!=\!3/2$ 
orbit). This means that the $\Lambda N$ contribution to the spacing should
be nearly the same. Table~\ref{tab:spacings} shows that the $\Lambda N$
contribution for \lam{10}{B} is actually slightly larger than for \lam{12}{C}.
Table~\ref{tab:spacings} also shows that the $\Lambda$-$\Sigma$ coupling 
increases the doublet separation in \lam{12}{C} while decreasing it
slightly in \lam{10}{B}. This is because the $\langle\sigma\tau\rangle$ 
matrix elements involving the lowest $3/2^-$ and $1/2^-$ states are of 
opposite sign for the two core nuclei. The coefficients of $\overline{V}'$ and 
$\Delta'$ for matrix elements involving the same core state are of opposite 
sign for the $1^-$ and $2^-$ states and the sign changes between \lam{10}{B} and
\lam{12}{C}. Although this is a substantial effect, it is lessened by that 
fact that the $1^-$ states in both hypernuclei are pushed down by a 
coupling to $\Sigma$ states that have a $1/2^-$ core state. 
 It is certainly possible to reduce the
spacing in \lam{10}{B} appreciably by changing the $\Lambda$-$\Sigma$ 
coupling interaction ~\cite{halderson08,millener10}. It has also been 
suggested that charge-symmetry breaking effects could lower the
transition energy in \lam{10}{B}~\cite{gal15}.

 Another way to try to measure the ground-state doublet spacing for the 
$A\!=\!10$ hypernuclei is to look for $\gamma$-rays from the $2^-$ and $3^-$ 
states in \lam{10}{Be} based on the 2.43-MeV $5/2^-$ state in $^9$Be via
the $^{10}$B$(K^-,\pi^0\gamma)$\lam{10}{Be} reaction~\cite{millener12}; 
this reference also considers \lamb{8}{Li} and \lamb{9}{Be} as  possible
sources of unassigned $p$-shell hypernuclear $\gamma$ rays. Unfortunately, 
the $2^-_2\to 2^-_1$ $\gamma$-ray branch is predicted to be only 13\% and
the $2^-_2\to 1^-_1$ and $3^-_1\to 2^-_1$ transitions could have very
similar energies. There is no chance to see the ground-state doublet 
transition itself because the B(M1) is proportional to 
$(g_c-g_\Lambda)^2$~\cite{dg78} ($g_c\!=\!-0.746$, $g_\Lambda\!=\!-1.226$)
leading to very long electromagnetic lifetime meaning that the $2^-$ level
will undergo weak decay.

 In the \piKg\ reaction on $^{11}$B, six $\gamma$-ray transitions with
energies of 264, 458, 505, 570, 1483, and 2477 keV have been identified
as transitions in \lam{11}{B}~\cite{miura05}. The 1483-keV transition is 
by far the most intense and is identified as coming from the $1/2^+$
level based on the 718-keV $1^+;0$ level of $^{10}$B and acts as a
collection point for $\gamma$-rays from strongly populated $3/2^+$ and
$1/2^+$ levels higher in the spectrum. A $3/2^+;1$ level based on the
5.16-MeV $2^+;1$ level of $^{10}$B should be the strongest and the
source of the 2477-keV $\gamma$-ray seen in the Doppler-corrected spectrum.
By making use of the relative intensities and lifetime limits for these
$\gamma$-rays a plausible decay scheme has been established by comparison
with shell-model calculations~\cite{millener08}. Assignments for the
lower part of the spectrum, shown in Fig.~\ref{fig:gamma-spectra},
have been confirmed from an analysis of the three $\gamma$-rays seen
following proton emission from \lam{12}{C}~\cite{ma10}. The main failing
of the shell-model calculation is that it does not produce high enough 
excitation energies for the \lam{11}{B} states based on the $1^+;0$ states of
$^{10}$B at 0.72 and 2.15 MeV~\cite{millener10}.

 The preceding discussion shows that one set of $p_Ns_Y$ parameters
is quite successful in reproducing data on the doublet spacings in the $p$
shell (with some adjustment for \lamb{7}{Li}). This statement refers to
$\Delta$, $S_\Lambda$, $T$ and the $\Lambda$-$\Sigma$ coupling parameters.
The parameter $S_N$ augments the nuclear spin-orbit interaction, gives
a substantial contribution to $B_\Lambda$ values in the $p$ 
shell~\cite{millener10}, and works in the right direction to reproduce 
the changes in spacing of doublet centroids from the spacing in the core
nucleus. However, a considerably larger value of $S_N$ is required to
reproduce the energies of excited-state doublets in \lam{11}{B}, \lam{12}{C},
and \lam{13}{C}. In terms of the $\Lambda N$ interaction alone, the small
value for $S_\Lambda$ means that the strengths of the symmetric and
antisymmetric spin-orbit interactions have to be very nearly equal. This
is not the case for effective interactions derived from free-space $YN$ models
and nor is the value for $S_N$ large enough~\cite{millener10}. However,
the double one-pion-exchange $\Lambda NN$ interaction 
(Gal, Soper, and Dalitz, 1971) is
independent of the $\Lambda$ spin and gives, when averaged over the 
$s_\Lambda$ wave function, an effective $NN$ interaction that operates in
the nuclear core. This interaction contains an antisymmetric spin-orbit
component that behaves rather like $S_N$ and has its largest effect
beyond the middle of the $p$ shell (Gal, Soper, and Dalitz, 1971). 
It may, in fact, be responsible for much of the empirical value of $S_N$ 
and should be reintroduced into $p$-shell hypernuclear calculations.

\begin{figure} 
\includegraphics[width=8cm]{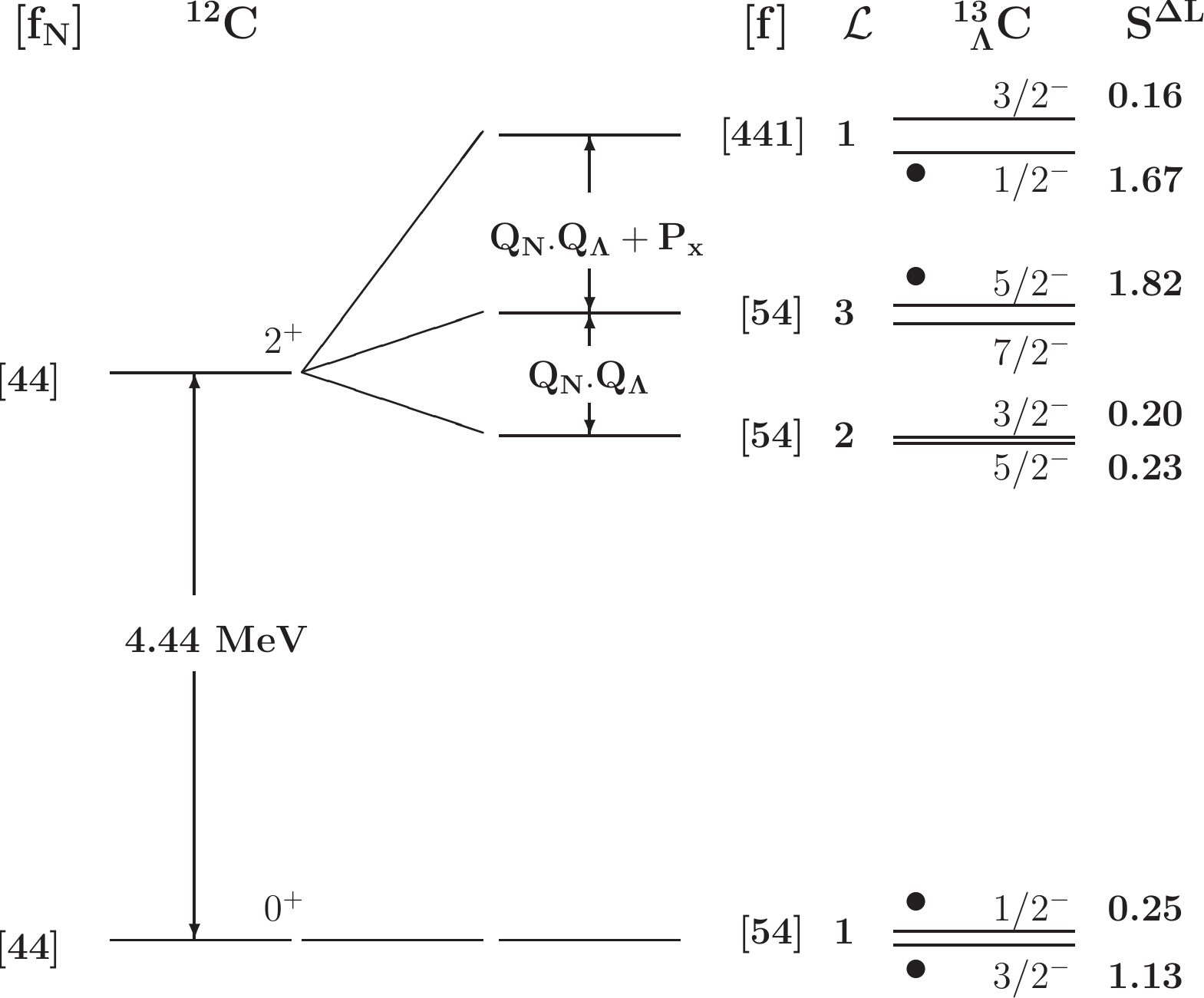}
\caption{ $p_\Lambda$ states in \lam{13}{C} based on the 
lowest $0^+$ and $2^+$ states of the $^{12}$C core. The spin-doublet 
structure is explained in the text and Eq.(\ref{eq:scriptl}). The states of 
the $2^+\times p_\Lambda$ multiplet are split by the quadrupole-quadrupole 
component of the $p_Np_\Lambda$ interaction. The states are labelled by 
their tendency towards a good supermultiplet symmetry $[f]$. The energy
of the uppermost doublet is sensitive to the space-exchange component
in the $\Lambda N$ interaction. The $S^{\Delta L}$ on the right are
structure factors governing the relative population of states in
the \Kpi\ reaction with no spin flip ($\Delta L\!=\!0$ for the $1/2^-$ states
and  $\Delta L\!=\!2$ for the others).}
\label{fig:lc13} 
\end{figure}

 In \lam{13}{C}, the $\Lambda$ threshold is the lowest particle-decay
channel and the $p_\Lambda$ orbit is just bound. As noted earlier, the
$\sim\!11$-MeV $\gamma$-rays from the lowest $3/2^-$ and $1/2^-$ states 
were measured using an array of NaI detectors and the separation of the 
states, $152\pm 54({\rm stat})\pm 36({\rm syst})$, was determined from
the shift in the peak with pion scattering angle~\cite{ajimura01,kohri02}.
Figure~\ref{fig:lc13} shows the $p^8p_\Lambda$ states  based on the lowest 
$0^+$ and $2^+$ states of the $^{12}$C core. From an older BNL
experiment~\cite{may83}, the separation between the two $1/2^-$ states
was determined to be $6.0\pm 0.4$ MeV while that of the $1/2^-_2$ and 
$5/2^-_2$ states was $1.7\pm 0.4$ MeV. The doublets are characterized
by the quantum number ${\cal L}$ and split by the spin-dependent
interactions where~\cite{auerbach81,auerbach83}
\begin{equation}
{\cal \bm{L}} = \bm{J}_c+\bm{l}_\Lambda\quad {\rm and}\quad
\bm{J} = {\cal \bm{L}}+\bm{s}_\Lambda .
\label{eq:scriptl}
\end{equation}
The spectrum, including $\Lambda$-$\Sigma$ coupling, can be calculated from
the Gaussian or Yukawa representations of the G matrices derived from the 
free $YN$ interaction model. Beause the $p_\Lambda$ states are only bound by
about 0.8 MeV, the calculation is performed using Woods-Saxon wave functions 
for this binding energy. One can also use an interaction obtained by
adjusting the strengths in the various $\Lambda N$ channels to reproduce
the $p_Ns_\Lambda$ matrix elements in Eq.~(\ref{eq:param11}). There are
20 independent $p_Np_\Lambda$ matrix elements and pieces of the
interactions such as the even-state tensor interaction enter. Furthermore,
a $Q_N\cdot Q_\Lambda$ multipole component of the interaction is active
as compared to just the spatial monopole for $p_Ns_\Lambda$. It is this
quadrupole component that splits the ${\cal L}\!=1,2,3$ states of the 
$2^+\times p_\Lambda$ multiplet in Fig.~\ref{fig:lc13}. This can involve 
strong mixing of the $p_{1/2}$ and $p_{3/2}$ $\Lambda$ states to make
states with good ${\cal L}$~\cite{auerbach83}.

 For $p_Ns_\Lambda$,
there is no way to separate the contributions from the even- and odd-state
central interactions. However, for $p_Np_\Lambda$ different strengths 
in the even- and odd-state central interactions give rise to a space-exchange
interaction that will separate states with different spatial symmetries.
Coupling a $p_\Lambda$ to the dominantly [44] states of $^{12}$C leads to
[54] and [441] symmetries for the nine $p$-shell baryons. These are not very
good quantum numbers for the hypernuclear states. Nevertheless, the uppermost
doublet in Fig.~\ref{fig:lc13} tends towards [441] symmetry; note the large
structure factor for the substitutional $1/2^-$ state reached via 
$\Delta L\!=\!0,\Delta S\!=\!0$ from the $^{13}$C ground state in the
\Kpi\ reaction. The excitation energy of this doublet is indeed sensitive
to the space-exchange interaction. For example, the NSC97f interaction has
repulsion in both singlet- and triplet-odd states producing a too
large separation of $\sim 6.9$ MeV from the lower ${\cal L}\!=\!1$ doublet
and a separation of $\sim 2.2$ MeV from the ${\cal L}\!=\!3$ doublet.
On the other hand, the ESC04 model~\cite{rijken06a} has repulsion in the 
singlet-odd channel and attraction in the triplet-odd channel giving 6.0 
and 1.2 MeV for the two separations. We note that the $^{12}$C ground state
has a considerable $L\!=\!1,S\!=1$ component that allows various
spin-dependent components of the $\Lambda N$ interaction to contribute
to the spacing of the lowest $1/2^-$ and $3/2^-$ states, in contrast
to the situation for the $3\alpha +\Lambda$ model~\cite{hiyama00}. The
tensor interaction and the $\Lambda$-$\Sigma$ coupling both work to
put the $1/2^-$ state below the $3/2^-$ state.

 The $p^np_\Lambda$ shell-model calculations were performed~\cite{auerbach83} 
to understand \Kpi\ reaction data coming from CERN and BNL. While these
calculations have been updated to include  $\Lambda$-$\Sigma$ coupling 
and the use of realistic radial wave functions, they need to be extended
to full $1\hbar\omega$ calculations that include an $s_\Lambda$ coupled to
$1\hbar\omega$ states of the core nucleus. These states are mixed with
the $p^np_\Lambda$ states both by the $\Lambda N$ interaction and by
the requirement that the physical  $1\hbar\omega$ states are free from spurious
center-of-mass components. The need for such calculations is apparent in
the extra structure near the $p_\Lambda$ peak in Fig.~\ref{fig:specb12l}
and the fact that a number of $p$-shell hypernuclear $\gamma$-rays are
seen in \lamb{9}{Be}, \lam{11}{B}, and \lam{15}N following proton emission 
from the primary hypernucleus. In the latter case, the $p^{n-1}(sd)s_\Lambda$ 
component in the wave function gives the $(sd)$ proton spectroscopic factor 
that controls the relative population of states in the daughter hypernucleus.

 The Nijmegen baryon-baryon interactions have continued to evolve
with a variety of ESC04~\cite{rijken06a} and ESC08 (Rijken, Nagels, and 
Yamamoto, 2010; Nagels, Rijken, and Yamamoto, 2015b) models becoming 
available. The improvements cover many aspects from strangeness 0 to -4 
(Rijken, Nagels, and Yamamoto, 2013). As far as $p$-shell spectra are
concerned, it is found that ESC04a and ESC04b do a reasonable job while
ESC04c and ESC04d do not~\cite{halderson08}. In addition, the tensor 
interaction is too weak (wrong ordering of the ground-state doublet in
\lam{16}{O}) and the $\Lambda N$-$\Sigma N$ coupling potentials have an
unusual radial behavior. For the ESC08 models, the strength of the
$\Lambda$-spin-dependent spin-orbit interaction has been reduced with
respect to earlier models (Yamamoto, Motoba, and Rijken, 2010) as demanded 
by the data. However, the ordering of many doublets in the $p$-shell 
hypernuclei are inverted because the combination of attractive triplet-even 
and triplet-odd central interactions makes the triplet interaction
stronger than the singlet ($\Delta < 0$). As noted in the section on
$s$-shell hypernuclei, all of the models are missing something. In
practice, empirical adjustments to the derived G-matrix interactions are 
made to fit the available data. Of course, these fits also cover for
the missing three-body interactions, the effect of which is likely to
be mostly on the absolute binding energies and on vector (SLS and ALS)
interactions in the core nuclei (represented phenomenologically by $S_N$).

\subsection{Weak decays of $\Lambda$ hypernuclei} 
\label{subsec:weakdecay}

\subsubsection{Mesonic decays} 
\label{subsubsec:mesonicdecay}

$\Lambda$ hypernuclei are unstable to weak decays of the $\Lambda$ 
hyperon. In free space, the $\Lambda$ weak-interaction lifetime 
$\tau_{\Lambda}=\hbar/\Gamma_{\Lambda}^{\rm free}=2.632 \times 10^{-10}$~s 
is dominated ($99.7\%$) by nonleptonic, mesonic two-body decay \cite{pdg14}: 
\begin{equation} 
\Lambda \rightarrow p + \pi^{-} + 38~{\rm MeV}~~~(63.9\pm 0.5)\%,  
\label{eq:Gamma_pi-} 
\end{equation} 
\begin{equation} 
\Lambda \rightarrow n + \pi^{0} + 41~{\rm MeV}~~~(35.8\pm 0.5)\%. 
\label{eq:Gamma_pi0} 
\end{equation} 
The ratio 
$\Gamma_{\Lambda\to p+\pi^-}^{\rm free}/
\Gamma_{\Lambda\to n+\pi^0}^{\rm free}$ 
for these branches is close to 2, in agreement with the $\Delta I = 1/2$ 
rule \cite{boyle13} which is also satisfied to a level of a few percent by 
all other known strangeness-changing nonleptonic weak decays, e.g. in kaon 
decays. In contrast, a purely $\Delta I = 3/2$ rule would give a branching 
ratio 1/2. The effective $\Lambda \to N \pi$ weak decay Lagrangian is written 
as 
\begin{equation} 
\label{eq:wdlagr} 
\mathcal{L}^W_{\Lambda N \pi} = -iG_F{m_{\pi}^2}{\bar \psi}_N
(A+B\gamma_5)\bm{\tau}\cdot\bm{\phi}_{\pi}\psi_{\Lambda}, 
\end{equation} 
where $G_F{m_{\pi}^2}=2.211 \times 10^{-7}$, and $A=1.06,~B=-7.10$ are 
fixed by the measured free-space $\Lambda$ decay parameters. The isospin  
operator $\bm{\tau}$ imposes the $\Delta I = 1/2$ rule once the $\Lambda$ 
hyperon is assigned a fictitious isospin state $(I,I_z)=(1/2,-1/2)$. 
The nonrelativistic approximation to the free $\Lambda$ decay width yields 
\begin{eqnarray} 
\label{eq:gammafree1} 
\nonumber 
\Gamma^\text{free}_{\alpha}&=&c_{\alpha}(G_F{m_{\pi}^2})^2\int{\frac{d^3\bm{q}}
{(2\pi)^3 2\omega(\bm{q})}2\pi \delta(m_{\Lambda}-\omega(\bm{q})-E_N)} \\ 
& &\mbox{} \times (S^2+\frac{P^2}{m_{\pi}^2}\bm{q}^2), 
\end{eqnarray} 
where $c_{\alpha}=1,2$ for $\alpha=\Lambda\to n \pi^0,~\Lambda\to p \pi^-$, 
respectively, $S=A,~P/m_{\pi}=B/(2m_N)$, and $E_N$ and $\omega(\bm{q})$ are 
the total energies of the emitted nucleon and $\pi$ meson, respectively. 
This leads to the following expression for the total free-space decay width: 
\begin{equation} 
\label{eq:gammafree2} 
\Gamma_{\Lambda}^{\rm free}=\frac{3}{2\pi}(G_F{m_{\pi}^2})^2 
\frac{m_N q_{\rm c.m.}}{m_{\Lambda}} (S^2+\frac{P^2}{m_{\pi}^2}q_{\rm c.m.}^2)~, 
\end{equation} 
with $q_{\rm c.m.} \approx 100$ MeV/c for the pion momentum in the 
center-of-mass frame. 

The empirical $\Delta I = 1/2$ rule \cite{boyle13} is not well understood. 
However here a key question is whether, and to what extent, it is satisfied 
by {\it in medium} $\Lambda$ weak decays. There has been no unambiguous 
experimetal test of the validity of this rule in hypernuclei. One reason 
is the difficulty to resolve two-body exclusive decay channels in the 
continuum, where a combination of several isospin values for the residual 
nucleus washes out the effect of the primary $\Delta I = 1/2$ weak decay. 
For example, the total mesonic decay widths of $^4_\Lambda$He given in 
Table~\ref{tab:WD1} naively suggest that a $\Delta I = 3/2$ rule holds. 
However, realizing the dominance of the {\it two-body} decay 
$^4_{\Lambda}{\rm He} \rightarrow \pi^0 + {^4{\rm He}}$, and the impossibility 
of a $\pi^- + {^4{\rm He}}$ two-body final state owing to charge conservation, 
the reversal of the $\pi^-/\pi^0$ ratio from close to 2 in the free-space 
decay to close to 1/2 in $^4_{\Lambda}{\rm He}$ decay only reflects the 
dominance of the $^4$He ground-state branch. A similar trend is also seen in 
the $\pi^-/\pi^0$ ratio of \lam{12}{C} total mesonic decay widths listed 
in the table. On the other hand, the $\pi^-/\pi^0$ ratio for $^5_\Lambda$He is 
close to the free-space ratio, reflecting the difficulty to divert sufficient 
kinetic energy to break up the $^4$He core in the quasi-free decays 
$^5_\Lambda{\rm He} \to {^4{\rm He}}+N+\pi$. The systematics of the 
$\pi^-/\pi^0$ ratio, owing to the nuclear structure of $p$-shell 
$\Lambda$ hypernuclei, was discussed by  Motoba \textit{et al.} (1988).  

\begin{table}[t]
\caption{Measured total pionic decay widths of selected hypernuclei in 
units of $\Gamma_{\Lambda}^{\rm free}$.} 
\label{tab:WD1} 
\begin{ruledtabular} 
\begin{tabular}{lccl} 
$^A_\Lambda$Z & $\Gamma_{\pi^-}$ & $\Gamma_{\pi^0}$ & 
\multicolumn{1}{c}{Reference} \\
\hline 
$^4_\Lambda$He & $0.289 \pm 0.039$ & $0.604 \pm 0.073$ & Parker \textit{et al.}
(2007) \\  
$^5_\Lambda$He & $0.340 \pm 0.016$ & $0.201 \pm 0.011$ & Kameoka \textit{et al.}
(2005), \\
 & & & ~~Okada \textit{et al.} (2005) \\ 
$^{12}_{~\Lambda}$C & $0.123\pm 0.015$ & $0.165\pm 0.008$ & Kameoka \textit{et al.}
(2005), \\
 & & & ~~Okada \textit{et al.} (2005) \\ 
$^{28}_{~\Lambda}$Si & $0.046 \pm 0.011$ &--& Sato \textit{et al.} (2005) \\
$_\Lambda$Fe & $\leq 0.015~ (90\%~{\rm CL})$ &--& Sato \textit{et al.} (2005) \\
\end{tabular}
\end{ruledtabular}
\end{table}

Another reason for the difficulty of testing the $\Delta I = 1/2$ rule 
in mesonic decays of hypernuclei is the rapid decrease of the pionic decay 
width $\Gamma_{\pi} = \Gamma_{\pi^-} + \Gamma_{\pi^0}$ as a function of 
hypernuclear mass number $A$. This is shown in Table \ref{tab:WD1} where 
some of the latest determinations of $\pi^-$ decay widths in hypernuclei 
for $A \geq 11$ are listed \cite{sato05}. The pionic decay widths fall off 
from about $0.9~\Gamma_{\Lambda}^{\rm free}$ in $^4_{\Lambda}{\rm He}$ to 
few percent in $_\Lambda$Fe. This had been anticipated from the low momentum 
$q \approx 100$~MeV/c, $q<p_F$, of the nucleon recoil in the pionic decay 
and was indeed confirmed quantitatively by detailed calculations of mesonic 
decay of $\Lambda$ hypernuclei. Equation (\ref{eq:gammafree1}) for the 
free-space decay width is replaced in hypernuclei by 
\begin{eqnarray} 
\label{eq:gammamed} 
\nonumber 
\Gamma_{\alpha}&=& c_{\alpha}(G_F{m_{\pi}^2})^2 \sum_f \int{\frac{d^3\bm{q}} 
{(2\pi)^3 2\omega(\bm{q})}2\pi\delta(E_{\Lambda}-\omega(\bm{q})-E_N^f)} 
\nonumber \\
& & \mbox{} \times \left( S^2\mid \int{d^3\bm{r}\phi_{\Lambda}(\bm{r})
\phi_{\pi}(\bm{r};\bm{q}) \phi_f^*(\bm{r})}\mid^2 \right. \nonumber \\
& & \mbox{}+ \left. \frac{P^2}{m_{\pi}^2}\mid \int{d^3\bm{r}\phi_{\Lambda}(\bm{r})
\bm{\nabla}\phi_{\pi}(\bm{r};\bm{q})\phi_f^*(\bm{r})}\mid^2 \right),
\end{eqnarray} 
where the sum extends over the unoccupied nucleon states $f$, and the pion 
wavefunction $\phi_{\pi}(\bm{r};\bm{q})$ is a solution of the Klein Gordon 
equation in the presence of a pion-nuclear optical potential $V_{\rm opt}$: 
\begin{equation} 
\label{eq:KG} 
\left\{ \bm{\nabla}^2 - m_{\pi}^2 -2\omega(\bm{q})V_{\rm opt}(\bm{r}) + 
\left[ \omega - V_c(\bm{r})\right]^2 \right\} \phi_{\pi}(\bm{r};\bm{q}) = 0\, . 
\end{equation} 
The free-space Eq.~(\ref{eq:gammafree1}) is recovered from 
Eq.~(\ref{eq:gammamed}) by extending the sum over occupied nucleon states 
as well, neglecting the pion-nuclear final-state interaction, i.e., 
$\phi_{\pi}^{\rm free}(\bm{r};\bm{q}) = \exp(i\bm{q}_\text{c.m.}\cdot \bm{r})$, 
and using closure. The reduction of the mesonic decay width in hypernuclei 
by several orders of magnitudes as $A$ increases is due to limiting the 
sum to unoccupied nucleon states. In realistic calculations, however, the 
final-state nuclear interaction of the emitted pion plays a significant 
role, providing {\it enhancement} of the decay rate in heavy hypernuclei 
by one to two orders of magnitude over what a plane-wave impulse
approximation calculation (using $\phi_{\pi}^{\rm free}(\bm{r};\bm{q})$) would 
give (Oset and Salcedo, 1985; Itonaga, Motoba, and Band\={o}, 1988; Nieves and
Oset, 1993; Motoba and Itonaga, 1994). 

\begin{table*}[t]
\caption{Hypernuclear spin assignments provided by pionic weak-decay studies.} 
\label{tab:WD2} 
\begin{ruledtabular} 
\begin{tabular}{lccll} 
$^A_\Lambda$Z & $J^{\pi}$ & Decay branch & \multicolumn{1}{c}{Theory} & 
\multicolumn{1}{c}{Experiment} \\
\hline 
$^3_\Lambda$H & ${\frac{1}{2}}^+$ & $\pi^- + ^3$He & Dalitz and Liu (1959)
 & Ammar, Dunn, and Holland (1962), \\
 & & & & ~~Block \textit{et al.} (1964), and Bertrand \\
 & & & & ~~\textit{et al.} (1970) \\
$^4_\Lambda$H & $0^+$ & $\pi^- + ^4$He &  Dalitz and Liu (1959) & Ammar
 \textit{et al.} (1961), Block, Lendinara, \\
 & & & & ~~and Monari (1962), Block \textit{et al.} (1964), \\
 & & & & ~~and Bertrand \textit{et al.} (1970) \\
$^4_\Lambda$He & $0^+$ & $\pi^0 + {\rm all}$ & Dalitz and Liu (1959)
 & Block \textit{et al.} (1964) and Fetkovich \\
 & & & & ~~\textit{et al.} (1972) \\
$^7_\Lambda$Li & ${\frac{1}{2}}^+$ & $\pi^ - + ^7$Be$^*(429$~keV) & 
Motoba \textit{et al.} (1988) & Sasao \textit{et al.} (2004) \\ 
 & & & ~~Motoba and Itonaga (1994) & \\
$^8_\Lambda$Li & $1^-$ & $\pi^- + ^4$He $+ ^4$He & 
Dalitz (1963a) & Davis, Levi Setti, and Raymund (1963) \\ 
$^{11}_{~\Lambda}$B & ${\frac{5}{2}}^+$ & $\pi^ - + ^{11}$C$^*(6.48$~MeV) & 
Ziemi\'{n}ska (1975) & Juri\v{c} \textit{et al.} (1973) \\ 
$^{12}_{~\Lambda}$B & $1^-$ & $\pi^ - + ^4$He $+ ^4$He $+ ^4$He & 
Ziemi\'{n}ska and Dalitz (1975) & Kielczewska  \textit{et al.} (1975) \\ 
 & & & ~~and  Kielczewska, Ziemi\'{n}ska, & \\
 & & & ~~and Dalitz (1975) & \\
$^{15}_{~\Lambda}$N & ${\frac{3}{2}}^+$ & $\pi^ - + ^{15}$O$_{\rm g.s.}$ &   
Gal (2009) & Agnello \textit{et al.} (2009) \\ 
\end{tabular}
\end{ruledtabular}
\end{table*} 

A weak $\pi^+$ decay branch with width of order $0.02~\Gamma_{\Lambda}^{\rm free}$ 
was observed in the decay of $^4_\Lambda$He in emulsion studies \cite{bohm69} 
and in helium bubble chambers \cite{fetkovich72}. Weaker evidence exists for 
$\pi^+$ decay of $^7_\Lambda$Be observed in emulsion. The {\it rare} $\pi^+$
branch was initially studied theoretically by Dalitz and von Hippel 
\cite{dalitz64b,vonhippel64} who observed that it required an intermediate 
strong-interaction step to occur through, e.g. (i) $\Lambda \to n + \pi^0$ 
followed by ($\pi^0,\pi^+$) charge exchange in the final state, or 
(ii) $\Lambda p \to \Sigma^+ n$, in order to generate a virtual $\Sigma^+$ 
component in the initial $\Lambda$ hypernuclear wavefunction followed by 
$\Sigma^+ \to n + \pi^+$. The pion charge-exchange mechanism was recalculated 
by Ciepl{\'{y}} and Gal (1997) where its rate was found larger than in the 
original calculation \cite{dalitz64b}, but still short by about a factor of 
two with respect to the observed rate. Gibson and Timmermans (1998) argued 
that relatively large $\Sigma^+$ admixtures were unique to $^4_\Lambda$He and 
could explain the large $\pi^+$ rates observed.  

\begin{figure}[t]
\includegraphics[width=8cm]{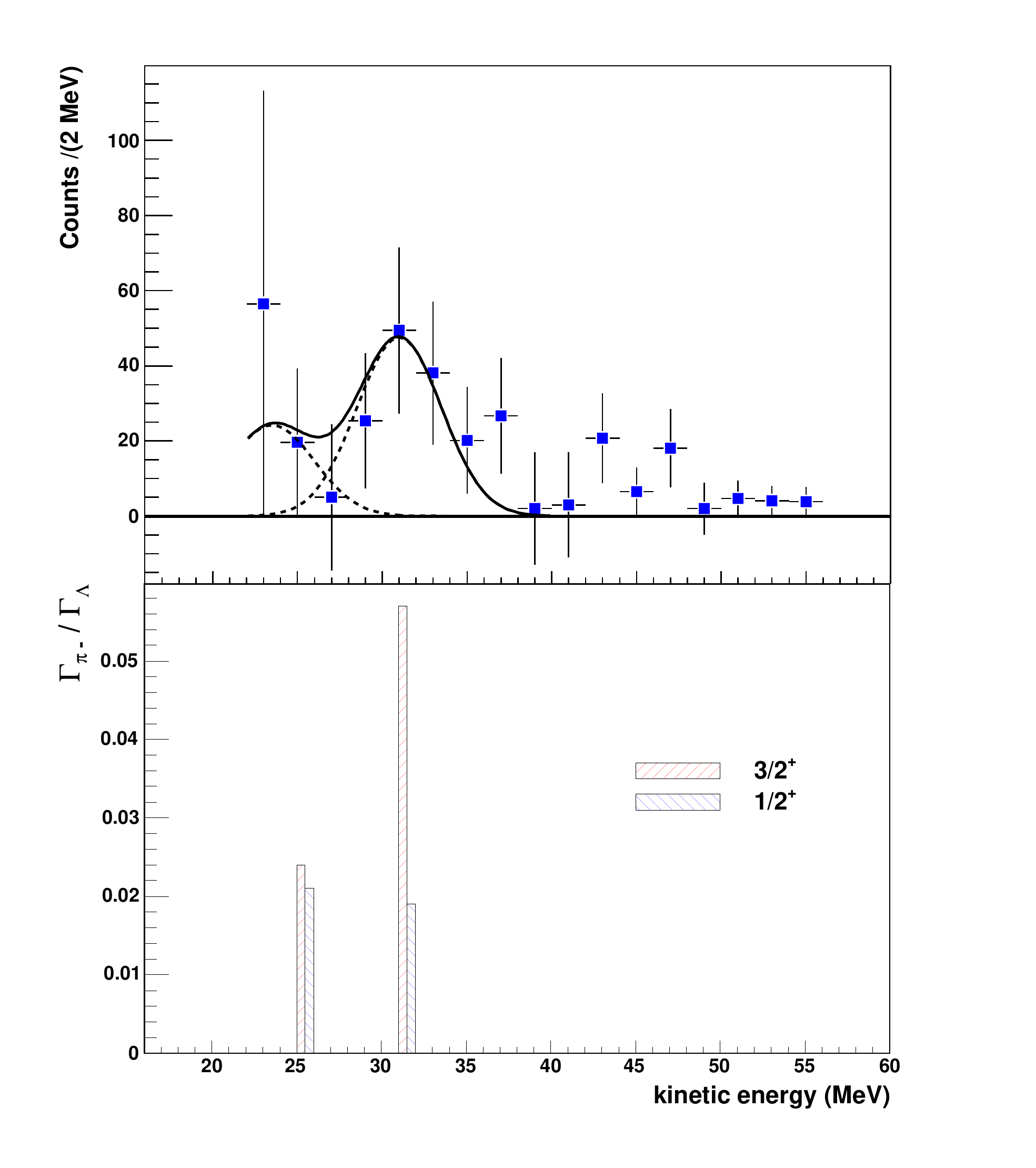}
\caption{Mesonic weak-decay spectrum of 
\lam{15}{N}\,$\to \pi^- + ^{15}$O 
(upper panel) observed at DA$\Phi$NE by the FINUDA Collaboration, compared to 
calculations (lower panel) for the two possible spin values of the decaying 
$\Lambda$ hypernucleus \cite{gal09} which show preference for a \lam{15}{N} 
g.s. spin $3/2^+$. Adapted from Agnello \textit{et al.}, 2009.}
\label{fig:fingal09} 
\end{figure} 

The study of exclusive two-body pionic weak decays of light hypernuclei 
has yielded valuable information on the ground-state spins of several 
species, as summarized in Table~\ref{tab:WD2}. These pionic weak decays 
show selectivity to the spin of the hypernuclear ground state owing to the 
dominance ($88\%$) of the $s$-wave, parity-violating $\Lambda \to N \pi$ 
amplitude [$A$ term in Eq.~(\ref{eq:wdlagr})]. This is demonstrated in 
Fig.~\ref{fig:fingal09}, taken from a recent FINUDA work \cite{agnello09}, 
showing a $\pi^-$ weak-decay spectrum for \lam{15}{N}, with a preference for 
a g.s. spin $3/2^+$ for \lam{15}{N} \cite{gal09}. 
In terms of nuclear-core spin $J_c$ values the derived hypernuclear spins 
$J$ satisfy $J=J_c - \frac{1}{2}$ in the $s$ shell and $p_{\frac{3}{2}}$ 
subshell, and $J=J_c + \frac{1}{2}$ for \lam{15}{N} in the $p_{\frac{1}{2}}$ 
subshell, all consistent with the $\Lambda N$ spin-singlet interaction being 
stronger than the spin-triplet interaction.

\begin{figure}[t]
\includegraphics[width=8.0cm]{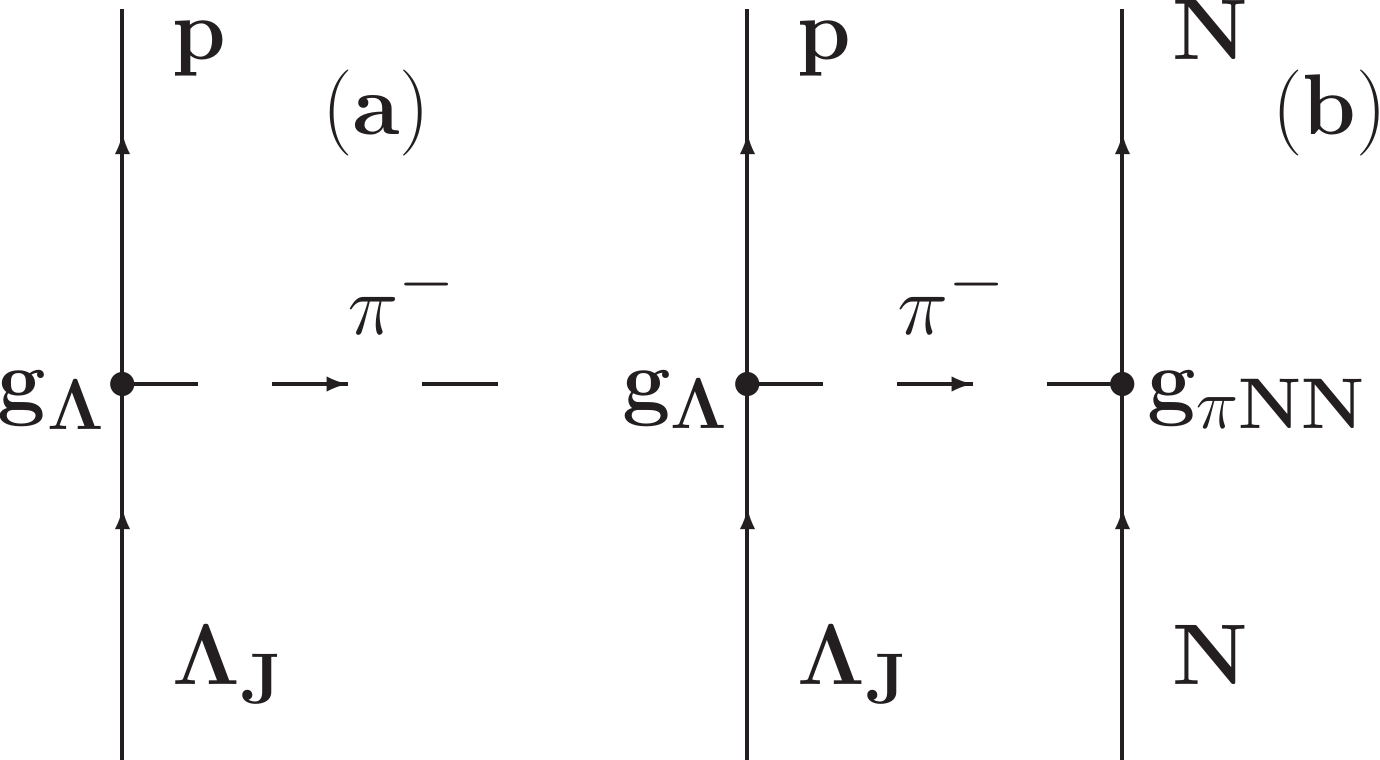} 
\caption{Graph (a) is for mesonic $\Lambda_J \to p \pi^-$ decay, where 
$\Lambda_J$ denotes a $\Lambda$ hyperon of total spin $J$. Graph (b) depicts 
nonmesonic de-excitation for a $\Lambda_J$ hyperon in nuclear matter. 
 Adapted from Dalitz, 2005.}
\label{fig:wdalitz} 
\end{figure} 

\subsubsection{Nonmesonic decays}

$\Lambda$ hypernuclear total decay widths $\Gamma_{\Lambda}$ are known to 
remain close to the free-$\Lambda$ decay width $\Gamma_{\Lambda}^{\rm free}$, 
in spite of the rapid decrease as a function of $A$ of the $\Lambda \to N \pi$ 
mesonic weak decay (MWD) widths $\Gamma_{\pi}$, as demonstrated in 
Table~\ref{tab:WD1}. A new mode of nonmesonic weak decay (NMWD), predicted by 
Cheston and Primakoff (1953), emerges upon increasing $A$ through the 
absorption of a weak-decay, virtual pion on one or more nucleons, as 
illustrated in Fig.~\ref{fig:wdalitz}. Other weak-decay virtual mesons may 
also mediate these NMWD modes. Historically, Karplus and Ruderman (1956) used 
the observed rates of the nonmesonic weak decay of $\Lambda$ hypernuclei to 
argue that the spin of the $\Lambda$ hyperon was consistent with 
$J_{\Lambda}=1/2$, and that there was no need to ascribe the relatively long 
lifetimes of strangeness weak decays to an exceptionally large value of 
$J_{\Lambda}$. 

\begin{figure}[t]
\includegraphics[width=8cm]{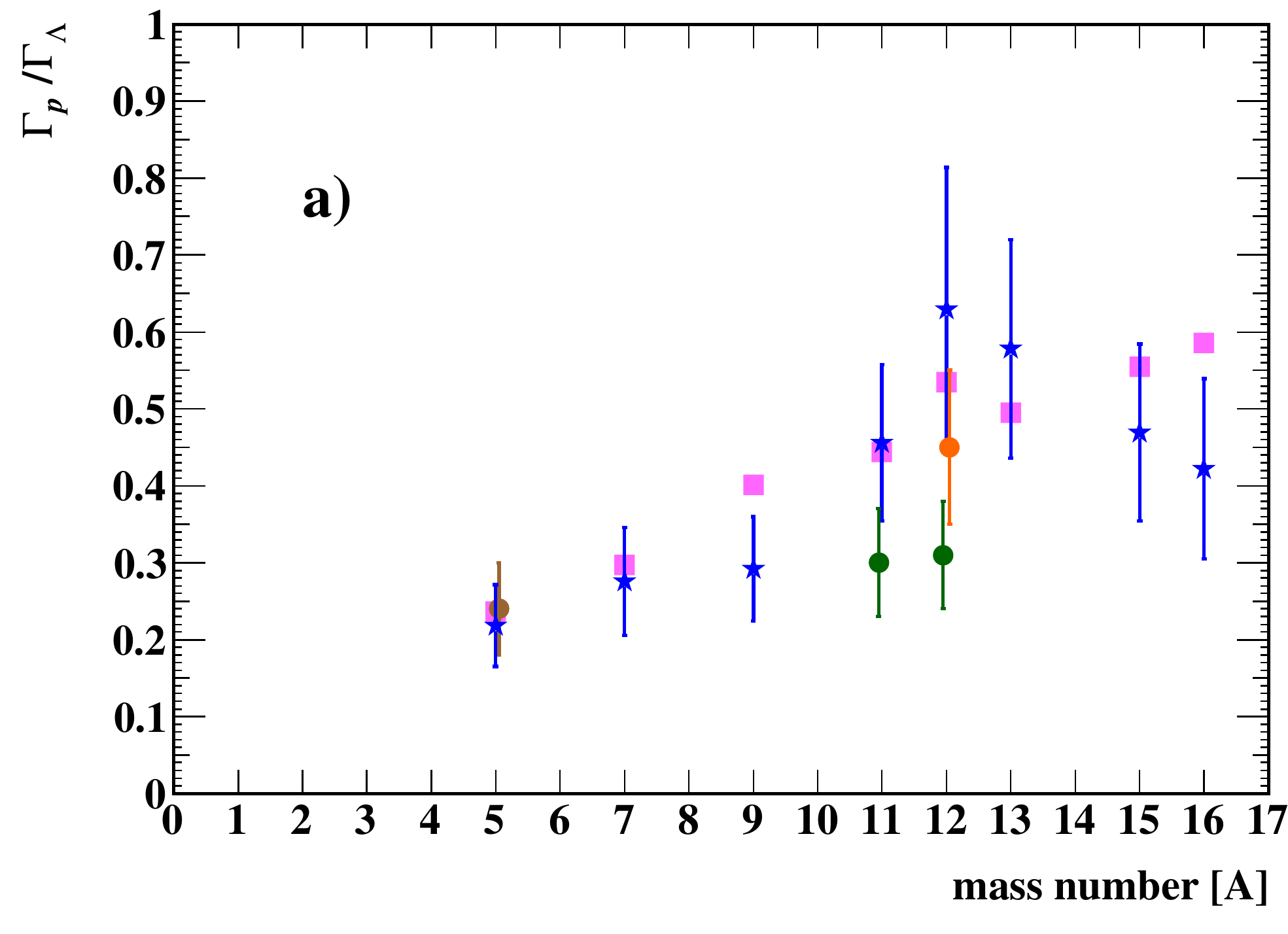}
\includegraphics[width=8cm]{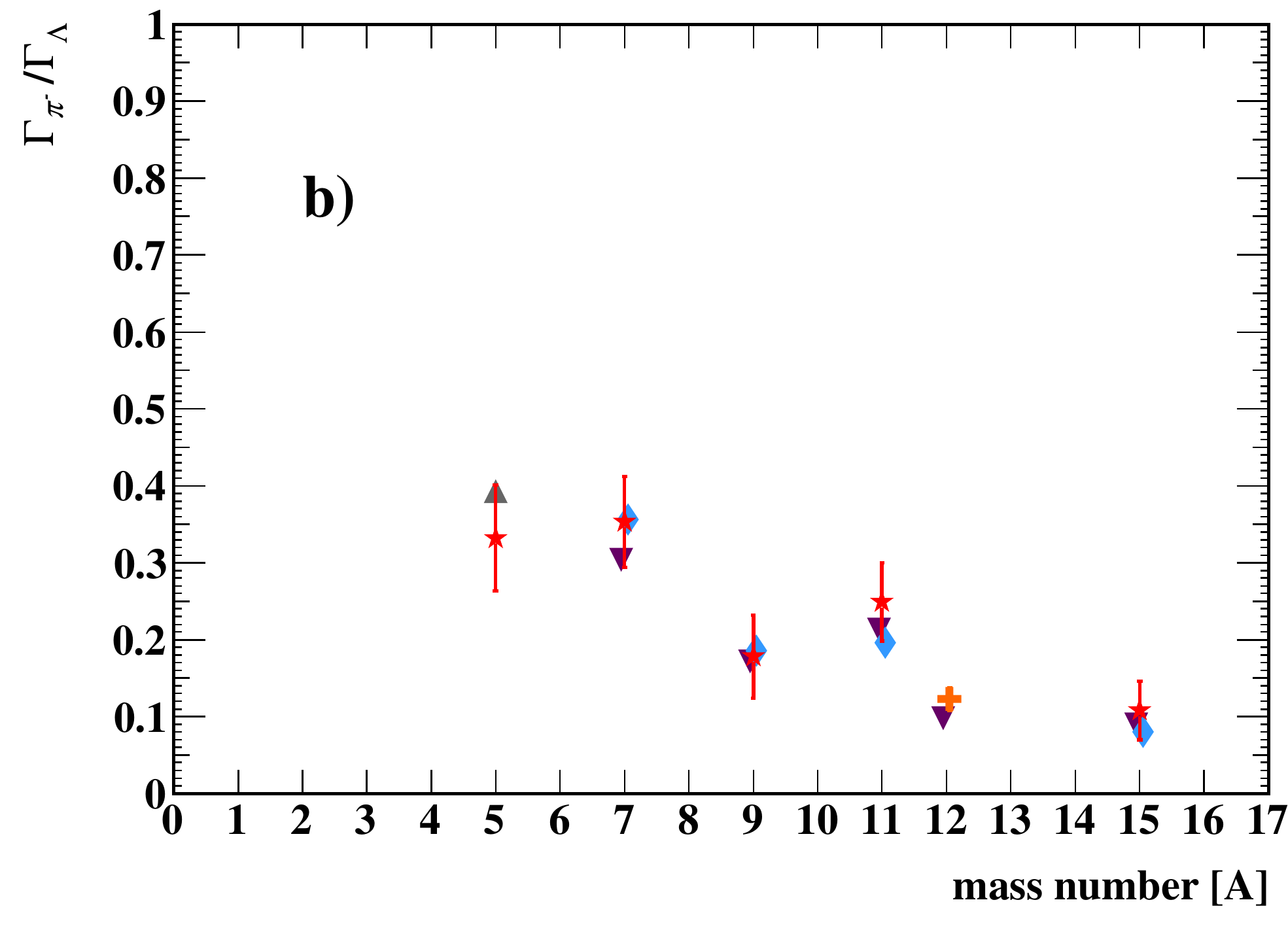}
\caption{$\Gamma_p$ (blue stars, upper panel) and 
$\Gamma_{\pi^-}$ (red stars, lower panel), in units of the free $\Lambda$ 
decay width, as a function of $A$ from measurements and analysis reported 
by the FINUDA Collaboration \cite{agnello09,agnello14}. Other experimental 
results and theoretical calculations are also marked, see caption to the 
original Fig.~3 in Agnello \textit{et al.}, 2014. \label{fig:botta14}} 
\end{figure} 

The dominant NMWD modes are believed to involve one nucleon 
in the initial state:  
\begin{equation} 
\Lambda + p \to n + p + 176~{\rm MeV}~~(\Gamma_p), 
\label{eq:Gamma_p} 
\end{equation} 
\begin{equation} 
\Lambda + n \to n + n + 176~{\rm MeV}~~(\Gamma_n), 
\label{eq:Gamma_n} 
\end{equation} 
having a summed width $\Gamma_1 = \Gamma_p + \Gamma_n$. 
Two-nucleon ($2N$) modes are also possible \cite{alberico91}, 
\begin{equation} 
\Lambda + N + N \to n + N + N + 176~{\rm MeV}~~(\Gamma_2). 
\label{eq:Gamma_2} 
\end{equation}     
A conservative estimate given by Alberico \textit{et al.} (1991) 
for these decays is $\Gamma_2/\Gamma_1 \sim 0.2$. The total hypernuclear 
weak-decay width, $\Gamma_{\Lambda}=\Gamma_{\pi}+\Gamma_{\rm nm}$, 
is a sum of the MWD width $\Gamma_{\pi}$ and the NMWD width, denoted by 
$\Gamma_{\rm nm} = \Gamma_1 + \Gamma_2 + \cdots$. The dots stand for more 
involved multinucleon decay modes. Very little is known about multinucleon 
decay modes beyond the two-nucleon mode as most experimental and theoretical 
studies of $\Lambda$ hypernuclear weak decay have focused on the one-nucleon 
modes, Eqs.~(\ref{eq:Gamma_p}) and (\ref{eq:Gamma_n}). The branching ratio of 
the $2N$ NMWD contribution to the total \lam{12}{C} NMWD width has been 
determined in KEK \cite{kim09} and in DA$\Phi$NE \cite{agnello11b} experiments,
with values given by 
\begin{equation} 
\frac{\Gamma_2}{\Gamma_{\rm nm}} =~~0.29\pm 0.13,~~~0.21\pm 0.08 , 
\label{eq:gamma2} 
\end{equation} 
respectively. The latter value was derived from analysis of several NMWD 
spectra, assuming that this branching ratio is constant in the $p$ shell. 
The $2N$ NMWD mode was observed recently through a complete kinematical 
reconstruction of a three-nucleon final state in two \lamb{7}{Li}~$\to 
{^4{\rm He}}+n+n+p$ decay events at DA$\Phi$NE \cite{agnello12b}, 
as demonstrated earlier in Fig.~\ref{fig:2Nfinuda}. 

NMWD dominates the $\Lambda$-hypernuclear decay in all but the lightest 
hypernuclei. This is demonstrated in Fig.~\ref{fig:botta14} where $\Gamma_p$, 
the largest contributor to NMWD, and $\Gamma_{\pi^-}$, the largest contributor 
to MWD, are shown as a function of $A$ along the $p$ shell as determined by 
FINUDA and in comparison to various calculations. It is seen clearly that 
$\Gamma_p$ rises roughly by a factor of 2, whereas $\Gamma_{\pi^-}$ decreases 
roughly by a factor of 3 from \lamb{5}{He} to \lam{15}{N}, with the ratio 
$\Gamma_p/\Gamma_{\pi^-}$ reaching a value somewhat larger than 4 at the end 
of the $p$ shell. NMWD is the only practical way to study the four-fermion, 
weak-decay interaction. The relatively large momentum transfer, 
$\approx$420~MeV/c in free space, could mean that sub-nucleon degrees of 
freedom are important, but at the present level of experimental data there 
seems no advantage to invoke explicitly sub-nucleon models. The status of 
models that consider direct quark (DQ) processes, in addition to meson 
exchanges, is summarized by Sasaki, Izaki, and Oka (2005). DQ models offer a 
natural 
theoretical framework for departing from the $\Delta I =1/2$ rule. However, 
there is no compelling evidence so far that this rule is not satisfied in 
$\Lambda$ hypernuclear NMWD. The models reviewed here are hadronic models 
that are built upon meson exchanges for which the $\Delta I =1/2$ rule is 
assumed to hold. A common approximation is that NMWD occurs dominantly from 
$s$-wave $\Lambda N$ states owing to the short range nature of these decays. 
The possible $\Lambda+N\to N+N$ transitions are listed in Table~\ref{tab:WD3} 
as taken from Block and Dalitz (1963), together with the spin dependence of the 
corresponding matrix elements. Thus, for capture from $^1S_0$ states, parity 
nonconservation in the weak interactions allows both the parity-conserving 
(PC) $^1S_0\to{^1S_0}$ as well as the parity-violating (PV) $^1S_0\to{^3P_0}$ 
transitions. Of the six amplitudes listed, those with $a$, $c$, and $d$ are PC 
and those with $b$, $e$, and $f$ are PV; those with $c$, $d$, and $e$, leading 
to  $I=0$ $NN$ states, are unique to $\Lambda p\to np$ whereas for the $a$, $b$,
and $f$ amplitudes, which lead to $I=1$ $NN$ states, both $nn$ and $np$ final 
states are possible with $a_n=\sqrt{2}a_p,~b_n=\sqrt{2}b_p,~f_n=\sqrt{2}f_p$ 
satisfying the $\Delta I =1/2$ rule. 

\begin{table}[ht]
\caption{$\Lambda + N \to N + N$ amplitudes \cite{block63}. The Pauli spin 
operator, $\bm{\sigma}_{\Lambda}$, acts on the initial $\Lambda$ particle 
and the final neutron. The final neutron momentum is $\bm{q}$, and 
$\bm{Q} \equiv \bm{q}/m_N$.} 
\label{tab:WD3} 
\begin{ruledtabular} 
\begin{tabular}{lccc} 
Transition & Operator & $I_{NN}$ & Rate \\ \hline 
$^1S_0 \to {^1S_0} $ & $\frac{a}{4}(1- \bm{\sigma}_{\Lambda} \cdot 
\bm{\sigma}_N)$ & 1 & ${\mid a \mid}^2$ \\ 
$^1S_0 \to {^3P_0} $ & $\frac{b}{8}(\bm{\sigma}_{\Lambda}-\bm{\sigma}_N) 
\cdot {\bm Q}~(1- \bm{\sigma}_{\Lambda} \cdot \bm{\sigma}_N)$ & 1 & 
${\mid b \mid}^2 Q^2$ \\ 
$^3S_1 \to {^3S_1} $ & $\frac{c}{4}(3+\bm{\sigma}_{\Lambda} \cdot 
\bm{\sigma}_N)$ & 0 & ${\mid c \mid}^2$ \\ 
$^3S_1 \to {^3D_1} $ & $\frac{3d}{\sqrt 2}(\bm{\sigma}_{\Lambda} \cdot 
\bm{Q}~\bm{\sigma}_N \cdot \bm{Q} -\frac{1}{3}\bm{\sigma}_{\Lambda} 
\cdot \bm{\sigma}_N~Q^2) $ & 0 & ${\mid d \mid}^2 Q^4$ \\
$^3S_1 \to {^1P_1} $ & $\frac{e{\sqrt 3}}{8}(\bm{\sigma}_{\Lambda}-
\bm{\sigma}_N)\cdot \bm{Q}~(3+ \bm{\sigma}_{\Lambda} \cdot 
\bm{\sigma}_N)$ & 0 & ${\mid e \mid}^2 Q^2$ \\ 
$^3S_1 \to {^3P_1} $ & $\frac{f{\sqrt 6}}{4}(\bm{\sigma}_{\Lambda} + 
\bm{\sigma}_N)\cdot {\bf Q} $ & 1 & ${\mid f \mid}^2 Q^2$  \\ 
\end{tabular}
\end{ruledtabular}
\end{table}

\begin{figure}[t]
\includegraphics[width=8.0cm]{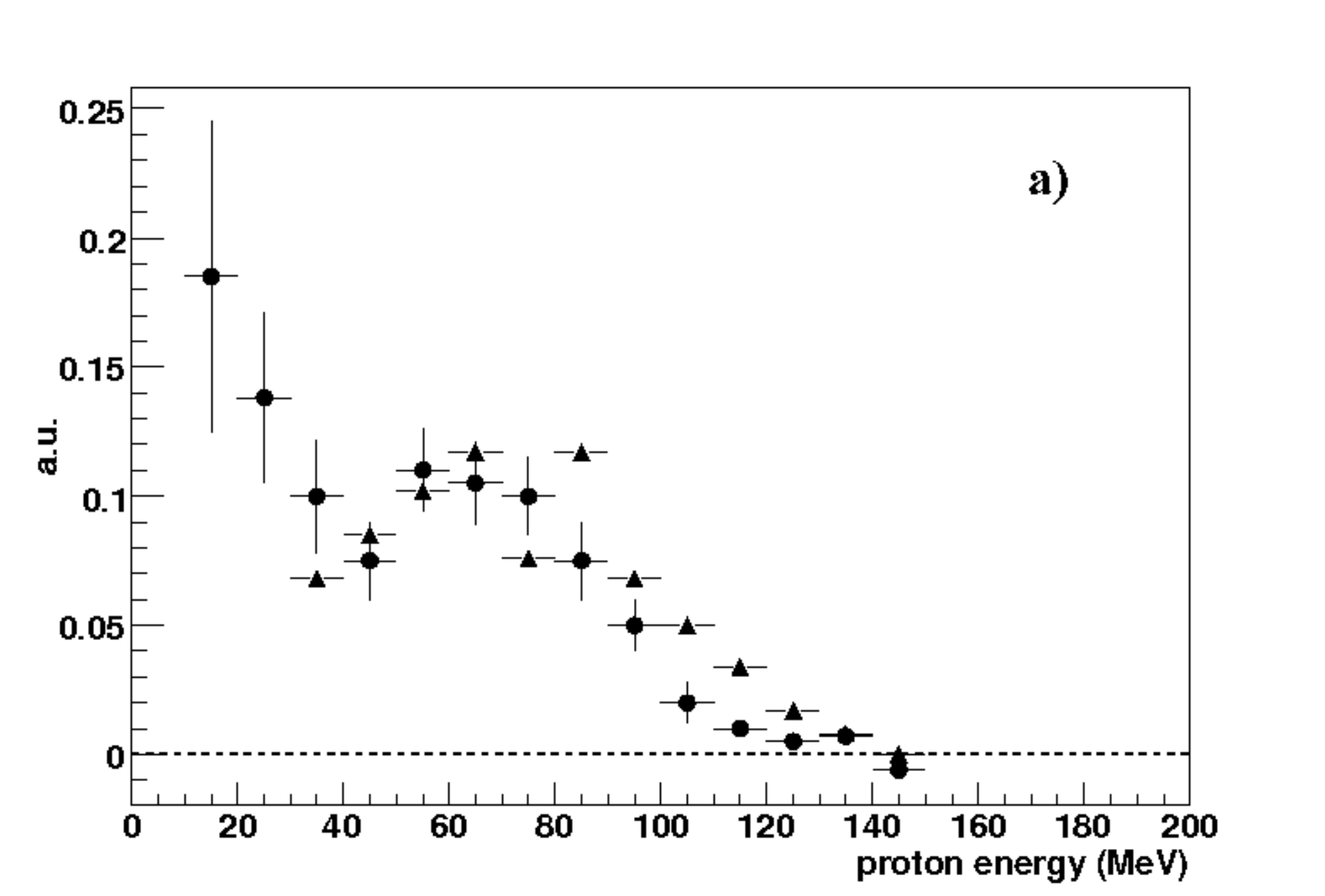} 
\includegraphics[width=8.0cm]{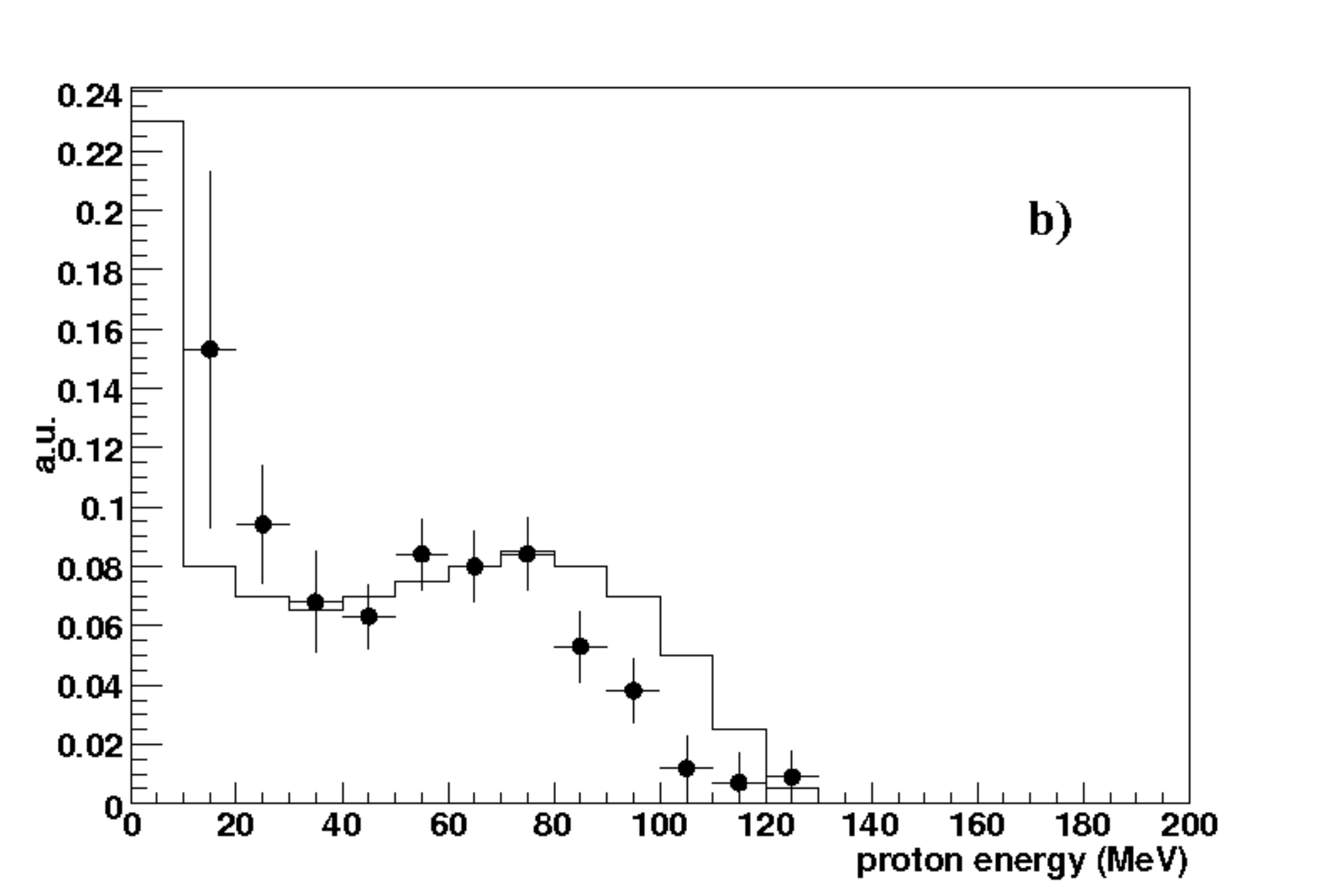} 
\caption{Upper panel: proton-energy spectrum from \lamb{5}{He} nonmesonic weak 
decay measured by FINUDA (circles) and at KEK (triangles). The two spectra 
were normalized beyond 35 MeV (threshold of the KEK spectrum). Lower panel: 
comparison between the FINUDA proton-energy spectrum (circles) from the upper 
panel and the INC calculation (histogram) of \cite{garbarino04}. The two 
spectra were normalized beyond 15 MeV (threshold of the FINUDA spectrum). 
Adapted from Agnello \textit{et al.}, 2008.} 
\label{fig:wdfinuda} 
\end{figure} 

\begin{table*}[t]
\caption{Measured and calculated NMWD widths and related entities for selected 
hypernuclei in units of $\Gamma_{\Lambda}^{\rm free}$.} 
\label{tab:WD4} 
\begin{ruledtabular} 
\begin{tabular}{llcc} 
Entity & \multicolumn{1}{c}{Method} & \lamb{5}{He} & \lam{12}{C} \\  
\hline 
$\Gamma_n/\Gamma_p$ & Emulsion ($_\Lambda$B,~$_\Lambda$C,~$_\Lambda$N) 
\cite{montwill74} &  & $0.59 \pm 0.15 $ \\ 
 & KEK E462/E508 \cite{kang06,kim06} & $0.45 \pm 0.11 \pm 0.03$ 
 & $0.51 \pm 0.13 \pm 0.05$  \\
 & OME+$2\pi+2\pi/\sigma$ \cite{chumillas07} & 0.415 & 0.366 \\
 & OME+$2\pi/\sigma+a_1$ (Itonaga \textit{et al.}, 2008; Itonaga and Motoba,
 2010) & 0.508 & 0.418 \\[4pt] 
$\Gamma_{nm}$&KEK E462/E508 \cite{okada05}& $0.406 \pm 0.020$ & 
$0.953 \pm 0.032$\\ 
 & OME+$2\pi+2\pi/\sigma$ \cite{chumillas07} & 0.388 & 0.722 \\ 
 & OME+$2\pi/\sigma+a_1$ (Itonaga \textit{et al.}, 2008; Itonaga and Motoba,
 2010)  & 0.358 & 0.758 \\[4pt] 
$\Gamma_{\Lambda}$ & KEK E462/E508 \cite{kameoka05} & $0.947 \pm 0.038$ & 
$1.242 \pm 0.042$ \\[4pt]
$a_{\Lambda}$ & KEK E462/E508 \cite{maruta07} & $0.07 \pm 0.08 +0.08$ & 
$-0.16 \pm 0.28 +0.18$ \\ 
 & OME \cite{chumillas07,chumillas08} & $-0.590$ & $-0.698$ \\ 
 & With final-state interactions &  $-0.401$ & $-0.340$ \\
 & OME+$2\pi+2\pi/\sigma$ \cite{chumillas07,chumillas08} & $+0.041$ &
 $-0.207$ \\
 & With final-state interactions  & $+0.028$ & $-0.126$ \\ 
 & OME+$2\pi/\sigma+a_1$ (Itonaga \textit{et al.}, 2008; Itonaga and Motoba,
 2010) & $+0.083$ & $+0.044$ \\ 
\end{tabular}
\end{ruledtabular}
\end{table*}

It is instructive to show the structure of the one-pion-exchange (OPE) 
transition potential generated by the diagram of Fig.~\ref{fig:wdalitz}b. 
To this end, the weak-interaction Lagrangian Eq.~(\ref{eq:wdlagr}) is 
augmented by a strong-interaction component  
\begin{equation} 
{\mathcal L}^S_{NN\pi} = - ig_{NN\pi}{\bar \psi}_N \gamma_5 
\bm{\tau}\cdot\bm{\phi}_{\pi}\psi_N , 
\label{eq:silagr} 
\end{equation} 
where $g_{NN\pi}=13.2$ is the strong-interaction coupling constant. 
Including the pion propagator between the two vertices given by 
Eqs.~(\ref{eq:wdlagr}) and (\ref{eq:silagr}) and applying a nonrelativistic 
reduction, one obtains the OPE momentum-space transition potential 
\begin{eqnarray} 
V_\text{OPE}(\bm{q}) & = &-G_F{m_{\pi}^2}\frac{g_{NN\pi}}{2m_N}
\left(A+\frac{B}{2m_{\rm av}}
\bm{\sigma}_{\Lambda}\cdot\bm{q}\right) \nonumber \\
& & \mbox{}\times \frac{\bm{\sigma}_N\cdot\bm{q}}
{\bm{q}^2+m_{\pi}^2}\bm{\tau}_{\Lambda}\cdot\bm{\tau}_N , 
\label{eq:Vope} 
\end{eqnarray} 
where $m_\text{av}=(m_N+m_{\Lambda})/2$. The OPE potential, owing to the 
sizable momentum transfer involved, is dominated by the tensor component, 
amplitude $d$ of Table~\ref{tab:WD3}. For this amplitude the final $NN$ 
state has isospin $I=0$, which is allowed for $np$ but forbidden for $nn$. 
Thus, the full OPE transition potential calculations produce a small value 
for $\Gamma_n/\Gamma_p \leq 0.1$. This is considerably smaller than the 
range of values, $\Gamma_n/\Gamma_p \sim 0.5$, deduced from old nuclear 
emulsion work \cite{montwill74} and from the most recent KEK experiments 
\cite{kang06,kim06}, indicating that OPE is insufficient to describe 
quantitatively NMWD.

In a semiclassical description of the hypernuclear $\Lambda + N \to n + N$ 
decay, the energy of each one of the two outgoing nucleons should peak at 
roughly 80 MeV which, assuming equal sharing of the released energy, is 
about half of the energy available in the decay. 
A proton-energy spectrum, taken by the FINUDA Collaboration \cite{agnello08} 
from nonmesonic weak decay of $^5_{\Lambda}$He produced on thin Li targets, 
is shown in the upper part of Fig.~\ref{fig:wdfinuda} (circles) in comparison 
with a proton spectrum taken at KEK \cite{okada04} (triangles). The two 
spectra were normalized above 35 MeV which is the KEK proton-energy threshold. 
A peak around 60-90 MeV is clearly observed, with a low-energy rise due to 
final state interactions (FSI), and perhaps also due to multinucleon induced 
weak decay. The FINUDA proton spectrum is compared in the lower part 
of Fig.~\ref{fig:wdfinuda} with the theoretical spectrum calculated by 
Garbarino, Parre\~{n}o, and Ramos (2004) using an intranuclear cascade 
(INC) code. The two spectra 
were normalized above 15 MeV which is the FINUDA proton-energy threshold. 
The agreement between experiment and theory is only qualitative. A more 
refined methodology to extract NMWD information from the FINUDA measured 
proton spectra has been presented recently by Agnello \textit{et al.} (2014). 
Neutron-energy spectra were reported by the KEK-PS Experiments 462/508 
\cite{okada04}, with a shape similar to that of the proton spectrum shown 
here, and with a similar rise at low energies. We note that the proton and 
neutron yields, $N_p$ and $N_n$ respectively, when properly normalized are 
related to the one-nucleon widths by 
\begin{equation} 
N_p=\Gamma_p~, \quad N_n=\Gamma_p+2\Gamma_n .  
\label{eq:1N} 
\end{equation} 
These expressions disregard FSI and multinucleon stimulated decays.

In the KEK experiments, the number of $np$ pairs, $N_{np}$, and $nn$ 
pairs, $N_{nn}$, corresponding to back-to-back final-state kinematics were 
identified and determined. Assuming that FSI has a negligible effect on the 
ratio $N_{nn}/N_{np}$, the ratio $\Gamma_n/\Gamma_p$ was approximated by 
$N_{nn}/N_{np}$ and the reported values for \lamb{5}{He} and \lam{12}{C} 
are listed in Table~\ref{tab:WD4}. For \lam{12}{C} the KEK result agrees 
within error bars with the old emulsion value. A recent 
reevaluation of the KEK spectra by Bauer \textit{et al.} (2010), accounting 
also for FSI, leads to a 
value of $\Gamma_n/\Gamma_p=0.66\pm 0.24$, in agreement with the emulsion and 
KEK values cited in the table. Previous determinations of $\Gamma_n/\Gamma_p$ 
from single-nucleon spectra gave considerably higher values, often in the 
range $1-2$, but are understood at present to have been subject to strong and 
unaccounted for FSI effects. This caveat refers, in principle, also to the 
value cited in the table from emulsion work, which was obtained by matching 
the experimentally observed fast ($T_p>30$~MeV) proton spectrum with 
appropriately weighted spectra from Monte-Carlo INC simulations of both 
proton and neutron FSI processes (recall that neutrons are not observed 
directly in emulsion). However, the emulsion estimate of $\Gamma_n/\Gamma_p$ 
appears to agree with the result of the more refined KEK analysis. Finally, 
two recent calculations using one-meson exchanges (OME) beyond OPE are listed 
in the table \cite{chumillas07,itonaga08}. These calculations reproduce 
satisfactorily the $\Gamma_n/\Gamma_p$ values deduced from the experiments 
listed in the table. They include also two-pion-exchange processes, with or 
without coupling the $\Lambda N$ system to $\Sigma N$, plus the two-pion 
($J^\pi\!=\!0^+,I\!=\!0$) resonance known as $\sigma$ and the axial 
vector meson $a_1$ considered as a $\rho-\pi$ resonance. The addition of 
$\sigma$ and $a_1$ exchanges does not effectively change the 
$\Gamma_n/\Gamma_p$ ratio, but proves to be significant in the calculation 
of the $\Lambda$ asymmetry parameter as discussed below. Earlier 
calculations by Jido, Oset, and Palomar (2001), using a chiral-interaction 
EFT approach, gave a very similar result, $\Gamma_n/\Gamma_p=0.53$ in 
\lam{12}{C}. 

Shown also in Table~\ref{tab:WD4} are experimentally deduced, as well 
as calculated values of the total NMWD width $\Gamma_{nm}$ for \lamb{5}{He} 
and \lam{12}{C}. The deduced NMWD width more than doubles between 
\lamb{5}{He} and \lam{12}{C} and is already close to saturation for $A=12$. 
Both calculations reproduce well the deduced NMWD width in \lamb{5}{He}, 
but fall short of it in \lam{12}{C}, perhaps due to the increased role of 
the $2N$ branch which was not included in the calculation. However, earlier 
calculations using the same exchanges, but with somewhat different couplings 
and with different prescriptions for the short-range behavior of the OME 
exchanges, were able to produce values $\Gamma_{\rm nm}({^{12}_{~\Lambda}{\rm 
C}})\sim (1.0-1.2)\,\Gamma_{\Lambda}^{\rm free}$ (Itonaga, Ueda, and Motoba, 2002;
Barbero \textit{et al.}, 2003). On the other hand, a more recent calculation 
by Bauer and Garbarino (2010), considering 
g.s. short-range correlations and including consistently a $2N$ branch, 
$\Gamma_2/\Gamma_{\rm nm}=0.26$, obtained a value $\Gamma_{\rm nm}=0.98\,
\Gamma_{\Lambda}^{\rm free}$, in very good agreement with the KEK deduced 
NMWD width. The saturation of the NMWD width for large values of $A$ is 
demonstrated in Table~\ref{tab:lifetimes} where \textit{total} hypernuclear 
decay lifetimes measured to better than $10\%$ accuracy are displayed. Recall 
from Table~\ref{tab:WD1} that for $A=56$ the mesonic decay width is 
no more than few percent of the nonmesonic width, hence the total width 
(lifetime) agrees to this accuracy with the nonmesonic width (lifetime). 

In the $\Lambda + N \to n + N$ two-body reactions, each of the final-state 
nucleons receives a momentum (energy) of order 400 MeV/c (80 MeV), which is 
well above the Fermi momentum (energy). This large value of momentum transfer 
justifies the use of semiclassical estimates for inclusive observables, such 
as the total nonmesonic decay rate of $\Lambda$ hypernuclei. 
Denoting a properly spin-isospin averaged nonmesonic decay width on a bound 
nucleon in nuclear matter by ${\bar \Gamma}_\Lambda$, the total hypernuclear 
rate is given in the local density approximation by 
\begin{equation} 
\frac{{\bar\Gamma}_\Lambda}{\rho_0}\int{\rho_\Lambda(r)\rho_N(r){{\rm d}^3}r}, 
\label{eq:sat} 
\end{equation} 
where $\rho_\Lambda(r)$ and $\rho_N(r)$ are the $\Lambda$ and the nucleon 
densities, normalized to 1 and to $A$, respectively, $\rho_0$ denotes 
nuclear-matter density, and zero range was implicitly assumed for the 
$\Lambda + N \to n + N$ amplitudes. Approximating the nucleon density 
$\rho_N(r)$ by $\rho_0$ for values of $r$ over which the $0s_\Lambda$
density $\rho_\Lambda(r)$ is localized, Eq.~(\ref{eq:sat}) reduces to 
${\bar \Gamma}_\Lambda$, 
independently of $A$. For nuclei with $N\neq Z$, the limiting value 
${\bar\Gamma}_\Lambda$ is replaced by 
\begin{equation} 
{\bar \Gamma}_\Lambda^0 + {\bar \Gamma}_\Lambda^1~\frac{N-Z}{A} 
= \Gamma_n \frac{N}{A} + \Gamma_p \frac{Z}{A}~, 
\label{eq:sat*} 
\end{equation} 
where ${\bar \Gamma}_\Lambda^0=(\Gamma_n+\Gamma_p)/2$ and 
${\bar \Gamma}_\Lambda^1=(\Gamma_n-\Gamma_p)/2$. Equation~(\ref{eq:sat*}) 
provides the leading term in a systematic expansion in powers of the 
neutron excess parameter $(N-Z)/A$. Finally, accepting that mesonic partial 
decay widths become negligible in medium- and heavier-weight hypernuclei 
and the total decay widths are essentially given by the nonmesonic decay 
widths, the total nonmesonic decay rate is expected to saturate in heavy 
hypernuclei, as was demonstrated in Table~\ref{tab:lifetimes}. 

The last item in Table~\ref{tab:WD4} concerns the $\Lambda$ intrinsic 
asymmetry parameter $a_{\Lambda}$ in the nonmesonic weak decay 
Eq.~(\ref{eq:Gamma_p}) of polarized $\Lambda$ hypernuclei. 
The angular distribution of the decay protons is given by 
\begin{equation} 
W(\theta) =W_0(1+a_{\Lambda}{\cal P}_{\Lambda}\cos{\theta}) .
\label{eq:asymm1} 
\end{equation} 
where ${\cal P}_{\Lambda}$ is the polarization of 
the $\Lambda$ spin in the decaying hypernucleus [as produced, e.g., in 
($\pi^+,K^+$) reactions] and $\theta$ is the emission angle of the protons 
with respect to the polarization axis. The asymmetry arises from the 
interference between PC and the PV weak-decay amplitudes. The values 
of $a_{\Lambda}$ deduced from experiment and listed in the table are 
close to zero, in strong disagreement with OME calculations; see, e.g., 
Parre\~{n}o, Ramos, and Bennhold (1997) and Parre\~{n}o and Ramos (2001).
A more recent representative example for such 
calculations is shown in Table~\ref{tab:WD4}. This long-standing problem was 
recently resolved with the introduction of a scalar-isoscalar ($0^+,0$) 
exchange which reduces the size of the negative and large asymmetry parameter 
produced in the OME calculations (Sasaki, Izaki, and Oka, 2005; Barbero and 
Mariano, 2006). These studies were motivated by the EFT approach adopted by 
Parre\~{n}o, Ramos, and Holstein (2004, 2005) where the largest contact 
term necessary for fitting the weak-decay rates and asymmetries was found to 
be spin- and isospin-independent; see also the review by Parre\~{n}o (2007).
A careful consideration of scalar-isoscalar two-pion exchange, in terms of a 
dynamically generated $\sigma$ resonance plus uncorrelated pion exchanges, was 
shown to resolve the $a_{\Lambda}$ puzzle, as listed in Table~\ref{tab:WD4}, 
without spoiling the agreement with experimental values of $\Gamma_{\rm nm}$ and 
$\Gamma_n/\Gamma_p$ \cite{chumillas07}. In contrast, Itonaga \textit{et al.} 
(2008) and Itonaga and Motoba (2010), using perhaps a less microscopic 
version of $\sigma$-meson degrees of freedom, have claimed that a satisfactory 
resolution of the $a_{\Lambda}$ puzzle requires a consideration of the 
axial-vector $a_1$, the chiral partner of the $\rho$ meson, in terms of 
$\rho-\pi$ and $\sigma-\pi$ correlated exchanges. Their results are also 
listed in Table~\ref{tab:WD4}. A similarly small and positive value for 
\lam{12}{C}, $a_{\Lambda}=0.069$, has also been calculated recently by Bauer and
Garbarino (2012). 

\section{$\Sigma$ Hypernuclei} 

\subsection{Overview} 

Evidence for relatively narrow $\Sigma$-hypernuclear continuum excitations 
in $^6$Li, $^9$Be, $^{12}$C and $^{16}$O, with widths of order few MeV, 
was suggested during the 1980s from $(K^-,\pi^{\pm})$ experiments in flight 
at CERN \cite{bertini80,bertini84,bertini85} and at BNL \cite{piekarz82} 
using $K^-$ beams with incident momentum $p_{\rm lab}=450-720$ MeV/c, 
and with stopped $K^-$ mesons at KEK \cite{yamazaki85}. 
Supporting evidence for $\Sigma$-nuclear attraction, of order $25-30$ MeV 
at central nuclear densities, existed from the ``old'' analysis of $\Sigma^-$ 
atom level shifts and widths \cite{batty79}, but the same analysis also 
yielded estimates of order $20-30$ MeV for the $\Sigma$-nuclear widths at 
the central nuclear densities expected in $\Sigma$ hypernuclei. The strength 
of the $\Sigma^- p \to \Lambda n$ reaction, deduced from cross section data 
at low energies, was shown to be in agreement with this width estimate 
\cite{gal80}. In this, and in other calculations (Dover, Millener, and Gal, 
1989), the $\Sigma N \to \Lambda N$ one-pion-exchange transition was perceived 
to provide the underlying mechanism for $\Sigma$ hypernuclear widths. No sound 
theoretical calculation was able to reproduce the narrow structures suggested 
by the reported $\Sigma$ hypernuclear spectra. These spectra, however, 
typically consisted of a small number of events of questionable statistical 
significance above the kaon decay background. Subsequent $(K^-,\pi^{\pm})$ 
experiments at BNL, with improved statistics, failed to confirm the existence 
of narrow $\Sigma$ hypernuclear structures (Tang \textit{et al.}, 1988;
Bart \textit{et al.}, 1999), particularly on the same targets ($^6$Li and 
$^9$Be) and in the same reactions for which previous claims of quasibound 
states were made. The new BNL experimental spectra showed somewhat broad 
continuum enhancements which indicated a very shallow, or even repulsive 
$\Sigma$ nuclear potential, as had been already argued (Dover, Millener, and 
Gal, 1989). This was verified by calculations \cite{dabrowski99} 
of the pion spectrum in the $(K^-,\pi^+)$ reaction on $^9$Be \cite{bart99}. 

\begin{figure} 
\includegraphics[width=8.0cm]{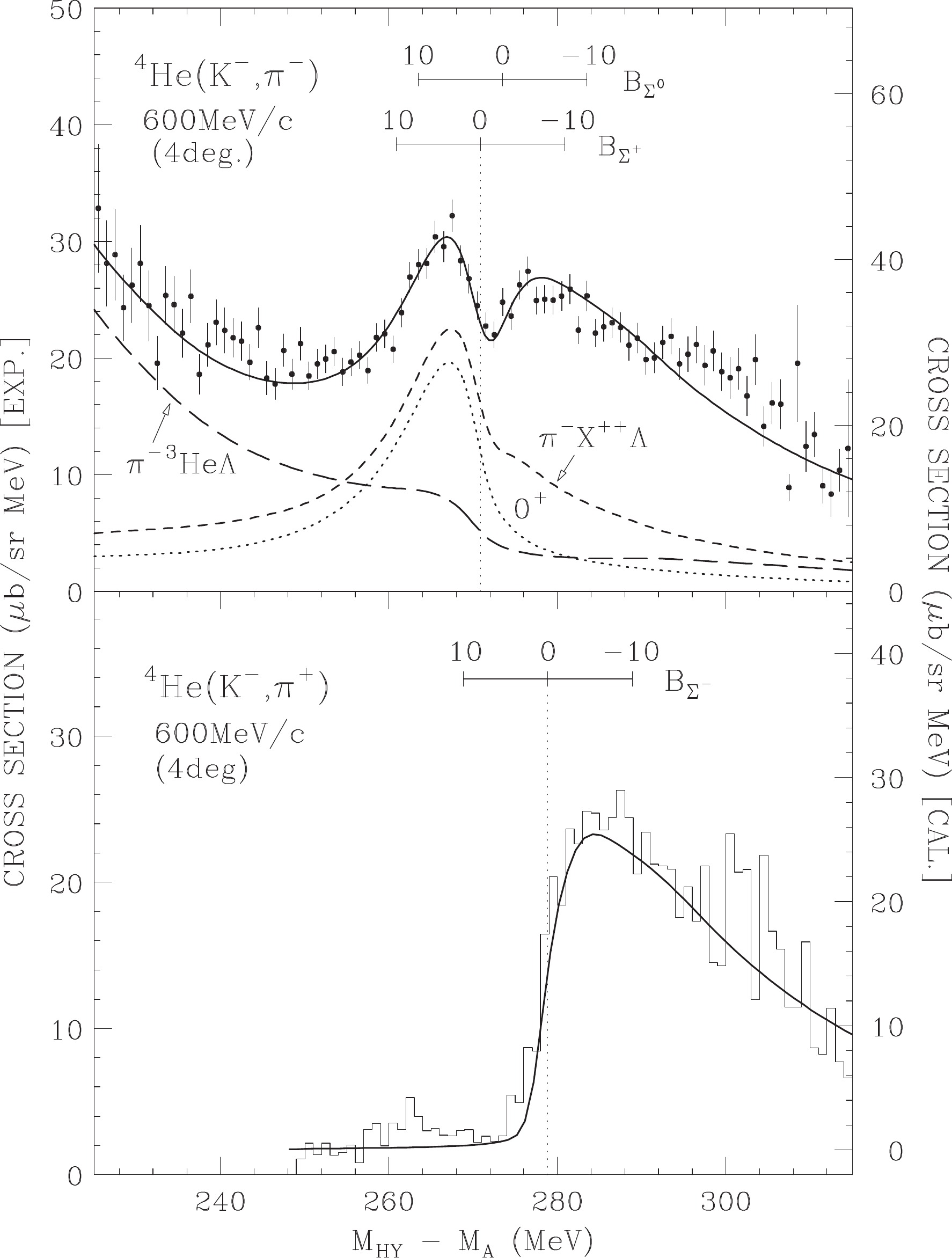}
\caption{$^4{\rm He}(K^-,\pi^{\pm})$ spectra measured at BNL \cite{nagae98} 
and as calculated by Harada (1998), providing evidence for a $^4_\Sigma$He 
$I=1/2$ quasibound state in the $\pi^-$ channel, with fitted values of 
binding energy $B_{\Sigma^+} =4.4 \pm 0.3 \pm 1$ MeV and width 
$\Gamma=7.0 \pm 0.7^{+1.2}_{-0.0}$ MeV. Adapted from Harada, 1998.} 
\label{fig:Sig1} 
\end{figure} 

A notable exception is provided by $^4_\Sigma$He, where a quasibound state 
below the $\Sigma^+$ threshold was discovered in a ($K^{-}_{\rm stop},\pi^-$) 
experiment on $^4$He at KEK~\cite{hayano89}. This quasibound state was 
confirmed in a ($K^-,\pi^-$) in-flight experiment, with $p_{\rm lab} = 600$ 
MeV/c, at BNL~\cite{nagae98}; see Fig.~\ref{fig:Sig1}. No evidence was found 
for quasibound states in the companion ($K^-, \pi^+$) experiment on $^4$He. 
For this reason the $^4_\Sigma$He quasibound state was assigned an isospin 
value $I=1/2$. Comparison of the two spectra in the figure suggests a strong 
isospin dependence of the $\Sigma$ nuclear potential. This dependence was 
taken into account in coupled-channel calculations (Harada \textit{et al.},
1990; Harada, 1998) which used $^3{\rm He}+\Lambda$, $^3{\rm He}+\Sigma^0$ and 
$^3{\rm H}+\Sigma^+$ channels for ($K^-,\pi^-$) and $^3{\rm H}+\Sigma^-$ 
for ($K^-,\pi^+$). A similar isospin dependence is also clearly seen in the 
CERN data \cite{bertini84} on $^{12}$C, as deduced by Dover, Gal, and
Millener (1984), and in the BNL measurements \cite{bart99} on $^6$Li and 
$^9$Be, as deduced by Dabrowski (1999). The strong isospin dependence may 
be parameterized in terms of a strong Lane term $V_1^{\Sigma}$ of the $\Sigma$ 
nuclear potential: 
\begin{equation} 
\label{eq:lane} 
V_{\Sigma}(r) = \left(V_0^{\Sigma}+\frac{1}{A}~V_1^{\Sigma}~\bm{T}_A{\cdot}
\bm{t}_{\Sigma}\right)\,\frac{\rho(r)}{\rho_0}\, , 
\end{equation} 
where $\bm{t}_{\Sigma}$ is the $\Sigma$ isospin operator and $\bm{T}_A$ 
is the nuclear isospin operator with z projection $(Z-N)/2$. Owing to 
the smallness of $A$ ($A=4$), the Lane term in the case of $^4_\Sigma$He, 
with a large and positive value of $V_1^{\Sigma}$, provides sufficient 
attraction to generate a quasibound state, whereas the relatively small 
width is due to the isoscalar repulsion (Harada \textit{et al.},
1990; Harada, 1998, 2001). A large value, $V_1^{\Sigma}\approx 80$~MeV, had been 
predicted by Dover, Gal, and Millener, (1984) from the ($K^-,\pi^{\pm}$) CERN 
data on $^{12}$C \cite{bertini84}. 

Recent measurements at KEK of the $\Sigma^-$ spectrum in the $(\pi^-,K^+)$ 
reaction on targets across the periodic table \cite{noumi02,noumi03,saha04} 
have established that the $\Sigma$ nuclear interaction is strongly 
repulsive. This was subsequently confirmed in DWIA calculations by 
Harada and Hirabayashi (2005, 2006) as reviewed in Sec.~\ref{subsec:pimkp} . 
In parallel, density-dependent analyses of $\Sigma^-$-atom data in the early 
1990s led to the conclusion that the \textit{nuclear} interaction of $\Sigma$'s 
is dominated by repulsion (Batty, Friedman, and Gal, 1994a, 1994b, 1997;
Mare\v{s} \textit{et al.}, 1995). Based on the various 
analyses discussed above, A reasonable estimate of the $\Sigma$ isoscalar 
repulsion, based on the various analyses discussed above, is $V_0^{\Sigma}
\approx 30\pm 20$~MeV, a value listed in Table~\ref{tab:Sig1}. 
The repulsion of $\Sigma^-$ in nuclear matter, and also in neutron matter, 
has important repercussions for the balance of strangeness in the inner crust 
of neutron stars, primarily by delaying to higher densities, or even aborting 
the appearance of $\Sigma^-$ hyperons \cite{balberg97}. 

\begin{table} 
\caption{Representative values of isoscalar and isovector 
$\Sigma$--nuclear potential depths (in MeV) [see Eq.~(\ref{eq:lane})] 
taken from Gal (2010) for Nijmegen soft-core potentials (Rijken, Nagels, and
Yamamoto, 2010), and from Haidenbauer and Mei{\ss}ner (2015) for EFT 
potentials with cutoff parameter 600~MeV.} 
\label{tab:Sig1} 
\begin{ruledtabular}
\begin{tabular}{lcccccc} 
 & NSC97f & ESC04d & ESC08b & LO & NLO & Phenomenology \\
\hline 
$V_0^{\Sigma}$ & $-$13.9 & $-$26.0 & $+$20.3 & $+$22.1 & $+$14.8 & 
$+$30$\pm$20 \\
$V_1^{\Sigma}$ & $-$30.4 & $+$30.4 & $+$85.2 & $+$58.1 & $+$67.8 & 
$\approx$80 \\ 
\end{tabular} 
\end{ruledtabular}    
\end{table} 

Values of $V_0^{\Sigma}$ and $V_1^{\Sigma}$ are listed in Table~\ref{tab:Sig1} 
for several representative Nijmegen soft-core potentials and recent EFT 
calculations, in comparison with phenomenological values derived from several 
sources of data analyses. Of the hard-core, earlier Nijmegen potentials, only 
Model F provided isoscalar repulsion and a sizable ``attractive'' Lane term 
($V_1^{\Sigma}>0$), both of which are required to fit the data, as shown by 
Dabrowski (1999). For the soft-core Nijmegen models, it is worth noting that 
the widely used NSC97 models, and the J\"{u}lich model \cite{haidenbauer05}, 
produced attractive isoscalar $\Sigma$-nuclear potentials and ``repulsive'' 
isovector potentials, just opposite of what phenomenology demands 
(as marked in the last column of the table). Subsequent Nijmegen potentials 
have removed this discrepancy by imposing a strongly repulsive $T=3/2$ 
${^3S_1}-{^3D_1}$ $\Sigma N$ interaction on their parameter fit. This was 
motivated by the SU(6) quark-model, resonating-group method calculations by the
 Kyoto-Niigata group \cite{kohno00}, reviewed by Fujiwara, Suzuki, and 
Nakamoto (2007), in which 
a strong Pauli repulsion appears in this $\Sigma N$ channel; see also recent 
calculations of hyperon-nucleus potentials by Kohno and Fujiwara (2009) 
and Kohno (2010). The latest EFT potentials, LO (Polinder, Haidenbauer, and
Mei{\ss}ner, 2006) and NLO \cite{haidenbauer13a}, also impose repulsion in 
this particular $\Sigma N$ channel. An earlier SU(3) chiral perturbation 
calculation by Kaiser (2005) 
yielded repulsion of order $V_0^{\Sigma}\approx 60$~MeV.

\subsection{$\Sigma$ nuclear potentials from fits to 
$\Sigma^-$ atoms} 

\begin{figure}
\includegraphics[width=8.0cm]{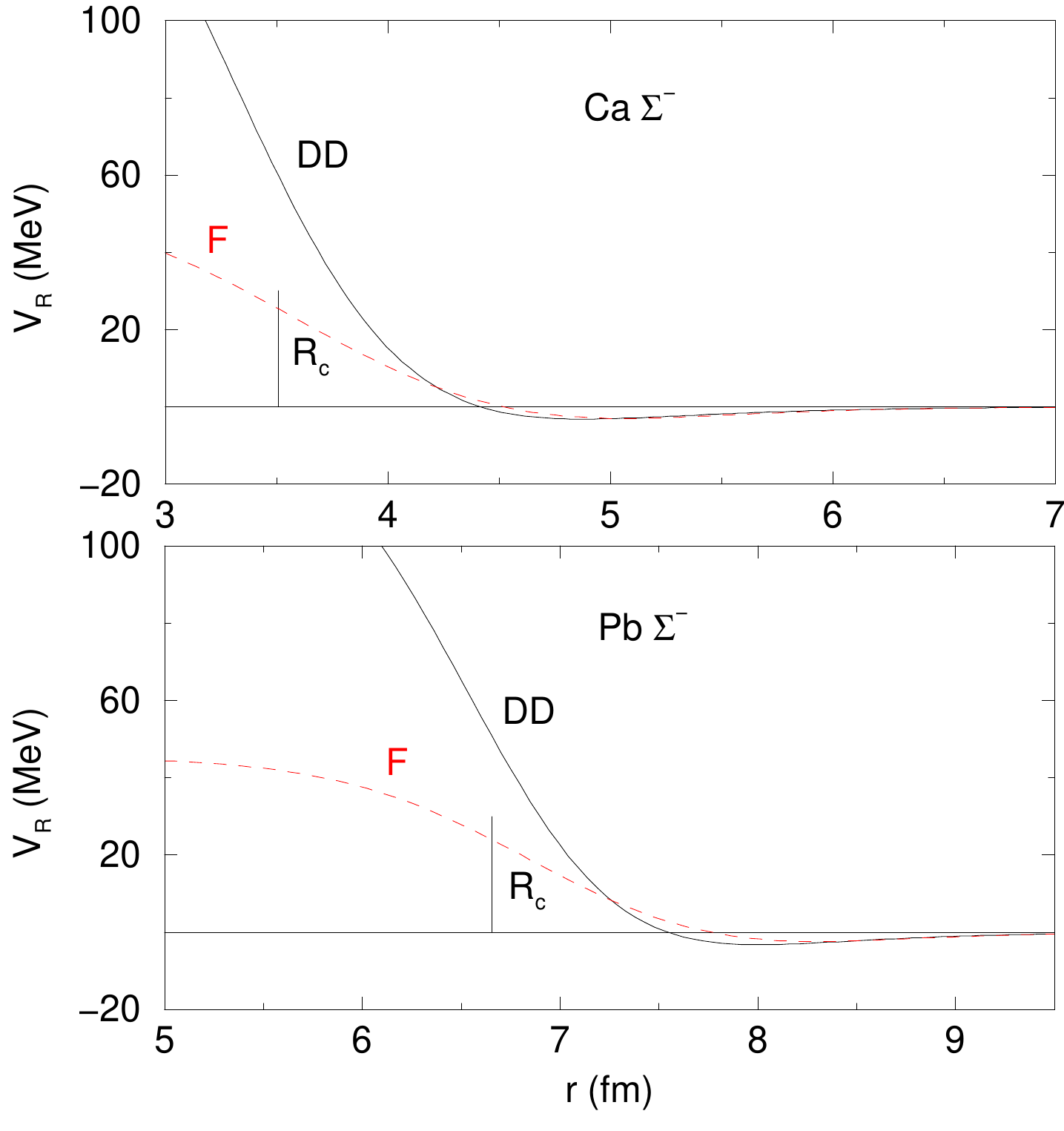} 
\caption{Re $V_{\rm opt}$ ($V_R$) for two different 
parameterizations of the $\Sigma^-$ nuclear potential, DD (solid) and F 
(dashed), fitted to $\Sigma^-$ atomic data. Vertical bars indicate the 
half-density radius of the nuclear charge distribution. 
From Friedman and Gal, 2007.} 
\label{fig:Sig2} 
\end{figure}

$\Sigma^-$ nuclear potentials resulting from two fits to the full set of 
$\Sigma^-$ atomic data, with different parameterizations for the density 
dependence of $V_\Sigma$, are shown in Fig.~\ref{fig:Sig2}. The data consist 
of 23 strong-interaction level shifts, widths, and yields. A phenomenological 
density-dependent isoscalar potential, DD, was introduced by the 
form (Batty, Friedman, and Gal, 1994a, 1994b) 
\begin{equation}
\label{eq:DD}
V_\Sigma(r) \sim \left\{ b_0 + B_0 \left[{\rho(r)/\rho(0)}\right]^\alpha \right\}
\rho(r) , \qquad \alpha > 0  
\end{equation} 
and a ``geometrical'' potential, F, was introduced by the form (Mare\v{s},
  Friedman, and Gal, 2006)
\begin{equation}
\label{eq:F} 
V_\Sigma(r) \sim \left\{ b_0 \left[1-F(r)\right] + B_0 F(r)\right\}\rho(r) . 
\end{equation} 
In these expressions 
\begin{equation}
\label{eq:Fx} 
F(r)=\frac{1}{e^x+1}, \quad x=\frac{r-R_x}{a_x} , 
\end{equation}  
with $R_x = R_{x0} A^{1/3}+\delta_x$ close to the radius of the nucleus, 
and $a_x \approx 0.5$~fm close to accepted values of the nuclear 
diffusivity. Greatly improved fits with respect to fitted $t \rho(r)$ type 
potentials are obtained by fitting the parameters $b_0, B_0$, and $\alpha$, 
for DD, and $b_0, B_0$, and $R_{x0}, \delta_x, a_x$, for F. Isovector 
components are readily included, but are found to have a marginal effect. 
The fit to the data is equally good in the two models, with a $\chi ^2$ 
per degree of freedom of 1.0 for DD and 0.9 for F. 
The half-density radius of the charge distribution $R_c$ is indicated in
Fig.~\ref{fig:Sig2}. The figure demonstrates that the transition from
outward attraction to inward repulsion occurs well outside $R_c$, a property
supported also by other types of fits to $\Sigma^-$ atomic 
data~\cite{friedman07}. The precise magnitude and shape of the repulsive 
component within the nucleus is not determined by the $\Sigma^-$ atomic 
data. Although both models show weak attraction at large radii, this is 
too weak to support bound states. The conclusion is in agreement with the 
experimental results from BNL \cite{bart99} showing the absence of $\Sigma$ 
hypernuclear quasibound peaks beyond He. 

\begin{figure*} 
\includegraphics[width=12cm]{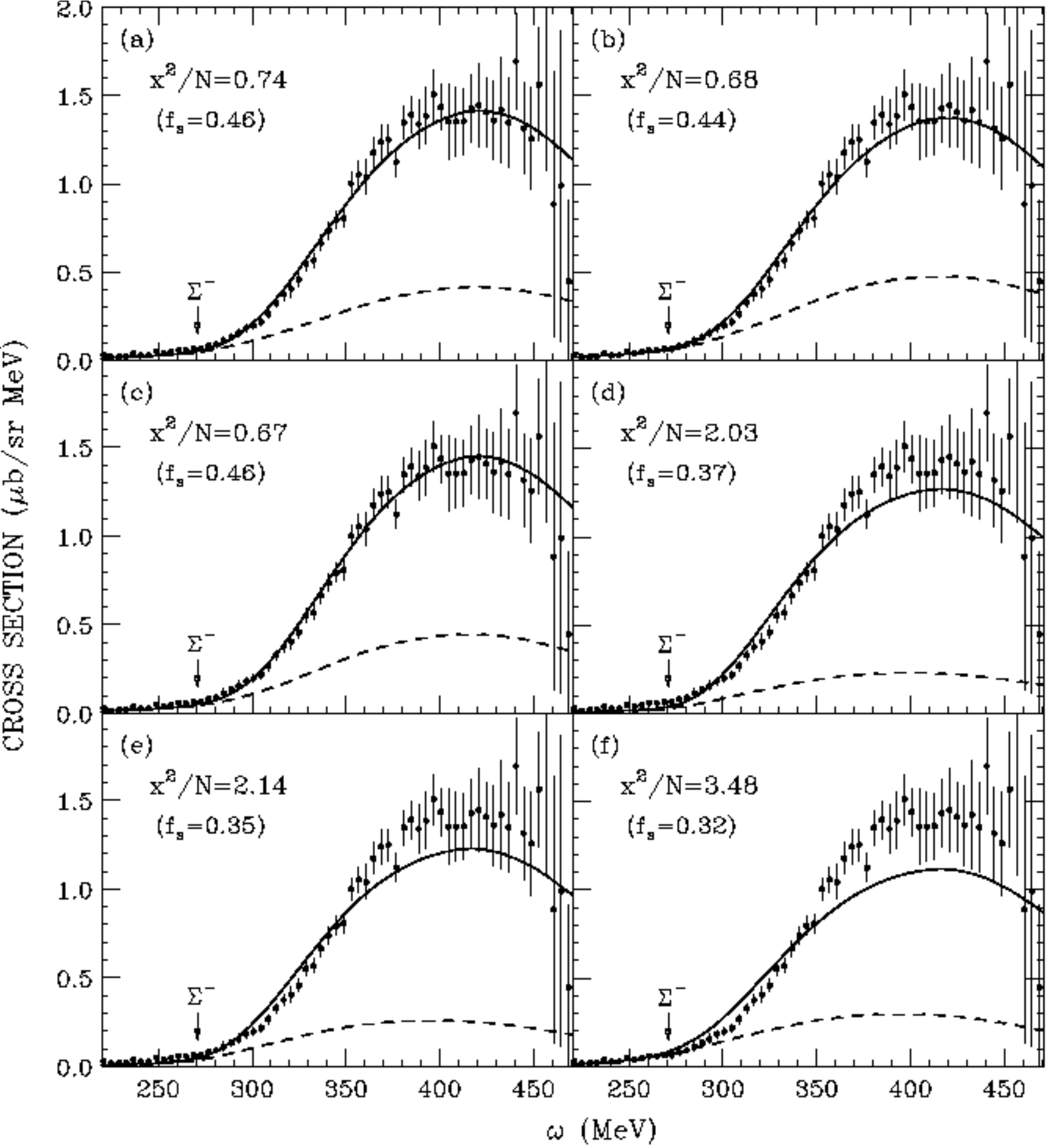} 
\caption{Comparison between DWIA calculations~{\cite{harada05,harada06}} 
using six $\Sigma$-nucleus potentials, (a)-(c) with inner repulsion, 
(d)-(f) fully attractive, and the measured $^{28}{\rm Si}(\pi^-,K^+)$ 
spectrum~{\protect\cite{saha04}}. The solid and dashed curves denote the 
inclusive and $\Lambda$ conversion cross sections, respectively. 
Each calculated spectrum was normalized by a fraction $f_s$. The arrows 
mark the ${\Sigma^-} - {^{27}}{\rm Al}$(g.s.) threshold at 
$\omega = 270.75$~MeV. From Harada and Hirabayashi, 2005.} 
\label{fig:Sig3} 
\end{figure*} 

\subsection{Evidence from $(\pi^-,K^+)$ spectra}
\label{subsec:pimkp}

More straightforward information on the nature of the $\Sigma$-nuclear 
interaction has been provided by recent measurements of inclusive 
$(\pi^-,K^+)$ spectra on medium to heavy nuclear targets at 
KEK~\cite{noumi02,noumi03,saha04}. These spectra were fitted using Woods-Saxon 
potentials with depths $V_0 \approx 100$ MeV for the \textit{repulsive} real 
part and $W_0=-40$ MeV for the imaginary part. There is less sensitivity to 
the imaginary (absorptive) component. The repulsive potential in this analysis 
is of the same order of magnitude as obtained for the DD potential in the 
nuclear surface region, Fig.~\ref{fig:Sig2}. 

\begin{figure*} 
\includegraphics[width=15cm]{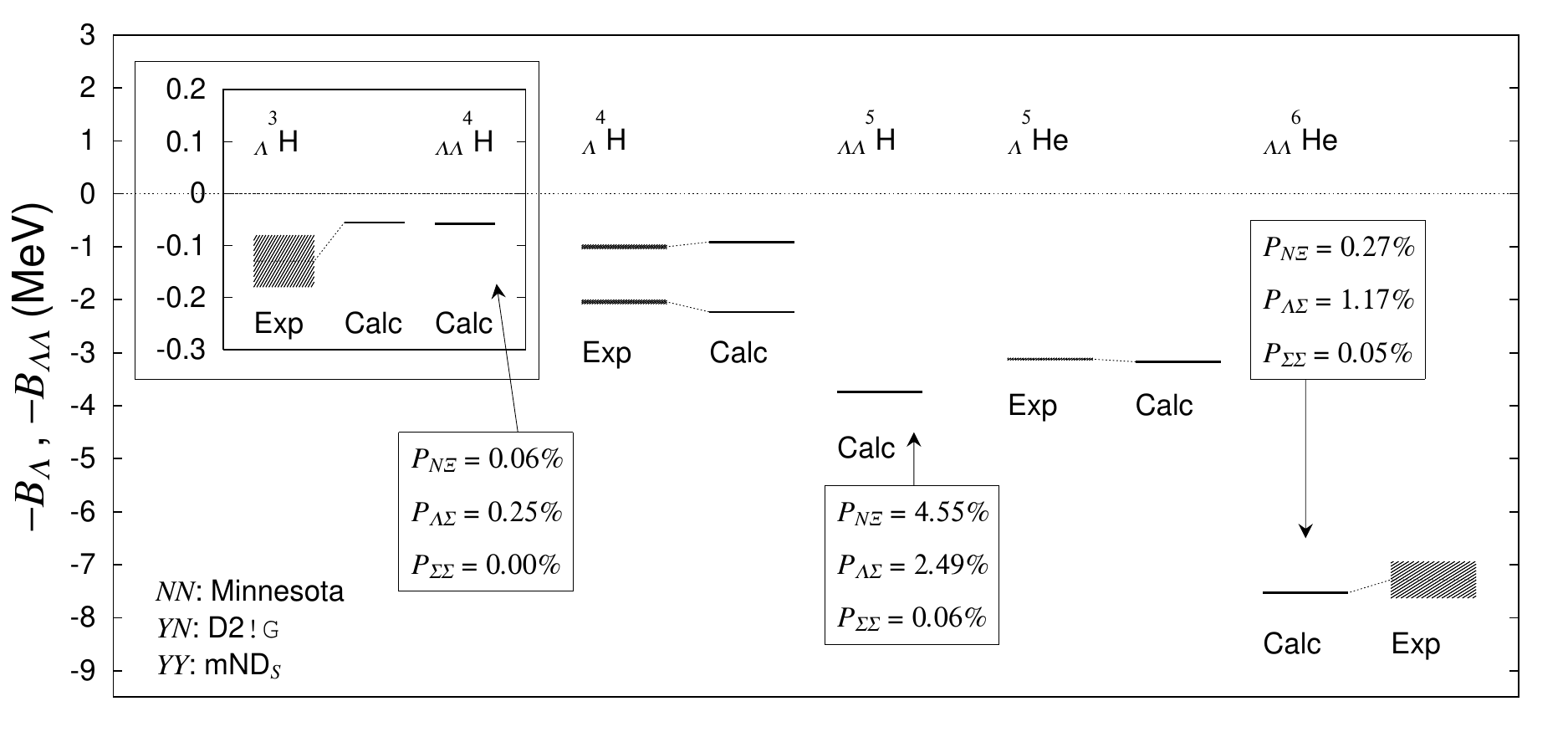} 
\caption{Calculated $\Lambda$ and $\Lambda\Lambda$ separation energies of 
$s$-shell hypernuclei. From Nemura \textit{et al.}, 2005.} 
\label{fig:nemura} 
\end{figure*} 

More sophisticated theoretical analyses of these KEK $(\pi^-,K^+)$ spectra 
(Kohno \textit{et al.}, 2004, 2006; Harada and Hirabayashi, 2005, 2006) have 
also concluded that the $\Sigma$-nuclear potential is repulsive within the 
nuclear volume, although they yield a weaker repulsion in the range of 
$10-40$ MeV. An example of a recent analysis of the Si spectrum is shown in 
Fig.~\ref{fig:Sig3} from  Harada and Hirabayashi (2005), where six different 
$\Sigma$-nucleus potentials are tested for their ability to reproduce the 
measured $^{28}{\rm Si}(\pi^-,K^+)$ spectrum~\cite{saha04} within the DWIA. 
This particular DWIA version was tested on the well-understood 
$^{28}{\rm Si}(\pi^+,K^+)$ quasifree $\Lambda$ hypernuclear spectrum which was 
also taken at KEK with pions of the same incident momentum, 
$p_{\rm lab} = 1.2$~GeV/c. The potential of Fig.~\ref{fig:Sig3}(a) is the DD, 
type A$^\prime$ potential of Batty, Friedman, and Gal (1994a, 1994b), that of 
Fig.~\ref{fig:Sig3}(b) is one of the RMF potentials of Mare\v{s} 
\textit{et al.} (1995), with $\alpha_{\omega} = 1$, and 
that of Fig.~\ref{fig:Sig3}(c) is a local-density approxiamation 
version of a $G$ matrix constructed from the Nijmegen model F. These three 
potentials are repulsive within the nucleus but differ considerably 
from each other. The potentials of Figs.~\ref{fig:Sig3}(d)-(f) are all 
attractive within the nucleus, with that of of Fig.~\ref{fig:Sig3}(f) being 
of a $t \rho$ form. All of the six potentials are attractive 
outside the nucleus, as required by fits to the ``attractive'' $\Sigma^-$ 
atomic level shifts. The figure shows clearly that fully attractive potentials 
are ruled out by the data, as deduced from $\chi^2$ fits, and that only the 
``repulsive'' $\Sigma$-nucleus potentials reproduce the spectrum, although 
without preference to any of these repulsive potentials.

\section{$\Lambda-\Lambda$ Hypernuclei}

Until 2001 only three emulsion events had been considered serious candidates 
for $\Lambda\Lambda$ hypernuclei: $_{\Lambda\Lambda}^{~10}$Be (Danysz 
\textit{et al.}, 1963a, 1963b), $_{\Lambda\Lambda}^{~~6}$He \cite{prowse66} and 
$_{\Lambda\Lambda}^{~13}$B \cite{aoki91}. The $\Lambda\Lambda$ binding 
energies deduced from these emulsion events indicated that the 
$\Lambda\Lambda$ interaction was quite attractive in the $^{1}S_0$ 
channel (Dalitz \textit{et al.}, 1989; Dover \textit{et al.}, 1991; Yamamoto,
Takaki, and Ikeda, 1991), with a $\Lambda\Lambda$ excess binding energy 
$\Delta B_{\Lambda\Lambda} \sim 4.5$ MeV. However, it was realized that the 
binding energies of $_{\Lambda\Lambda}^{~10}$Be and $_{\Lambda\Lambda}^{~~6}$He were 
inconsistent with each other (Bodmer, Usmai, and Carlson, 1984; Wang, Takaki,
and Band\={o}, 1986). Here, the $\Lambda\Lambda$ excess binding energy is 
defined by 
\begin{equation} 
\label{eq:delB} 
\Delta B_{\Lambda\Lambda} (^{~A}_{\Lambda \Lambda}Z) 
= B_{\Lambda\Lambda} (^{~A}_{\Lambda \Lambda}Z) 
- 2{\overline B}_{\Lambda} (^{(A-1)}_{~~~~\Lambda}Z) , 
\end{equation} 
where $B_{\Lambda\Lambda}(^{~A}_{\Lambda \Lambda}Z)$ is the $\Lambda\Lambda$ 
binding energy of the hypernucleus $^{~A}_{\Lambda \Lambda}Z$ and ${\overline 
B}_{\Lambda} (^{(A-1)}_{~~~~\Lambda}Z)$ is the (2$J$+1)-average of $B_{\Lambda}$ 
values for the $^{(A-1)}_{~~~~\Lambda}Z$ hypernuclear core levels in the g.s. 
doublet, as appropriate to a spin-zero $(1s_{\Lambda})^2$ configuration 
of the double-$\Lambda$ hypernucleus $^{~A}_{\Lambda \Lambda}Z$. The 
unambiguous observation of $_{\Lambda\Lambda}^{~~6}$He \cite{takahashi01} 
from the KEK hybrid-emulsion experiment E373 lowered the accepted $\Delta 
B_{\Lambda\Lambda}$ value substantially from the value deduced from the older, 
dubious event \cite{prowse66}, down to $\Delta B_{\Lambda\Lambda}(_{\Lambda
\Lambda}^{~~6}{\rm He})=0.67\pm 0.17$~MeV \cite{ahn13}. With this new value 
of $\Delta B_{\Lambda\Lambda}$, it is natural to inquire where the onset of 
$\Lambda\Lambda$ binding occurs. From the very beginning it was recognized 
that the $\Lambda\Lambda$ system \cite{dalitz63b} and the three-body 
$\Lambda\Lambda N$ system were unbound \cite{tang65}; if $\Lambda\Lambda N$ 
were bound, the existence of a bound $nn\Lambda$ would follow and 
$_{\Lambda\Lambda}^{~~6}$He would most likely become overbound \cite{gal13a}. 
The existence of a $_{\Lambda\Lambda}^{~~4}$H bound state was claimed by 
the AGS experiment E906 \cite{ahn01a}, studying correlated weak-decay 
pions emitted sequentially from $\Lambda\Lambda$ hypernuclei apparently 
produced in a $(K^-,K^+)$ reaction on $^9$Be, but this interpretation is 
ambiguous \cite{randeniya07}. 

The issue of $^{~~4}_{\Lambda\Lambda}$H binding was addressed in several 
subsequent studies. A Faddeev-Yakubovsky (FY) four-body calculation 
\cite{filikhin02a} found no bound state when using an $s$-wave 
$V_{\Lambda\Lambda}$ fitted to $B_{\Lambda\Lambda}(^{~~6}_{\Lambda\Lambda}$He) 
and a $V_{\Lambda N}$ partially fitted to $B_{\Lambda}(^3_{\Lambda}$H). 
However, when fitting a $\Lambda d$ potential to the low-energy parameters 
of the $s$-wave Faddeev calculation for $\Lambda pn$ and solving 
the $s$-wave Faddeev equations for a $\Lambda\Lambda d$ model of 
$^{~~4}_{\Lambda\Lambda}$H, a $1^+$ bound state was obtained. Disregarding 
spin it can be shown, for essentially an attractive $\Lambda\Lambda$ 
interaction and for a static nuclear core $d$, that a two-body $\Lambda d$ 
bound state implies binding for the three-body $\Lambda\Lambda d$ system. 
Nevertheless, for a \textit{non-static} nuclear core $d$ (made of a $pn$ 
interacting pair), a $\Lambda d$ bound state does not necessarily imply 
binding for the $\Lambda\Lambda d$ system. 

This $^{~~4}_{\Lambda\Lambda}$H no-binding conclusion was challenged by 
Nemura, Akaishi, and Myint (2003) and Nemura \textit{et al.} (2005) 
who showed that $\Lambda N$-$\Sigma N$ coupling, 
which is so important for the quantitative discussion of light $\Lambda$ 
hypernuclei, is capable of inducing appreciable $\Xi N$ admixures into light 
$\Lambda\Lambda$ hypernuclei via the $\Sigma \Lambda - \Xi N$ coupling. 
This is shown in Fig.~\ref{fig:nemura} along with all other bound $\Lambda$ 
and $\Lambda\Lambda$ $s$-shell hypernuclei. Although in their calculation the 
second $\Lambda$ in $^{~~4}_{\Lambda\Lambda}$H is bound by $0-0.07$ MeV, no 
firm conclusion can be made regarding the particle stability of this species 
since in their $^{~~6}_{\Lambda\Lambda}$He calculation the second $\Lambda$ 
is overbound by 0.22 MeV. Thus, the issue of the onset of $\Lambda\Lambda$ 
binding, in particular whether or not $^{~~4}_{\Lambda\Lambda}$H is 
particle stable, is still unresolved. Further experimental work is needed 
to decide whether the events reported in the AGS experiment E906 correspond 
to $^{~~4}_{\Lambda\Lambda}$H \cite{ahn01a,randeniya07}, and also in view of 
subsequent conflicting theoretical analyses (Kumagai-Fuse and Okabe, 2002;
Kahana, Kahana, and Millener, 2003). 

\begin{figure}[t] 
\includegraphics[width=8.0cm]{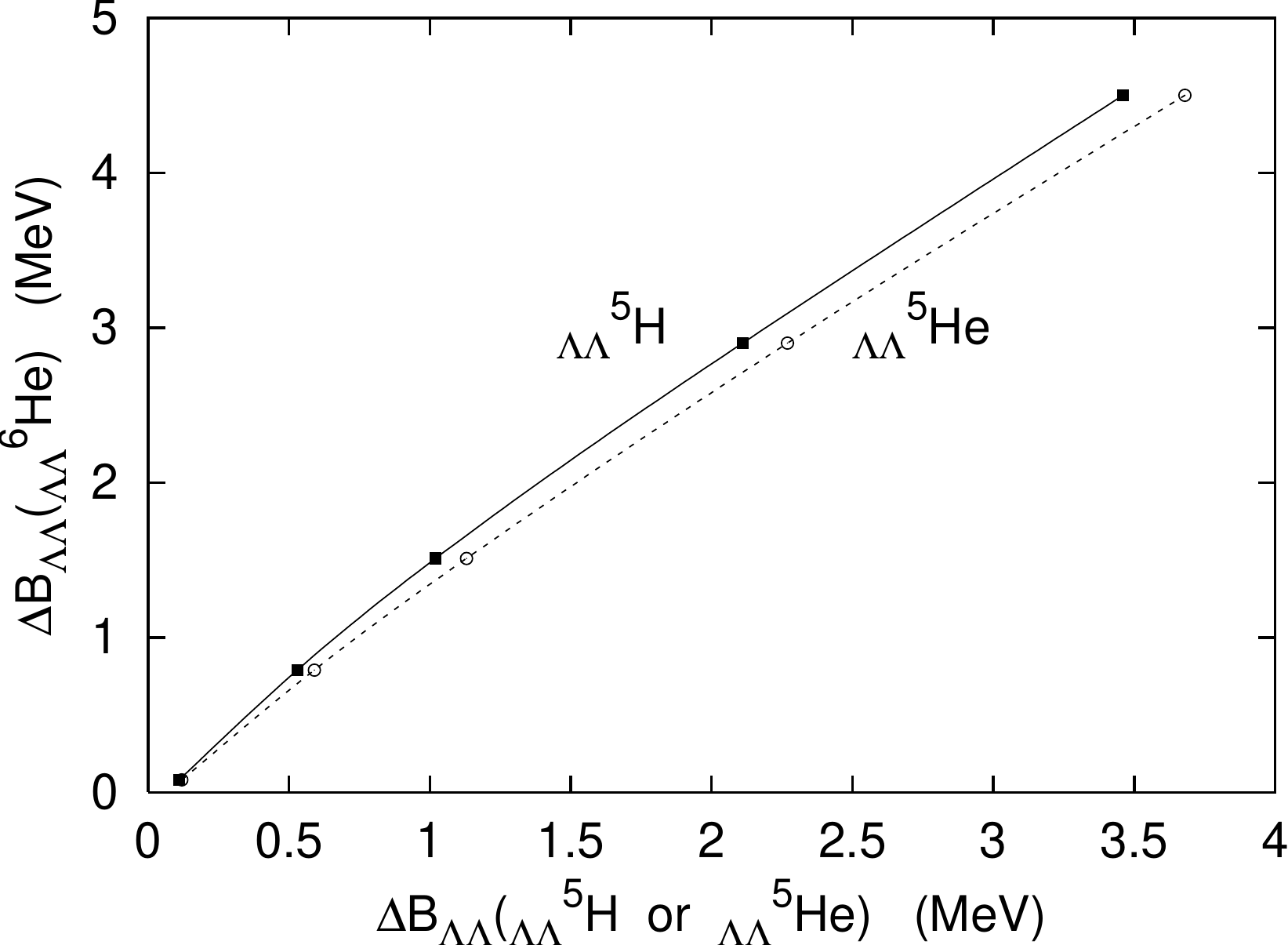} 
\caption{Faddeev calculations of $\Delta B_{\Lambda\Lambda}$ for 
$^{~~6}_{\Lambda\Lambda}$He vs. Faddeev calculations for the mirror $\Lambda
\Lambda$ hypernuclei $^{~~5}_{\Lambda\Lambda}$H and $^{~~5}_{\Lambda\Lambda}$He. 
The points mark results obtained for various assumptions 
on $V_{\Lambda\Lambda}$. From Filikhin and Gal, 2002a} 
\label{fig:filikhin} 
\end{figure} 

\begin{table*}[t] 
\caption{$B_{\Lambda\Lambda}$ values (in MeV) from KEK experiments E176 
\cite{aoki09} and E373 \cite{ahn13}, and as calculated in cluster models 
\cite{hiyama02,hiyama10} and in the shell model \cite{gal11}. $B_{\Lambda
\Lambda}(^{~~6}_{\Lambda\Lambda}$He) serves as input in both types of 
calculations. The E176 entries offer several assignments to the \textit{same} 
single emulsion event observed. \label{tab:LL}} 
\begin{ruledtabular} 
\begin{tabular}{lcccccc} 
Event & $_{\Lambda\Lambda}^{~~\rm A}{\rm Z}$ & 
${\overline B}_{\Lambda}({_{~~~\Lambda}^{\rm A-1}{\rm Z}})$ &
$B_{\Lambda\Lambda}^{\rm exp}$ & $B_{\Lambda\Lambda}^{\rm CM}$ & 
$B_{\Lambda\Lambda}^{\rm SM}$ \\  \hline 
E373-Nagara & $_{\Lambda\Lambda}^{~~6}{\rm He}$ & $3.12\pm 0.02$ & 
$6.91\pm 0.16$ & $6.91\pm 0.16$ & $6.91\pm 0.16$ \\ 
E373-DemYan & $_{\Lambda\Lambda}^{~10}{\rm Be}$ & $6.71\pm 0.04$ & 
$14.94\pm 0.13$ & $14.74\pm 0.16$ & $14.97\pm 0.22$\footnotemark[1] \\ 
E176-G2 & $_{\Lambda\Lambda}^{~11}{\rm Be}$ & $8.86\pm 0.11$ & 
$17.53\pm 0.71$ & $18.23\pm 0.16$ & $18.40\pm 0.28$ \\ 
E373-Hida & $_{\Lambda\Lambda}^{~11}{\rm Be}$ & $8.86\pm 0.11$ & 
$20.83\pm 1.27$ & $18.23\pm 0.16$ & $18.40\pm 0.28$ \\ 
E373-Hida & $_{\Lambda\Lambda}^{~12}{\rm Be}$ & $10.02\pm 0.05$ & 
$22.48\pm 1.21$ & -- & $20.72\pm 0.20$ \\ 
E176-E2 & $_{\Lambda\Lambda}^{~12}{\rm B}$ & $10.09\pm 0.05$ & 
$20.02\pm 0.78$ & -- & $20.85\pm 0.20$ \\ 
E176-E4 & $_{\Lambda\Lambda}^{~13}{\rm B}$ & $11.27\pm 0.06$ & 
$23.4\pm 0.7$ & -- & $23.21\pm 0.21$ \\ 
\end{tabular} 
\end{ruledtabular}    
\footnotetext[1] {$B_{\Lambda\Lambda}^{\rm SM}({_{\Lambda\Lambda}^{~10}
{\rm Be}})= 2~{\overline B}_{\Lambda}({_{\Lambda}^{9}{\rm Be}})+
4~[\overline{V}({_{\Lambda}^{9}{\rm Be}})-\overline{V}_{\rm average}]~+
\langle V_{\Lambda\Lambda}\rangle_\text{SM}$, see Eq.~(\ref{eq:deltaBLL}).}   
\end{table*} 

Regardless of whether $^{~~4}_{\Lambda\Lambda}$H is particle-stable 
or not, there is a general consensus that the mirror $\Lambda\Lambda$ 
hypernuclei $^{~~5}_{\Lambda\Lambda}$H and  $^{~~5}_{\Lambda\Lambda}$He 
are particle-stable, with $\Delta B_{\Lambda\Lambda} \sim 0.5 - 1$~MeV 
(Filikhin and Gal, 2002a; Filikhin, Gal, and Suslov, 2003; Lanskoy and
Yamamoto, 2004; Nemura \textit{et al.}, 2005). This is demonstrated 
in Fig.~\ref{fig:filikhin} where calculated $\Delta B_{\Lambda\Lambda}(A=5)$ 
values, for several potentials $V_{\Lambda\Lambda}$ with different strengths, 
are shown to be correlated with calculated $\Delta B_{\Lambda\Lambda}(A=6)$ 
values. A minimum value of $\Delta B_{\Lambda\Lambda}(A=5)\approx 0.1$ is seen 
to be required for getting $\Delta B_{\Lambda\Lambda}(A=6)>0$, and for the 
actual value of $\Delta B_{\Lambda\Lambda}(A=6)=0.67\pm 0.17$~MeV the $A=5$ 
$\Lambda\Lambda$ hypernuclei come out safely bound. It was also argued that 
$\Lambda\Lambda -\Xi N$ coupling is particularly important for the binding of 
the $A=5$ $\Lambda\Lambda$ hypernuclei, increasing $\Delta B_{\Lambda\Lambda}$ 
for these systems above the corresponding value of 1 MeV in $^{~~6}_{\Lambda
\Lambda}$He, with the Nijmegen model ESC04d giving as much as 2 MeV 
\cite{yamamoto08}. In addition, substantial charge-symmetry breaking effects 
are expected in these systems, resulting in a higher binding energy of 
$^{~~5}_{\Lambda\Lambda}$He by up to 0.5 MeV with respect to $^{~~5}_{\Lambda
\Lambda}$H \cite{lanskoy04,yamamoto08}. 

Whereas the assignment of $_{\Lambda\Lambda}^{~~6}{\rm He}$ to the KEK E373 
emulsion event \cite{takahashi01} is unique, because it has no particle-stable 
excited states and the daughter \lamb{5}{He} hypernucleus has 
no particle-stable excited states to be formed in sequential $\pi^-$ weak 
decays, the assignment of other, heavier $\Lambda\Lambda$ hypernuclei to the 
few emulsion events reported by the KEK E176 and KEK E373 experiments is 
plagued by ambiguities resulting from the presence of particle-stable excited 
states in which a $\Lambda\Lambda$ hypernucleus may be formed or to which it 
may weakly decay. In fact, the $B_{\Lambda\Lambda}^{\rm exp}$ value listed in 
Table~\ref{tab:LL} for the KEK E373 Demachi-Yanagi event \cite{ahn01b} 
assumes that $^{~10}_{\Lambda\Lambda}{\rm Be}$ was formed in its $2^+$ first 
excited state \cite{filikhin02c,hiyama02}, whereas the earlier observation of 
$^{~10}_{\Lambda\Lambda}{\rm Be}$ \cite{danysz63} was interpreted as involving 
the weak decay of $^{~10}_{\Lambda\Lambda}{\rm Be}_{\rm g.s.}$ to the 
excited doublet levels ($3/2^+,5/2^+$) in \lamb{9}{Be} \cite{danysz63a}. 
The $\approx$3~MeV unobserved $\gamma$-ray de-excitation energy has to 
be accounted for in each one of these scenarios, and the $\approx$6~MeV 
difference between the $B_{\Lambda\Lambda}^{\rm exp}$ values originally 
claimed for these two events of $^{~10}_{\Lambda\Lambda}{\rm Be}$ is 
consistent (6=3+3) with the reinterpretations offered here. Other scenarios, 
involving production neutrons or decay neutrons which are unobserved 
in emulsion, have also been considered \cite{davis05}. Similarly, the 
$B_{\Lambda\Lambda}^{\rm exp}$ value assigned in the table to $^{~13}_{\Lambda
\Lambda}{\rm B}$ also assumes an unobserved $\gamma$ ray $E_{\gamma}\approx
$4.8~MeV from the electromagnetic decay of the excited doublet levels 
($3/2^+,5/2^+$) in \lam{13}{C} formed in the weak decay 
$^{~13}_{\Lambda\Lambda}{\rm B}\to$\,\lam{13}{C}($3/2^+,5/2^+$).  
 
Table~\ref{tab:LL} provides a comprehensive listing of candidate $\Lambda
\Lambda$-hypernuclear emulsion events, along with $\Lambda\Lambda$ binding 
energy values derived from these events, with caveats explained earlier for 
$^{~10}_{\Lambda\Lambda}{\rm Be}$ and $^{~13}_{\Lambda\Lambda}{\rm B}$. 
The table also lists calculated $\Lambda\Lambda$ binding energies using (i) 
few-body cluster models \cite{hiyama02,hiyama10}, and (ii) shell-model 
evaluations \cite{gal11}. The table makes it clear that the shell-model 
methodology is able to confront any of the reported $\Lambda\Lambda$ species, 
whereas cluster models have been limited so far to 3-,4- and 5-body 
calculations. For those $\Lambda\Lambda$ hypernuclei where a comparison 
between the two models is possible, the calculated binding energies are 
remarkably close to each other. The shell-model (SM) estimate for 
$B_{\Lambda\Lambda}$ in the nuclear $p$ shell is given simply by 
\begin{equation} 
B_{\Lambda\Lambda}^{\rm SM}(_{\Lambda\Lambda}^{~~\rm A}{\rm Z})=
2{\overline B}_{\Lambda}({_{~~~\Lambda}^{\rm A-1}{\rm Z}}) + 
\langle V_{\Lambda\Lambda}\rangle_{\rm SM} , 
\label{eq:BLL} 
\end{equation} 
where $\langle V_{\Lambda\Lambda}\rangle_{\rm SM}$ is a $\Lambda\Lambda$ interaction 
matrix element identified with $\Delta B_{\Lambda\Lambda}(_{\Lambda
\Lambda}^{~~6}{\rm He})=0.67\pm 0.17$~MeV. In cluster-model (CM) calculations 
\cite{hiyama10}, $\langle V_{\Lambda\Lambda}\rangle_{\rm CM} \equiv B_{\Lambda\Lambda}
(V_{\Lambda\Lambda}\neq 0)-B_{\Lambda\Lambda}(V_{\Lambda\Lambda}=0)$ assumes 
similar values: 0.54, 0.53 and 056~MeV for $_{\Lambda\Lambda}^{~~6}{\rm He}$, 
$_{\Lambda\Lambda}^{~10}{\rm Be}$ and $_{\Lambda\Lambda}^{~11}{\rm Be}$, 
respectively. To apply Eq.~(\ref{eq:BLL}), 
${\overline B}_{\Lambda}({_{~~~\Lambda}^{\rm A-1}{\rm Z}})$ is derived from the 
shell-model calculations outlined in Sec.~\ref{subsubsec:pshell} on $p$-shell 
single-$\Lambda$ hypernuclei. Apart from the spin dependence of the $\Lambda 
N$ interaction, which is fully constrained by the $\gamma$-ray measurements and 
their shell-model analyses, the validity of a uniform shell-model description 
of hypernuclei throughout the whole $p$ shell depends on the constancy of the 
$\Lambda N$ spin-independent matrix element $\overline{V}$ in the mass range 
considered. Indeed, excluding $^{9}_\Lambda{\rm Be}$ which deviates 
substantially from the 
other species, a common value $\overline{V}^{\rm SM}\!=\!-1.06\pm 0.03$ MeV 
can be assigned. In $^{9}_\Lambda{\rm Be}$, the $\Lambda$ hyperon is attached 
to a somewhat loose $\alpha-\alpha$ structure, but in 
$_{\Lambda\Lambda}^{~10}{\rm Be}$ the second $\Lambda$ is bound with respect 
a normal \lamb{5}{He}--$\alpha$ structure. This suggests an extension of the 
validity of Eq.~(\ref{eq:BLL}) also to $_{\Lambda\Lambda}^{~10}{\rm Be}$ by 
adding to its right-hand side a correction term $\delta B_{\Lambda\Lambda}^{\rm SM}$ 
due to the normally bound second $\Lambda$: 
\begin{equation} 
\delta B_{\Lambda\Lambda}^{\rm SM}(_{\Lambda\Lambda}^{~~\rm A}{\rm Z})=(A-6)\,
\left[\,\overline{V}({_{~~~\Lambda}^{\rm A-1}{\rm Z}})-\overline{V}^{\rm SM}\,
\right], 
\label{eq:deltaBLL} 
\end{equation} 
where $\Lambda\!-\!\Sigma$ contributions $\lesssim 0.1$ MeV were disregarded. 
Cluster models, on the other hand, are able to treat the $^8$Be core in terms 
of a loose $\alpha-\alpha$ structure, as well as $^{9}_\Lambda{\rm Be}$ and 
$_{\Lambda\Lambda}^{~10}{\rm Be}$ as $\alpha\alpha n$ and $\alpha\alpha nn$ 
clusters, respectively, but they encounter difficulties in consistently
evaluating spin-dependent $\Lambda N$ interaction contributions.  

Inspection of Table~\ref{tab:LL} shows that the binding energies of both 
$_{\Lambda\Lambda}^{~10}{\rm Be}$ and $_{\Lambda\Lambda}^{~13}{\rm B}$ are 
well reproduced by the shell model, thereby confirming the interpretations 
of the corresponding emulsion events discussed earlier. Of the other 
$\Lambda\Lambda$ hypernuclear candidates, the E373-Hida event \cite{ahn13} 
does not fit any reasonable assignment as $_{\Lambda\Lambda}^{~11}{\rm Be}$ 
or $_{\Lambda\Lambda}^{~12}{\rm Be}$. Regarding the species listed in the 
table as due to E176, they all correspond to different assignments of the 
\textit{same} event, for which the $_{\Lambda\Lambda}^{~13}{\rm B}$ assignment 
is statistically preferable \cite{aoki09}.

\section{$\Xi$ Hypernuclei}

Very little is established 
experimentally or phenomenologically on the interaction of $\Xi$ hyperons 
with nuclei. Dover and Gal (1983), analyzing old emulsion data which were 
interpreted as due to $\Xi^-$ hypernuclei, obtained an attractive $\Xi$-nucleus
interaction with a nuclear potential well depth of $-V_{0}^{\Xi}=21-24$~MeV. 
This range of values agreed well with the theoretical 
prediction~\cite{dover84b} for $\Xi$ in nuclear matter, using 
the early hard-core model D of the Nijmegen group (Nagels, Rijken, and de Swart,
1977) to describe baryon-baryon interactions in a SU(3)$_{\rm f}$ framework. 
However, this is in contrast with the $\Xi$-nucleus repulsion obtained using 
the other hard-core model, model F (Nagels, Rijken, and de Swart,
1979). Predictions made subsequently using more detailed $G$-matrix studies 
(Yamamoto \textit{et al.}, 1994; Yamamoto, 1995a, 1995b) 
spanned a whole range of $\Xi$-nucleus well depths by varying the hard-core 
radius in these Nijmegen models. The confidence in the predictive power of 
model D in strangeness $-$2 hypernuclear physics was due, to a large extent, 
to its success in yielding the substantial attractive $\Lambda\Lambda$ 
interaction that was deemed necessary to reproduce the three known $\Lambda
\Lambda$ binding energies before 2001. This picture has changed since then 
for several reasons.  

\begin{itemize} 

\item Inclusive $(K^-,K^+)$ spectra taken at KEK and at BNL on $^{12}$C 
\cite{fukuda98,khaustov00}  yield more moderate 
values for the attractive $\Xi$ well depth, $-V_{0}^{\Xi} \sim 15$~MeV 
when fitted near the $\Xi^-$-hypernuclear threshold.  

\item The uniquely identified $_{\Lambda \Lambda}^{~~6}$He hypernucleus 
\cite{takahashi01} implies a considerably weaker $\Lambda \Lambda$ interaction 
than produced by the original version of the Nijmegen hard-core Model D. 
The Nijmegen soft-core potentials NSC97~\cite{stoks99} and extended soft-core 
potentials ESC04~\cite{rijken06b} provide a more realistic framework for 
the weaker $\Lambda \Lambda$ interaction. The NSC97 potentials slightly 
underestimate $\Delta B_{\Lambda \Lambda}(_{\Lambda \Lambda}^{~6}{\rm He})$, 
whereas the ESC04 potentials overestimate it, occasionally by about 0.5 MeV, 
and the ESC08 potentials only by up to 0.3 MeV (Yamamoto, Motoba, and Rijken,
2010). 

\end{itemize} 

Representative values of isoscalar, $V_0^{\Xi}$, and isovector, $V_1^{\Xi}$, 
$\Xi$ potential depths and width, $\Gamma_{\Xi}$, from $G$-matrix 
calculations at nuclear-matter density ($k_F=1.35~{\rm fm}^{-1}$) using 
the Nijmegen extended soft-core models ESC04d and ESC08c, are listed in 
Table~\ref{tab:Xi1}. The isovector (Lane) potential $V_1^{\Xi}$ is defined by 
Eq.~(\ref{eq:lane}) where $\bm{t}_{\Sigma}$ is replaced by $\bm{t}_{\Xi}$.
The isoscalar potential comes out repulsive in ESC04a,b and attractive 
in ESC04c,d, whereas it is attractive in all ESC08 versions. The focus 
in Table~\ref{tab:Xi1} on attractive $\Xi$-nucleus isoscalar potentials, 
$V_0^{\Xi}<0$, is motivated by the experimental hints from KEK \cite{fukuda98} 
and BNL \cite{khaustov00} mentioned above. Both ESC04d and ESC08c $\Xi N$ 
potentials are attractive in the isospin $I=0,1$ $^3S_1-{^3D_1}$ channels, 
which might lead to $\Xi N$ bound states, while the $^1S_0$ channels 
are repulsive. The models give rise to a positive isovector potential 
depth $V_1^{\Xi}$. The predictions of spin-flavor SU(6) quark models  
(Fujiwara, Kohno, and Suzuki, 2007; Fujiwara, Suzuki, and Nakamoto, 2007) 
differ in detail, but the overall picture for the isoscalar $\Xi$-nuclear 
potential depths is similar, with a slightly attractive isoscalar potential, 
$V_0^\Xi<0$, and a positive isovector potential depth, $V_1^\Xi>0$. In both 
approaches, however, the $\Xi-\alpha$ system will not bind, but $3N-\Xi$ bound 
states are predicted depending on the spin-isospin two-body model dependence. 

\begin{table} 
\caption{Isoscalar, $V_0^{\Xi}$, and isovector, $V_1^{\Xi}$, $\Xi$ 
nuclear-matter potential depths, and widths $\Gamma_{\Xi}$, all in MeV, 
in recent extended soft core (ESC) Nijmegen potentials, ESC04 \cite{rijken06b} 
and ESC08 (Nagels, Rijken, and Yamamoto, 20015a).}  
\label{tab:Xi1} 
\begin{ruledtabular} 
\begin{tabular}{cccc} 
Potential & $V_0^{\Xi}$ & $V_1^{\Xi}$ & $\Gamma_{\Xi}$  \\  
\hline 
% ESC04a  & $+$15.1 & $+$32.5 &  --  \\ 
  ESC04d  & $-$18.7 & $+$50.9 & 11.4 \\ 
% ESC04d* & $-$12.1 & $+$51.5 & 12.7 \\ 
  ESC08c  & $-$7.0  & $+$21.6 & 4.5  \\
\end{tabular} 
\end{ruledtabular}    
\end{table} 

If the interaction of $\Xi$ hyperons with nuclei is sufficiently attractive 
to cause binding as has been repeatedly argued since the original work of 
\cite{dover83}, then a rich source of spectroscopic information would become 
available and the properties of the in-medium $\Xi N$ interaction could be 
extracted. Few-body cluster model calculations using the ESC04d model have 
been reported recently \cite{hiyama08}. Bound states of $\Xi$ hypernuclei 
would also be useful as a gateway to form double-$\Lambda$ hypernuclei 
(Dover, Gal, and Millener, 1994; Ikeda \textit{et al.}, 1994; 
Millener, Dover, and Gal, 1994; Yamamoto \textit{et al.}, 1994). Finally, a 
minimum strength of about 15 MeV for $-V_{0}^{\Xi}$ is required to realize the 
exciting possibility of ``strange hadronic matter''~\cite{schaffner00}, 
where protons, neutrons, $\Lambda$'s and $\Xi$'s are held together to form 
a system which is stable against strong-interaction decay. 

\section{Strange Dense Matter} 

\subsection{Strange hadronic matter} 
\label{subsec:SHM}

Bodmer (1971), and more specifically Witten (1984), 
suggested that strange quark matter, with roughly equal composition of 
$u$, $d$ and $s$ quarks, might provide an absolutely 
stable form of matter. Metastable strange quark matter was studied by 
Chin and Kerman (1979). Jaffe and collaborators (Farhi and Jaffe, 1984;
Berger and Jaffe, 1987) 
subsequently charted the various scenarios possible for the stability 
of strange quark matter, from absolute stability down to metastability 
due to weak decays. Finite strange quark systems, so called strangelets, 
have also been considered \cite{farhi84,gilson93}. 

\begin{figure} 
\includegraphics[width=8.5cm]{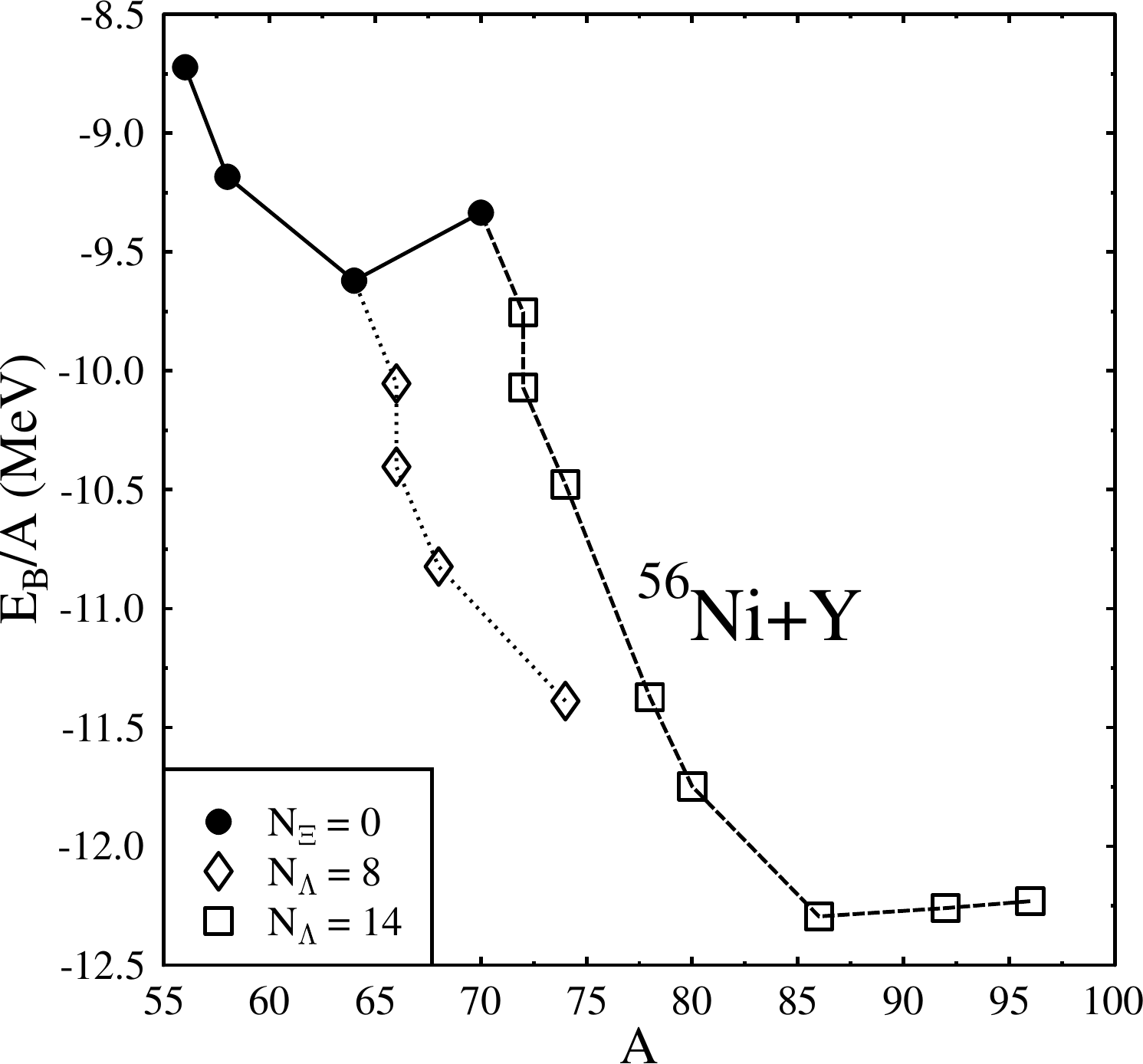} 
\caption{Calculated binding energy of multi-strange nuclei of $^{56}{\rm Ni}$ 
plus $\Lambda$ and $\Xi$ hyperons, as function of baryon number $A$.  
From Schaffner \textit{et al.}, 1993.} 
\label{fig:Ni} 
\end{figure}

Less known is the suggestion \cite{schaffner93,schaffner94} that metastable 
strange systems with similar properties, i.e., a strangeness fraction 
$f_{S} \equiv -S/A \approx 1$ and a charge fraction 
$f_{Q} \equiv Z/A \approx 0$, might also exist in hadronic form at moderate 
values of density, between 2 and 3 times nuclear matter density. 
These strange systems are made of $N$, $\Lambda$ and $\Xi$ baryons. 
The metastability (i.e., stability with respect to strong interactions, 
but not to $\Delta S \not= 0$ weak-interaction decays) of these strange 
hadronic systems was established by extending relativistic mean field (RMF) 
calculations from ordinary nuclei ($f_{S} = 0$) to multi-strange nuclei 
with $f_{S}\not= 0$. Although the detailed pattern of metastability, 
as well as the actual values of the binding energy, depend specifically 
on the partly unknown hyperon potentials in dense matter, the 
predicted phenomenon of metastability turned out to be robust in these 
calculations (Balberg, Gal, and Schaffner, 1994). A conservative example is given in 
Fig.~\ref{fig:Ni}, assuming a relatively weakly attractive hyperon-hyperon 
interaction. The figure shows the calculated binding energy of 
$^{56}{\rm Ni} + N_{\Lambda}\Lambda$ multi-$\Lambda$ hypernuclei for 
$N_{\Lambda}=0,2,8,14$ and how it becomes energetically favorable to 
add $\Xi$ hyperons when $N_{\Lambda}$ exceeds some fairly small threshold 
value. As soon as the $\Lambda$ $p$-shell is filled, $\Xi$ hyperons 
may be placed in their $s$-shell owing to Pauli blocking of the  
strong-interaction conversion process $\Xi N \to \Lambda\Lambda$ which in 
free space releases about 25 MeV.   

\begin{figure}[t]
\includegraphics[width=8.0cm]{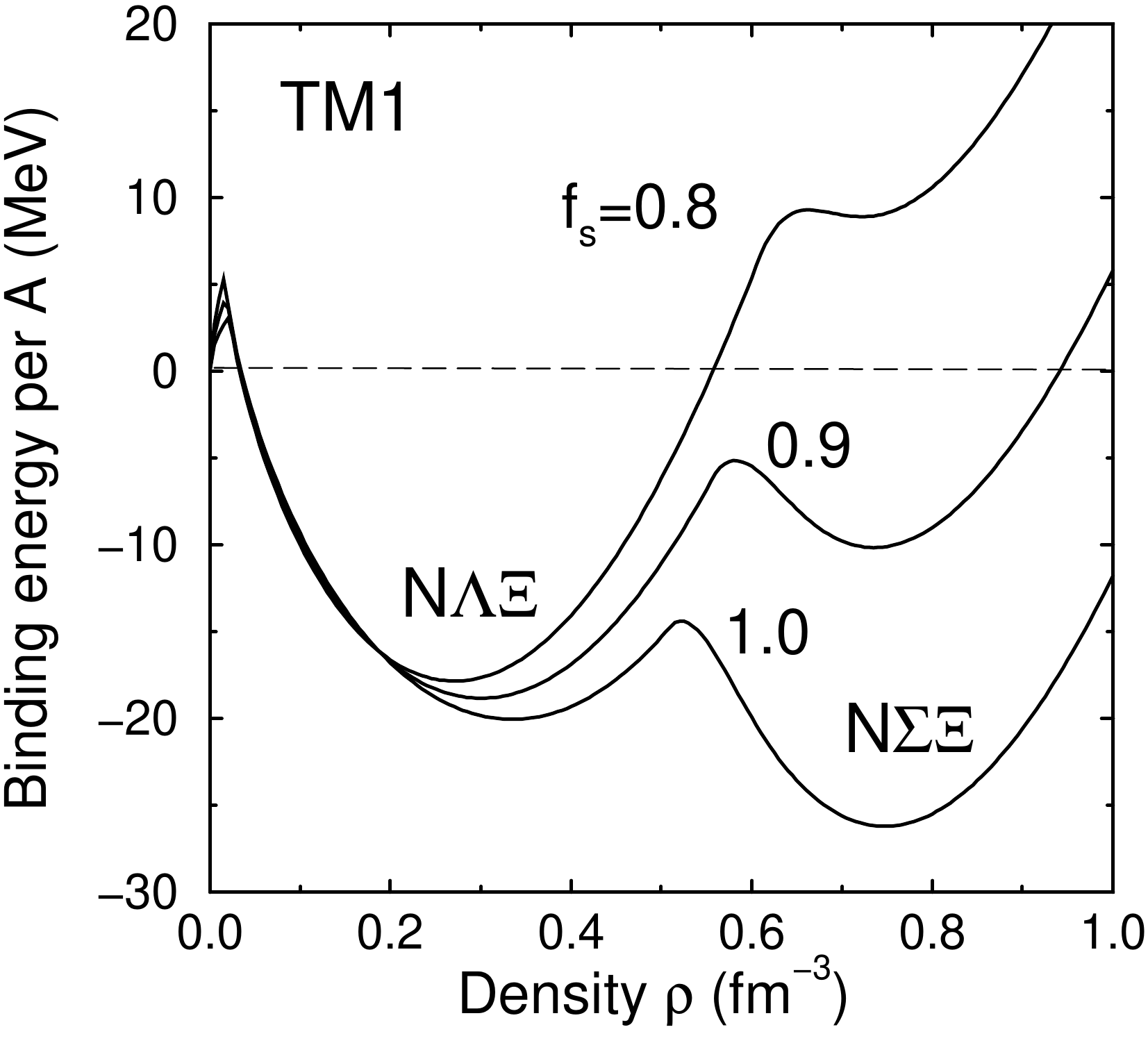} 
\caption{Transition from $N\Lambda\Xi$ to $N\Sigma\Xi$ matter 
upon increasing the strangeness fraction $f_{S}$. From Schaffner-Bielich
and Gal, 2000.} 
\label{fig:shm1} 
\end{figure} 

A less conservative example is provided by applying the Nijmegen soft-core 
model NSC97 \cite{stoks99} which predicts strongly attractive $\Xi \Xi$, 
$\Sigma \Sigma$ and $\Sigma \Xi$ interactions, but fairly weak 
$\Lambda \Lambda$ and $N \Xi$ interactions that roughly agree with existing 
phenomenology. Schaffner-Bielich and Gal (2000) found that strange hadronic 
matter (SHM) is comfortably metastable for any allowed value of $f_{S} > 0$. 
However for $f_{S} \geq 1$, $\Sigma$'s replace $\Lambda$'s due to the 
exceptionally strong $\Sigma \Sigma$ and $\Sigma \Xi$ interactions in this 
model. A first-order phase transition occurs from 
$N\Lambda\Xi$ dominated matter for $f_{S} \leq 1$ to $N\Sigma\Xi$ dominated 
matter for $f_{S} \geq 1$, as shown in Fig.~\ref{fig:shm1} where the binding 
energy is drawn versus the baryon density for several representative fixed 
values of $f_S$. At $f_S \approx 1.0$ a secondary minimum at higher baryon 
density becomes energetically favored. The system then undergoes a first-order 
phase transition from the low-density state to the high-density state. 

\begin{figure}[t]  
\includegraphics[width=8.0cm]{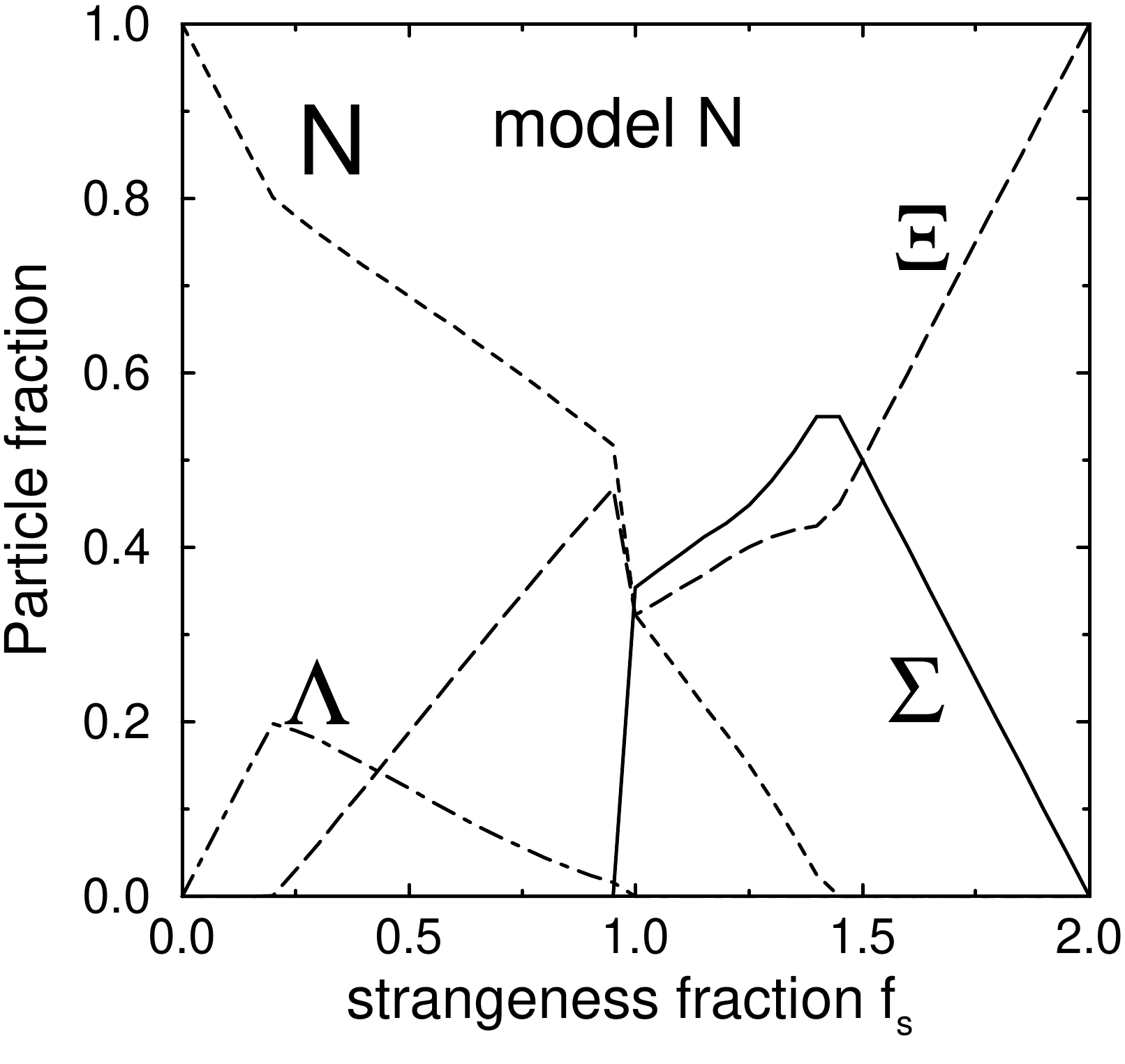}  
\caption{Strange hadronic matter composition as function of strangeness 
fraction $f_S$. From Schaffner-Bielich and Gal, 2000.} 
\label{fig:shm2} 
\end{figure} 

Figure~\ref{fig:shm2} demonstrates explicitly that the phase transition
involves transformation from $N\Lambda\Xi$ dominated matter to $N\Sigma\Xi$
dominated matter, by showing the calculated composition of SHM for this 
model (denoted $N$ for Nijmegen) as function of the strangeness fraction $f_S$.
The particle fractions for each baryon species change as function of
$f_S$. At $f_S=0$, one has pure nuclear matter, whereas at $f_S=2$ one has
pure $\Xi$ matter. In between, matter is composed of baryons as dictated by
chemical equilibrium. A change in the particle fraction may occur quite
drastically when new particles appear, or existing ones disappear. 
A sudden change in the composition is seen in Fig. \ref{fig:shm2} 
for $f_S=0.2$ when $\Xi$'s (long-dashed line) emerge in the medium,
or at $f_S=1.45$ when nucleons (short-dashed line) disappear. 
The situation at $f_S=0.95$ is a special one, as $\Sigma$'s (solid line) 
appear in the medium, marking the first-order phase transition observed 
in the previous figure. The baryon composition alters completely at that 
point, from $N\Xi$ baryons plus a rapidly vanishing fraction of $\Lambda$'s 
(dot-dashed line) into $\Sigma\Xi$ hyperons plus a decreasing fraction of 
nucleons. At the very deep minimum of the binding-energy curve (not shown 
here) SHM is composed mainly of $\Sigma$'s and $\Xi$'s with a very small 
admixture of nucleons. The phase transition demonstrated above has been 
discussed by the Frankfurt group \cite{schaffner02} in the context of a phase 
transition to hyperon matter in neutron stars. Unfortunately, it will be 
difficult to devise an experiment to determine the depth of the $\Lambda\Xi$, 
$\Xi\Xi$, $\Xi\Sigma$, $\Sigma\Sigma$ interaction potentials, which are so 
crucial to verify these results.  

\subsection{Neutron stars}
\label{subsec:NS} 

Neutron stars are gravitationally bound massive objects in $\beta$ equilibrium 
with radii of about 12 km and masses of about $(1-2)M_{\odot}$, perhaps up to 
$2.5M_{\odot}$. Here $M_{\odot}$ stands for a solar mass \cite{leahy11}. 
Although their composition at low density is dominated by neutrons, 
transmutation to hyperons, beginning at 2 to 3 times normal nuclear matter 
density $\rho_0=0.17$~fm$^{-3}$, would act to alleviate the Pauli pressure 
of nucleons and leptons. Matter in the core of neutron stars is further 
compressed to about $(5-6)\rho_0$. At these high densities strange hadronic 
matter, which may already be self bound at densities $(2-3)\rho_0$, 
could become stable even to weak decay \cite{schaffner02}. Such matter 
may perhaps form kaon condensates \cite{kaplan86} and even deconfine to 
quarks \cite{baym76}, forming strange quark matter. However, it is also 
possible that a star having a mixed phase of hyperons and quarks in its 
interior is produced. Because the star rapidly rotates, losing energy via 
radiation, the rotational inertia of the star changes, and the rotational 
frequency depends on its composition which is coupled to the rotational 
frequency. Obviously, while more astrophysical observations are needed, 
the only terrestrial handle on this physics comes from hypernuclei, 
particularly multi-strange hypernuclei. The physics of neutron stars 
was reviewed recently by Lattimer (2012). 

\begin{figure}[t] 
\includegraphics[width=8cm]{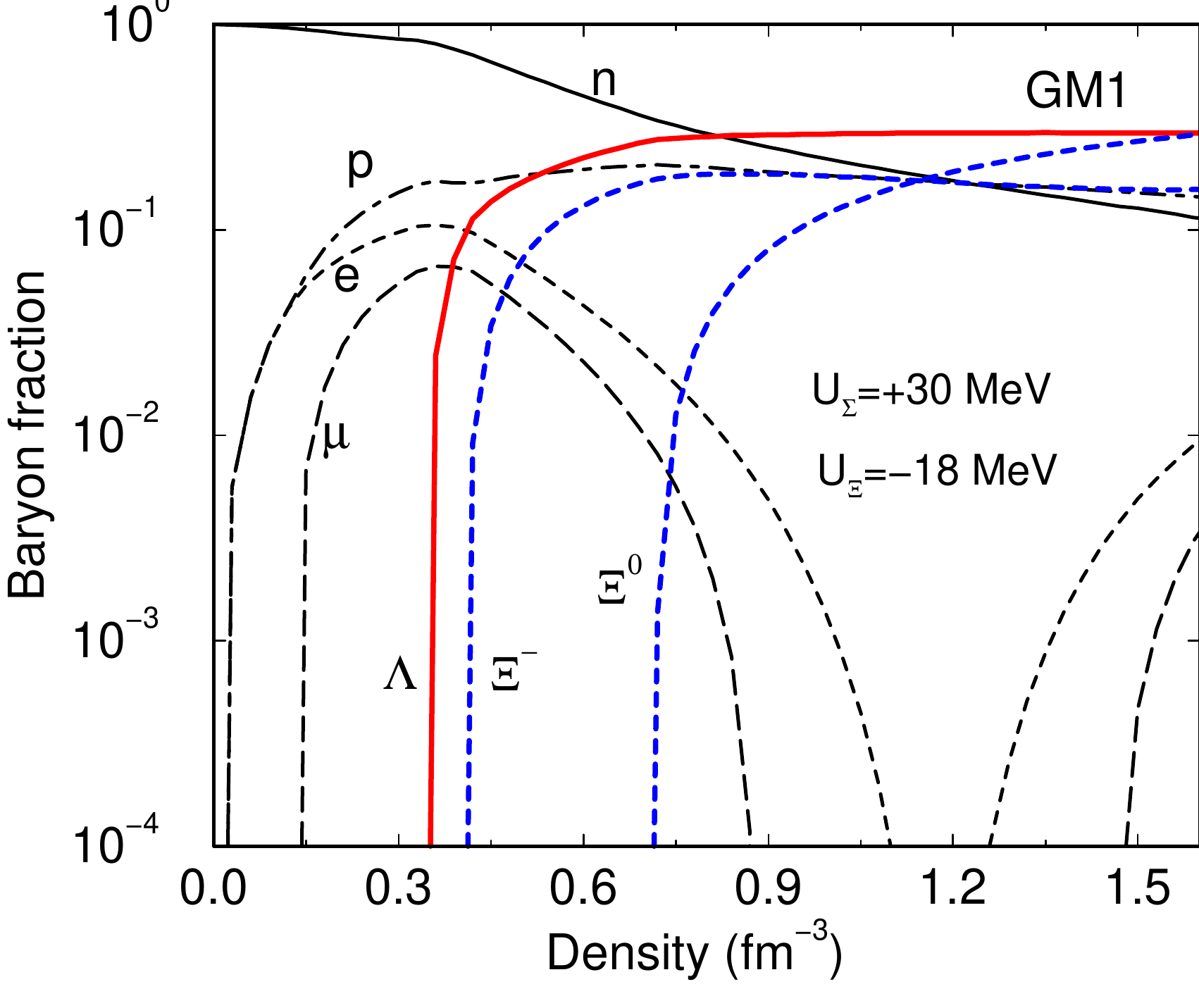}
\caption{Neutron star matter fractions of baryons and 
leptons, calculated as a function of density. From Schaffner-Bielich, 2008. 
\label{fig:n_star1}} 
\end{figure} 

It is important to recognize that hypernuclei, and, in particular, 
multistrange hypernuclei which were reviewed in Sec.~\ref{subsec:SHM}, are a 
low-density manifestation of strange hadronic matter. As such, studies of their 
interactions at normal nuclear density impact the construction of models of 
density-dependent interactions for use at higher densities. Thus, hyperon 
potentials in dense matter control the composition of dense neutron-star 
matter, as shown by a recent RMF calculation in Fig.~\ref{fig:n_star1}. 
As a function of density, the first hyperon to appear is the lightest one, 
the $\Lambda$ at about $2\rho_0$, by converting protons and electrons 
directly to $\Lambda$'s instead of neutrons, thereby decreasing the neutron 
Pauli pressure. It is reasonable to assume that this composition varies 
radially, perhaps having a crust and an atmosphere composed of neutrons. 
Among the negatively charged hyperons, the lightest one $\Sigma^-$ does not 
appear at all over the wide range of densities shown owing to its repulsion 
in nuclear matter, and most likely also in neutron matter \cite{balberg97}. 
Its potential role in reducing the Pauli pressure of the leptons 
($e^-$ and $\mu^-$) could be replaced by the heavier $\Xi^-$ hyperon, 
assuming overall $\Xi$-nuclear attraction. The specific calculation 
sketched by Fig.~\ref{fig:n_star1} predicts that the hyperon population 
overtakes the nucleon population for densities larger than about 
$6\rho_0$, where the inner core of a neutron star may be viewed as 
a giant hypernucleus \cite{glendenning85}. 

\begin{figure}[t] 
\includegraphics[width=8cm]{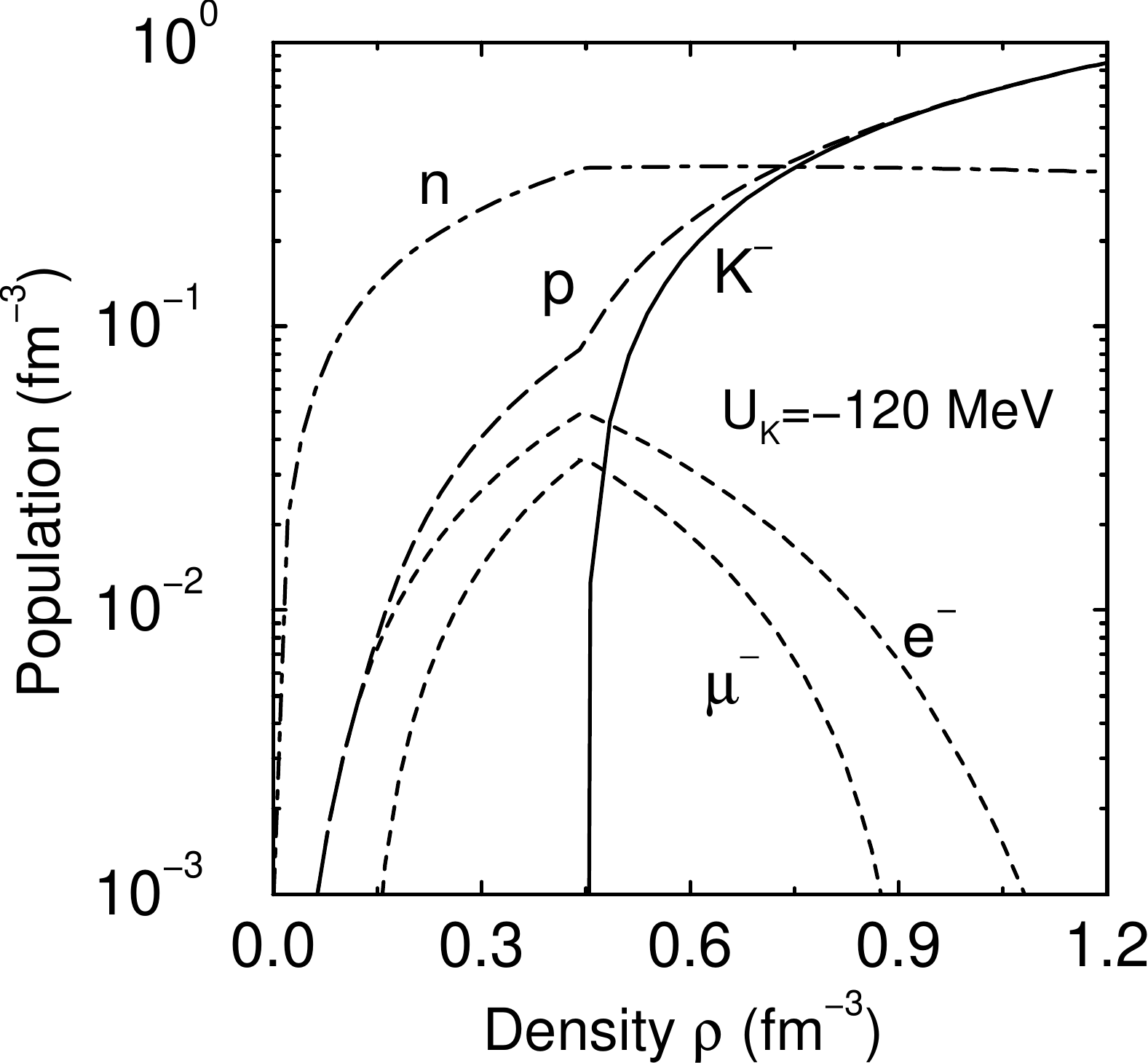}
\caption{Population of neutron star matter, allowing for kaon condensation, 
calculated as a function of nucleon density. From Glendenning and
Schaffner-Bielich, 1999. \label{fig:n_star2}} 
\end{figure}

Negative strangeness may also be injected into neutron-star matter by 
agents other than hyperons. Thus, a robust consequence of the sizable 
$\bar K$-nucleus attraction, as discussed in Sec.~\ref{sec:kbar}, is that 
$K^-$ condensation is expected to occur in neutron stars at a density about 
$3\rho_0$ in the absence of hyperons, as shown in Fig.~\ref{fig:n_star2} 
for a RMF calculation using a strongly attractive $K^-$ nuclear potential 
$U_{\bar K}(\rho_0)=-120$~MeV. Since it is more favorable to produce kaons 
in association with protons, the neutron density shown in the figure stays 
nearly constant once kaons start to condense, while the lepton populations 
decrease as the $K^-$ provides a new neutralizing agent via the weak 
processes $\ell^- \to K^- + \nu_{\ell}$. However, including negatively 
charged hyperons in the equation of state (EoS) of neutron-star matter 
defers $K^-$ condensation to higher densities \cite{knorren95,glendenning01} 
where the neutron-star maximum mass $M_{\rm max}$ is lowered by only 
$\approx 0.01M_{\odot}$ below the value reached through the inclusion 
of hyperons (Knorren, Prakhash, and Ellis, 1995).

\begin{figure}[t] 
\includegraphics[width=8.0cm,clip]{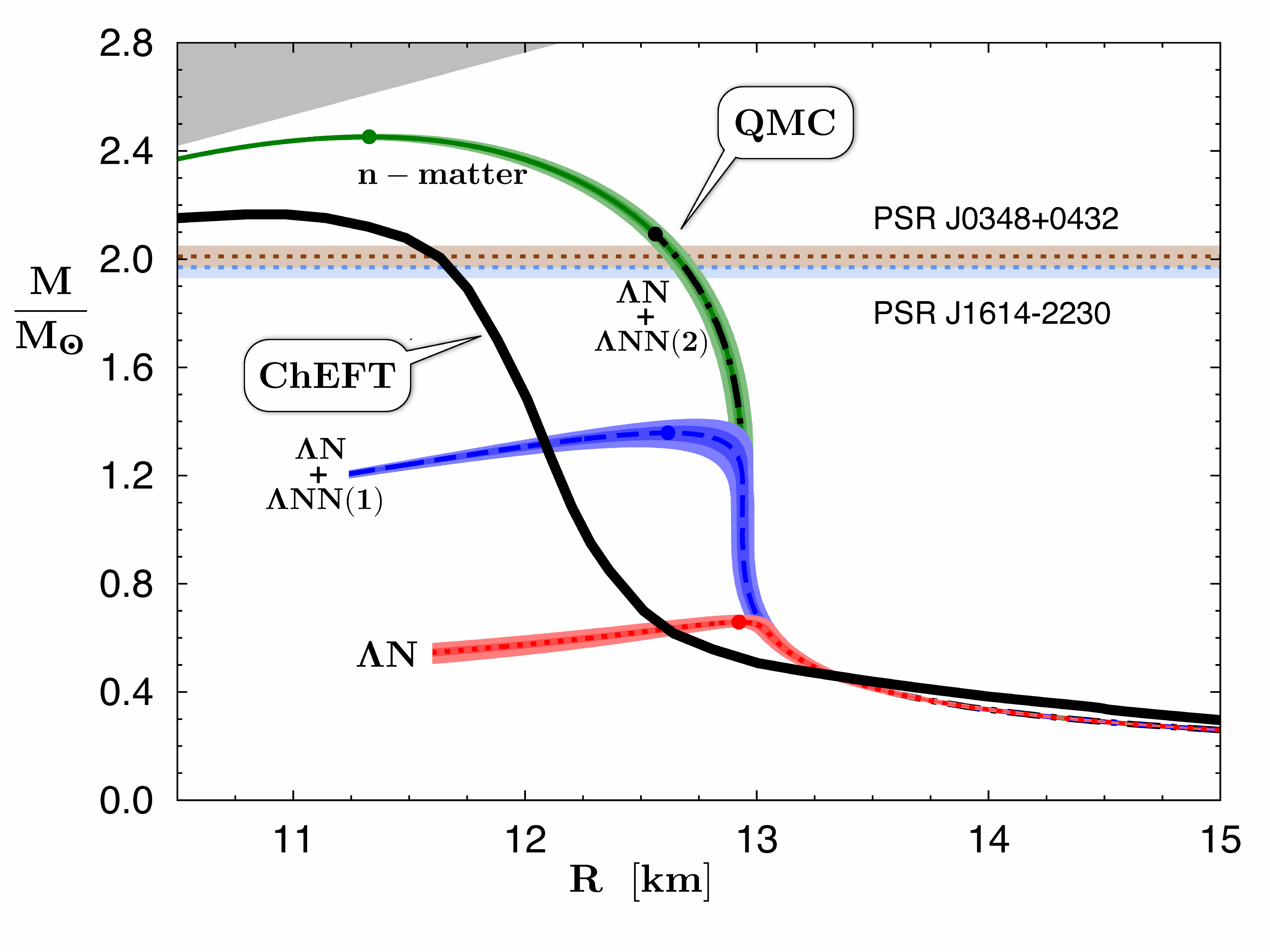} 
\caption{Mass-radius relationship for various EoS scenarios of neutron stars, 
including nucleons and leptons only \cite{weise14} as well as upon including 
$\Lambda$ hyperons \cite{lonardoni15}. From Weise, 2015. 
\label{fig:n_star3}} 
\end{figure} 

Given the high matter density expected in a neutron star, a phase transition 
from ordinary nuclear matter to some exotic mixtures cannot be ruled out. 
Whether a stable neutron star is composed dominantly of hyperons, quarks, 
or some mixture thereof, and just how this occurs, is not clear as both the 
strong and weak interactions, which operate on inherently different time 
scales, are in play. The EoS of any possible composition constrains the 
mass-radius relationship for a rotating neutron star. Thus, the 
maximum mass $M_{\rm max}$ for a relativistic free neutron gas is given 
by $M_{\rm max}\approx 0.7M_{\odot}$ \cite{tolman39,oppenheimer39}, 
whereas higher mass limits are obtained under more realistic EoS assumptions. 
Without strangeness, but for interacting nucleons (plus leptons) $M_{\rm max}$ 
comes out invariably above $2M_{\odot}$, as shown by the curves marked 
$n$ matter from Quantum Monte Carlo (QMC) calculations \cite{lonardoni15} and 
chiral effective field theory ($\chi$EFT) \cite{weise14} in Fig.~\ref{fig:n_star3}. 
$M_{\rm max}$ values of up to $2M_{\odot}$ are within the reach of hybrid (nuclear
plus quark matter) star calculations in which strangeness materializes via 
non-hadronic degrees of freedom \cite{alford05}. In the hadronic basis, adding 
hyperons softens the EoS, thereby lowering $M_{\rm max}$ in RMF calculations to 
the range $(1.4-1.8)M_{\odot}$ (Knorren, Prakhash, and Ellis, 1995; Glendenning,
2001), also if and when a phase transition occurs to SHM \cite{schaffner02}. 
More recent Hartree-Fock and Bruckner-Hartree-Fock calculations using the 
NSC97, ESC08 and $\chi$EFT $YN$ interactions find values of $M_{\rm max}$ lower 
than $1.4M_{\odot}$ (Schulze \textit{et al.}, 2006; Djapo, Schaefer, and 
Wambach, 2010; Schulze and Rijken, 2011), while the inclusion of several of the 
$YY$ interactions from the Nijmegen ESC08 model appears to increase 
$M_{\rm max}$ by $0.3M_{\odot}$ to about $1.65M_{\odot}$ \cite{rijken16}.   

\begin{figure} 
\includegraphics[width=8cm,clip]{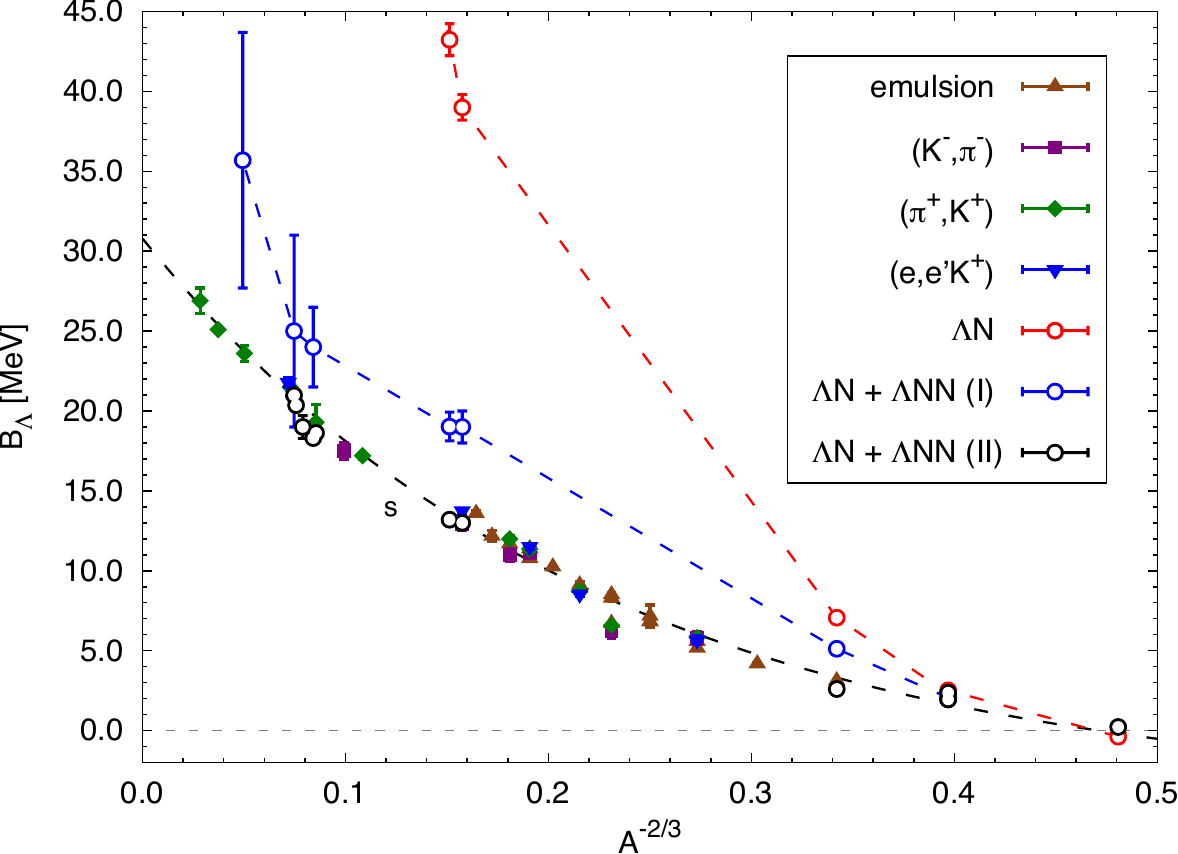} 
\caption{QMC calculations of $\Lambda$ hypernuclear binding energies for 
purely two-body $\Lambda N$ interactions and for two versions of adding 
repulsive $\Lambda NN$ interactions. Adapted from Gandolfi and Lonardoni, 
2015.} 
\label{fig:n_star4} 
\end{figure} 

Until recently, the neutron-star mass distribution for radio binary pulsars 
was given by a narrow Gaussian with mean and width values $(1.35\pm 0.04)
M_{\odot}$ \cite{thorsett99}, somewhat below the Chandrasekhar limit 
of $1.4M_{\odot}$ for white dwarfs, above which these objects become 
gravitationally unstable. However, there is now some good evidence from 
x-ray binaries classified as neutron stars for masses about and greater than 
$2M_{\odot}$ (Barret, Olive, and Miller, 2006). The highest \textit{accepted} 
value of neutron-star mass is provided at present by the precise mass 
measurements of the pulsars PSR J1614-2230 \cite{demorest10} and PSR J0348+0432 
\cite{antoniadis13}, marked by horizontal lines in Fig.~\ref{fig:n_star3}. 
These yield nearly $2M_{\odot}$ and thereby exclude several ``soft'' EoS 
scenarios for dense matter \cite{freire09,lattimer12}. The figure demonstrates 
how the gradual introduction of repulsive $\Lambda NN$ interactions 
\cite{lonardoni15}, from version 1 to version 2, leads to a corresponding 
increase of the calculated $M_{\rm max}$ value by increasing the matter density 
$\rho_{\rm min}$ at which $\Lambda$ hyperons appear first in neutron-star matter 
to higher values, until this $\rho_{\rm min}$ exceeds the value $\rho_{\rm max}$ 
corresponding to $M_{\rm max}$. When this happens, for version 2, the 
mass-radius dotted curve overlaps with the purely ``$n$-matter'' green curve 
below the point marked in the figure for the value of $M_{\rm max}$ reached. 
This scenario in which hyperons are excluded from the EoS of neutron stars 
exclusively by strongly repulsive $YNN$ forces, thereby resolving the 
``hyperon puzzle'', requires further study.

\begin{figure}[t] 
\includegraphics[width=8cm]{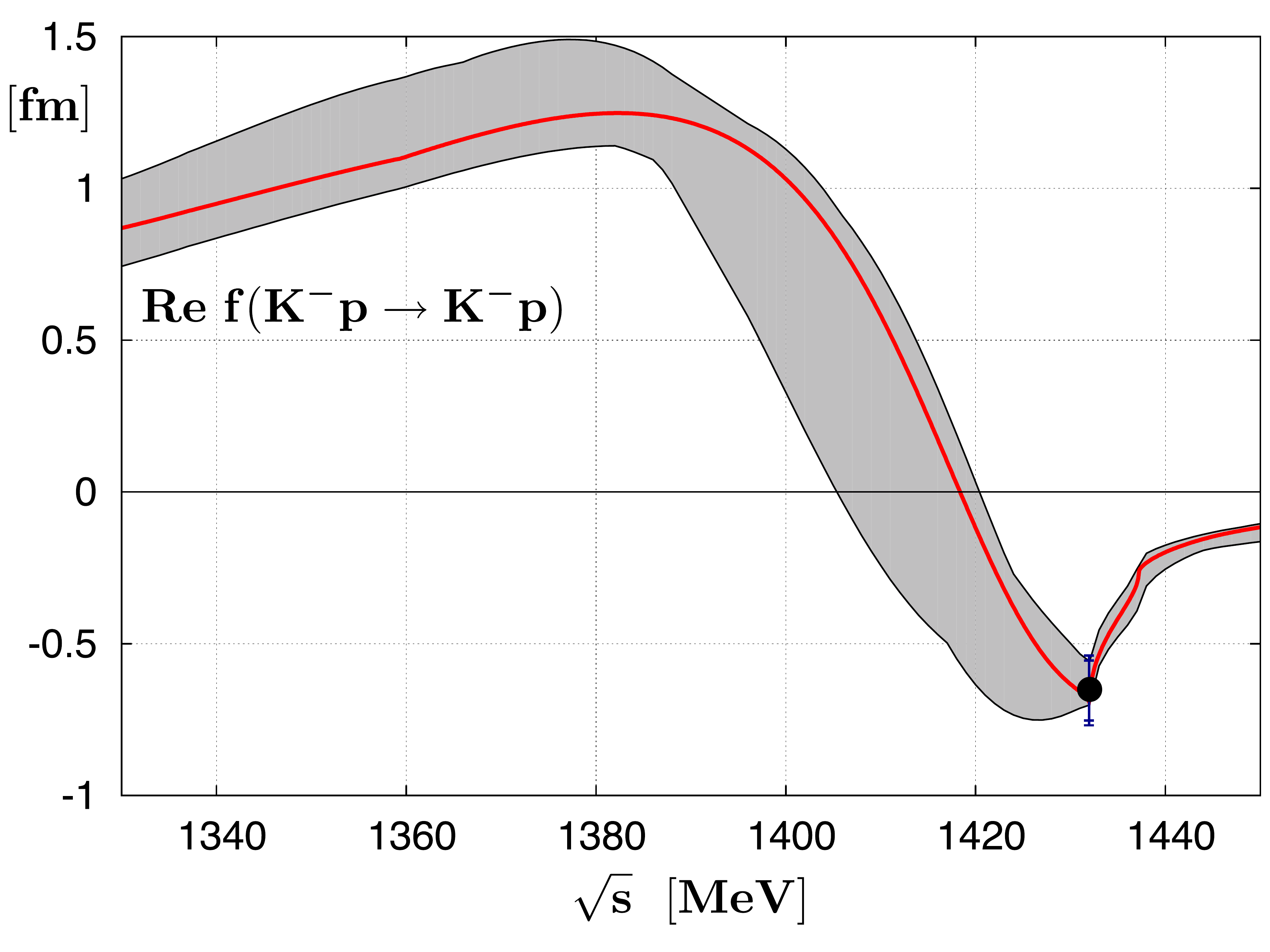} 
\includegraphics[width=8cm]{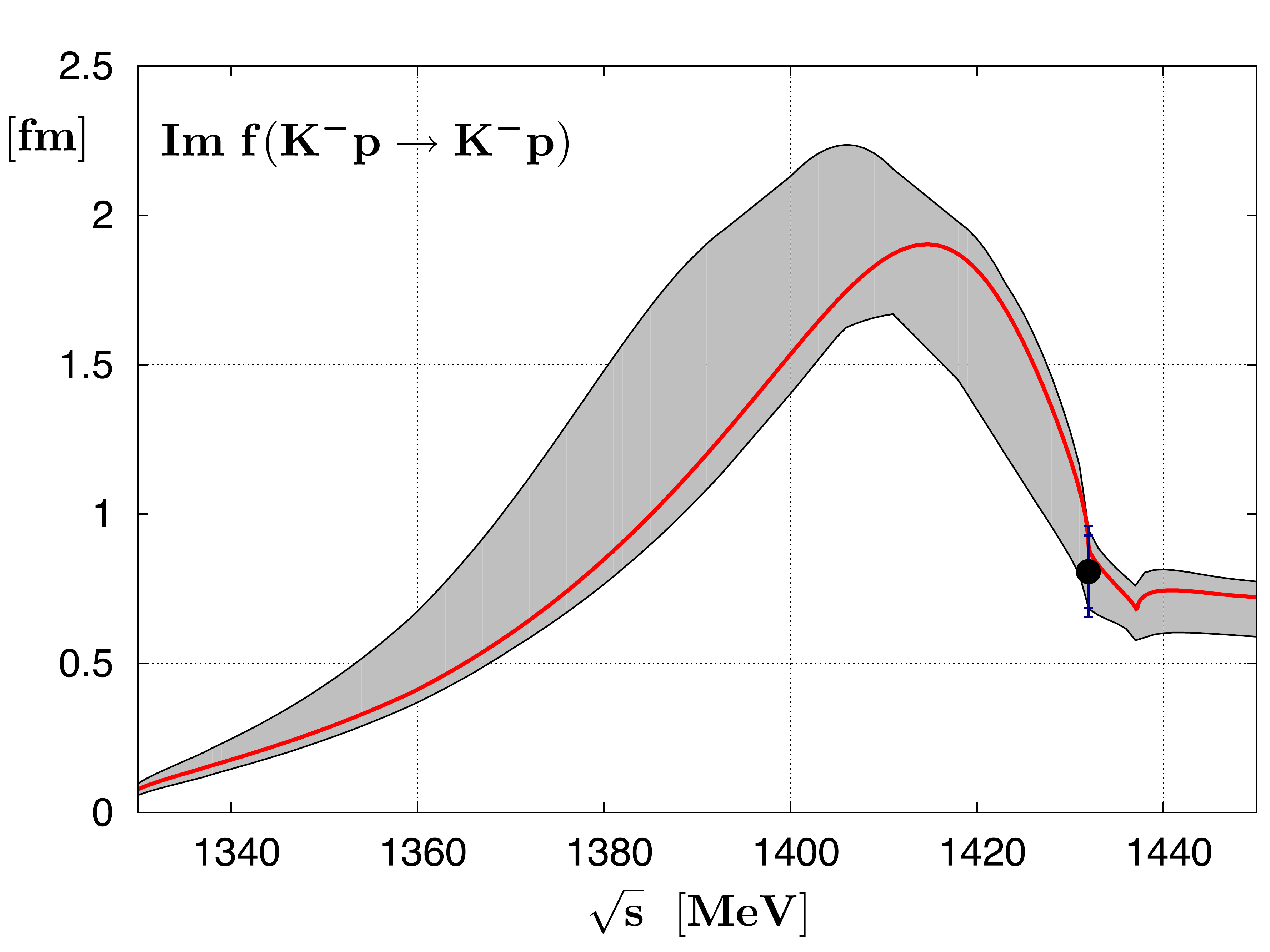} 
\caption{NLO chiral-model calculation of the real and imaginary parts of the 
$K^-p$ cm scattering amplitude, denoted IHW in the text (Ikeda, Hyodo, and 
Weise, 2012). The pole position of the $\Lambda(1405)$ resonance is at 
1424$-i$26 MeV. The $K^-p$ threshold values marked by solid dots follow from 
the SIDDHARTA measurement of kaonic hydrogen $1s$ level shift and 
width \cite{bazzi11,bazzi12}. Adapted from Ikeda, Hyodo, and Weise, 2012.
\label{fig:IHW}} 
\end{figure} 

In this context, Fig.~\ref{fig:n_star4} shows how the introduction of 
repulsive $\Lambda NN$ interactions within QMC calculations relieves 
the over-binding of $\Lambda$ hypernuclei which arises progressively 
with increasing the mass number $A$ (corresponding to smaller values of 
$A^{-2/3}$ in the figure) upon using microsocopically constructed purely 
two-body $\Lambda N$ interactions dominated by attraction. In particular, 
the same version ``$\Lambda N$+$\Lambda NN$~(II)'' that according to 
Fig.~\ref{fig:n_star3} resolves the hyperon puzzle also resolves, 
according to Fig.~\ref{fig:n_star4}, the ``$B_{\Lambda}$ over-binding'' problem. 
It is worth noting, however, that the purely two-body $\Lambda N$ interaction 
of version `$\Lambda N$' overbinds heavy $\Lambda$ hypernuclei substantially 
beyond the $\Lambda N$ two-body contribution $D^{(2)}_{\Lambda}\sim 60$~MeV 
to the $\Lambda$-nucleus potential well depth derived from the $A$ dependence 
of the ($\pi^+,K^+$)-measured $\Lambda$ binding energies (Millener, Dover,
and Gal, 1988). This excessive overbinding is then compensated in Lonardoni,
Pederiva, and Gandolfi (2014) by a similarly excessive $\Lambda NN$ repulsion 
which makes the neutron-star matter EoS so stiff as to exclude hyperons from 
appearing in neutron-star matter. In other, phenomenological models that 
introduce softer repulsive $\Lambda NN$ interactions in a more controlled way, 
values of $M_{\rm max}$ in the range $(1.6-1.7)M_{\odot}$ are obtained 
\cite{balberg97,vidana11}, short however of resolving the hyperon puzzle. 
Nevertheless, it is possible to reach values of $M_{\rm max}\ge 2M_{\odot}$ by 
introducing in addition to moderately repulsive $\Lambda NN$ interactions also 
phenomenological repulsive $NNN$ interactions that have not been tested yet in 
nuclear structure calculations \cite{yamamoto13,yamamoto14,yamamoto16}. 
Obviously, more work is required in this direction to make sure whether 
or not the hyperon puzzle is indeed resolved; see Chatterjee and
Vida\~{n}a (2016) for a comprehensive review of related works.

\section{$\bar K$-Nuclear Interactions and Bound States} 
\label{sec:kbar} 

The $\bar K N$ interaction near and below threshold is attractive 
in models which dynamically generate the $\Lambda(1405)$ subthreshold 
resonance. This motivates a search for $K^-$ quasibound states in 
nuclei \cite{hyodo13,gal13a}. The $\Lambda(1405)$ was predicted as early 
as 1959 \cite{dalitz59b} by analyzing the available data on the strong 
interactions of $K^-$ mesons with protons above threshold, and was discovered 
two years later in the Berkeley hydrogen bubble chamber \cite{alston61} as 
an $I=0$ $\pi\Sigma$ resonance by studying the reaction $K^-p\to\Sigma+3\pi$ 
for several charge states. The proximity of this $\pi\Sigma$ resonance to 
the ${\bar K}N$ threshold, at 1432 MeV for $K^-p$, suggested that it can be 
dynamically generated by ${\bar K}N-\pi\Sigma$ interhadron forces. This was 
subsequently shown (Dalitz, Wong, and Rajasekaran, 1967) to be possible within 
a dynamical model of SU(3)-octet vector-meson exchange. The model provides a 
concrete physical mechanism for the Tomozawa-Weinberg leading term in the 
chiral expansion of the meson-baryon Lagrangian \cite{tomozawa66,weinberg66}. 

\begin{table}[t] 
\caption{Calculated $K^-pp$ binding energies $B$ \& widths $\Gamma$. 
DHW stands for Dot\'{e}, Hyodo, and Weise (2008, 2009), BGL for Barnea, Gal,
and Liverts (2012), IKS for Ikeda, Kamano, and Sato (2010), RS for 
R\'{e}vai and Shevchenko (2014), YA for Yamazaki and Akaishi (2002), 
WG for Wycech and Green (2009), SGM for Shevchenko, Gal and Mare\v{s}
(2007) and Shevchenko \textit{et al.} (2007), and IS 
for Ikeda and Sato (2007, 2009).} 
\begin{ruledtabular} 
\begin{tabular}{lcccc} 
\multicolumn{5}{c}{Energy dependent meson-baryon interactions} \\ 
 & \multicolumn{2}{c}{Variational} & \multicolumn{2}{c}{Faddeev} \\ 
(MeV) & DHW & BGL & IKS & RS \\  
\hline 
$B$ & 17--23 & 16 & 9--16 & 32 \\ 
$\Gamma$ & 40--70 & 41 & 34--46 & 49 \\[5pt] 
\multicolumn{5}{c}{Energy independent meson-baryon interactions} \\ 
 & \multicolumn{2}{c}{Variational} & \multicolumn{2}{c}{Faddeev} \\ 
(MeV) & YA & WG & SGM & IS \\ 
$B$ & 48 & 40--80 & 50--70 & 60--95 \\ 
$\Gamma$ & 61 & 40--85 & 90--110 & 45--80 \\ 
\end{tabular} 
\end{ruledtabular}
\label{tab:kpp} 
\end{table} 

\begin{figure}[t] 
\includegraphics[width=8cm]{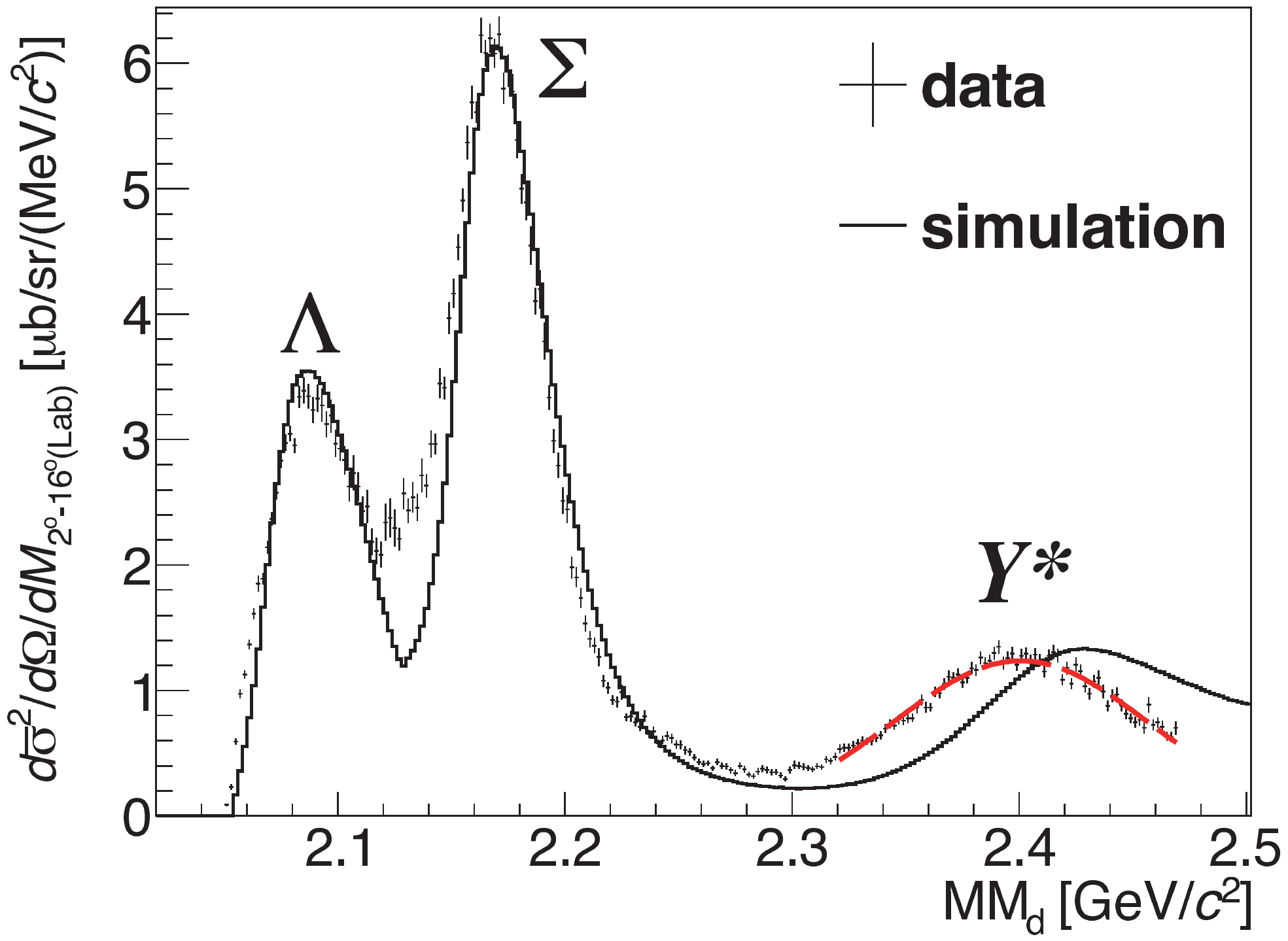} 
\caption{Missing-mass spectrum (MM$_{\rm d}$) of the $d(\pi^+,K^+)$ reaction in 
the J-PARC E27 experiment at forward angles. A phase-space simulated spectrum 
is shown by a solid line. Adapted from Ichikawa \textit{et al.}, 2014.}  
\label{fig:E27a} 
\end{figure}  

A NLO chiral-model calculation of the $K^-p$ center-of-mass (c.m.) scattering 
amplitude $f_{K^-p}$ is shown in Fig.~\ref{fig:IHW}. This NLO amplitude agrees 
qualitatively with leading-order $K^-p$ amplitudes derived in the mid 1990s 
[see, e.g., Kaiser, Siegel, and Weise (1995) and Oset and Ramos (1998)], 
the main quantitative improvement arising from the threshold value constraint 
provided by the SIDDHARTA measurement of kaonic hydrogen $1s$ level shift 
and width \cite{bazzi11,bazzi12}. The large positive values of Re~$f_{K^-p}$, 
which exceed 1~fm in the subthreshold region, indicate a strong attraction. 
Although all NLO models agree above threshold, because of fitting to the 
same $K^-N$ low-energy scattering and reaction data, a non-negligible model 
dependence below threshold can be deduced by comparing to other NLO chiral 
calculations; see, e.g., Guo and Oller (2013). However, it is the 
subthreshold region that is needed in bound-state calculations, which is also 
true for kaonic atoms where the kaon energy is essentially at threshold 
\cite{gal14b}. Fortunately, the two $K^-N$ scattering amplitudes used in the 
most recent atomic and nuclear quasi-bound-state calculations, IHW (Ikeda, 
Hyodo, and Weise, 2011, 2012) of Fig.~\ref{fig:IHW} and NLO30 \cite{cieply12} 
shown in a later figure, are also similar in the subthreshold region despite 
the different methodologies involved in their derivations. 

The lightest $\bar K$-nuclear quasibound state is expected to be $K^-pp$. 
Such a $\bar K NN$ state would have isospin $I=\frac{1}{2}$ and spin-parity 
$J^{\pi}=0^-$, dominated by $I_{NN}=1$ and $s$ waves. A representative 
compilation of recent few-body calculations of this system is given in 
Table~\ref{tab:kpp}. These calculations suggest robust binding for $K^-pp$, 
but the calculated widths are all large (of order 50 MeV). The table shows 
that chiral-model calculations using energy-dependent $\bar KN$ interactions 
give weaker binding than those calculated when disregarding the energy 
dependence away from the $\bar KN$ threshold. Since the $K^-pp$ quasi-bound 
state may be regarded as $\Lambda(1405)N$ bound state (Uchino, Hyodo, and
Oka, 2011), this difference partly reflects the higher $\Lambda(1405)$ mass 
obtained in chiral models [see the caption to Fig.~\ref{fig:IHW} for the 
$\Lambda(1405)$ pole position in that calculation].

\begin{figure}[t] 
\includegraphics[width=8.5cm]{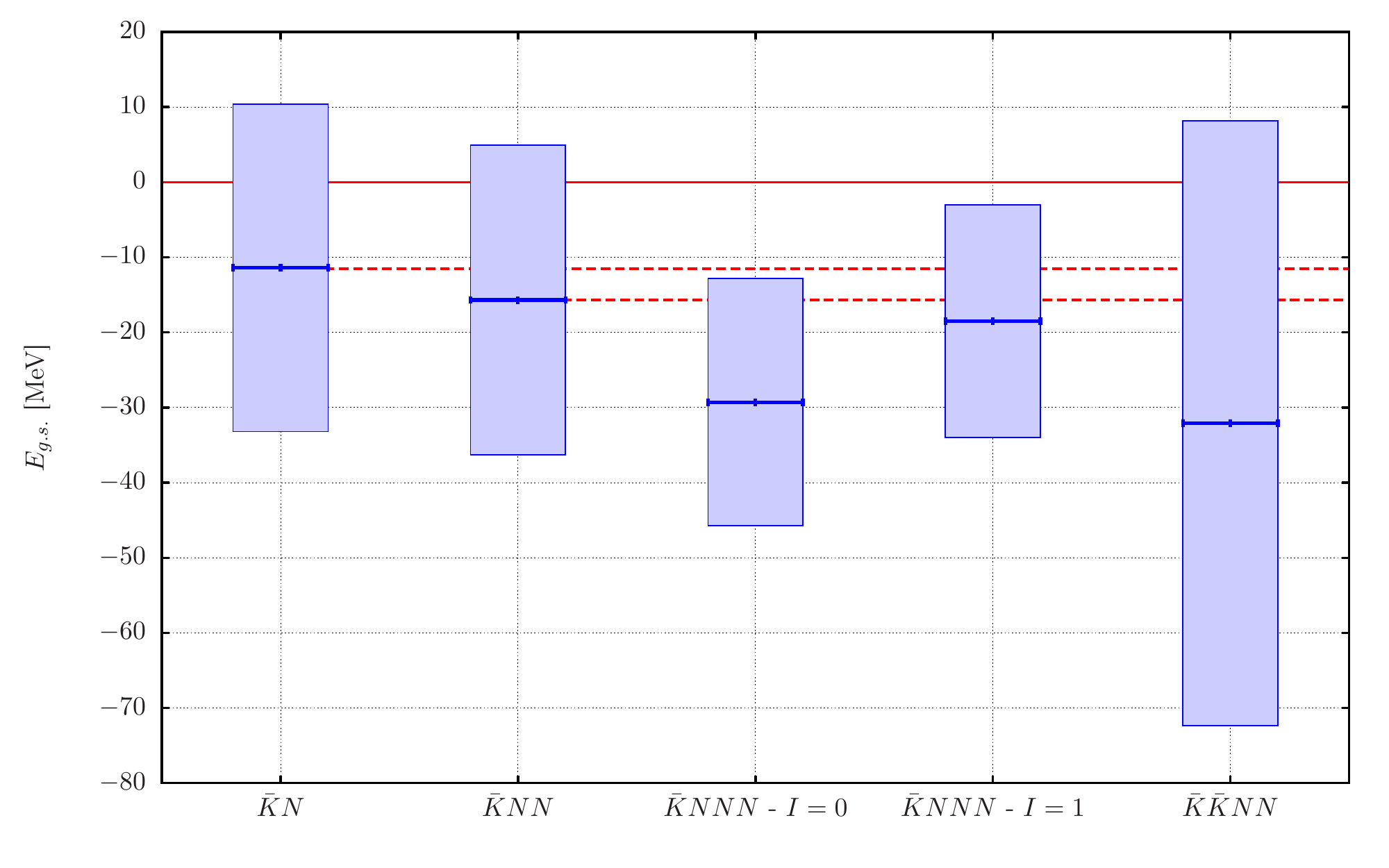} 
\caption{Binding energies and widths, $\Gamma(\bar K N\to\pi Y)$, of $\bar K$ 
and $\bar K\bar K$ few-body quasibound states (in MeV) calculated by 
Barnea, Gal, and Liverts (2012). Horizontal lines denote particle-stability 
thresholds. Widths are represented by vertical bars. A possible 
$I=\frac{1}{2}$, $J^{\pi}={\frac{1}{2}^+}$ $\bar K\bar K N$ quasibound state 
\cite{shevchenko15} is not shown in this figure. Figure courtesy of N. Barnea.} 
\label{fig:BGL} 
\end{figure} 

While several experiments have suggested evidence for a $K^-pp$ quasibound 
state with somewhat conflicting binding energy, there seems to be no consensus 
on this matter and it awaits further experimentation. In Fig.~\ref{fig:E27a}, a 
missing-mass spectrum is shown for the $d(\pi^+,K^+)$ reaction at 1.69 GeV/c 
taken at J-PARC \cite{ichikawa14}. The main features of this spectrum are the 
quasifree $\Lambda$, $\Sigma$ and $Y^{\ast}$ components. The latter rests 
on a broad phase-space structure. As for dynamical structures aside from 
the expected $\Sigma N$ cusp structure around 2.13 GeV/c$^2$, one observes 
a 20--30 MeV downward shift of the broad bump representing the 
$Y^{\ast}$ component. This indicates attraction for the $Y^{\ast}N$ system. 
Unfortunately, in this kinematical region the contributions of $\Sigma(1385)$ 
and $\Lambda(1405)$ overlap and are indistinguishable. A $\Sigma(1385)N$ 
quasibound realization of such a structure was previously discussed by 
Gal and Garcilazo (2013) as a possible $I=\frac{3}{2}$, $J^{\pi}=2^+$ $\pi YN$ 
resonance near the $\pi\Sigma N$ threshold (about 100 MeV below the $\bar K NN$ 
threshold). The main attraction in this ``pion-assisted dibaryon'' comes from 
the $p_{3/2}$-wave pion-baryon interactions, where $\bar KNN$ admixtures play 
a negligible role. 

Of the $K^-pp$ calculations listed in Table~\ref{tab:kpp}, we chose to review 
the hyperspherical-basis variational calculations including also four-body 
bound states (Barnea, Gal, and Liverts, 2012). The energy dependence of the 
$\bar KN$ interaction in this calculation is treated self consistently. 
The binding energies are shown in Fig.~\ref{fig:BGL} for three- and four-body 
kaonic bound states. $\Gamma(\bar K N\to\pi Y)$ width estimates are plotted as 
vertical bars, given by 
\begin{equation} 
\frac{\Gamma}{2}\approx\langle \,\Psi_\text{g.s.} |
- \text{Im}\,{\cal V}_{\bar{K}N}\, | \, \Psi_\text{g.s.} \, \rangle , 
\label{eq:Gamma} 
\end{equation} 
where ${\cal V}_{\bar{K}N}$ consists of all pairwise $\bar{K}N$ interactions. 
Equation~(\ref{eq:Gamma}) provides a good approximation because $|\text{Im}\,
{\cal V}_{\bar{K}N}|\ll |\text{Re}\,{\cal V}_{\bar{K}N}|$ \cite{hyodo08}. 
The calculated binding energies (widths) typically are found to be 10 
(10 to 40) MeV lower than when one uses threshold values as input, 
due to the self-consistency requirement which results in weaker $\bar{K}N$ 
interactions below threshold. In particular, the $I=\frac{1}{2}$ $\bar{K}NN$ 
g.s. ($K^-pp$) lies only 4.3 MeV below the 11.4 MeV centroid of the $I=0$ 
$\bar KN$ quasibound state. The latter value differs substantially from 
the 27 MeV binding energy traditionally assigned to the $\Lambda(1405)$ 
resonance used in nonchiral calculations. The $\bar{K}N\to\pi Y$ widths 
are of order 40 MeV for single-$\bar{K}$ clusters and twice that for 
double-$\bar{K}$ clusters. Additional $\bar{K}NN\to YN$ contributions of up 
to $\sim$10~MeV in $K^-pp$ (Dot\'{e}, Hyodo, and Weise, 2009) and $\sim$20~MeV 
in the four-body systems (Barnea, Gal, and Liverts, 2012) are likely. 

\begin{figure}[t] 
\includegraphics[width=6.5cm]{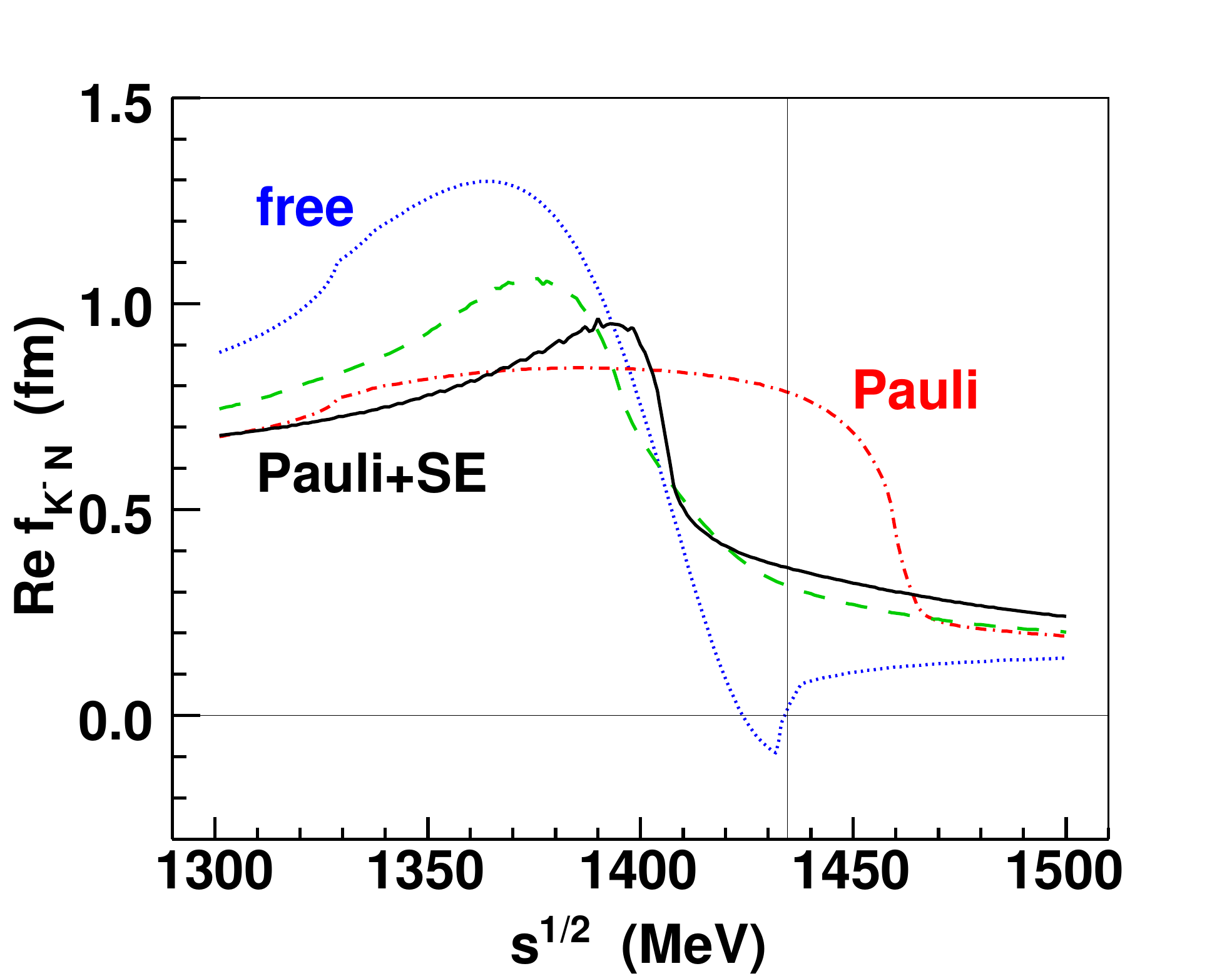}
\includegraphics[width=6.5cm]{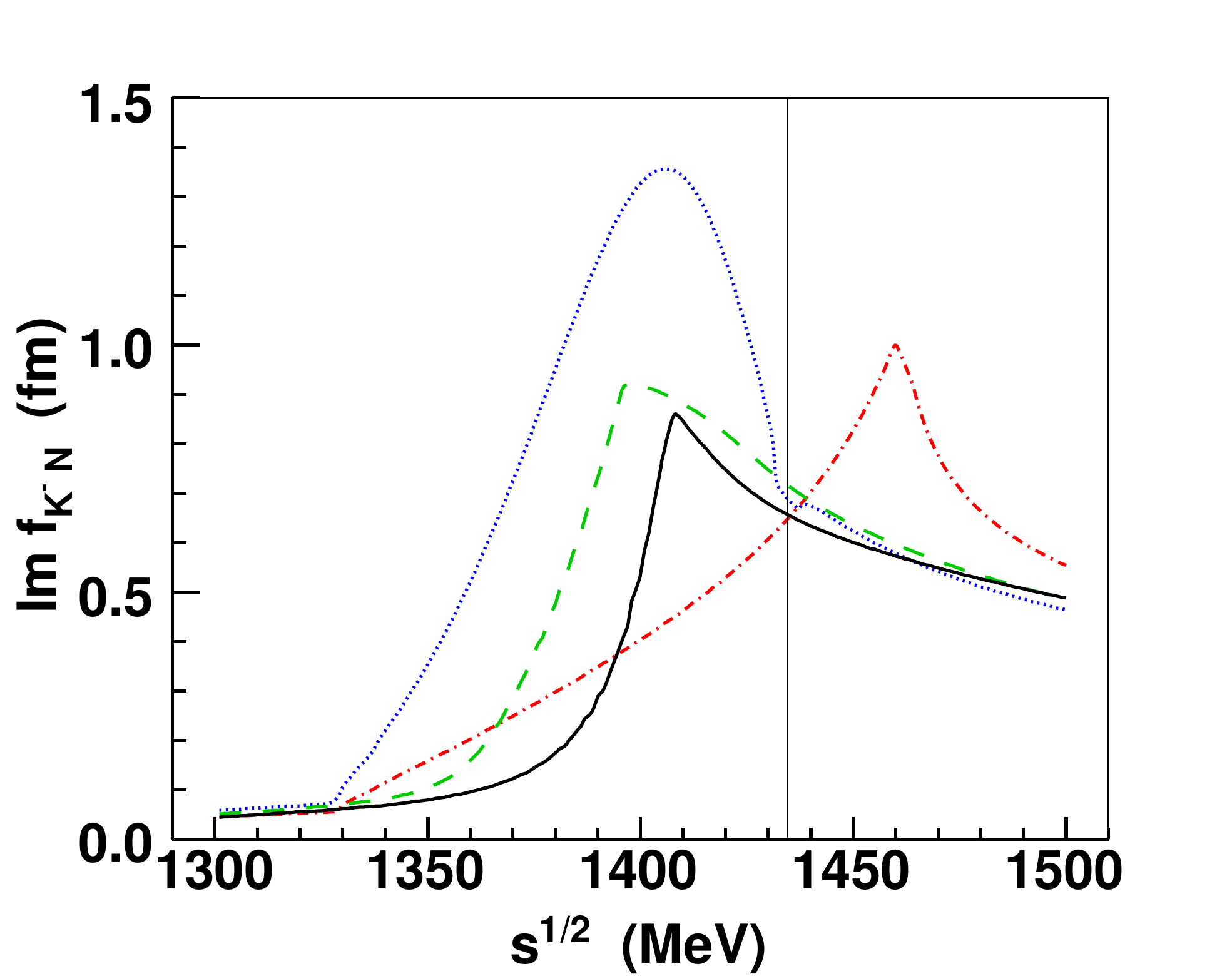}
\caption{Near-threshold energy dependence of $K^-N$  center-of-mass
scattering amplitudes in model NLO30 \cite{cieply12} for 
free-space (dotted) and Pauli-blocked amplitudes at $\rho=\rho_0$ with 
(solid) and without (dot-dashed) meson and baryon self-energies (SE). 
The dashed curves show Pauli-blocked amplitudes with SE at $\rho=0.5\rho_0$. 
The $K^-N$ threshold is marked by a thin vertical line. Figure courtesy
of A. Ciepl{\'{y}}.} 
\label{fig:CS} 
\end{figure} 

For calculations involving heavier single-$\bar{K}$ nuclear systems one needs 
in-medium $\bar{K}N$ scattering amplitudes. The in-medium $K^-N$ isoscalar 
amplitudes obtained from the chirally motivated coupled-channel model of 
Ciepl{\'{y}} and Smejkal (2012), and denoted NLO30 in the text, are shown in 
Fig.~\ref{fig:CS} above and below threshold. The real part of the subthreshold 
amplitude, which is relevant to $K^-$ atomic and nuclear states, is strongly 
attractive ($\sim$1~fm) and similar to that of the IHW subthreshold amplitude. 
This implies that $K^-$ quasibound states are likely to exist. Note that the 
attraction as well as absorption (expressed by the imaginary part of the 
amplitude) become moderately weaker for $\rho\geq 0.5\rho_0$, as demonstrated 
by comparing the solid ($\rho=\rho_0$) and dashed curves ($\rho=0.5\rho_0$). 

\begin{table}[b] 
\caption{Self-consistently calculated \cite{gazda12} binding energies $B_K$ 
and widths $\Gamma_K$ (in MeV) of $K^-$ quasi-bound states in Ca using 
a static RMF Ca density and NLO30 in-medium $K^-N$ subthreshold amplitudes 
\cite{cieply12}.} 
\begin{ruledtabular}
\begin{tabular}{lcccccc} 
 & \multicolumn{2}{c}{NLO30} & \multicolumn{2}{c}{+ $p$ wave} & 
\multicolumn{2}{c}{+ $2N$ abs.} \\ 
 & $B_K$ & $\Gamma_K$ & $B_K$ & $\Gamma_K$ & $B_K$ & $\Gamma_K$ \\
\hline 
$1s_K$ & 70.5 & 14.9 & 73.0 & 14.8 & 68.9 & 58.9 \\ 
$1p_K$ & 50.6 & 18.0 & 53.1 & 17.9 & 49.2 & 53.6 \\ 
$1d_K$ & 28.8 & 30.3 & 32.1 & 29.3 & 27.7 & 59.7 \\ 
$2s_K$ & 23.9 & 33.8 & 26.3 & 34.2 & 21.6 & 67.1 \\ 
\end{tabular} 
\label{tab:GM12} 
\end{ruledtabular} 
\end{table} 

\begin{figure}[t] 
\begin{center} 
\includegraphics[width=6.5cm]{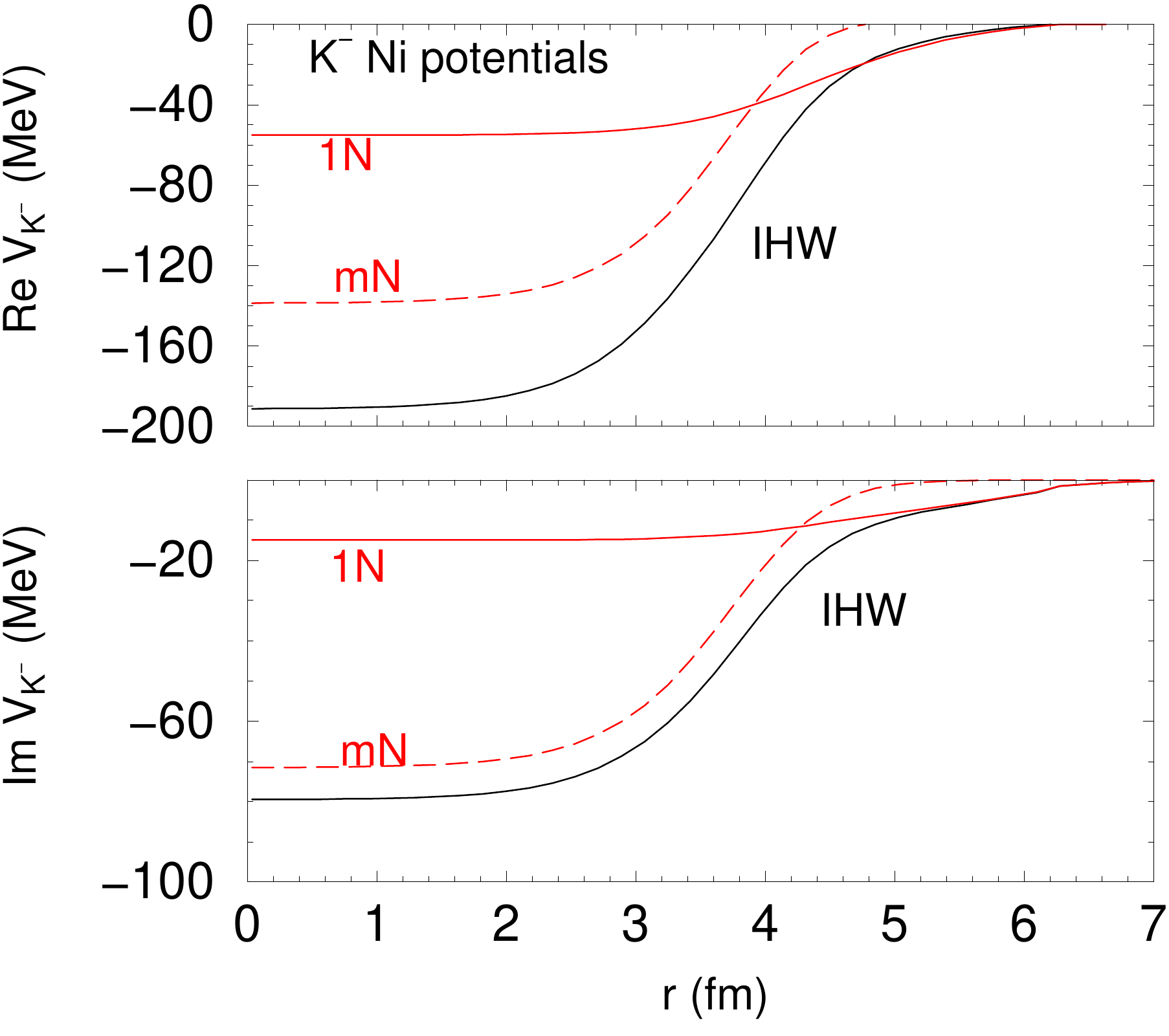} 
\caption{A self-consistent $K^-$ nuclear potential $V_{K^-}$ 
for $K^-$ atoms of Ni derived from global fits \cite{friedman13a} based on 
in-medium IHW one-nucleon ($1N$) amplitudes, together with its $1N$ and 
multinucleon ($mN$) components.} 
\label{fig:FG13} 
\end{center} 
\end{figure} 

The NLO30 in-medium $\bar KN$ $s$-wave scattering amplitudes shown in 
Fig.~\ref{fig:CS} were used by Gazda and Mare\v{s} (2012) to evaluate 
self-consistently $K^-$ quasibound states using RMF nuclear-core densities 
across the periodic table. Calculated $K^-$ binding energies, $B_K$, and 
widths, $\Gamma_K$, in Ca are listed in Table~\ref{tab:GM12} for several 
choices of input interactions. Listed in the table are also values of $B_K$ 
and $\Gamma_K$ derived by adding a $\Sigma(1385)$-motivated $p$-wave $K^-N$ 
interaction from Weise and H\"{a}rtle (2008). 
This marginally increases $B_K$ by a few MeV and modifies $\Gamma_K$ by less 
than 1 MeV. By adding a two-nucleon ($2N$) $K^-NN$$\to$$YN$ absorption term 
estimated from fitting to kaonic atoms, a $\lesssim$2~MeV decrease of $B_K$ 
results, but the width substantially increases to $\Gamma_K\sim (50-70)$~MeV. 
Given these large widths, it is unlikely that distinct quasibound states can 
be uniquely resolved, except perhaps in very light $K^-$ nuclei.

The hierarchy of widths listed in Table~\ref{tab:GM12} is also worth noting. 
One expects a maximal width in the lowest, most localized 1$s_K$ states for 
energy-independent potentials, which gradually decreases in excited states 
since these are less localized within the nucleus. The reverse is observed 
here, particularly when excluding $2N$ absorption. This is a corollary of the 
required self consistency; the more excited a $K^-$ quasibound state, the 
lower nuclear density it feels and thus a smaller subthreshold downward 
shift it experiences. Since Im$\,f_{K^-N}(\rho)$ decreases strongly below 
threshold (see Fig.~\ref{fig:CS}) the contribution to the calculated width 
gets larger as the excitation energy of the quasibound state increases. 

\begin{figure}[t] 
\begin{center} 
\includegraphics[width=6.5cm]{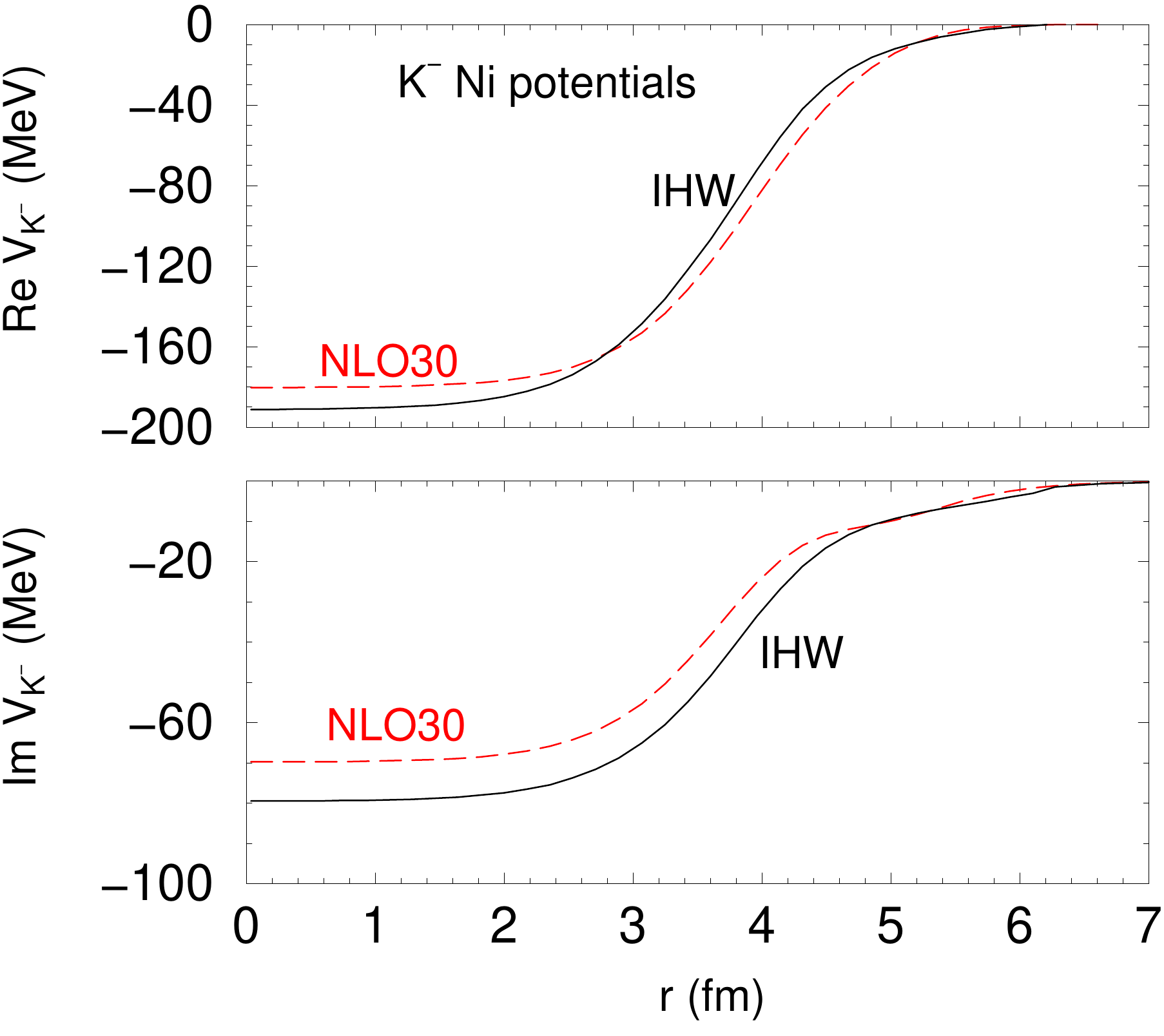} 
\caption{Self-consistent $K^-$ nuclear potentials $V_{K^-}$ for $K^-$ atoms 
of Ni derived from global fits \cite{friedman13a} based on the in-medium IHW 
$1N$ amplitudes (solid curves) (see Fig.~\ref{fig:FG13}) and as based on the 
in-medium NLO30 $1N$ amplitudes (dashed curves).} 
\label{fig:FG12} 
\end{center} 
\end{figure}

$K^-$ nucleus optical potential fits to kaonic-atom data across the periodic 
table reveal that the in-medium IHW-based, or NLO30-based one-nucleon ($1N$) 
amplitude input to $V_{K^-}$ fails to reproduce, even qualitatively, the $K^-$ 
atomic level shifts and widths. This is demonstrated in Fig.~\ref{fig:FG13} 
by the considerably stronger component, attributed to multinucleon ($mN$) 
processes with $m=2,3,{\cdots}$, of the fitted $V_{K^-}$. The composition 
of the imaginary part of the potential is of particular interest. 
It indicates that the $mN$ component, which is sizable in the nuclear interior, 
becomes negligible about half a fermi outside the half-density radius. 
This has implications for optimally choosing the kaonic-atom candidates 
where widths of two atomic levels can be measured \cite{friedman13b} to 
substantiate the $1N$ vs $mN$ pattern observed in global fits 
\cite{friedman12,friedman13a}. 
Finally, Fig.~\ref{fig:FG12} demonstrates that both IHW and NLO30 
energy-dependent in-medium amplitude inputs to $V_{K^-}$ lead to practically 
the same strongly attractive and absorptive nuclear-matter potential 
$V_{K^-}(\rho_0)$.

It is worth noting that the strong $K^-$ nuclear attraction forces the atomic 
$K^-$ wavefunction to overlap appreciably with the nuclear density down to 
almost 90\% of the central nuclear density $\rho_0$ \cite{friedman07,gal13a}. 
This does not hold for the shallower optical potentials $V_{K^-}$ 
based on $1N$ energy-independent $f_{K^-N}$ input consisting of threshold 
values (Baca, Garcia-Recio, and Nieves, 2000). Such potentials do not 
penetrate significantly beyond 10\% of $\rho_0$ and also do not provide 
equally good atomic fits as shown in Fig.~22 of Friedman and Gal (2007). 
In this context, a reaction that discriminates 
between deep and shallow attractive $K^-$ nuclear potentials is the formation 
of $\Lambda$ hypernuclear states localized within the nuclear interior in 
$K^-$ capture at rest. The calculated formation rates show sensitivity to 
how far the relevant $K^-$ atomic wavefunctions penetrate into the nucleus 
\cite{cieply11}. Formation rates of several $p$-shell hypernuclear ground 
states, available from FINUDA experiments \cite{agnello11a} and analyzed 
by Ciepl{\'{y}} \textit{et al.} (2011), favor deep $K^-$ nuclear potentials 
to shallow ones. 

\begin{figure}[t] 
\begin{center} 
\includegraphics[width=6.5cm]{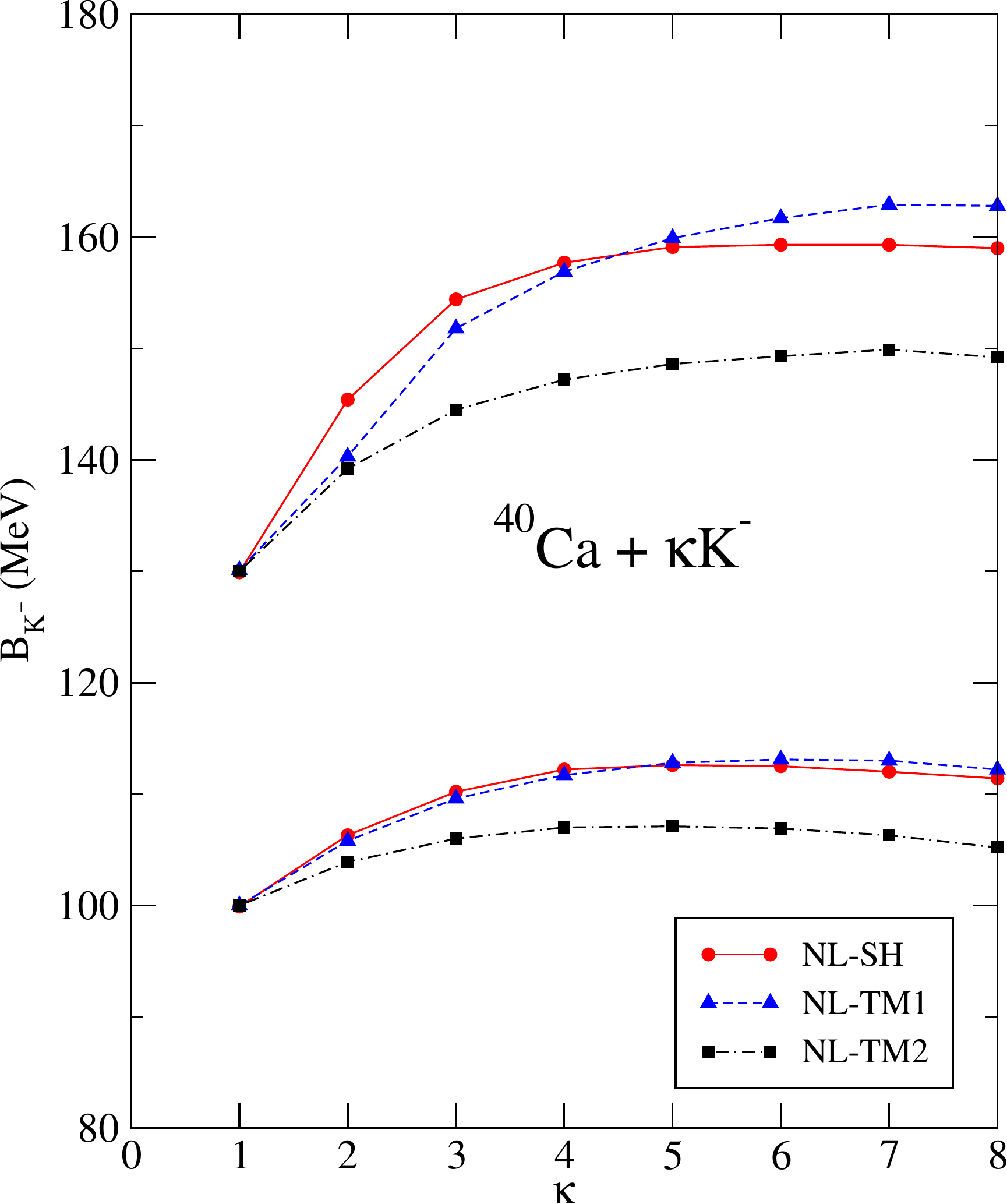} 
\caption{Saturation of $1s_{K^-}$ separation energies $B_{K^-}$ as calculated 
in multi-$K^-$ ($^{40}{\rm Ca}+\kappa K^-$) nuclei \cite{gazda08} for several 
versions of RMF input marked in the inset. The lower (upper) group of 
curves was constrained to produce $B_{K^-}=100~(130)$ MeV for $\kappa=1$.} 
\label{fig:GFGM08} 
\end{center} 
\end{figure} 

One might expect increased binding in multi-$K^-$ nuclei when calculated 
using strongly attractive $K^-$ nuclear potentials, which are fitted to $K^-$ 
atom data, since the bosonic nature of kaons allows them to occupy the same 
high-density central region of nuclei. This turns out not to be the case, as 
demonstrated by the RMF calculations of Gazda \textit{et al.} (2008) shown in 
Fig.~\ref{fig:GFGM08}. The difference between the various curves representing 
a given starting value of $B_{K^-}$, originates from the balance of the 
RMF inputs between the vector fields which generate ${\bar K}{\bar K}$ 
repulsion and the $\sigma$ scalar field which generates overall attraction. 
The separation energies, $B_{K^-}$, saturate as a function of the number of 
$K^-$ mesons, $\kappa$, such that $B_{K^-}(\kappa\to\infty)\ll (m_K+M_N-M_{
\Lambda})\approx 320$~MeV. This implies that antikaons do not replace 
$\Lambda$ hyperons in the ground-state realization of multistrange hadronic 
systems. Stated differently, antikaons do not condense in a finite self-bound 
hadronic system. 

\section{Future Experiments and Directions}
\label{sec:future} 

Even though SU(3)$_{\rm f}$ symmetry is badly broken, it is a useful way to 
organize the discussion of strangeness within a nucleus. Thus techniques in, 
and knowledge of, traditional nuclear physics may readily be applied. 
As examples, spectroscopy that resolves the spin structure, and the weak 
decay mechanisms that operate within the nuclear interior illuminate new 
features of the hadronic many-body problem. 

Because the $\Lambda$N interaction is weak, hypernuclear spectroscopy can be 
represented by a superposition of particle-hole states resulting in 5-10 MeV 
spaced $\hbar\omega$ structures, and these can be resolved, as previously 
discussed, by experiments with 1--2 MeV resolution. However, it is more 
difficult to extract levels which involve nuclear-core excitations, or to 
resolve $\Lambda$ spin-flip excitations within the enclosing $\hbar\omega$ 
structures. Indeed, direct observation of $\Lambda$ spin-doublet structure 
in many instances requires resolutions approaching 100 keV or better, and 
thus well beyond the capabilities of present magnetic spectroscopy. 
Still, resolution of nuclear-core excitations at the $\le$ 500 keV level 
carry substantial physics interest, and are accessible with modern, 
continuous-beam electron accelerators \cite{nakamura13b}, and perhaps 
also with meson beams at the J-PARC (Japan Proton Accelerator Research Complex) 
50 GeV proton synchrotron \cite{takahashi13}. 

In addition to spectroscopy, nonmesonic weak decays provide information on 
the local nuclear environment, including for example $NN$ correlations. Also by 
comparing energy shifts between charge-symmetric hypernuclei, information on 
the dynamical behavior of the nuclear core and the admixture of other hyperons 
in the ground-state wave function can be obtained. Finally, multihyperon 
states provide information on hyperon-hyperon interactions that is needed 
to extend $SU(3)_{\rm f}$ symmetry and develop a better understanding of 
nuclear matter at high density in astrophysical objects.    

Future programs will be driven by the new proton accelerator at J-PARC, 
the continuous electron accelerators at Jlab and Mainz, and the antiproton 
facility at FAIR. Not only do these facilities have infrastructure designed 
for hypernuclear research, but the experiments will be able to take advantage 
of new, innovative detectors and electronics that will allow higher rates, 
better energy resolution, and better particle and signal identification. 
It is anticipated that this field will remain interesting and fertile to 
new exploration.

\subsection{Spectroscopy using meson beams} 

\subsubsection{Hyperon production and hyperon-nucleon interactions} 

As discussed in Secs.~\ref{sec:intro} and \ref{sec:lamhyp}, the mainstays 
of hypernuclear research have been the \Kpi\ and \piK\ mesonic reactions. On 
the other hand, studies of heavy hypernuclear systems may prove difficult. 
Therefore, it is important to 
undertake better measurements of elementary hyperon production cross sections 
and, in particular, polarization observables may prove useful. Polarization 
is small at the forward angles where the  $\Lambda$ production 
amplitude is sufficient to be experimentally useful. However, polarization 
is crucial in experiments attempting to measure the weak-decay asymmetry. 
Although the residual polarization after hypernuclear production appears 
consistent with zero, polarization due to the large spin-flip amplitudes in 
the \Kpi\ reaction at 1.1 and 1.5 GeV/c has not been explored systematically. 
This may be more accessible with the intense kaon beams available at J-PARC, 
as indeed proved in the E13 experiment by populating the \lamb{4}{He}(1$^+$) 
level in the \Kpi\ reaction on $^4$He at $p_K=1.5$~GeV/c \cite{yamamoto15}. 

Most importantly, there should be a plan to systematically study the 
elementary hyperon-nucleon ($YN$) interaction. To date only approximately 
40 data points of $YN$ scattering cross sections are available from mostly 
old experiments that studied hyperon post-production secondary interactions. 
Some of the more recent $\Sigma N$ data were obtained using the SCIFI 
(scintillator fiber) active detector system of the 1990s. One approved 
experiment at J-PARC, E40 \cite{takahashi13}, will extend these measurements. 
Such new and improved data are particularly important from a theoretical 
standpoint in constructing $YN$ potential models for use in hypernuclear 
structure applications. We recall from Table~\ref{tab:Sig1} that 
successive Nijmegen ESC potentials, the latest of which is ESC08 
(Nagels, Rijken, and Yamamoto, 2015a, 2015b), have led to 
increasingly repulsive $\Sigma$-nucleus $G$-matrix potentials, in agreement 
with deductions made from $\Sigma$ hypernuclear production experiments. 
Therefore, it would be useful to enhance the $YN$ data base of these models 
by new and more precise $\Sigma N$ cross section data in order to confirm the 
validity of these deductions. Similarly, it would be useful to 
enhance the $S=-2$ baryon-baryon data base by new and more precise $\Xi N$ 
cross section data, particularly by remeasuring and extending the poorly 
measured $\Xi^- p\to\Lambda\Lambda$ reaction cross sections. This input is 
crucial for confirming that the $S=-2$ baryon-baryon interactions are fairly 
weak, as suggested by the absence of a particle-stable $H$ dibaryon and by the 
accurately known $B_{\Lambda\Lambda}({_{\Lambda\Lambda}^{~6}{\rm He}})$ value, 
and in agreement with a recent NLO $\chi$EFT study by 
Haidenbauer, Mei{\ss}ner, and Petschauer (2016). 

\subsubsection{Reaction spectroscopy with mesons} 

The absence of a modern hadron accelerator, providing intense beams of 
energetic kaons and pions, has hindered the exploration of hypernuclear 
experiments, particularly those involving the study of doubly-strange 
nuclear systems. This impediment is being resolved with the introduction 
of experiments at J-PARC \cite{takahashi13}. The 30 GeV proton beam at 
J-PARC is operative, producing various high-intensity beams of secondary 
pions and kaons. Two beamlines are initially available, with high-resolution 
magnetic spectrometers that are able to reach missing-mass resolution 
of somewhat less than 2~MeV at best. A proposed high-resolution \piK\ 
spectrometer for use in a future extension of the hadron facility 
should achieve missing mass resolutions for hypernuclear spectroscopy of 
$\le 500$~keV. So far, the spectroscopy of single-$\Lambda$ hypernuclei has 
been addressed in brief running periods of experiments E10, search for 
\lamb{6}{H} \cite{sugimura14}, and E13, $\gamma$-ray studies in the $s$, $p$, 
and $sd$ shells \cite{tamura13}, with the latter observing a 1.41~MeV 
$1^+\to 0^+$ $\gamma$ transition in \lamb{4}{He} \cite{yamamoto15}. 
Also high on the hypernuclear agenda is experiment E05 which is a search 
for the $^{12}_{~\Xi}$Be hypernucleus via $^{12}$C$(K^{-},K^{+})^{12}_{~\Xi}$Be 
\cite{nagae13}. In this experiment, the overall energy resolution in the 
$\Xi^-$ bound-state region is expected to be in the range of 1.5--3~MeV 
at FWHM. 

\subsubsection{Experiments using emulsion detectors} 

As described earlier, nuclear emulsion was the first detection system used to 
investigate hypernuclear events. The advantage of emulsion is its excellent 
position and energy resolution, which allows detailed investigation 
of a reaction and its decay products. Coupling counters with emulsion, 
although somewhat clumsy, can still provide needed information under certain 
experimental conditions. Indeed, this technique was crucial in the KEK E373 
determination of the binding energy of $_{\Lambda \Lambda}^{~6}{\rm He}$ 
(Takahashi \textit{et al.}, 2001; Ahn \textit{et al.}, 2013). A coupled 
counter and emulsion detector is proposed for the study of $\Lambda\Lambda$ 
systems at J-PARC. In this experiment E07 $\Xi^{-}$ are produced in a diamond 
target upstream of the emulsion and are tracked as they recoil into, and stop, 
in the emulsion \cite{takahashi13}. Particle emission from the stopping vertex 
is then analyzed for various reactions, including the production of 
$S\,=\,-2$ systems.

\subsubsection{Spectroscopy using electromagnetic transitions}

While the energy resolution using direct spectroscopy to specific states 
with magnetic spectrometers and meson beams is presently limited to no 
better than a few hundred keV, the energy of electromagnetic transitions 
between states can be measured to a few keV. Thus, measurement of 
electromagnetic transitions is a powerful tool for hypernuclear spectroscopy. 
This requires a dedicated beam line to tag the formation of a specific 
hypernucleus, and large acceptance, high resolution Ge detectors. The photon 
detectors to be used have high photo-peak efficiency and rate handling 
capabilities. The system at J-PARC is called Hyperball-J \cite{tamura13} 
and consists of 28 mechanically-cooled Ge detectors having 60\% relative 
efficiency. Each Ge crystal is enclosed by 2 cm thick lead tungstate (PWO) 
counters to suppress Compton scattering and $\gamma$ rays from $\pi^{0}$ 
decays. The readout requires special electronics for high counting rate and 
large dynamic range of the signals. 

J-PARC has tested and mounted equipment to undertake a study of $\gamma$
emission from excited levels in \lamb{4}{He}, \lam{10}{Be}, \lam{11}{Be}, 
and \lam{19}{F} \cite{tamura13}. A first result for \lamb{4}{He} has been 
obtained~\cite{yamamoto15}. Lifetimes can be measured using the Doppler shift 
attenuation method (DSAM) that was first used to extract the lifetime 
of the $5/2^+$ state of \lamb{7}{Li}, and thus its electromagnetic E2 
transition strength B(E2) value~\cite{tanida01}. Also, the lifetime of the 
lowest $1/2^+;T\!=\!1$ state in \lam{15}{N} has been measured~\cite{ukai08}. 
Perhaps with the higher intensities provided at J-PARC, the $\Lambda$ magnetic 
moment in the nuclear medium might also be inferred from measuring the 
lifetime of M1 transitions between ground-state hypernuclear doublet levels, 
such as the ($3/2^+\to 1/2^+$) $\gamma$ ray in \lamb{7}{Li}~\cite{tamura13}.  
In the weak-coupling limit the strength of the electromagnetic M1 transition 
B(M1) is proportional to $(g_c-g_{\Lambda})^2$, where $g_c$ is the 
core $g$-factor and $g_{\Lambda}$ is the $\Lambda$ $g$-factor (for the 
$0s_{\Lambda}$ orbit in this example). For the simple $\Lambda$-hypernuclear 
configurations considered here, the in-medium $\Lambda$ $g$-factors could 
deviate from their corresponding free-space single-particle Schmidt values 
by at most 10\%  (Dover, Feshbach, and Gal 1995; Saito, Oka, and Suzuki, 1997). 
The lifetime measurement accuracy required to test a few-percent departure of
$g_{\Lambda}$ from its Schmidt value can be reached at J-PARC \cite{tamura13}.  

As the target mass increases to heavier systems the number of both nuclear 
and hypernuclear $\gamma$ rays increases while the yield to specific 
hypernuclear states decreases. Although the Doppler shift of in-flight 
hypernuclear transitions can discriminate between at-rest nuclear transitions, 
it still becomes more difficult to assign observed $\gamma$ rays to a 
particular hypernuclear level scheme. Thus, coincident $\gamma$ decays, as well 
as better resolution of the tagging spectrometer, becomes more important. 

The first $\gamma\gamma$ coincidence observation was reported \cite{ukai06}, 
but $\gamma$ coincidences cannot be a widely used tool until production rates are 
substantially improved. Note that an increase in yield involves more than 
increasing beam flux, because $\gamma$ detectors are sensitive to backgrounds 
of all types, and resolution is degraded by rate-dependent electronic pileup. 

In addition to $\gamma\gamma$ coincidence measurements, a coincidence 
between a $\gamma$ and a weak decay can be used to extract information 
about hypernuclear structure. For hypernuclei with masses up to the middle 
of the $p$ shell, mesonic, as opposed to non-mesonic, weak decay is 
sufficiently probable that detection of mono-energetic $\pi^{-}$ emission 
can be used as a coincidence to tag a specific  hypernucleus. If the 
hypernucleus can be uniquely identified from its mesonic decay, then 
detection and missing-mass analysis of the production reaction would 
not be necessary, and the observation of $\gamma$ rays from hyperfragments in 
coincidence with their $\pi^{-}$ decay would increase the efficiency of 
a $\gamma$-ray experiment. The technique also gives access to hypernuclei 
which could only be produced by fragmentation or nucleon emission; see also 
the discussion of the Mainz program in Sec.~\ref{subsubsec:MAMI}. 

\subsection{Spectroscopy with electron accelerators}

\subsubsection{Electroproduction at Mainz (MAMI)} 
\label{subsubsec:MAMI}

\begin{figure}[t] 
\includegraphics[width=8.0cm]{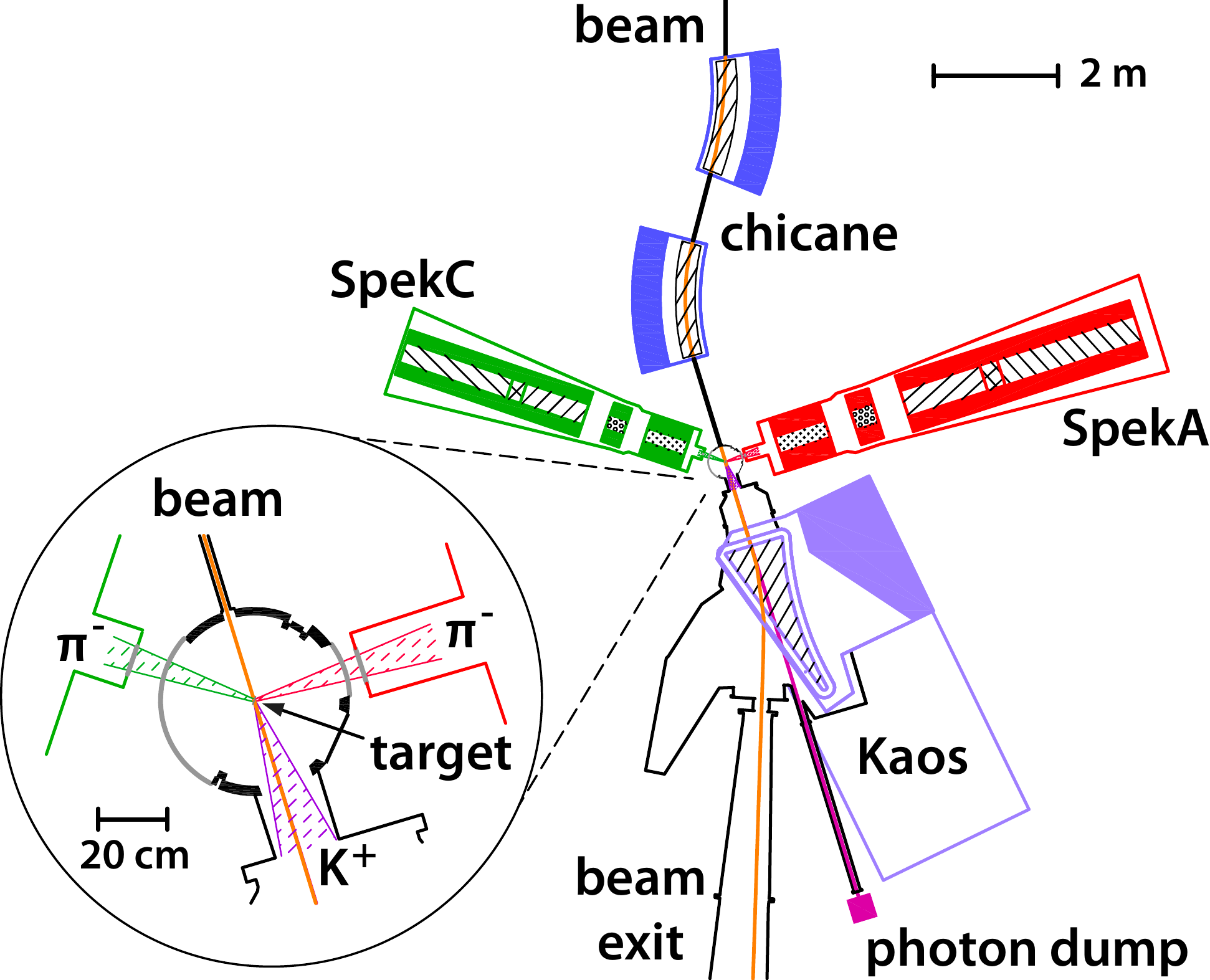} 
\caption{The spectrometer system at the Mainz microtron designed to observe 
the pion decay of light hypernuclei formed by fragmentation of heavier 
hypernuclear systems formed in kaon electroproduction. The KAOS spectrometer 
detects the kaons emitted in the \eeK\ reaction, and spectrometers A and C 
detect the decay pions. From Esser \textit{et al.}, 2015.} 
\label{fig:geom} 
\end{figure}

An ongoing program at the Mainz microtron (MAMI) involves studying 
the mesonic weak decay of light hypernuclei formed by fragmentation of 
excited hypernuclear levels reached in electroproduction. This interesting, 
unexplored, technique uses counters, not emulsion. The microtron energy 
of 1.5 GeV allows experiments to determine ground-state masses of light 
hypernuclei by measuring the pion weak decays following the fragmentation 
of heavier hypernuclear systems reached in kaon electroproduction. 
For example, \lamb{4}{H} is strongly produced in $K^-$ absorption on a 
number of $p$-shell targets and can be identified by the monochromatic
$\pi^-$'s with $p_{\pi}=133$~MeV/c from the two-body decay  
\lamb{4}{H}$\to {^4}$He\,+$\pi^-$ \cite{tamura89}. In fact, the \lamb{4}{H}
line has been studied recently \cite{esser15} using the setup shown in
Fig.~\ref{fig:geom} [see also \cite{esser13}] with a $^9$Be target. 
The kaon spectrometer (KAOS) detects kaon production with the kaons 
identified by time-of-flight and an 
aerogel Cherenkov detector. Spectrometers A and C detect the decay pions 
(spekC for the high-momentum \lamb{4}{H} line). The binding 
energy value of $B_{\Lambda}$(\lamb{4}{H})=2.12$\pm$0.01$\pm$0.09~MeV
was obtained and is consistent with the old emulsion value 2.04$\pm$0.04~MeV 
(cf. Table~\ref{tab:blambda}). This is an important result
given the importance of establishing precisely the degree of charge-symmetry
breaking in the $A\!=\!4$ hypernuclei and the fact that the emulsion
values for the binding energies were derived from three-body decays because
there was no calibration for long-ranged pions in emulsion.  
For the lower momentum pions typical of $p$-shell hypernuclei, problems 
certainly exist in assigning the observed pion decay spectrum to specific 
hypernuclear states. Nevertheless, because the decay of these hypernuclei 
can be determined by 2-body kinematics, the assignment of masses and binding 
energies is potentially possible. However, note that mesonic decays from 
hypernuclear ground states do not necessarily end up in the corresponding 
daughter-nuclei ground states (Motoba and Itonaga, 1994; Randeniya and 
Hungerford, 2007; Gal, 2009).

\begin{figure}[t] 
\includegraphics[width=8.5cm]{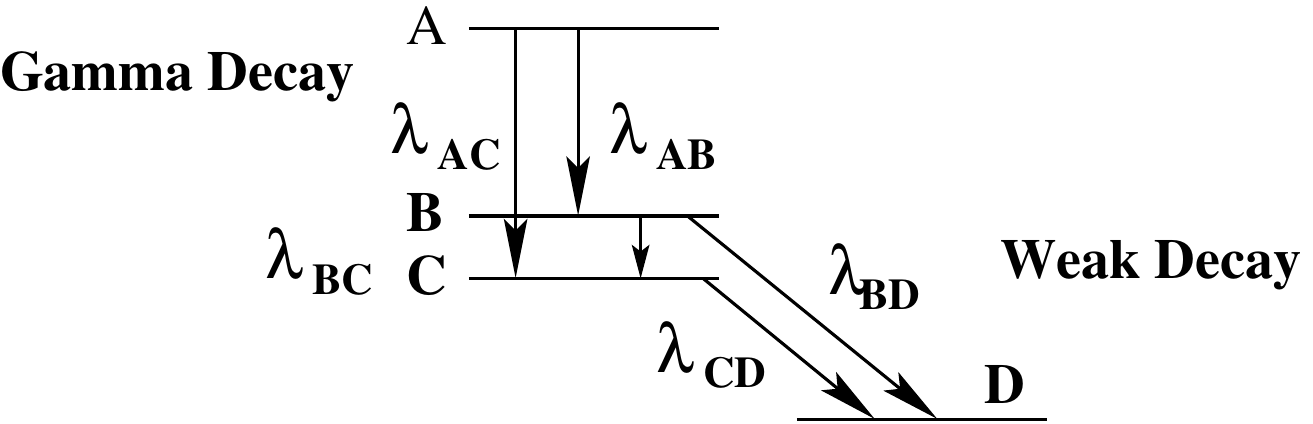} 
\caption{A schematic illustration of $\gamma$ and weak decay between 
hypernuclear levels with a ground state doublet (B,C) having energy 
spacing $\le$ 100 keV.} 
\label{fig:level_decay} 
\end{figure}

The use of a $\gamma$-weak decay coincidence has also been proposed to obtain 
the lifetime of hypernuclear levels that have $\gamma$ lifetimes comparable 
to those of weak decay (200 ps). This could be used, for example, to measure 
the $\gamma$ lifetime of the upper level of a hypernuclear ground-state 
doublet, where the $\gamma$ decay of the upper level competes with weak decay. 
This generally 
occurs for high multipolarity transitions of low transition energy, $\le$ 100 
keV. A simultaneous fit to the coincidence times between the weak decays 
of the doublet levels and the $\gamma$ transitions from A to B and B to C as 
shown in the level diagram of Fig.~\ref{fig:level_decay} would provide the 
lifetimes of the B and C levels. Such a program fits into a potential program 
at Mainz, but the hypernuclei are electroproduced and will be accompanied by
significant gamma backgrounds that may preclude $\gamma$-pion 
coincidence experiments. 

\subsubsection{Electroproduction at Jlab}

There is substantial, new electroproduced hyperon data from the CLAS 
detector Collaboration at JLab, particularly polarization and spin transfer 
data \cite{carman09,mcCracken10,dey10}, providing a consistent data base for 
partial-wave $YN$ amplitude analysis. The electroproduction of hyperons 
is a complicated process involving a number of overlapping strange and 
non-strange resonances \cite{bydzovsky13,skoupil16}. Whereas $s$-channel 
diagrams 
are found to be most important at low energy, $t$-channel/Reggeon exchange 
dominates when $W>2$ GeV (i.e., above the resonance region). More data 
is expected from CLAS and and also from LEPS at SPring-8 \cite{niiyama13}.

Jlab will be upgraded to a higher energy with more intense beams. 
The new large solid-angle spectrometers drawn in Fig.~\ref{fig:layout}, 
HKS and HES, with a new splitting magnet (SPL), will be available. 
Previously \eeK\ hypernuclear programs were undertaken in both Hall A and Hall 
C. When Jlab transitions to 12 GeV electron beams, hypernuclear experiments 
will take place in only one Hall. If this is Hall A, a plan exists to design 
two new septum magnets and move the HKS and the HES from Hall C into Hall A 
behind the target station. A waterfall target (H$_{2}$O) will be retained and 
could be used to further study the elementary electroproduction amplitude at 
forward angles and for spectrometer calibrations. By carefully selecting the 
scattering geometry, bremsstrahlung and M\"{o}ller backgrounds can be reduced 
and the luminosity increased to obtain rates of several 10's per hour to 
specific states. This allows electromagnetic production of hypernuclei 
through the $sd$ shell with perhaps resolutions approaching 300 keV. 
Proposals have been made for improved energy resolution experiments, 
after the 12 GeV upgrade, aiming at the electroproduction of $\Lambda$ 
hypernuclei beyond the $p$-shell hypernuclei explored so far in Halls A 
and C \cite{garibaldi13,tang14}. 

\subsection{Experiments at PANDA} 

The PANDA collaboration using antiprotons at the FAIR future facility in 
Darmstadt, proposes to produce double-$\Lambda$ hypernuclei, followed by 
high-resolution $\gamma$-spectroscopy study, in order to provide for the 
first time precise information on their bound-state spectra \cite{esser13}. 
The PANDA detector is to be set up at the high-energy storage ring (HESR) 
that produces high-intensity phase-space cooled antiprotons with momenta 
between 1.5 and 15 GeV/c. The antiprotons from the storage ring are extracted 
and allowed to interact on a nuclear target at $p_{\rm lab}\approx$3 GeV/c 
\cite{pochodzalla05}, 
\begin{equation} 
{\bar p}~+~p ~\rightarrow~ \Xi^- ~+~{\bar \Xi}^+ \,, ~~~
{\bar p}~+~n ~\rightarrow~ \Xi^- ~+~{\bar \Xi}^0 \,. 
\label{eq:xiprod} 
\end{equation} 
The trigger for these reactions will be based on the detection of 
high-momentum $\bar \Xi$ anti-hyperons at small angles or on $K^{+}$ mesons 
produced by the absorption of anti-hyperons in the primary target nuclei. 
Produced $\Xi^-$, with typical momenta between 0.5 to 1~GeV/c, are decelerated 
in a secondary target. The slow $\Xi^-$ are then either directly absorbed by 
the nucleus or are captured into an atomic orbit, cascading downward through 
the $\Xi^-$ atom levels until absorbed in the $\Xi^- p\to \Lambda\Lambda$ 
reaction, thereby partially forming a double-$\Lambda$ hypernucleus. 
X-ray deexcitation between $\Xi$ atomic states, and $\gamma$ deexcitation 
between states in the $\Lambda\Lambda$ hypernuclei which may be formed, 
are to be studied with an array of Ge detectors (Pochodzalla, 2005; Esser
\textit{et al.}, 2013). One expects to identify approximately 3000 stopped 
$\Xi^{-}$ hyperons per day; see the simulation by Ferro \textit{et al.} 
(2007). $\Xi^{-}$ capture yields, 
associated fragmentation mass spectra, and production cross sections of 
double-$\Lambda$ hypernuclei have been estimated in two recent works 
\cite{gaitanos12,gaitanos14} using transport in-medium calculations. 

\subsection{Weak decay of hypernuclei} 
\label{subsec:WDfuture}

\subsubsection{mesonic decays} 

Mesonic decays of hypernuclei have been studied since the beginning of 
hypernuclear experimentation, first in emulsion and more recently in counter 
experiments at BNL, KEK and by the FINUDA Collaboration at DA$\Phi$NE, 
Frascati \cite{botta12}. A wealth of binding energies 
and spin-parity values of light $\Lambda$ hypernuclei were deduced in these 
studies. The well-understood mesonic decay of the $\Lambda$ can be used as a 
tool to explore nuclear structure when strangeness is injected into the nuclear
medium. The pion-decay spectroscopy program at Mainz \cite{esser13}, which was 
reviewed in Sec.~\ref{subsubsec:MAMI}, is poised to develop this tool, 
primarily by improving the momentum resolution in detecting the emitted pion. 

The limitation of mesonic-decay studies to light hypernuclei is due to the low 
momentum of the recoiling nucleon in the $\Lambda \to N + \pi$ decay, which 
is well below the nuclear Fermi momentum $p_F$ for $A\geq 6$. However, the 
$\Lambda$ mesonic-decay rate in the nuclear medium is extremely sensitive 
to pion distortion effects from in-medium nuclear and electromagnetic 
interactions. The inclusion of pion-nuclear distortion allows the recoiling 
nucleon to assume momentum values greater than $p_F$, enhancing both $\pi^{0}$ 
and $\pi^{-}$ emission, while Coulomb distortion is expected to raise the 
$\pi^{-}$ decay rates to measurable levels for the heaviest hypernuclei. 
Indeed, the prediction is that the ratio of the in-medium to free rate 
saturates at about $10^{-2}$ \cite{motoba94c}. However, another calculation, 
which predicts somewhat similar behavior, results in a rate about a factor 
of 10 lower in the case of $^{208}$Pb \cite{oset94}. There are no available 
experimental data.
 
Hypernuclei generally deexcite by $\gamma$ emission to the ground state where 
they undergo weak decay. In situations where the ground state belongs to 
a spin doublet based on the nuclear core g.s., weak decay from the upper 
level can successfully compete with the M1 doublet transition when the 
transition energy is lower than typically 100~keV; see 
Fig.~\ref{fig:level_decay}. This may occur in the case of the 
[$1^{-}$(g.s.),$2^{-}$] doublet in $^{10}_{~\Lambda}$B where no $\gamma$ ray 
between these two levels has been seen \cite{chrien90,tamura05}. 
Of the two levels, only the $2^-$ is expected to have been populated in 
the non-spin-flip production reactions used in these experiments. Therefore, 
in \lam{10}{B} either the doublet splitting is less than 100 keV, thereby 
hindering the $\gamma$ transition with respect to weak decay, or the level 
ordering of the spin-doublet members is reversed. 

Furthermore, the $\pi^{-}$ decay spectrum is substantially different for 
weak decays from each member of the doublet \cite{gal09}, providing a way to 
identify the decay sequence. However, in general one might expect a mixture 
of weak decays from the doublet levels, and a more detailed analysis would 
be required to extract the decay ratios and determine the ordering. Note that 
an energy resolution of $\le$ 100~keV is required to measure the $\pi^{-}$ 
transition energy shifts in the decays. This may be possible if excellent 
resolution and sufficient statistics are available. Nevertheless, comparison 
of the observed pion decay to one calculated for various spin possibilities 
should allow the level order to be determined. 

\subsubsection{nonmesonic decays} 

Of the various observables studied so far, data on nonmesonic weak-decay 
asymmetries are scarce. Asymmetry and coincident weak-decay experiments are 
difficult, requiring thick targets, with low yields. A definitive asymmetry 
experiment would require a substantial increase in intensity and/or 
polarization, as well as the determination of the polarization of the 
hypernuclear ground state from which the decay occurs. Better missing-mass 
resolution to tag ground-state production and the use of a polarizing reaction 
such as $(\pi^{+},K^{+})$ at an angle $>10^{\circ}$ would help, but this 
requires higher beam intensity. 

It would also be important to measure the neutron and proton simulated decays 
from $^{4}_{\Lambda}$H compared to the same decays from $^{4}_{\Lambda}$He. 
This comparison would significantly help to resolve the question as to whether 
the $\Delta I = 1/2$ rule applies in nonmesonic weak decay $\Lambda+N\to N+N$ 
transitions. However, the production of $^{4}_{\Lambda}$H requires a charge 
exchange as well as a strangeness exchange reaction when using a $^{4}$He 
target. Photoproduction is a possibility as well as the $(K^{-},\pi^{0})$ 
reaction. High beam intensity and large solid angle detectors would be 
required. A test of the $\Delta I = 1/2$ rule requires that the final $NN$ 
states have isospin $I_f(NN)=1$, which is reached by the $a$, $b$, and $f$ 
amplitudes defined in Table~\ref{tab:WD3}. This practically leads to the 
requirement that the initial $\Lambda N$ state is a purely $^1S_0$. In this 
case the $\Delta I=1/2$ rule predicts that 
\begin{equation} 
\Gamma_n(^4_\Lambda{\rm He})~=~2\,\Gamma_p(^4_\Lambda{\rm H})~, 
\label{eq:test4} 
\end{equation} 
which may be tested in the nonmesonic hypernuclear decays of the $A=4$ 
hypernuclei. The value of the left-hand side $\Gamma_{n}(^{4}_{\Lambda}$He) 
has been determined to be very small, 
$\Gamma_n(^4_\Lambda{\rm He})/\Gamma^{\rm free}_\Lambda \leq 0.035$ 
\cite{parker07}, whereas the value of $\Gamma_{p}(^{4}_{\Lambda}$H) 
is unknown. This will be studied in the J-PARC E22 experiment. 

Another area of interest for nonmesonic weak decays would be to study 
exclusive decay modes, in analogy to the exclusive, two-body mesonic-decay 
modes of $\Lambda$ hypernuclei that have provided valuable 
information on spins of $\Lambda$-hypernuclear levels; see Table~\ref{tab:WD2}. 
The study of exclusive decay modes in \textit{nonmesonic} weak decays could 
yield valuable information on the $\Lambda+N\to N+N$ amplitudes of 
Table~\ref{tab:WD3}. Examples of such modes in light nuclei are 
\begin{equation} 
^5_\Lambda{\rm He} \to n{^4{\rm He}},~ddn,~nn{^3{\rm He}},~pn{^3{\rm H}}, 
\label{eq:WD5} 
\end{equation} 
\begin{equation} 
^4_\Lambda{\rm He} \to p{^3{\rm H}},~n{^3{\rm He}},~dd,~dpn. 
\label{eq:WD4} 
\end{equation} 
Rates for some of these decays were measured in bubble chambers and emulsion 
\cite{coremans70}. In passing we mention that the $\Lambda$-hypernuclear 
program at J-PARC also includes a search for multinucleon emission in the 
weak decay of hypernuclei, experiment E18 \cite{takahashi13}. 

\subsubsection{$\Lambda$ hypernuclear lifetimes} 
\label{subsubsec:lifetimes2}

\begin{table}[t] 
\caption{\lamb{3}{H} lifetime (in ps): measurements vs. theory. The free 
$\Lambda$ lifetime is ($263 \pm 2$)~ps \cite{pdg14}. The first marked error 
is statistical, the second one is systematic. BC denotes a bubble-chamber
measurement.} 
\label{tab:L3H} 
\begin{ruledtabular}
\begin{tabular}{ccccc} 
BC\footnotemark [1] & STAR\footnotemark [2] & HypHI\footnotemark [3] & 
ALICE\footnotemark [4] & Theory\footnotemark [5]  \\  \hline 
$246^{+62}_{-41}$ & $182^{+89}_{-45}\pm 27$ & $183^{+42}_{-32}\pm 37$ & 
$181^{+54}_{-39}\pm 33$ & 256  \\ 
\end{tabular} 
\end{ruledtabular}
\footnotetext [1] {Keyes \textit{et al.} (1973).}
\footnotetext [2] {Abelev \textit{et al.} (2010).}
\footnotetext [3] {Rappold \textit{et al.} (2013a).}
\footnotetext [4] {Adam \textit{et al.} (2016a).}
\footnotetext [5] {Kamada \textit{et al.} (1998).}
\end{table} 

Accurate measurements of $\Lambda$-hypernuclear lifetimes in heavy systems 
beyond $A=56$, as listed in Table~\ref{tab:lifetimes}, could confirm the 
saturation of the nonmesonic decay width, Eq.~(\ref{eq:sat*}), as well as 
provide a check on the $\Gamma_n/\Gamma_p$ ratio systematics as a function of 
$A$. Previously, lifetime measurements in delayed fission triggered by proton 
and antiproton reactions on heavy nuclei, were interpreted as due to the 
production of $\Lambda$ hypernuclei and their subsequent weak decay. The 
latest and most accurate measurements of this kind yielded lifetimes 
[Cassing \textit{et al.} (2003), Kulessa \textit{et al.} (1998), and 
Armstrong \textit{et al.} (1993), respectively], 
\begin{equation} 
\tau_{\Lambda}(p+{\rm Au})=(145 \pm 11)~ \text{ps}  \,, 
\label{eq:Au} 
\end{equation} 
\begin{equation}
\tau_{\Lambda}(p+{\rm Bi})=(161 \pm 7 \pm 14)~ \text{ps}  \,,
\label{eq:Bi} 
\end{equation} 
\begin{equation} 
\tau_{\Lambda}({\bar p}+{\rm U})=(125 \pm 15)~ \text{ps}  \,. 
\label{eq:U} 
\end{equation} 
These are considerably shorter than values extrapolated from 
Table~\ref{tab:lifetimes}, and taken at face value, imply unreasonably 
large values for $\Gamma_n/\Gamma_p$ for heavy hypernuclei. 
Finally, we would like to focus attention again to recent measurements 
of the \lamb{3}{H} lifetime in heavy-ion experiments. As reviewed in 
Sec.~\ref{subsubsec:lifetime}, the \lamb{3}{H} lifetime was measured at
several heavy-ion facilities using the time dilation of a Lorentz boost to 
a recoiling hypernucleus produced in a heavy-ion reaction. Lifetimes deduced 
by the STAR Collaboration at BNL-RHIC, by the HypHI Collaboration at GSI and 
very recently by the ALICE Collaboration at CERN-LHC (see Fig.~\ref{fig:alice} 
in Sec.~\ref{subsubsec:lifetime}) are listed in Table~\ref{tab:L3H} 
together with a \lamb{3}{H} lifetime derived in bubble-chamber  
studies \cite{keyes70,keyes73}. The \lamb{3}{H} lifetime values deduced from 
measurements made at the heavy-ion facilities are about 25\% shorter than 
the free $\Lambda$ lifetime, and about 20\% shorter than the value measured 
in a bubble chamber. Note that the bubble-chamber measurement does not suffer 
from the uncertainty incurred in emulsion by a possible in-flight Coulomb 
dissociation of \lamb{3}{H} \cite{bohm70c}. A recent statistical analysis of 
all the reported \lamb{3}{H} lifetime measurements gives an average value 
$\tau$(\lamb{3}{H})=(216$^{+19}_{-16})$~ps \cite{rappold14}.  A realistic 
calculation of the lifetime \cite{kamada98} derives a lifetime shorter by 
only 3\% than the free $\Lambda$ lifetime $\tau_{\Lambda}$=(263$\pm$2)~ps, in 
agreement with Rayet and Dalitz (1966) that marks 
the first correct calculation of $\tau$(\lamb{3}{H}). The discrepancy between 
the lifetimes measured in heavy-ion collisions and the lifetime prescribed 
by theory is disturbing, posing a major problem for the understanding of 
\lamb{3}{H}, the lightest and hardly bound hypernucleus. More work is 
necessary to understand the heavy-ion lifetime results. We note that 
$\tau$(\lamb{4}{H}) is also considerably shorter than $\tau_{\Lambda}$, with 
a world average of $\tau$(\lamb{4}{H})=192$^{+20}_{-18}$~ps \cite{rappold14}, 
but this is theoretically anticipated and well understood. 

\subsection{Multi-strange systems}

Nuclear systems with $S\!=\!-2$ are essential to experimentally access the 
hyperon-hyperon interaction. While several light double-$\Lambda$ hypernuclei 
have been observed, and their phenomenology is fairly well understood 
\cite{gal11}, bound $\Xi$ hypernuclei have yet to be observed. Light $\Xi$ 
hypernuclear systems are predicted to be bound by several MeV, and with 
sufficiently narrow widths to provide spectroscopy \cite{hiyama08}. Intense 
$K^{-}$ beams are required for their investigation. The E05 experiment 
searching for the $^{12}_{~\Xi}$Be hypernucleus \cite{nagae07} is high on the 
agenda of J-PARC. The proposal is to use the $^{12}\mbox{C}(K^{-},K^{+})$ 
reaction to obtain the 1.5 MeV (FWHM) resolution~\cite{takahashi13} that should 
be sufficient to observe any quasibound structure. $\Xi^-$ hypernuclear 
$0^\circ$ production cross sections in the bound-state region, using targets in 
this mass range, are estimated to be a fraction of a 
$\mu$b/sr~(Dover and Gal, 1983; Dover, Gal, and Millener, 1994; Ikeda
\textit{et al.}, 1994; Shyam, Tsushima, and Thomas, 2012).

A similar experimental setup is also capable of producing $\Lambda\Lambda$ 
hypernuclei, either directly or by the conversion $\Xi\, N\,\rightarrow\, 
\Lambda\,\Lambda$. Identification of a $\Lambda \Lambda$ hypernucleus could 
occur either through direct production or by observation of the decay 
products. In direct production, one would observe the missing mass in 
a $(K^{-},K^{+})$ reaction. In this case, $0^\circ$ cross sections are small, 
a few nb/sr at most (Baltz, Dover, and Millener, 1983; Harada, Hirabayashi, 
and Umeya, 2010), due to the fact that the reaction requires a multistep 
interaction on two nucleons. On the other hand, detection in light hypernuclei
could occur by observing sequential monoenergetic $\pi^{-}$ decays of the 
embedded $\Lambda$'s. In either case, good energy resolution and tracking is 
important. All experiments will be difficult because production rates are not 
expected to be high. A particularly important task would be to settle the 
question as to whether $_{\Lambda\Lambda}^{~~4}$H is bound 
(Filikhin and Gal, 2002c; Nemura, Akaishi, and Myint, 2003). 
Interest in $_{\Lambda\Lambda}^{~~4}$H arises as it may be the least 
bound double-$\Lambda\Lambda$ system. A previous experimental claim for 
the observation of $_{\Lambda\Lambda}^{~~4}$H \cite{ahn01a} is probably 
incorrect, as shown by a reanalysis of the data \cite{randeniya07}. 

A possibly strong $\Lambda-\Xi$ attraction in the NSC97 model was pointed out 
by Filikhin and Gal (2002c). Here the $S=-3$ hypernucleus $_{\Lambda\Xi}^{~~6}$He, 
or $_{\Lambda\Lambda\Xi}^{~~~~7}{\rm He}$, may provide the onset of $\Xi$ 
stability in nuclear mater. This observation, and the repulsive nature of 
the $\Sigma$-nucleus potential, are relevant to the composition of neutron 
stars, as discussed in Sec.~\ref{subsec:NS}. 

\subsection{Experiments at heavy-ion facilities} 
\label{subsec:heavy-ion} 

Collisions between heavy nuclei ($A \gg 1$) at relativistic energies produce 
copiously hadrons and antihadrons, including hyperons and strange mesons. The 
formation of exotic nuclear systems and their study in relativistic heavy-ion 
collisions was suggested by Kerman and Weiss (1973). This was further developed 
by more quantitative evaluations using a variety of production mechanisms 
(Baltz \textit{et al.}, 1994; Pop and Gupta, 2010; Andronic \textit{et al.},
2011; Steinheimer \textit{et al.}, 2012). Following collision, 
the local hadron density produced in the ``fireball" stabilizes in times of 
order 60 fm/c, resulting in the formation of hadronic clusters. These clusters 
potentially include strange dibaryons, hypernuclei, and other multistrange 
hadrons. Predictions of production rates use two kinds of models: (i) thermal 
models in which entropy conservation governs the resulting production yields, 
following chemical freeze-out at a limiting temperature $T\approx$160~MeV 
\cite{andronic11}, and (ii) coalescense models which apply internuclear 
cascade simulations of particle collisions and captures, based on particle 
overlaps in both coordinate and momentum phase space \cite{steinheimer12}. 

\begin{figure}[t] 
\includegraphics[width=8.5cm]{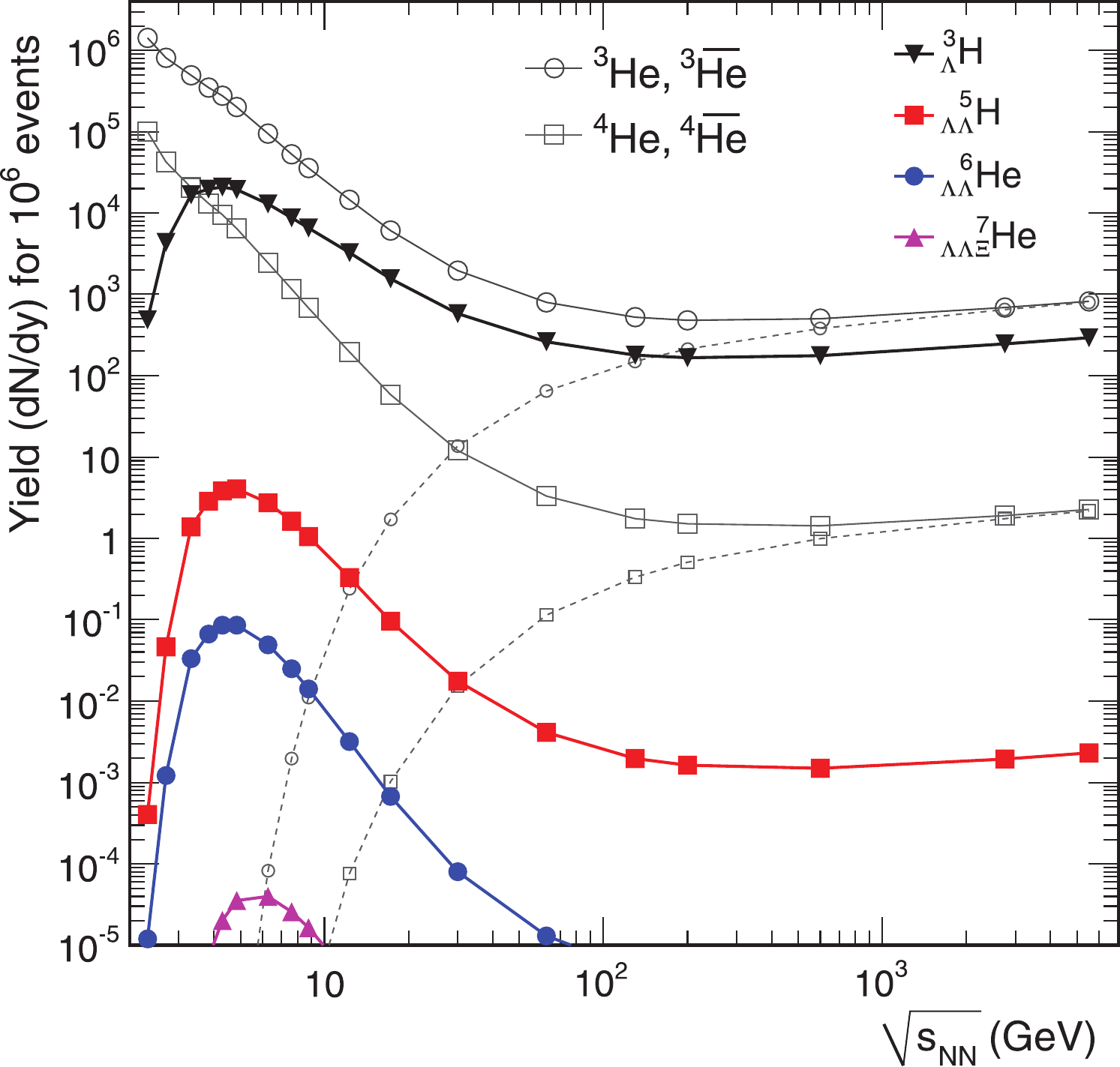} 
\caption{Energy dependence of predicted yields for several multistrange 
isotopes of hydrogen and helium at midrapidity for $10^{6}$ heavy-ion 
central collisions. Predicted yields for two non-strange helium isotopes 
and their anti-isotopes are also plotted for comparison. From Andronic 
\textit{et al.}, 2011.} 
\label{fig:hi_production} 
\end{figure} 

Somewhat surprisingly, the predicted production yields of hypernuclei are 
model independent above an approximate collision energy of 10~$A$~GeV, 
and both types of models predict saturation of the yield at beam energies 
$\approx$15~$A$~GeV (Andronic \textit{et al.}, 2011; Botvina, Gudima, and
Pochodzalla, 2013). Dibaryon production, however, 
is found to be strongly model dependent. These simulation studies demonstrate 
that $(10-20)$A~GeV is the optimal energy for hypernuclear production. 
Observation of hypernuclear production in relativistic heavy-ion collisions 
is difficult, and except for light systems, present-day detectors are not 
really designed to identify and investigate hypernuclear systems of unknown 
mass and binding energies. The development of a hypernuclear research program 
using ion beams of lower mass, e.g., C, with energies of approximately 
$(10-20)$A~GeV would seem appropriate, and can be pursued at the FAIR and NICA 
facilities \cite{botvina15}. Figure~\ref{fig:hi_production} illustrates yield 
predictions for the production of light multistrange hypernuclei at 
midrapidity per $10^{6}$ central collisions. These thermal-model predictions 
were constrained by fitting to RHIC hadron production yields at 200~GeV. 

Focusing on the lightest $A\!=\!3,4$ hypernuclei, which are essentially the 
only ones studied so far in relativistic heavy-ion collisions, the BNL-AGS E864 
Collaboration \cite{armstrong04} reported the observation of \lamb{3}{H} in 
central Au+Pt collisions at an energy per $NN$ collision of 
$\sqrt{s_{NN}}=11.5$~GeV. Subsequent work by the STAR Collaboration at the 
BNL-RHIC collider \cite{abelev10} identified both \lamb{3}{H} and its 
antihypernucleus, $\overline{^{3}_{\Lambda}{\rm H}}$, in Au+Au collisions at 
$\sqrt{s_{NN}}=200$~GeV. This was followed recently at the CERN-LHC facility 
by the ALICE Collaboration \cite{adam16a} in Pb+Pb collisions at 
$\sqrt{s_{NN}}=2.76$~TeV. The \lamb{3}{H} lifetime measurements reported by 
these heavy-ion experiments were listed and discussed in Table~\ref{tab:L3H} 
and in the related text. 

Searches for exotic nuclear states such as $\Lambda\Lambda$ and 
$\overline{\Lambda n}$ bound states were also undertaken by the ALICE 
Collaboration~\cite{adam16b}, thereby placing upper limits that are typically 
smaller by one order of magnitude than yields anticipated from thermal models 
for the production of such states. Another ALICE Collaboration experiment 
studied the low-energy $\Lambda$-$\Lambda$ interaction, producing useful 
constraints on the scattering length and effective range: 
$a_{\Lambda\Lambda}=-1.10\pm 0.37^{+0.68}_{-0.08}$~fm and $r_{\Lambda\Lambda}
=8.52\pm 2.56^{+2.09}_{-0.74}$~fm \cite{adamczyk15}. This result suggests 
a relatively weak $\Lambda$-$\Lambda$ interaction, in accord with other 
existing experimental and theoretical estimates summarized recently by 
Morita, Furumoto, and Ohnishi (2015). 

A program somewhat similar to that of the HypHI Collaboration at GSI 
\cite{rappold15} was proposed for the under-construction nuclotron-based ion 
collider facility (NICA) at Dubna, using an approximate 3 GeV/nucleon $^{6}$Li 
beam incident on a $^{\rm nat}$C target. A more sophisticated trigger would 
be based on identifying the recoiling hypernuclei by using a new magnetic 
spectrometer to measure the momentum of their two-body pionic decays. 
The pions and residual particles from the decays would be detected with 
multiwire proportional chambers placed behind the spectrometer magnet 
to reconstruct the hypernuclei from their decay products, which were 
presumed to be $^{A}_{\Lambda}{\rm H}\to {^{A}{\rm He}}+\pi^-$ or 
$^{A}_{\Lambda}{\rm He}\to {^{A}{\rm Li}}+\pi^-$ \cite{averyanov08}. 
The main interest in this program would be the potential production of light, 
neutron-rich hypernuclei inaccessible by other reactions. However, obtaining 
lifetimes of heavy hypernuclei, where mesonic decay is suppressed and 
essentially unobservable, is more compelling at present. 

\begin{table*}
\caption{J-PARC scheduled experiments related to strangeness nuclear physics.
DCX stands for double-charge exchange and TES for transition-edge sensor.} 
\label{tab:JPARC} 
\begin{ruledtabular} 
\begin{tabular}{lll} 
Exp. & \multicolumn{1}{c}{Title} & \multicolumn{1}{c}{Status} \\ 
\hline
E03 & X rays from $\Xi^-$ atoms &  \\
E05 & $^{12}{\rm C}(K^-,K^+){_{~\Xi}^{12}{\rm Be}}$ & Day-1 experiment \\
E07 & S=-2 emulsion-counter studies & \\ 
E10 & DCX studies of neutron-rich $_\Lambda^AZ$ & Negative result for 
\lamb{6}{H} \\
E13 & $\gamma$-ray spectroscopy of $\Lambda$ hypernuclei & Day-1 experiment, 
\lamb{4}{He} $\gamma$ ray observed \\  
E15 & Search for $K^-pp$ in $^3{\rm He}(K^-,n)$ & Day-1 experiment, 
shallow $K^-pp$ \\
    &       & ~~bound state suggested \\ 
E18 & $_{~\Lambda}^{12}$C weak decays & \\  
E19 & Search for $\Theta^+$ pentaquark in $\pi^- p \to K^- X$ & Day-1 
experiment, upper bound established \\  
E22 & Weak interactions in $_{\Lambda}^4{\rm H}-{_{\Lambda}^4{\rm He}}$ & \\ 
E27 & Search for $K^-pp$ in $d(\pi^+,K^+)$ & Deeply-bound ``$K^-pp$-like''\\
    &   & ~~bound state claimed \\ 
E31 & Study of $\Lambda(1405)$ by in-flight $d(K^-,n)$ & \\ 
E40 & Measurement of $\Sigma p$ scattering & \\ 
E42 & Search for $H$-dibaryon in $(K^-,K^+)$ nuclear reactions & \\ 
E62 & Precision spectroscopy of X-rays from kaonic atoms with TES &
Supersedes old day-1 experiment E17 \\
\end{tabular}
\end{ruledtabular}
\end{table*}

\subsection{$\overline{K}$-nucleus bound-state searches} 
\label{subsec:kbarexp} 

The topic of $K^{-}$-nuclear bound states has generated much heat and perhaps 
little illumination. Experimental searches for these states using stopped 
kaon reactions with outgoing neutrons, at KEK, or protons, at DA$\Phi$NE, 
at first suggested bound-state structure at more than 100~MeV below threshold. 
However, the KEK observation \cite{suzuki04,suzuki05} of a $\bar K NNN$ 
structure is now believed to be an experimental artifact, and at least 
a large part of the FINUDA Collaboration observation of a $K^{-}pp$ structure 
at DA$\Phi$NE \cite{agnello05b} must be due to final-state interactions 
\cite{magas06}. Yet the theoretical prediction of a $K^{-}pp$ bound state 
is reasonably robust, with microscopic preference for shallow binding of few 
tens of MeV \cite{gal13a}. Recent searches by the HADES Collaboration using 
the $pp\to \Lambda p K^+$ reaction at GSI and performing a complete background 
evaluation \cite{epple15} have refuted earlier claims for a deeply bound 
$K^{-}pp$ state based on a DISTO Collaboration analysis of older proton-beam 
data \cite{yamazaki10}. In addition, the LEPS Collaboration at SPring-8 also 
published upper limits, although less significant than with meson beams, for 
the production of a $K^-pp$ bound state via the $d(\gamma,K^+\pi^-)$ reaction 
at photon energy $E_{\gamma}= 1.5 - 2.4$~GeV \cite{tokiyasu14}. 

Ongoing experiments at J-PARC using meson beams reach contradictory results. 
E27 claims to have observed a deeply bound $K^{-}pp$-like structure in the 
$d(\pi^+,K^+)$ reaction at $p_{\pi}=1.69$~GeV/c \cite{ichikawa15}, whereas 
E15 presented upper limits in the $^3$He$(K^-,n)$ reaction at $p_{K}=1$~GeV/c 
\cite{hashimoto15} that appear to rule out a $K^-pp$ bound state with binding 
energy similar to that claimed by E27. However, E15, by focusing on the 
detection of $\Lambda p$ pairs, now suggests a broad $K^{-}pp$ bound-state 
structure at just 15 MeV below threshold \cite{sada16}. This ambiguity 
in identifying broad $\bar K$-nuclear bound-state structures reflects an 
experimental difficulty to directly access the formation and decay of such 
kaonic bound states. In particular, the detector used in such experiments 
must have good resolution, particle identification, and large angular 
acceptance. Further, improved experimentation searching for $\bar K$-nucleus 
bound-state structures is required to settle this issue. 

\section{Summary} 

Strangeness nuclear physics has been invesitgated since the first hyperon 
(the $\Lambda$) was observed in cosmic rays. Progress in this field has not 
been rapid but continuous, with its development critically dependent on both 
the experimental and theoretical tools to fully exploit the physics. The 
previous sections reviewed the production mechanisms with which $\Lambda$ and 
$\Sigma$ hyperons are injected into the nuclear medium. In addition, 
multistrangeness and the hyperon puzzle in neutron stars were reviewed, along 
with the strong interaction of $\bar K$ mesons in and with nuclei, including 
the possibility to form $\bar K$-nuclear quasibound states. The nonmesonic weak
decay of hypernuclei offers the unique opportunity to study the four-fermion 
weak interaction, and in particular, the fundamental origin (if any) of the 
empirical $\Delta$I = 1/2 rule. A number of potential experimental areas 
which seem critical for further advances in this field were pointed out. 

To highlight obvious achievements in strangeness nuclear physics and 
outstanding problems facing this field of research for the coming years, 
a brief, perhaps subjective list follows: 

\begin{itemize} 

\item With the $\Lambda N$ hypernuclear spin dependence largely deciphered
     via $\gamma$-ray studies, why is the $\Lambda$-nuclear spin-orbit
     splitting so small?

\item What is the role of three-body $\Lambda NN$ interactions in hypernuclei 
      and at neutron-star densities? 

\item The $\Sigma$-nuclear interaction is established as being repulsive,
     but how repulsive?  

\item Where is the onset of $\Lambda\Lambda$ binding: 
  ${_{\Lambda\Lambda}^{~~4}{\rm H}}$ 
   or ${_{\Lambda\Lambda}^{~~5}{\rm H}}$ and ${_{\Lambda\Lambda}^{~~5}{\rm He}}$? 

\item Do $\Xi$ hyperons bind in nuclei and how broad are the
      single-particle levels given the 
      $\Xi N\to\Lambda\Lambda$ strong decay channel? 

\item Where is the onset of $\Xi$ stability: $_{\Lambda\Xi}^{~~6}{\rm He}$ or 
$_{\Lambda\Lambda\Xi}^{~~~7}{\rm He}$?  

\item Although no $\bar K$ condensation occurs in self-bound stable matter, 
 can one observe $\bar K$ bound states in spite of the expected large 
 widths $\Gamma\geq 50$~MeV (for example,  $K^-pp$)? 

\item Is strange hadronic matter, made of roughly equal amounts of nucleons, 
 $\Lambda$ and $\Xi$ hyperons, likely to provide the ground state of strange 
 matter? 

\end{itemize}

The field is now poised to begin exploiting the new programs proposed at 
J-PARC, MAMI, FAIR, and at the upgraded JLab. These programs take advantage 
of new detection and electronic technologies which allow higher rates and 
coincidence experiments. To demonstrate the richness of the experimental 
programs we list in Table~\ref{tab:JPARC} the J-PARC scheduled experiments 
which, obviously, are limited to meson beams but still cover a broad spectrum 
of strangeness nuclear physics topical issues. 

\section*{Acknowledgments}

This work was supported by the U.S. DOE under Contract Nos. DE-SC0012704
(D. J. M.) and DE-SC0011598 (E. V. H.). 

\bibliographystyle{apsrmp}
\bibliography{hyprmp} 

\end{document}